%% file: home1.tex
\documentclass[11pt,twoside]{report}
\usepackage{epsfig}
\usepackage{amscd}
\usepackage{amssymb}
\usepackage{amsfonts}
\usepackage{amsmath}
\usepackage{fancyheadings}
\usepackage{graphicx}
\usepackage{graphics}
\usepackage[mathscr]{euscript}
\usepackage{psfrag}
\usepackage{pstricks}
\usepackage{pst-node}
\makeatletter
\@addtoreset{equation}{section}
\makeatother

\renewcommand{\chaptermark}[1]%
               {\markboth{#1}{#1}}
\renewcommand{\sectionmark}[1]%
               {\markright{\it{\thesection\ #1}}}
\begin{document}
\thispagestyle{empty}
\begin{center}
{\large Universit\`{a} degli Studi di Trieste} \\
\vspace{.4cm}
{\large Dottorato di Ricerca in Fisica - XV Ciclo} \\
\vspace{.4cm}
{\large Anno Accademico 2001/2002} \\
\vspace{5cm}
{\scshape\Huge
\begin{tabular}{cc}
\bf{Topics}\\
\bf{in}\\
\bf{Koopman-von Neumann}\\
\bf{Theory}
\end{tabular}
}\\

\bigskip
\bigskip

{\large{Ph.D. Thesis}}\\
\vspace{6cm}
{\Large Dottorando: Danilo Mauro} \\
\vspace{1cm}
{\Large Relatore: prof. Ennio Gozzi, Universit\`a di Trieste}\\
\vspace{1cm}
{\Large Coordinatore: prof. M. Francesca Matteucci, Universit\`a di
Trieste}
\end{center}
\newpage

\thispagestyle{empty}
$\qquad\quad$

\newpage
\setcounter{page}{1}

\pagestyle{fancy}
\markboth{\it{Contents}}{\it{Contents}}
\pagenumbering{roman}

\tableofcontents
\markboth{\it{Contents}}{\it{Contents}}
\newpage

\pagenumbering{arabic}
%%%%%%%%%%%%%%%%%%%%%%%%%%%%%%%%%%%%%%%%%%%%%%%%%%%%%%%%%%%%%%%%%%%%

%-------------------------- Thesis     ---------------------

%\mainmatter
\thispagestyle{empty}
\setcounter{page}{1}

\input{introduction}

 \newpage \thispagestyle{plain} 
 \newpage
\part*{PART I: \\Koopman-von Neumann Theory without Forms}
\addcontentsline{toc}{part}{{\Large{\underline{PART I}:\smallskip\\
Koopman-von Neumann Theory without Forms}}}

\input{chapter1}

 \newpage \thispagestyle{plain}
\input{chapter2}

 \newpage \thispagestyle{plain}
 $\qquad$
 \newpage

\part*{PART II: \\Koopman-von Neumann Theory with Forms}
\addcontentsline{toc}{part}{{\Large{\underline{PART II}:\smallskip\\
Koopman-von Neumann Theory with Forms}}}

\input{chapter3}

 \newpage \thispagestyle{plain}

\input{chapter4}

 \newpage \thispagestyle{plain}
\input{chapter5}

 \newpage $\qquad\qquad$
 \thispagestyle{plain}
\input{conclusions}

\newpage \thispagestyle{plain}

\pagestyle{fancy}
\markboth{\it{Bibliography}}{\it{Bibliography}}

\end{document}

%% file: introduction.tex
\pagestyle{fancy}
\markboth{{\it{Introduction}}}{\it{Introduction}}
\setcounter{page}{1}
\chapter*{Introduction}
\addcontentsline{toc}{chapter}{\numberline{}Introduction}
\def \HT{{\mathcal H}}
\def \LT{{\mathcal L}}
\def \ET{{\widetilde{\mathcal E}}}
\def \HCT{\hat{\mathcal H}}
\def \s{\scriptscriptstyle}
\raggedbottom

It is well-known that classical mechanics (CM) and quantum mechanics (QM) 
are formulated using completely different mathematical tools.
The first one (CM) uses the phase space and the Poisson brackets 
while the second one (QM) uses the Hilbert space on which suitable 
states and operators are defined. In the literature a lot of attempts have been made in 
order to overcome these differences and to get a better understanding of the
interplay between CM and QM. These attempts can be divided
in two sectors. One is the reformulation of
quantum
mechanics in a ``classical" language by replacing wave functions with suitable 
distributions in phase space (Wigner functions \cite{Wigner}) and by replacing
the 
commutators of the theory with some suitable brackets (Moyal brackets \cite{Moyal}) 
which are a deformation of the Poisson brackets.  
Another, even older, direction is to reformulate CM in an operatorial
language by using a Hilbert space of square integrable functions on the phase space and
by replacing the Poisson brackets with some suitable classical commutators. This
is what has been done in the 30's by Koopman and von Neumann (KvN) \cite{Koopman}\cite{von Neumann}.
More recently a functional (or path integral) approach to the KvN method has been 
proposed in
\cite{Goz89}.  

In this thesis we have analysed, clarified and explained some aspects of
the 
operatorial KvN
approach to CM and of its path integral counterpart. 
The starting point of KvN is the introduction of a Hilbert space of square
integrable
and complex functions $\psi(q,p)$ whose modulus square are just the usual 
probability densities in phase space $\rho(q,p)=|\psi(q,p)|^2$. A topic which
was not analysed 
by KvN is the fact that the introduction of these Hilbert space elements
requires an 
enlargement of the set of observables of the theory. If we stick to the accepted
wisdom 
that in CM the observables are only the functions of $\varphi\equiv(q,p)$ then a
superselection
mechanism is triggered and this limits the Hilbert space to be made of just the
Dirac deltas 
centered on a single point of phase space. In order to have $\psi(\varphi)$,
which are linear 
superpositions of the Dirac deltas mentioned above with complex coefficients, 
we have to prevent the superselection mechanism to set in 
and we do that by enlarging the space of observables to functions not only of 
$\varphi$ but also of $\displaystyle \frac{\partial}{\partial \varphi}$. In this
way we can build a Hilbert space made up of square integrable functions with phases. 

In QM the phases of the wave 
functions are crucial to explain phenomena like, for example, the interference
effects in
two slit-experiments. So one can ask which is the role of the phases in the KvN 
approach to CM. In this respect a crucial difference between CM and QM is given
by the form
of the operator of evolution. In fact KvN postulated that the evolution of the  
$\psi(\varphi)$ must be given by the Liouvillian
which, containing only first order derivatives, 
evolves also the probability densities $\rho(\varphi)$, differently than what happens 
in QM.
The presence of only first order derivatives in the Liouvillian has a lot 
of peculiar consequences for what concerns the role of phases in KvN theory. 
For example, differently from what happens in QM, the equations of evolution
for the phase and the modulus of the $\psi$ are completely decoupled: the phase do not enter
the equations
of the modulus and vice versa. One could then think that 
phases are completely useless in CM and that it is sufficient to consider real
instead
of complex wave functions in order to describe all the physics.
This is true only if we limit ourselves to consider the particular
representation in which both the positions 
and the momenta of the theory are given by multiplicative operators. If we want
to remain
free to consider also other kinds of representations it is possible
to prove that we have to consider complex wave functions. 
In principle the classical phases of KvN functions could be measured by using observables depending 
both on $\widehat{\varphi}$ and on $\displaystyle \frac{\partial}{\partial\varphi}$ which 
generalize the usual ones of CM which depend only on $\widehat{\varphi}$.
From these last ones it is impossible to extract information on phases, as we have proved
in detail both by using the superselection principle mentioned above and by implementing
a classical analog of the double-slit experiment via the KvN formalism.
These classical phases could appear in a regime at the interface between CM and QM.
For further details on this part see paper \cite{5P}.

Phases in QM appear also in other settings like for example the Aharonov-Bohm
experiment.
In QM it is possible to show \cite{Sak} that the effect of these phases (or
better of the magnetic
field) is felt by the spectrum of the system even if the magnetic field
is confined to regions inaccessible to the particle.  
In CM this cannot happen because the motion of a charged
particle is determined uniquely by the Lorentz force and this is identically zero if
the magnetic field
is zero. So a Aharonov-Bohm-like effect must be absent also in the KvN operatorial
approach for CM.
Before checking this we had to understand which are the analog in KvN theory of the minimal 
coupling rules of QM. Having established all this, it was then quite easy to show that magnetic fields \`a la 
Aharonov-Bohm cannot change the spectrum of the Liouville operator, differently 
than what happens in QM. All this analysis, which had never been performed before in the literature,
made us understand and clarify concepts like the gauge invariance in the KvN formalism. For further 
details on this part see paper \cite{2P}. This concludes the first part of the thesis which is entirely
devoted to the study of the original KvN approach for CM.

In the second part we have analysed the path integral counterpart \cite{Goz89} 
of the KvN approach that we have indicated with the acronym CPI (for Classical
Path Integral).
The CPI is not only the functional
counterpart of the operatorial 
formulation of CM but it provides also an interesting generalization. In fact
the CPI gives the kernel
of evolution not only for the KvN states $\psi(q,p)$ but also for more
general objects $\psi(q,p,c^q,c^p)$ 
depending not only on $q,p$ but also
on some Grassmann variables $c^q,c^p$. These variables make their appearance in
the theory in a 
very natural way and, 
from a geometrical point of view, they can be interpreted as a basis for the differential 
forms associated to the phase space manifold \cite{Marsd}. To study the geometrical 
and the physical role of these forms is the main aim of the second part of the thesis.
Consequently we have called this part ``Koopman-von Neumann Theory with Forms" and the first 
part, where such differential structures are not present, ``Koopman-von Neumann Theory without Forms".

The identification of the Grassmann variables with a basis for differential forms
brings into the CPI
a lot of beautiful geometry. For example all the tensors manipulations on
symplectic manifolds (the
so-called Cartan calculus) can be reformulated in terms of the Grassmann
variables and structures
present in the CPI: it is possible to reproduce exterior derivatives, interior
contractions, Lie derivatives, Lie brackets and all their generalizations,
like the Schouten-Nijenhuis, the Fr\"olicher-Nijenhuis and the
Nijenhuis-Richardson brackets using the structures present in the CPI. Further details
can be found in the paper \cite{1P}.

We are aware that some readers might be not familiar with
Grassmann variables, but one can easily prove that every Grassmann
algebra can
be realized via tensor products of Pauli matrices and that the functions
$\psi(q,p,c^q,c^p)$ can
be replaced by ordinary vectors in suitable tensor products of KvN spaces. 
As a consequence, while in the original CPI all the Cartan calculus
was performed via Grassmann variables,
now, by replacing these variables with matrices, 
we have all the Cartan calculus reproduced via tensor products of Pauli
matrices. For further details on this part see paper \cite{8P}.

Even if the CPI is richer than the original KvN theory from a geometrical point
of view, the introduction
of the Grassmann variables creates some problems. In fact at the level of KvN
theory without forms 
it was easy to
construct a scalar product on the Hilbert space of the $\psi(\varphi)$. 
According to this scalar product all the norms of the states
were positive definite and the Liouvillian was a Hermitian operator.
Consequently the evolution turned
out to be unitary and all the probabilistic interpretation of the function
$\psi(\varphi)$ turned out
to be consistent. In fact the norm of $\psi(\varphi)$ was the total probability
of finding a particle
in a point of the phase space and this had to be a conserved quantity. Things
become more difficult 
when we introduce the auxiliary Grassmann variables. In this case one can
construct a lot of different
inner products but all of them satisfy the following no-go theorem. For all the
scalar products which are
positive definite the operator of evolution $\widehat{\mathcal{H}}$ is not Hermitian while in all
the scalar products
for which $\widehat{\mathcal{H}}$ is Hermitian there are states with negative
norms. This no-go theorem is due to the Grassmannian character of the auxiliary
variables but it is also a 
peculiar feature of CM. It was not true for example in the supersymmetric
quantum mechanical models developed in \cite{Witten}\cite{Salomonson} where
there was no 
problem in having both the unitary of the evolution and the positivity of the norm
of the states, even if there were states containing Grassmann variables. From
a physical point of view the non hermiticity of the operator of evolution in CM
is somehow due to
the fact that the norm of the Grassmannian wave functions can be associated with
the length of the Jacobi 
fields. Now the length of a Jacobi field can change during the time evolution
(for example it diverges exponentially in the case of chaotic systems).
Therefore the evolution of the 
Jacobi fields, which is given by the operator $\widehat{\mathcal{H}}$, cannot be in general
unitary and this seems not to be 
a problem for CM. 

Anyhow, if the reader is uncomfortable with a non unitary evolution or negative
norm states,
we have considered two other different functional formulations of CM: in the first
one the auxiliary variables
of the CPI are all bosonic and they belong to the vector representation of the
symplectic group; in the second
one we have an infinite number of Grassmann variables belonging to the spinor
representation of the
metaplectic group, which is the covering group of the symplectic one. In both
cases it is possible to endow 
the associated Hilbert space with a positive definite scalar product and to
describe the dynamics via
a Hermitian Hamiltonian. The price we have to pay for this is twofold: on one hand the
geometrical richness
of the original CPI can be obtained only by introducing in the theory some
further structures from outside, like a
complicated tensor product structure, on the other hand a real physical
understanding of these approaches 
is lacking. For further details on this part see the papers \cite{6P} 
and \cite{7P}.

This thesis is organized as follows.

In Chapter {\bf 1} we will briefly review the main features of KvN theory without forms, underlying the differences
and the similarities with ordinary QM. Then we try to understand the role of phases in this approach 
and in the different representations one can use to describe the theory. We will show that there is at least 
one representation in which phases and moduli do not interact with each other and we will prove how interference
effects are killed by the particular form of the classical evolution.

In Chapter {\bf 2} we will construct the ``minimal coupling" rules, i.e. the rules to
go from the description of a free particle to the description of a particle in interaction with an electromagnetic
field. We will study in detail how to implement the Abelian gauge invariance and we use the results
to analyse two particular applications of the formalism: the Landau problem and
the Aharonov-Bohm effect.

From Chapter {\bf 3} we begin to take into account the forms. 
After a brief review of the many geometrical structures of the CPI, we will prove
how it is possible to construct within the formalism some generalizations of the Lie brackets and how it 
is possible to replace the Grassmann variables with Pauli matrices without losing anything of the initial
geometrical richness. 

In Chapters {\bf 4} and {\bf 5} we will show that the enlarged ``Hilbert space" associated to the CPI cannot be 
endowed with a scalar product with positive definite norm states and unitary evolution. 
Nevertheless we will show that it is possible to construct other functional formulations of CM
satisfying both requirements.

In the Conclusions we will give a brief summary of what we have done and we will indicate which further 
lines of research we are currently pursuing with the results contained in this thesis.

%% file: chapter1.tex
\def \HT{{\mathcal H}}
\def \LT{{\mathcal L}}
\def \ET{{\widetilde{\mathcal E}}}
\def \HCT{\widehat{\mathcal H}}
\def \s{\scriptscriptstyle}

\pagestyle{fancy}
\chapter*{\begin{center}
1. Koopman-von Neumann Waves
\end{center}}
\addcontentsline{toc}{chapter}{\numberline{1}Koopman-von Neumann Waves}
\setcounter{chapter}{1}
\markboth{{\it{1. Koopman-von Neumann Waves}}}{}

\begin{quote}
{\it{
Ordinary mechanics must also be statistically formulated: the determinism
of classical physics turns out to be an illusion, it is an idol, not an ideal 
in scientific research.}}\medskip\\
-{\bf Max Born}, 1954 Nobel Prize Lecture.
\end{quote}

\bigskip

\noindent As we have said in the Introduction, the starting point of the KvN approach to CM 
\cite{Koopman}\cite{von Neumann} is the introduction of a Hilbert space of {\it complex} 
and {\it square integrable} functions $\psi(\varphi)$ such that $\rho(\varphi)\equiv|\psi(\varphi)|^2$ can be 
interpreted as the probability density
of finding a particle at the point $\varphi=(q,p)$ of the phase space. 

We know that in QM the complex character of the wave function is crucial: in fact
while the modulus of $\psi$ is the square root of the probability density $\rho$, also 
the phase of $\psi$ brings in
some physical information. For example it is related to the mean value
of the momentum operator $\widehat{p}$ and it gives origin to the well-known
interference effects.
In the following sections we want to study which is the role of the phases in the KvN waves.
Part of the content of this chapter has been already published in \cite{5P}.

\bigskip

\section{Operatorial Approach to Classical Mechanics}

We know that in classical statistical mechanics
the probability density $\rho$ has to evolve in time according to the well-known Liouville equation:
%%%
\begin{equation}
\displaystyle
i\frac{\partial}{\partial t}\rho(\varphi)=\widehat{L}\rho(\varphi) \label{ijmpa.first}
\end{equation}
%%%
where $\widehat{L}$ is the Liouville operator 
%%%
\begin{equation}
\widehat{L}=-i\partial_{p_i}H(\varphi)\partial_{q_i}+i\partial_{q_i}H(\varphi)\partial_{p_i} \label{ijmpa.ht}
\end{equation}
%%%
and $H(\varphi)$ is the Hamiltonian of the standard phase space.  
Koopman and von Neumann {\it postulated} the same evolution for the elements $\psi(\varphi)$
of the Hilbert space they had introduced:
%%%
\begin{equation}
\displaystyle
i\frac{\partial}{\partial t}\psi(\varphi)=\widehat{L}\psi(\varphi)\;\;\Longrightarrow\;\; \frac{\partial}{\partial t}\psi
=(-\partial_{p_i}H\partial_{q_i}+\partial_{q_i}H\partial_{p_i})\psi. \label{ijmpa.lio1} 
\end{equation}
%%%
We can think of (\ref{ijmpa.lio1}) as 
the fundamental equation governing the evolution of the vectors in the Hilbert space of CM: it is
the analog of the Schr\"odinger equation for QM. A space, in order to be defined as a Hilbert space,
must be equipped with a scalar product. KvN choose the following one:
%%%
\begin{equation}
\displaystyle \langle\psi|\tau\rangle=\int d\varphi\,\psi^*(\varphi)\tau(\varphi).
\end{equation}
%%%
The norm of the states $|\psi\rangle$ is then:
%%%
\begin{equation}
\displaystyle \|\psi\|^2=\int d\varphi\,\psi^*(\varphi)\psi(\varphi)=\int d\varphi\,\rho(\varphi).
\end{equation}
%%%
With this scalar product it is easy to show that 
$\langle\psi|\widehat{L}\tau\rangle=\langle\widehat{L}\psi|\tau\rangle$, i.e. 
the Liouvillian $\widehat{L}$ is a Hermitian operator; consequently the norm of 
the state $\displaystyle \langle\psi|\psi\rangle=\int d\varphi\,\psi^*(\varphi)\psi(\varphi)$ 
is conserved during
the evolution and we can consistently interpret $\rho(\varphi)=\psi^*(\varphi)\psi(\varphi)$ 
as the probability density of finding the particle in a point of the phase space.

Coming back to (\ref{ijmpa.lio1}) we note that no complex factor ``$i$" appears on its RHS
and that the operator of evolution $\widehat{L}$ contains only first order derivatives. 
These ingredients are crucial to check that (\ref{ijmpa.first}) can be derived from (\ref{ijmpa.lio1}).
In fact, if we take the complex conjugate of (\ref{ijmpa.lio1}), we obtain:
%%%
\begin{equation}
\displaystyle
\frac{\partial}{\partial t}\psi^*=(-\partial_{p_i}H\partial_{q_i}+\partial_{q_i}H\partial_{p_i})\psi^* \label{ijmpa.lio2}
\end{equation}
%%%
i.e. $\psi$ and $\psi^*$ satisfy the same equation.  Now, multiplying (\ref{ijmpa.lio1}) 
by $\psi^*$, (\ref{ijmpa.lio2}) by $\psi$ and summing the two resulting equations, we re-obtain (\ref{ijmpa.first}), 
i.e. the evolution of $\rho(\varphi)\equiv\psi^*(\varphi)\psi(\varphi)$ via the Liouvillian operator:
%%%
\begin{equation}
\displaystyle
\frac{\partial}{\partial t}\rho=(-\partial_{p_i}H\partial_{q_i}+\partial_{q_i}H\partial_{p_i})\rho
\;\;\Rightarrow\;\;  i\frac{\partial}{\partial t}\rho(\varphi)=\widehat{L}\rho(\varphi). \label{ijmpa.lio3}
\end{equation}
%%%
So we have derived the standard Liouville equation for $\rho(\varphi)$ as a 
consequence of having postulated (\ref{ijmpa.lio1}) for $\psi$. We notice from (\ref{ijmpa.lio3}) 
that the equation of motion of the modulus square of $\psi$, i.e. $\rho$, 
is not coupled with the phase of $\psi$. We will see in (\ref{decoupling}) that also the opposite is true,
i.e. the equation of motion of the phase is completely independent of the modulus of $\psi$.
This does not happen in QM where the analog of $\widehat{L}$ is given by the 
Schr\"odinger operator which contains second order derivatives in its kinetic term. In fact
if we start from the Schr\"odinger equation for the quantum wave function $\psi(q,t)$,
%%%
\begin{equation}
\displaystyle
i\hbar\frac{\partial\psi(q,t)}{\partial t}=\widehat{H}\psi(q,t)\;\;\Rightarrow\;\;i\hbar\frac{\partial\psi(q,t)}{\partial
t}=-\frac{\hbar^2}{2m}\nabla^2\psi(q,t) +V(q)\psi(q,t) \label{ijmpa.third}
\end{equation}
%%%
then we have that the probability density $\rho(q,t)=|\psi(q,t)|^2$ satisfies a continuity equation of the following form:
%%%
\begin{equation}
\displaystyle
\frac{\partial\rho}{\partial t}=-\textrm{div}\,{\mathbf j} \label{ijmpa.tre}
\end{equation}
%%%
where we have indicated with ${\mathbf j}$ the probability density current:
%%%
\begin{equation}
\displaystyle 
{\mathbf j}=-\frac{i\hbar}{2m}\biggl(\psi^*\nabla\psi
-\psi\nabla\psi^*\biggr).
\end{equation}
%%%
If we now write the wave function as $\psi=\sqrt{\rho}\,\textrm{exp}[iS/\hbar]$ we discover 
immediately that the phase $S$ enters explicitly into the expression of the current probability density ${\mathbf j}$:
%%%
\begin{equation}
{\mathbf j}=\frac{\rho\nabla S}{m}. \label{ijmpa.cinque}
\end{equation}
%%%
As a consequence the equation of evolution of $\rho$ (\ref{ijmpa.tre}) couples 
the phase $S$ and the modulus square $\rho$ of the wave functions \cite{Sak}. We 
can also notice from (\ref{ijmpa.tre})
that in QM, differently than in CM, the probability density $\rho$ does not evolve in time with the same
Schr\"odinger Hamiltonian $\widehat{H}$ which gives the evolution of the wave function $\psi$. 

The analysis presented above indicates some {\it crucial} differences between the KvN equation
(\ref{ijmpa.lio1}) and the Schr\"odinger one (\ref{ijmpa.third}). These differences have their origin in the fact that 
the Liouvillian contains only {\it first order} derivatives
while the Schr\"odinger operator contains {\it second order} derivatives. 
 
Before going on we want to briefly introduce the functional approach developed in \cite{Goz89}
which represents the path integral counterpart of the KvN operatorial formalism and which we will indicate 
with the acronym CPI (for Classical Path Integral).   
First of all let us ask ourselves: which is the {\it probability} of finding a particle at a point\break
$\varphi^a=(q^1,\dots,q^n;p_1,\dots,p_n)$ of the phase space at time
$t$ if it was at $\varphi^a_i$ at the initial time
$t_i$? This probability is one if $\varphi^a_i$ and $\varphi^a$ are connected with a {\it classical path} $\phi^a_{cl}$,
i.e. a path that solves the classical equations of motion, and it is zero in all the other cases. So we can write:
%%%
\begin{equation}
P(\varphi^a,t|\varphi^a_i,t_i)=\delta(\varphi^a-\phi^a_{cl}(t;\varphi_i)) \label{ijmpa.prob}
\end{equation}
%%%
where $\phi^a_{cl}(t;\varphi_i)$ is the classical solution of the Hamiltonian equations of motion with initial
conditions $(\varphi^a_i,t_i)$. 
Since $P(\varphi^a,t|\varphi^a_i,t_i)$ is a classical
probability, we can rewrite it as a sum over all the possible intermediate configurations:
%%%
\begin{equation}
\label{CPI3}
\begin{array}{rl}
P(\varphi^a,t|\varphi^a_i,t_i) & =\displaystyle\sum_{k_{i}} P(\varphi^a,t|k_{\scriptscriptstyle{N-1}})
P(k_{\scriptscriptstyle{N-1}}|k_{\scriptscriptstyle{N-2}})\cdot...\cdot 
P(k_{\scriptscriptstyle{1}}|\varphi_i^a,t_i)
\vspace{.2cm} \\
&=\displaystyle\prod^N_{j=1}\int
d\varphi_{j}~\delta[\varphi^{a}_{j}-\phi^{a}_{cl}(t_{j} 
|\varphi_{j-1},t_{j-1})] \vspace{.2cm}\\
\xrightarrow{N\rightarrow\infty} &=\displaystyle\int{\mathcal
D}\varphi~\widetilde{\delta}[\varphi^{a}-\phi^{a}_{cl}(t;\varphi_i)]
\end{array} 
\end{equation}
%%%
\noindent where in the first equality $k_i$ denotes an intermediate configuration $\varphi_{k_i}$ between 
$(\varphi^a_{i},t_i)$ and $(\varphi^a,t)$ and the symbol $\widetilde{\delta}({\ldots})$ represents a 
{\it functional} Dirac delta, that is a 
product of an infinite number of Dirac deltas, each one referring to a different time $t$ along the classical trajectory.

The functional delta in (\ref{CPI3}) can be rewritten as a delta 
on the Hamiltonian equations of motion $\dot{\varphi}^a=\omega^{ab}\partial_bH(\varphi)$ via the introduction 
of a suitable functional determinant:
%%%
\begin{equation}
\widetilde{\delta}[\varphi^a-\phi^a_{cl}(t;\varphi_i)]=\widetilde{\delta}(\dot{\varphi}^a-\omega^{ab}\partial_bH)
\textrm{det}(\partial_t\delta^a_b-\omega^{ac}\partial_c\partial_bH). \label{ijmpa.Diracdelta}
\end{equation}
%%%
Next let us make a Fourier transform of the Dirac delta on the RHS of (\ref{ijmpa.Diracdelta})
introducing $2n$ extra variables $\lambda_a$; moreover let us 
exponentiate the determinant using $4n$ anticommuting variables
$c^a,\bar{c}_a$. The final result is the following one:
%%%
\begin{equation}
\displaystyle
P(\varphi, t|\varphi_{i},t_i)=\int{\mathcal D}^{\prime\prime}\varphi{\mathcal D}\lambda{\mathcal D}c
{\mathcal D}\bar{c}\,
\textrm{exp}\biggl[i\int_{t_i}^tdt\,\LT\biggr] \label{ann.prob3}
\end{equation}
%%%
where with ${\mathcal D}^{\prime\prime}\varphi$ we indicate that the integration is over paths with 
fixed end points in $\varphi$ and the Lagrangian $\LT$ is given by:
%%%
\begin{equation}
\LT=\lambda_a\dot{\varphi}^a+i\bar{c}_a\dot{c}^a-\lambda_a\omega^{ab}\partial_bH-i\bar{c}_a\omega^{ad}
(\partial_d\partial_bH)c^b. \label{ann.suplag}
\end{equation}
%%%
From (\ref{ann.prob3}) and the kinetic part of the Lagrangian (\ref{ann.suplag}) we can deduce the form of the graded
commutators \cite{Goz89} of the associated operatorial theory:
%%%
\begin{equation}
[\widehat{\varphi}^a,\widehat{\lambda}_b]=i\delta^a_b,\qquad\qquad
[\widehat{c}^a,\widehat{\bar{c}}_b]=\delta_b^a. \label{ijmpa.comm}
\end{equation}
%%%
All the other commutators are identically
zero. In particular, differently from the quantum case, we have that $[\widehat{q},\widehat{p}]=0$ which implies that 
we can determine with
an arbitrary precision the position and the momentum of a classical particle, like it happens in the standard 
approach to CM. 
Associated to the Lagrangian (\ref{ann.suplag}) there is a Hamiltonian which is 
%%%
\begin{equation}
\HT=\lambda_a\omega^{ab}\partial_bH+i\bar{c}_a\omega^{ad}(\partial_d\partial_bH)c^b. \label{ann.supham}
\end{equation}
%%%
We notice that, instead of just the original $2n$ phase space coordinates $\varphi^a$, we now have $8n$ variables 
$(\varphi^a,\lambda_a,
c^a,\bar{c}_a)$ whose geometrical meaning has been studied in detail in Refs.
\cite{1P}\cite{Regini}
and will be reviewed in the second part of this thesis.
We will indicate with ${\cal M}$ the original phase space coordinatized by $\varphi^a$ and with $\widetilde{\cal M}$
the space coordinatized by
$(\varphi^a,\lambda_a,c^a,\bar{c}_a)$. 
This space can be endowed with some extended Poisson structures as follows. 
From the Lagrangian (\ref{ann.suplag}) one could derive the equations of motion for the $8n$-variables
$(\varphi^a,\lambda_a,c^a,\bar{c}_a)$ by the simple variational principle. These equations are:
%%%
\begin{eqnarray}
\label{ann.eq:otto}
\dot{\varphi}^{a}&=&\omega^{ab}\partial_{b}H \nonumber\\
\dot{c}^a &=&\omega^{ac}\partial_{c}\partial_{b}Hc^{b} \nonumber\\
\dot{\bar{c}}_b&=&-{\bar
c}_{a}\omega^{ac}\partial_{c}\partial_{b}H \\
\dot{\lambda}_b&=&-\omega^{ac}\partial_{c}\partial_{b}H\lambda_{a}-i{\bar c}_{a}\omega^{ac}
\partial_{c}\partial_{d}\partial_{b}H c^{d}\nonumber.
\end{eqnarray}
%%%
The same equations could be derived from the Hamiltonian $\HT$ if we introduce the following 
extended Poisson brackets structure (epb) in the extended space $(\varphi^a,\lambda_a,c^a,\bar{c}_a)$:
%%%
\begin{equation}
\left\{
\begin{array}{l}
\{\varphi^a,\lambda_b\}_{epb}=\delta_b^a \medskip \\
\{\bar{c}_b,c^a\}_{epb}=-i\delta_b^a \label{ann.epb}
\end{array}
\right.
\end{equation}
%%%
where all the other brackets are identically zero. Then we get that the equations of motion (\ref{ann.eq:otto})
can be written in a compact way as
%%%
\begin{equation}
\frac{dO}{dt}=\{O,\HT\}_{epb}
\end{equation}
%%%
where $O$ is any function of the variables $(\varphi^a,\lambda_a,c^a,\bar{c}_a)$.
More details can be found in Ref. \cite{Goz89}.

What is nice about the extended space $\widetilde{\cal M}$ is that all these $8n$ variables can be put 
together in a single object known in the literature on supersymmetry as superfield. 
In order to construct it we first enlarge the base space including, besides the standard time $t$, two Grassmannian
partners $\theta,\bar{\theta}$ and then we build in the superspace $(t,\theta,\bar{\theta})$ the following superfield:
%%%
\begin{equation}
\Phi^a(t,\theta,\bar{\theta})=\varphi^a(t)+\theta
c^a(t)+\bar{\theta}\omega^{ab}\bar{c}_b(t)+i\bar{\theta}\theta\omega^{ab}\lambda_b(t).
\label{ann.super1}
\end{equation}
%%%
The index ``$a$" in $\Phi^a$ indicates either the first $n$ configurational variables $q$ or the 
second $n$ momentum ones $p$. The explicit expression for the two different kinds of superfields is given by:
%%%
\begin{eqnarray}
&& q\;\longrightarrow \;\Phi^q=q+\theta c^q+\bar{\theta}\bar{c}_p+i\bar{\theta}\theta\lambda_p \label{ann.sup1}\\ 
&& p\;\longrightarrow \;\Phi^p=p+\theta c^p-\bar{\theta}\bar{c}_q-i\bar{\theta}\theta\lambda_q \label{ann.sup2}
\end{eqnarray}
%%%
where we have put the index $q$ (or $p$) on $c,\bar{c},\lambda$ just to indicate that we refer to 
the first $n$ (or the second $n$) of the $c,\bar{c},\lambda$ variables.
Let us take the usual Hamiltonian $H$ of CM and replace in its argument the standard phase space 
variables $\varphi$ with the superfields $\Phi$. Next let us make the expansion of $H(\Phi)$ in $\theta$ 
and $\bar{\theta}$. It is then straightforward to prove the following formula:
%%%
\begin{equation}
H(\Phi)=H(\varphi)+\theta N_{\s H}-\bar{\theta}\,\overline{N}_{\s H}+i\theta\bar{\theta}\HT \label{gg}
\end{equation}
%%%
where the precise form of $N_{\s H},\overline{N}_{\s H}$ is not necessary in this section and can be 
found in Ref. \cite{Goz89}.
From (\ref{gg}) it is easy to prove that the connection between the standard $H(\varphi)$ and the
Hamiltonian appearing in the weight of the CPI is:
%%%
\begin{equation}
i\int d\theta d\bar{\theta} \,H[\Phi]=\HT. \label{ann.superH}
\end{equation}
%%%
The same steps we did above for the Hamiltonian, i.e. to replace $\varphi$ with $\Phi$
and to expand $O(\Phi)$ in $\theta,\bar{\theta}$, can be done for any function $O(\varphi)$ of the 
phase space ${\cal M}$:
%%%
\begin{equation}
O(\Phi)=O(\varphi)+\theta N_{\s O}-\bar{\theta}\overline{N}_{\s O}+i\theta\bar{\theta}{\cal O} \label{ann.supero}
\end{equation}
%%%
where ${\cal O}=\lambda_a\omega^{ab}\partial_bO+i{\bar c}_{a}\omega^{ad}(\partial_{d}\partial_{b}O)c^{b}$. 

In the first part of this thesis we will be interested only in the non-Grassmannian set of variables 
$(\varphi^a,\lambda_a)$.
So the Lagrangian $\LT$ of (\ref{ann.suplag}) and the Hamiltonian $\HT$ of (\ref{ann.supham}) will be reduced 
to:
%%%
\begin{eqnarray}
&&\LT_{\s B}=\lambda_a\dot{\varphi}^a-\lambda_a\omega^{ab}\partial_bH\label{ann.lagbos}\\
&&\HT_{\scriptscriptstyle B}=\lambda_a\omega^{ab}\partial_bH \label{ann.hambos}
\end{eqnarray}
%%%
and the superfields (\ref{ann.sup1}) and (\ref{ann.sup2}) to:
%%%
\begin{eqnarray}
&&\Phi^q=q+i\bar{\theta}\theta\lambda_p \label{ann.supb1}\\
&&\Phi^p=p-i\bar{\theta}\theta\lambda_q. \label{ann.supb2}
\end{eqnarray}
%%%
The commutators $[\widehat{\varphi}^a,\widehat{\lambda}_b]=i\delta_b^a$ 
can be realized considering $\widehat{\varphi}^a$ as multiplicative operators and $\widehat{\lambda}_b$ 
as derivative ones:
%%%
\begin{equation}
\displaystyle
\widehat{\varphi}^a=\varphi^a,\qquad
\widehat{\lambda}_b=-i\frac{\partial}{\partial\varphi^b}\;\;\Longrightarrow\;\;
\widehat{\lambda}_{q_j}=-i\frac{\partial}{\partial q_j},\;\;
\widehat{\lambda}_{p_j}=-i\frac{\partial} {\partial p_j}. \label{ijmpa.operatorial}
\end{equation}
%%%
From now on we will indicate this representation as the {\it Schr\"odinger representation} of CM.
Using (\ref{ijmpa.operatorial}) the Hamiltonian $\HT_{\s B}$ becomes the following operator:
%%%
\begin{equation}
\widehat{\HT}_{\s B}=-i\omega^{ab}\partial_bH\partial_a=-i\partial_{p_i}H\partial_{q_i}+i\partial_{q_i}H\partial_{p_i} 
\label{ann.operat}
\end{equation}
%%%
that is exactly the Liouvillian $\widehat{L}$ of (\ref{ijmpa.ht}). This confirms that the path integral 
(\ref{ann.prob3}) is the correct 
functional counterpart of the KvN operatorial theory. 

In deriving (\ref{ann.prob3}) we have started from the transition {\it probability} 
$P(\varphi,t|\varphi_i,t_i)$ of going from $\varphi_i$ to $\varphi$
in the time interval $t-t_i$. So we can say that the path integral (\ref{ann.prob3}) gives a kernel of 
evolution for the classical
probability densities $\rho(\varphi,t)$, in the sense that if we know the probability density at the 
initial time $t_i$ we can
derive the probability density $\rho$ at any later time $t$ via the standard relation:
%%%
\begin{equation}
\rho(\varphi,t)=\int d\varphi_iP(\varphi,t|\varphi_i,t_i)\rho(\varphi_i,t_i).
\end{equation}
%%%
Let us remember that KvN postulated for $\psi(\varphi)$ the same equation of motion (\ref{ijmpa.lio1}) 
as for $\rho(\varphi)$. 
As a consequence the evolution of the KvN waves can be represented as:
%%%
\begin{equation}
\psi(\varphi,t)=\int d\varphi_iK(\varphi,t|\varphi_i,t_i)\psi(\varphi_i,t_i)
\end{equation}
%%%
where the kernel of evolution $K(\varphi,t|\varphi_i,t_i)$ has the same expression
as the kernel of evolution $P(\varphi,t|\varphi_i,t_i)$ for the densities $\rho(\varphi)$.
To show that there is no contradiction in having the same kernel propagating both $\psi(\varphi)$ and $|\psi(\varphi)|^2$ now we will work out in detail the case of a free particle. 
The kernel $P(\varphi, t|\varphi_i,t_i)$ of $\rho(\varphi)$ is just (\ref{ijmpa.prob}) which is a Dirac delta. So, since the kernel $K(\varphi,t|\varphi_i,t_i)$ of propagation of
$\psi(\varphi)$ has to be the same, we have for a free particle:
%%%
\begin{equation}
\displaystyle
K(\varphi,t|\varphi_i,t_i=0)=\delta\biggl(q-q_i-\frac{p_it}{m}\biggr)\delta(p-p_i).
\end{equation}
%%%
Let us now use this expression to check what we get for the kernel of $\rho$ knowing that $\rho(t)=|\psi(t)|^2$.
%%%
\begin{eqnarray}
\displaystyle
\rho(t)=\psi^*(t)\psi(t)\hspace{-0.2cm}&=&\hspace{-0.2cm}\int d\varphi_i\,K^*(\varphi,t|\varphi_i,0)\,\psi^*(\varphi_i,0)\cdot\int
d\varphi_i^{\prime}\,K(\varphi,t|\varphi_i^{\prime},0)\,\psi(\varphi_i^{\prime},0)=
\nonumber\\ 
\hspace{-0.2cm}&=&\hspace{-0.2cm}\int
dq_idp_i\,\delta\biggl(q-q_i-\frac{p_it}{m}\biggr)\delta(p-p_i)\,\psi^*(q_i,p_i,0)\cdot
\nonumber\\
\hspace{-0.2cm}&&\hspace{-0.2cm}\cdot\int
dq^{\prime}_idp^{\prime}_i\,\delta\biggl(q-q^{\prime}_i-\frac{p^{\prime}_it}{m}\biggr)\delta(p-p^{\prime}_i)
\,\psi(q^{\prime}_i,p^{\prime}_i,0). \label{ijmpa.diruno}
\end{eqnarray}
%%%
Now we can use the properties of the Dirac deltas to rewrite:
%%%
\begin{eqnarray}
\rho(t)\hspace{-0.2cm}&=&\hspace{-0.2cm}\int dq_idp_idq_i^{\prime}dp_i^{\prime}\,
\delta\biggl(q-q_i-\frac{p_it}{m}\biggr)\delta(p-p_i)
\delta(p_i-p_i^{\prime})\cdot\nonumber\\
\hspace{-0.2cm}&&\hspace{-0.2cm}\;\;\;\cdot\delta\biggl(q_i-q_i^{\prime}+(p_i-p_i^{\prime})\frac{t}{m}\biggr)\,
\psi^*(q_i,p_i,0)\psi(q_i^{\prime},
p_i^{\prime},0). \label{ijmpa.dirdue}
\end{eqnarray}
%%%
The integrals over the primed variables can be done explicitly:
%%%
\begin{equation}
\int dq_i^{\prime}dp_i^{\prime}\,\delta(p_i-p_i^{\prime})
\delta\biggl(q_i-q_i^{\prime}+(p_i-p_i^{\prime})\frac{t}{m}\biggr)\psi^*(q_i,p_i,0)\psi(q_i^{\prime},
p_i^{\prime},0)=\rho(\varphi_i,0). \label{ijmpa.dirtre}
\end{equation}
%%%
Substituting (\ref{ijmpa.dirtre}) into (\ref{ijmpa.dirdue}) we have finally:
%%%
\begin{equation}
\rho(\varphi,t)=\int dq_idp_i K(\varphi, t|\varphi_i,0)\rho(\varphi_i,0).
\end{equation}
%%%
From this relation we get that the kernel of propagation for the probability density 
$\rho$ is the same as the one for 
the KvN wave $\psi$ and this confirms that there is no contradiction in the KvN postulate. 

\bigskip

\section{Spreading and Phases of KvN Waves}

One of the most characteristic effects of QM is the spreading of the wave functions during their 
time evolution. Let us
consider for example a quantum free particle in one dimension with Hamiltonian
$\displaystyle \widehat{H}=-\hbar^2\frac{\partial^2}{\partial q^2}$ and let us
take as initial wave function the following one of Gaussian type:
%%%
\begin{equation}
\displaystyle
\psi(q,t=0)=\frac{1}{\sqrt{\sqrt{\pi}a}}\textrm{exp}\biggl(-\frac{q^2}{2a^2}+\frac{i}{\hbar}p_iq\biggr). \label{ijmpa.initial}
\end{equation}
%%%
It is easy to check that, for what concerns the initial position $q$, its mean value is equal to zero and the uncertainty 
in its measurement is: $\displaystyle \overline{(\Delta q)^2}=\frac{a^2}{2}$. 
At time $t$ the wave function will be:
%%%
\begin{equation}
\displaystyle
\psi(q,t)=N\cdot \textrm{exp}\biggl[-\frac{m}{2(ma^2+i\hbar
t)}\biggl(q-\frac{p_it}{m}\biggr)^2+\frac{i}{\hbar}\biggl(p_iq-\frac{p_i^2}{2m}t\biggr)\biggr]. \label{ijmpa.ending}
\end{equation}
%%%
where $N$ is a normalization factor. From (\ref{ijmpa.ending}) 
we can note how the coefficient $p_i$, which at time $t=0$ appeared only in the phase
of the wave function, at any time $t>0$ appears also in its modulus. 
As a consequence the expectation value of the position 
$q$ at any time $t>0$ depends explicitly on the initial phase factor $p_i$;
in fact we have:
%%%
\begin{equation}
\overline{q(t)}=\int dq\,q\; |\psi(q,t)|^2=p_it/m.
\end{equation}
%%%
So we can say that, during the evolution, the information about the mean value of $q$ is carried by terms 
appearing originally in the phase of $\psi$. For the mean square deviation of $q$ we get:
%%%
\begin{equation}
\displaystyle
\overline{(\Delta q(t))^2}=\overline{(q-\bar{q}(t))^2}=\frac{a^2}{2}\biggl(1+\frac{t^2\hbar^2}{m^2a^4}\biggr)
\label{ijmpa.aabb}
\end{equation}
%%%
from which we obtain that, for $t\to\infty$, the wave function is totally delocalized:
%%%
\begin{equation}
\lim_{t\to\infty}\overline{(\Delta q(t))^2}=+\infty, \;\;\;\forall a>0.
\end{equation}
%%%
This effect is present also if we prepare an initial state very sharply peaked around the origin,
in fact:
%%%
\begin{equation}
\displaystyle
\lim_{a\to 0}\overline{\Delta(q(t))^2}=\lim_{a\to 0}\biggl(\frac{a^2}{2}+\frac{t^2\hbar^2}{2m^2a^2}
\biggr)=+\infty, \;\;\; \forall t>0.
\label{ijmpa.chiama}
\end{equation}
%%%
Note that the previous limit is $+\infty$ because of the presence of the parameter $a^2$ in 
the denominator of (\ref{ijmpa.chiama}). This relation is not surprising: if we take $a\to 0$ at the beginning then we have a state perfectly localized in space, i.e.
$\overline{(\Delta q)^2}\to 0$ but, from Heisenberg uncertainty relations, we deduce that
$\overline{(\Delta p)^2}\to +\infty$. So in this case the initial momentum is completely undetermined and, consequently,
even after an infinitesimal time interval, also the position of the particle becomes completely undetermined. These are the
well-known quantum mechanical effects. 

What happens in the operatorial version of CM?
As we have seen in the previous section, the evolution in time of the KvN waves is generated by the Liouvillian itself 
$\widehat{L}=-i\partial_pH\partial_q+i\partial_qH\partial_p$
which, 
in the particular case of a one-dimensional free particle, has the following simplified form:
%%%
\begin{equation}
\widehat{L}=
-i\frac{\widehat{p}}{m}\frac{\partial}{\partial q}. \label{ijmpa.pmq}
\end{equation}
%%%
The free Liouvillian is essentially the product of two commutative operators: an operator of multiplication 
$\widehat{p}$ and a derivative operator $\displaystyle -i\frac{\partial}{\partial q}$. So if we want
to diagonalize the Liouvillian $\widehat{L}$ of (\ref{ijmpa.pmq}) we have to diagonalize simultaneously both $\widehat{p}$
and $\displaystyle -i\frac{\partial}{\partial q}$. The eigenstates of $\widehat{p}$ associated to an arbitrary
real eigenvalue $p_0$ are the Dirac deltas $\delta(p-p_0)$; the eigenstates of 
$\displaystyle -i\frac{\partial}{\partial q}$ associated to an arbitrary real
eigenvalue\footnote[1]{
We call $\lambda_q$ the eigenvalues of $\displaystyle -i\frac{\partial}{\partial q}$ since
$\displaystyle -i\frac{\partial}{\partial q}$ is just a representation of the abstract Hilbert space 
operator $\widehat{\lambda}_q$: see (\ref{ijmpa.operatorial}) and the next section.} $\lambda_q$ are instead the plane waves
$\displaystyle \frac{1}{\sqrt{2\pi}}\textrm{exp}[i\lambda_qq]$.
So the eigenstates of the Liouvillian for a free particle (\ref{ijmpa.pmq}) are just the product of the eigenstates of
$\widehat{p}$ and $\displaystyle -i\frac{\partial}{\partial q}$:
%%%
\begin{equation}
\tau_{\lambda_qp_0}(q,p)=\frac{1}{\sqrt{2\pi}}\textrm{exp}[i\lambda_qq]\delta(p-p_0) \label{ijmpa.eigenstates}
\end{equation}
%%%
and the associated eigenvalues are: ${\mathcal E}=\displaystyle \frac{\lambda_qp_0}{m}$. Now, let us take as initial wave
function the following double Gaussian in $q$ and $p$:
%%%
\begin{equation}
\psi(q,p,t=0)=\frac{1}{\sqrt{\pi a b}}\textrm{exp}\biggl(-\frac{q^2}{2a^2}-\frac{(p-p_i)^2}{2b^2}\biggl) \label{ijmpa.double}
\end{equation}
%%%
where $a$ and $b$ are related to our initial uncertainty in the knowledge of $q$ and 
$p$:\break
$\overline{(\Delta q)^2}=a^2/2,\;\overline{(\Delta p)^2}=b^2/2$. Note that, 
since in CM $\widehat{q}$ and $\widehat{p}$
commute, there is no uncertainty relation. As a consequence $a$ and $b$ in (\ref{ijmpa.double})
are two completely independent parameters and the product $\overline{(\Delta q)^2}\cdot\overline{(\Delta p)^2}$
can assume arbitrary small values. 

Now we can write the initial KvN wave (\ref{ijmpa.double}) 
as a superposition of the eigenstates (\ref{ijmpa.eigenstates}) of the Liouvillian
$\widehat{L}$ as:
%%%
\begin{equation}
\psi(q,p,0)=\int d\lambda_qdp_0\,c(\lambda_q,p_0)\tau_{\lambda_qp_0}(q,p)
\end{equation}
%%%
where the coefficients $c(\lambda_q,p_0)$ are given by:
%%%
\begin{eqnarray}
\displaystyle
c(\lambda_q,p_0)&=&\int
dqdp\,\tau^*_{\lambda_q,p_0}(q,p)\psi(q,p,t=0)=\nonumber\\
&=&\sqrt{\frac{a}{\pi b}}\,\textrm{exp}\biggl(-\frac{\lambda_q^2a^2}{2}-\frac{(p_0-p_i)^2}{2b^2}\biggr).
\end{eqnarray}
%%%
The KvN wave at $t$ is:
%%%
\begin{eqnarray}
\displaystyle
\psi(q,p,t)&=&\int d\lambda_qdp_0\,c(\lambda_q,p_0)\,\textrm{exp}[-i{\mathcal E} t]\,\tau_{\lambda_qp_0}(q,p)
\nonumber\\
&=&\frac{1}{\sqrt{\pi a b}}\textrm{exp}\biggl[-\frac{1}{2a^2}\biggl(q-\frac{p}{m}t\biggr)^2-\frac{(p-p_i)^2}{2b^2}\biggr].
\label{ijmpa.param}
\end{eqnarray}
%%%
Note that this $\psi(t)$ is related to the initial KvN wave $\psi(0)$ by the following equation:
%%%
\begin{equation}
\displaystyle
\psi(q,p,t)=\psi\biggl(q-\frac{pt}{m},p,0\biggr)=\psi\biggl(q-\frac{\partial H}{\partial p}t, p+
\frac{\partial H}{\partial q}t, 0\biggr). \label{ijmpa.psievolution}
\end{equation}
%%%
The previous relation could be inferred also from the CPI (\ref{ann.prob3}). 
In fact, as we have seen in the previous section, in the case of a free particle the resulting kernel
of propagation is correctly given by:
%%%
\begin{equation}
\displaystyle
K(\varphi,t|\varphi_i,0)=\delta\biggl(q-q_i-\frac{p_it}{m}\biggr)\delta(p-p_i) \label{ijmpa.kernel}
\end{equation}
%%%
from which we obtain immediately:
%%%
\begin{equation}
\displaystyle
\psi(q,p,t)=\int d\varphi_i\,K(\varphi,t|\varphi_i,0)\psi(\varphi_i,0)=\psi\biggl(q-\frac{pt}{m},p,0\biggr).
\end{equation}
%%%
If we calculate the modulus square of the KvN wave
$|\psi(q,p,t)|^2=\rho(q,p,t)$, i.e., the probability density of finding the particle in a certain point of the phase space, we have from (\ref{ijmpa.psievolution}) that:
%%%
\begin{equation}
\rho(q,p,t)=\rho\biggl(q-\frac{\partial H}{\partial p}t, p+\frac{\partial H}{\partial q}t, 0
\biggr).
\end{equation}
%%%
which is in perfect agreement with the Liouville theorem $\displaystyle
\frac{d}{dt}\rho=0$. 

Let us now go back to (\ref{ijmpa.param}) and calculate 
the mean values of the dynamical variables at a generic time $t$.
They are:
%%%
\begin{equation}
\displaystyle
\bar{q}=\int dqdp\, q\;|\psi(q,p,t)|^2=\frac{p_it}{m},\;\;\;\;\;\;\;\bar{p}=\int dqdp\,p\;|\psi(q,p,t)|^2=p_i.
\label{ijmpa.equno}
\end{equation}
%%%
Note that, differently from QM, the information on the mean value of $\widehat{p}$ is given by coefficients 
which appear in the modulus of the KvN wave (\ref{ijmpa.param}).  
The mean square deviations are:
%%%
\begin{equation}
\overline{(\Delta q(t))^2}=\frac{a^2}{2}+\frac{b^2}{2}\frac{t^2}{m^2},\;\;\;\;\;\;\; 
\overline{(\Delta p(t))^2}=\frac{b^2}{2}.
\label{ijmpa.var12}
\end{equation}
%%%
Therefore also in CM if $b\neq 0$ then we have $\displaystyle \lim_{t\to\infty}
\overline{(\Delta q(t))^2}=+\infty$ and the KvN wave becomes totally delocalized. This is not strange if we consider
that we are giving a statistical description of a set of particles with a momentum  
distributed in a Gaussian way around $p_i$. This means that we can have particles with momenta both greater
and smaller than $p_i$. These particles
cause a spreading in the wave  function and, consequently, in the distribution of 
probability around the mean value of $q$.

If we instead consider the motion of a single particle we can measure exactly its position and momentum 
at the initial time. In this
case the terms that parameterize the Gaussians (\ref{ijmpa.double}) and (\ref{ijmpa.param}) 
go to zero ($a\to 0, b\to 0$).
Therefore the variances are identically zero because, differently from the quantum case
(\ref{ijmpa.chiama}), 
in (\ref{ijmpa.var12}) the parameters $a^2$ and $b^2$ do not appear in the denominator:
%%%
\begin{equation}
\lim_{a,b\to 0}\overline{(\Delta q(t))^2}=\lim_{a,b\to 0}\biggl(\frac{a^2}{2}
+\frac{b^2}{2}\frac{t^2}{m^2}\biggr)=0,
\;\;\;\;\;\;\lim_{a,b\to 0}\overline{(\Delta p(t))^2}=\lim_{b\to 0}\frac{b^2}{2}=0.\label{ijmpa.deltadelta}
\end{equation}
%%%
Since the previous relations hold for every time $t$ we can say that the particle remains perfectly localized in the phase space at every time $t$.

We feel that even this very simple and pedagogical example can be used to underline some very important differences between
the quantum and the classical operatorial approaches which are:\\ 
{\bf 1)} in CM we can know with absolute precision $q$ and $p$ since $\widehat{q}$ and 
$\widehat{p}$ are commuting operators and so there is no
uncertainty relation between them;\\
{\bf 2)} the classical dynamics given by $\widehat{L}$ is such that, if we know with
absolute precision the position and the momentum at $t=0$, they remain perfectly determined at every instant of time
$t$ and there is no spreading;\\ 
{\bf 3)} the knowledge about the average momentum of the classical particle is 
brought by terms appearing in the modulus, and not in the phase, of the KvN wave. 

For a classical free particle it is easy to show that, even if we add a phase factor to an {\it arbitrary}
initial KvN wave, then this phase factor does not pass into the real part 
during the evolution differently from what happens in QM, see (\ref{ijmpa.initial})-(\ref{ijmpa.ending}). 
This can be proved
as follows: every
initial classical KvN wave 
%%%
\begin{equation}
\psi(q,p)=F(q,p)\,\textrm{exp}[iG(q,p)] \label{ijmpa.effegi}
\end{equation}
%%%
can always be written as a superposition of the eigenstates of the free Liouvillian 
(\ref{ijmpa.pmq}) in the following way:
%%%
\begin{equation}
\psi(q,p)=\int d\lambda_qdp_0\,c(\lambda_q,p_0)\tau_{\lambda_qp_0}(q,p)
\end{equation}
%%%
where the eigenstates $\tau_{\lambda_q,p_0}(q,p)$ are given by (\ref{ijmpa.eigenstates})
while the coefficients $c(\lambda_q,p_0)$ are basically the Fourier transforms of $\psi(q,p_0)$ 
with respect to the variable $q$:
%%%
\begin{equation}
\displaystyle
c(\lambda_q,p_0)=\frac{1}{\sqrt{2\pi}}\int dq \,e^{-i\lambda_q q}F(q,p_0)e^{iG(q,p_0)}.
\end{equation}
%%%
Now the free evolution of the KvN wave can be obtained in the usual way:
%%%
\begin{eqnarray}
\displaystyle
\label{ijmpa.psiqpt}
\psi(q,p,t)&\hspace{-0.2cm}=&\hspace{-0.2cm}\int d\lambda_qdp_0\,c(\lambda_q,p_0)\,
\textrm{exp}[-i{\mathcal E} t]\,\tau_{\lambda_qp_0}(q,p)=\nonumber\\
&\hspace{-0.2cm}=&\hspace{-0.2cm}\int d\lambda_q \,c(\lambda_q,p)\,\textrm{exp}
\biggl[i\lambda_q\biggl(q-\frac{pt}{m}\biggr)\biggr]=\\
&\hspace{-0.2cm}=&\hspace{-0.2cm} F\biggl(q-\frac{pt}{m},p\biggr)
\textrm{exp}\biggl[iG\biggl(q-\frac{pt}{m},p\biggr)\biggr] \nonumber
\end{eqnarray}
%%%
which again is in perfect agreement with the kernel of evolution (\ref{ijmpa.kernel}).
Since the phase $G$ remains a phase also during the evolution we have that, in the case of a free particle, 
for {\it every} initial KvN wave $\psi(q,p)=F(q,p)\textrm{exp}[iG(q,p)]$
the probability density 
$|\psi(q,p,t)|^2$ does not depend on the phase $G$ not only at the beginning, but also at any
later time; in fact from (\ref{ijmpa.psiqpt}) we have that $\displaystyle |\psi(q,p,t)|^2=F^2\biggl(
q-\frac{pt}{m},p\biggr)$. 
This has some consequences also on the expectation values of the observables. Usually one assumes that 
observables in CM are the functions of the phase space $O(\varphi)$. In operatorial terms they become 
the operators $O(\widehat{\varphi})$ and it is easy to check that their
expectation values do not depend on the phase $G$ of the KvN wave (\ref{ijmpa.effegi}):
%%%
\begin{equation}
\langle O\rangle=\int d\varphi F^*(\varphi)\,\textrm{exp}[-iG(\varphi)]O(\varphi)
F(\varphi)\textrm{exp}[iG(\varphi)]=\int d\varphi
F^*(\varphi)O(\varphi)F(\varphi).
\end{equation}
%%%
This is true for any time $t$ as it is clear from the form (\ref{ijmpa.psiqpt}) of the KvN wave. 
This independence from the phase $G$ would not happen if there were observables dependent also on $\lambda$ 
because $\widehat{\lambda}$ is a derivative operator and phases enter the expectation values of the 
derivative operators\footnote[2]{We will analyse this in more details in Sec. {\bf 1.4}.}.

These considerations can be
extended to the case of a physical system characterized by a generic Liouvillian $\widehat{L}$. 
In fact the solution
of the equation:
%%%
\begin{equation}
\displaystyle
i\frac{\partial}{\partial t}\psi(q,p,t)=\widehat{L}\psi(q,p,t)
\end{equation}
%%%
is given by \cite{Sudarshan}:
%%%
\begin{equation}
\psi(q,p,t)=\psi\Bigl(\bar{q}(q,p,t),\bar{p}(q,p,t)\Bigr) \label{ijmpa.trasc}
\end{equation}
%%%
where $\bar{q}$ and $\bar{p}$ are the solutions of the equations:
%%%
\begin{equation}
\displaystyle
\dot{\overline{q}}_j(q,p,t)=-\frac{\partial H(\bar{q},\bar{p})}{\partial\bar{p}_j},\;\;\;
\dot{\overline{p}}_j(q,p,t)=\frac{\partial H(\bar{q},\bar{p})}{\partial\bar{q}_j}
\end{equation}
%%%
with the initial conditions $\bar{q}_j(q,p,0)=q^0_j,\;\; \bar{p}_j(q,p,0)=p^0_j$. So,
according to (\ref{ijmpa.trasc}), the evolution of a classical system via the Liouvillian does not modify,
in the Schr\"odinger representation, the functional form of $\psi$. As an immediate consequence, we have that, if we
take a KvN wave without any phase at the initial time, then phases {\it cannot} be generated during the evolution:
%%%
\begin{equation}
\widehat{L}:\;\;\;\psi\,(\textrm{without phases} \;\, t=0) \;\,\longrightarrow\;\, \psi\,(\textrm{without phases} \;\, t).
\end{equation}
%%%
In QM instead, even if we start from a wave function that does not contain phases, these ones
will be created in general at later times via the operator $\widehat{H}$:
%%%
\begin{equation}
\widehat{H}:\;\;\;\psi\,(\textrm{without phases}\;\, t=0) \;\;\;\longrightarrow \;\;\;\psi\,(\textrm{with phases} \; \, t).
\end{equation}
%%%,
An example of this phenomenon is given by (\ref{ijmpa.initial}) and (\ref{ijmpa.ending}). Even if we start with no phase $p_i=0$,
at time $t$ we get that the wave function (\ref{ijmpa.ending}) becomes:
%%%
\begin{equation}
\displaystyle
\psi= N\cdot \textrm{exp}\biggl[-\frac{m}{2(ma^2+i\hbar t)}q^2\biggr]
\end{equation}
%%%
and it has a phase because of the term $i\hbar t$ in the denominator. All this is a consequence of the fact that the phase
and the modulus of the wave functions are coupled in QM as one can see from standard text books, \cite{Messiah}.
In fact, writing
%%%
\begin{equation}
\psi(q)=A(q) \textrm{exp}\biggl[\frac{i}{\hbar}S(q)\biggr]
\end{equation}
%%%
and equating the real and imaginary part of the Schr\"odinger equation (\ref{ijmpa.third}), 
we obtain the following two equations for $A(q)$ and $S(q)$:
%%%
\begin{equation}
\label{ijmpa.messiah}
\left\{
\begin{array}{l}
\displaystyle
\frac{\partial S}{\partial t}+\frac{1}{2m}\biggl(\frac{\partial S}{\partial q}\biggr)^2
+V=\frac{\hbar^2}{2mA}
\frac{\partial^2 A}{\partial q^2}\smallskip\\
\displaystyle m\frac{\partial A}{\partial t}+\frac{\partial A}{\partial q}\frac{\partial S}{\partial q}
+\frac{A}{2}\frac{\partial^2 S}{\partial q^2}=0.\\ 
\end{array}
\right.
\end{equation}
%%%
From (\ref{ijmpa.messiah}) it is easy to see that $S$ and $A$ are {\it coupled} by their equations of motion. 

In CM instead if we start from
%%%
\begin{equation}
\psi(q,p)=F(q,p)\,\textrm{exp}[iG(q,p)]
\end{equation}
%%%
we can insert it in (\ref{ijmpa.lio1}): $\displaystyle i\frac{\partial\psi}{\partial t}=\widehat{L}\psi$
and, equating the real and imaginary part, we get that both the modulus and the phase evolve with the Liouville equation: 
%%%
\begin{equation}
\displaystyle
\label{decoupling}
i\frac{\partial F}{\partial t}=\widehat{L} F, \;\;\;\;\;\;\;\;\;\;\;i\frac{\partial G}{\partial t}=\widehat{L} G.
\end{equation}
%%%
So we see that in CM the phase and the modulus {\it decouple} from each other and they do not interact at all. 
Summarizing all this discussion we can give the following aphorism:
``{\it What is QM? Quantum mechanics is
the  theory of the interaction of a phase with a modulus}". 

As we have just proved, in CM there
is at least one representation in which this interaction 
is completely lost and the
evolution of the modulus is completely decoupled from the evolution of the phase. 
One may then think that it is useless to deal
with complex KvN waves if their phases do not bring in any physical information.  
This is true only if we decide to work in
the Schr\"odinger representation. In the next section we shall show that, changing representation
and using the one where 
$\widehat{p}$ is realized as the derivative
with respect to $\lambda_p$, the mean value of $\widehat{p}$ is related to the phase of the KvN waves.
So, if we want to be as general as possible and not just stick to the Schr\"odinger representation, we have to assume
that the classical KvN waves are complex objects. 

\bigskip

\section{Abstract Hilbert Space and the $(q,\lambda_p)$ Representation}

In the previous section we have restricted ourselves to the Schr\"odinger representation in which both $\widehat{q}$ and 
$\widehat{p}$ are realized as multiplicative operators, and we have worked out everything in this frame. 
What we want to do now is to construct the {\it Hilbert space} of CM from an {\it abstract} point of view,
i.e. without considering any particular representation. 
We can start observing that $\widehat{q}$ and $\widehat{p}$  can be
considered as a complete set of commuting operators whose real eigenvalues form a continuous spectrum 
which includes all the values from $-\infty$ to $+\infty$:
%%%
\begin{equation}
\widehat{q}|q,p\rangle=q|q,p\rangle;\;\;\;\;\;\widehat{p}|q,p\rangle=p|q,p\rangle. \label{ann.diagmix0}
\end{equation}
%%%
The eigenstates $|q,p\rangle$ form an orthonormal and complete set which can be used as a basis for the Hilbert space
of KvN. The orthonormality and completeness relations are respectively given by:
%%%
\begin{equation}
\displaystyle 
\langle
q^{\prime},p^{\prime}|q^{\prime\prime},p^{\prime\prime}\rangle=
\delta(q^{\prime}-q^{\prime\prime})\delta(p^{\prime}-p^{\prime\prime}),\;\;\;\;\;
\int dq dp\,|q,p\rangle\langle q,p|=1. \label{ijmpa.fortysix}
\end{equation}
%%%
The connection between the abstract vectors $|\psi\rangle$ and the KvN waves $\psi(q,p)$ is through the relation 
$\langle q,p|\psi\rangle=\psi(q,p)$. In this basis the operators
$\widehat{q}$ and $\widehat{p}$ are diagonal:
%%%
\begin{eqnarray}
&&\langle
q^{\prime},p^{\prime}|\widehat{q}|q^{\prime\prime},p^{\prime\prime}\rangle=q^{\prime}\delta(q^{\prime}-q^{\prime\prime})
\delta(p^{\prime}-p^{\prime\prime});\nonumber\\
&&\langle
q^{\prime},p^{\prime}|\widehat{p}|q^{\prime\prime},p^{\prime\prime}\rangle=p^{\prime}\delta(q^{\prime}-q^{\prime\prime})
\delta(p^{\prime}-p^{\prime\prime})
\end{eqnarray}
%%%
while the operators $\displaystyle -i\frac{\partial}{\partial q}\biggl(\displaystyle -i\frac{\partial}{\partial p}\biggr)$
defined by the relations:
%%%
\begin{equation}
\displaystyle
\langle
q^{\prime},p^{\prime}\biggl|-i\frac{\partial}{\partial q}\biggl(-i\frac{\partial}{\partial
p}\biggr)\bigg|\psi\rangle=-i\frac{\partial}{\partial q^{\prime}}
\biggl(-i\frac{\partial}{\partial p^{\prime}}\biggr)\langle q^{\prime},p^{\prime}|\psi\rangle \label{ijmpa.fortyeight}
\end{equation}
%%%
are Hermitian. From (\ref{ijmpa.fortysix})-(\ref{ijmpa.fortyeight}) it is easy to check that:
%%%
\begin{equation}
\displaystyle
\langle q^{\prime},p^{\prime}\biggl|\biggl[\widehat{q},-i\frac{\partial}{\partial q}\biggr]\biggr|\psi\rangle=\langle
q^{\prime},p^{\prime}|i|\psi
\rangle,\;\;\;\;\;\;\langle q^{\prime},p^{\prime}\biggl|\biggl[\widehat{p},-i\frac{\partial}{\partial p}\biggr]\biggr|\psi\rangle=\langle
q^{\prime},p^{\prime}|i|\psi
\rangle \label{ijmpa.fortynine}
\end{equation}
%%%
while all the other commutators are zero. Because of
the completeness of $\langle q^{\prime},p^{\prime}|$ and the arbitrariness of the state $|\psi\rangle$ we have that
(\ref{ijmpa.fortynine}) can be turned into the purely operatorial relations:
$\displaystyle \biggl[\widehat{q},-i\frac{\partial}{\partial q}\biggr]=i$ and $\displaystyle
\biggl[\widehat{p},-i\frac{\partial}{\partial p}\biggr]=i$ and so, from (\ref{ijmpa.comm}), we can identify
$\displaystyle
\widehat{\lambda}_q=-i\frac{\partial} {\partial q}$ and 
$\displaystyle \widehat{\lambda}_p=-i\frac{\partial}{\partial p}$.
Now it is easy to show that the Liouville equation (\ref{ijmpa.lio1}) 
is nothing more than a particular representation of the abstract
Liouville equation:
%%%
\begin{equation}
\displaystyle
i\frac{\partial}{\partial t}|\psi,t\rangle=\widehat{\lambda}_a\omega^{ab}\partial_bH|\psi,t\rangle \label{ijmpa.ann}
\end{equation}
%%%
obtained using as basis the eigenfunctions of $\widehat{q}$ and $\widehat{p}$. 
If we consider the following Hamiltonian in the standard phase space: $\displaystyle H=\frac{p^2}{2m}+V(q)$, then 
the Liouville equation (\ref{ijmpa.ann}) becomes:
%%%
\begin{equation}
\displaystyle
i\frac{\partial}{\partial t}|\psi,t\rangle=\biggl[\widehat{\lambda}_q\frac{\widehat{p}}{m}-
\widehat{\lambda}_p\partial_qV(q)\biggr]|\psi,t\rangle.
\label{ijmpa.anndue}
\end{equation}
%%%
Projecting the previous equation onto the basis $\langle q,p|$ we easily obtain:
%%%
\begin{eqnarray}
\displaystyle
i\frac{\partial}{\partial t}\langle q,p|\psi,t\rangle&\hspace{-0.2cm}=&\hspace{-0.2cm}
\langle q,p\biggl|\widehat{\lambda}_q
\frac{\widehat{p}}{m}\biggr|\psi,t\rangle
-\langle q,p|\widehat{\lambda}_p \partial_qV(q)|\psi,t\rangle=\\
&\hspace{-0.2cm}=&\hspace{-0.2cm}-i\frac{\widehat{p}}{m}\frac{\partial}{\partial q}\langle q,p|\psi,t\rangle+i\partial_qV(q)\frac{\partial}{\partial p}\langle
q,p|\psi,t\rangle \nonumber
\end{eqnarray}
%%%
that is equivalent to the usual Liouville equation:
%%%
\begin{equation}
\displaystyle
\frac{\partial}{\partial t}\psi(q,p,t)=\biggl[-\frac{\widehat{p}}{m}\frac{\partial}{\partial q}+\partial_qV(q)\frac{\partial}
{\partial p}\biggr]\psi(q,p,t).
\end{equation}
%%%

The basis of the eigenstates $|q,p\rangle$ is not the only one for the Hilbert space of CM. A very important
representation
\cite{Abrikosov} is the one in which we consider the basis of the simultaneous eigenstates of $\widehat{q}$ and 
$\widehat{\lambda}_p$ which, according to (\ref{ijmpa.comm}), are commuting operators:
%%%
\begin{equation}
\widehat{q}\,|q,\lambda_p\rangle=q|q,\lambda_p\rangle;\;\;\;\;\;\;\;\;\;\;\;
\widehat{\lambda}_p|q,\lambda_p\rangle=\lambda_p|q,\lambda_p\rangle. \label{ijmpa.eq2}
\end{equation}
%%%
Sandwiching the second relation in (\ref{ijmpa.eq2}) with the bra $\langle q^{\prime},p^{\prime}|$ we obtain
the following differential equation:
%%%
\begin{equation}
\displaystyle
-i\frac{\partial}{\partial p^{\prime}}\langle q^{\prime},p^{\prime}|q,\lambda_p\rangle=\lambda_p\langle q^{\prime},
p^{\prime}|q,\lambda_p\rangle \label{ijmpa.fivefive}
\end{equation}
%%%
whose solution is:
%%%
\begin{equation}
\displaystyle
\langle q^{\prime},p^{\prime}|q,\lambda_p\rangle=\frac{1}{\sqrt{2\pi}}\delta(q-q^{\prime})e^{ip^{\prime}\lambda_p}.
\label{ijmpa.fiftysix}
\end{equation}
%%%
Also the states $|q,\lambda_p\rangle$ form a complete set of orthonormal eigenstates, 
i.e. another possible basis for the
vectors of the KvN classical Hilbert space. In this basis we have:
%%%
\begin{equation}
\langle q,\lambda_p|\psi\rangle=\int dq^{\prime} dp\langle q,\lambda_p|q^{\prime},p\rangle\langle q^{\prime},p|\psi\rangle 
\end{equation}
%%%
which, via (\ref{ijmpa.fiftysix}), gives:
%%%
\begin{equation}
\displaystyle
\psi(q,\lambda_p)=\frac{1}{\sqrt{2\pi}}\int dp \,e^{-ip\lambda_p}\,\psi(q,p). \label{ijmpa.Fou}
\end{equation}
%%%
This means that the KvN waves in the $(q,\lambda_p)$ representation and in the Schr\"odinger one
are related via a Fourier transform\footnote[3]{We indicate the wave
functions in the new basis with the same symbol $\psi$ for notational simplicity.}. In this
new representation we have for the $\widehat{p}$ operator:
%%%
\begin{eqnarray}
\displaystyle
\langle q,\lambda_p|\widehat{p}|\psi\rangle &=&\int dq^{\prime}dp^{\prime}\langle q,\lambda_p|\,\widehat{p}\,|q^{\prime},
p^{\prime}\rangle\langle q^{\prime},p^{\prime}|\psi\rangle=\frac{1}{\sqrt{2\pi}}
\int dp^{\prime}\,p^{\prime}e^{-ip^{\prime}\lambda_p}
\psi(q,p^{\prime})=\nonumber\\
&=&\frac{1}{\sqrt{2\pi}}i\frac{\partial}{\partial\lambda_p}\int dp^{\prime}e^{-ip^{\prime}\lambda_p}\langle
q,p^{\prime}|\psi
\rangle=i\frac{\partial}{\partial\lambda_p}\langle q,\lambda_p|\psi\rangle \label{ijmpa.fiftynine}
\end{eqnarray}
%%%
while $\widehat{\lambda}_p$ is simply a multiplicative operator:
%%%
\begin{equation}
\langle q,\lambda_p|\widehat{\lambda}_p|\psi\rangle=\lambda_p\langle q,\lambda_p|\psi\rangle. \label{ijmpa.sixty}
\end{equation}
%%%
Summarizing (\ref{ijmpa.eq2})-(\ref{ijmpa.sixty}), we can say that, with respect to the Schr\"odinger 
representation, in the $(q,\lambda_p)$ one we have to consider 
$\widehat{p}$ as a derivative operator and $\widehat{\lambda}_p$ 
as a multiplicative one:
%%%
\begin{equation}
\left\{
	\begin{array}{l}
	\displaystyle q \longrightarrow \widehat{q}\smallskip\\
          \displaystyle \lambda_p \longrightarrow \widehat{\lambda}_p\smallskip\\
          \displaystyle \lambda_q  \longrightarrow -i\frac{\partial}{\partial q}\smallskip\\
          \displaystyle p \longrightarrow i\frac{\partial}{\partial \lambda_p}. \label{ann.mixrep}
	\end{array}
	\right.
\end{equation}
%%%
This is simply a different realization of the usual commutation relations: 
$[\widehat{\varphi}^a,\widehat{\lambda}_b]=i\delta_b^a$. 
Using $\langle q,\lambda_p|$ we get that the abstract Liouville equation (\ref{ijmpa.anndue}) becomes:
%%%
\begin{equation}
\displaystyle
i\frac{\partial}{\partial t}\psi(q,\lambda_p,t)=\frac{1}{m}\frac{\partial}{\partial
q}\frac{\partial}{\partial\lambda_p}\psi(q,\lambda_p,t) -\lambda_p\partial_qV(q)\psi(q,\lambda_p,t).
\end{equation}
%%%

We shall now show that a lot of the results of the 
previous section were strongly dependent on the particular kind of representation we used. 
In fact in the $(q,\lambda_p)$ representation, since the momentum $\widehat{p}$ has become a derivative operator,
we have that the information about its mean value is brought in by the phase of the KvN wave
similarly to what happens in quantum 
mechanics. 
For example, using the Fourier transform (\ref{ijmpa.Fou}), we have that, in the $(q,\lambda_p)$ representation, the 
double Gaussian state (\ref{ijmpa.double}) becomes the following one:
%%%
\begin{equation}
\displaystyle
\psi(q,\lambda_p,t=0)=\sqrt{\frac{b}{\pi a}}\textrm{exp}\Biggl(-\frac{q^2}{2a^2}\Biggr)
\textrm{exp}\Biggl(-\frac{\lambda_p^2b^2}{2}-ip_i\lambda_p\Biggr). \label{ijmpa.fed1}
\end{equation}
%%%
We immediately note that the KvN wave, which was real in the Schr\"odinger
representation, has become complex. The mean values of $\widehat{q}$ and $\widehat{p}$ are obviously the same
as before:
%%%
\begin{equation}
\displaystyle
\bar{q}=\langle\psi|\widehat{q}|\psi\rangle=0,\;\;\;\;\;\;\bar{p}=\langle\psi|\widehat{p}|\psi\rangle=
\langle\psi\biggl|i\frac{\partial}{\partial\lambda_p}\biggr|\psi\rangle=p_i \label{ijmpa.sixtre}
\end{equation}
%%%
but now we see that elements appearing in the phase of the KvN waves, like $p_i$ in (\ref{ijmpa.fed1}), begin to play an important role since 
they are linked with the mean values of physical observables like $\widehat{p}$. 

Let us now make the evolution of (\ref{ijmpa.fed1}) under the Liouvillian for a free particle.
This Liouvillian in the $(q,\lambda_p)$ representation is given by:
%%%
\begin{equation}
\widehat{L}=\frac{1}{m}\frac{\partial^2}{\partial q\partial\lambda_p}.
\end{equation}
%%%
Its eigenstates are:
%%%
\begin{equation}
\displaystyle
\tau_{\lambda_q,p}(q,\lambda_p)=\frac{1}{2\pi}\textrm{exp}[i\lambda_qq-i\lambda_pp].
\end{equation}
%%%
while the associated eigenvalues are $p\lambda_q/m$.
Writing the KvN wave (\ref{ijmpa.fed1}) 
as a superposition of the eigenstates $\tau_{\lambda_q,p}$ above and making the evolution of the system, we obtain
at time $t$:
%%%
\begin{equation}
\displaystyle
\psi(q,\lambda_p,t)=N\cdot
\textrm{exp}\Biggl[-\frac{q^2}{2a^2}-\frac{p_i^2}{2b^2}-\frac{1}{2}\frac{(\lambda_pma^2b^2
+iqtb^2+ip_ima^2)^2}{a^2b^2(m^2a^2+t^2b^2)}\Biggr] \label{ijmpa.fed2}
\end{equation}
%%%
where $N$ is the normalization factor.
From the previous formula we see how the factor $p_i$ which, according to (\ref{ijmpa.fed1}), 
at time $t=0$ entered only the phase factor,
at time $t$ has passed also into the real
part of the KvN wave, exactly as in the quantum case
we studied before. The expectation values and the variances of $q$ and $p$ are
still given by (\ref{ijmpa.equno})-(\ref{ijmpa.var12}):
%%%
\begin{equation}
\displaystyle
\bar{q}=\frac{p_it}{m},\;\;\;\;\;\;\bar{p}=p_i
\end{equation}
%%%
and:
%%%
\begin{equation}
\displaystyle
\overline{(\Delta q)^2}=\frac{a^2}{2}+\frac{b^2}{2}\frac{t^2}{m^2},\;\;\;\;\;\;\overline{(\Delta p)^2}=\frac{b^2}{2}.
\end{equation}
%%%
This is true because they are observable quantities and, consequently, they have to be 
independent of the representation we are using.
Anyhow to prepare a KvN wave of the form (\ref{ijmpa.fed1})
well-localized both in $q$ and in $\lambda_p$, we have to send $a\to 0,b\to\infty$. In this
limit we have that $\overline{(\Delta q)^2}\to\infty$ at every time $t>0$ and so there is an instantaneous
spreading of the KvN wave. This is not surprising. In fact if the initial KvN wave is very peaked around
$q=\lambda_p=0$ then we know precisely the values of $q$ and $\lambda_p$, instead of the values of $q$ and $p$. From the
commutator $[\widehat{p},\widehat{\lambda}_p]=i$  we can derive the following uncertainty relation: $\displaystyle
\Delta p\cdot\Delta\lambda_p\ge 1/2$, where $\Delta p$ and $\Delta\lambda_p$ are the square roots of the 
mean square deviations.
So if, by sending $b\to\infty$, we determine with absolute precision
$\lambda_p$, the value of $p$ is completely undetermined. Consequently also the position $q$ at every instant $t>0$ 
is completely undetermined, because $q$ and $p$ are linked by the classical equations of motion 
$\dot{q}=p/m$. 
An immediate consequence of this is the spreading of $q$ and the complete delocalization 
of the KvN wave at every instant of time following the
initial one\footnote[4]{The usual mechanics of the single particle can be reproduced also in this representation but we have to take
the limit
$a\to 0,b\to 0$, i.e. we have to use for $\lambda_p$ a plane wave of the type $\textrm{exp}(-ip_i\lambda_p)$.}.

Another aspect that we can study is the continuity equation
which in the Schr\"odinger representation was nothing more than the usual Liouville equation for
the probability density $\rho$, as we have seen in Sec. {\bf 1.1}. 
What happens in the $(q,\lambda_p)$ representation? According to what we have already
seen, the Liouville equation for $\psi$ becomes:
%%%
\begin{equation}
\displaystyle
i\frac{\partial}{\partial t}\psi(q,\lambda_p,t)=\frac{1}{m}\frac{\partial}{\partial
q}\frac{\partial}{\partial\lambda_p}\psi(q,\lambda_p,t) -\lambda_pV^{\prime}(q)\psi(q,\lambda_p,t). \label{ijmpa.seinove}
\end{equation}
%%%
Taking the complex conjugate we obtain the equation for $\psi^*(q,\lambda_p)$:
%%%
\begin{equation}
\displaystyle
-i\frac{\partial}{\partial t}\psi^*(q,\lambda_p,t)=\frac{1}{m}\frac{\partial}{\partial q}\frac{\partial}{\partial\lambda_p}
\psi^*(q,\lambda_p,t)-\lambda_pV^{\prime}(q)\psi^*(q,\lambda_p,t). \label{ijmpa.settezero}
\end{equation}
%%%
From (\ref{ijmpa.seinove}) and (\ref{ijmpa.settezero}) we get that the equation for $\rho(q,\lambda_p)=\psi^*(q,\lambda_p)
\psi(q,\lambda_p)$ is of the form:
%%%
\begin{equation}
\displaystyle
\frac{\partial}{\partial t}\rho(q,\lambda_p,t)+{\mathbf J}=0
\end{equation}
%%%
where:
%%%
\begin{equation}
\displaystyle
{\mathbf J}=\frac{i}{m}\bigg(\psi^*\frac{\partial}{\partial q}\frac{\partial}{\partial\lambda_p}\psi-
\psi\frac{\partial}{\partial q}\frac{\partial}{\partial\lambda_p}\psi^*\bigg).
\end{equation}
%%%
So in this case $\rho(q,\lambda_p)$ does not evolve with 
the Liouville equation and there is no manner to write ${\mathbf J}$ in terms only of $\rho$.
In fact, if we write 
$\psi(q,\lambda_p)=\sqrt{\rho}\,\textrm{exp}[iS(q,\lambda_p)]$,
the phase $S(q,\lambda_p)$ will enter explicitly into ${\mathbf J}$ and the equation of $\rho$. This creates 
a situation very similar to the quantum one,
where phases and moduli are coupled in the equations of motion. 

Another aspect that the $(q,\lambda_p)$ representation of CM has in common with the standard quantum
one is the following: even if we prepare a real KvN wave of $q$ and $\lambda_p$ at the initial time $t=0$, phases
will be created in general by $\widehat{L}$ during the evolution. This can be seen by means of our usual example. In fact, if we put
$p_i=0$, we have from (\ref{ijmpa.fed1}) and (\ref{ijmpa.fed2}) that:
%%%
\begin{equation}
\left\{
	\begin{array}{l}
	\displaystyle\sqrt{\frac{b}{\pi a}}\textrm{exp}\biggl(-\frac{q^2}{2a^2}-\frac{\lambda_p^2b^2}{2}\biggr)
        \;\longrightarrow\; N\cdot
	\textrm{exp}\biggl[-\frac{q^2}{2a^2}-\frac{1}{2}\frac{(\lambda_pma^2b^2+iqtb^2)^2}{a^2b^2(m^2a^2+t^2b^2)}
	\biggr]\bigskip\nonumber\\
	\displaystyle\widehat{L}: \;\;\;\psi(q,\lambda_p,t=0)\;\; \textrm{without phases}\;
        \longrightarrow\;\psi(q,\lambda_p,t) \;\;\textrm{with phases}. 
	\end{array}
	\right.
\end{equation}
%%%
Since the $(q,\lambda_p)$ representation has all these features in common with QM it is to be expected
that this representation turns out to be \cite{Abrikosov} the one where the process of quantization is best 
understood.

\bigskip

\section{Superselection Rules and Observables}

In the previous section we have given some reasons why the elements of the Hilbert space
of KvN must be chosen as complex. Now we want first to analyse in detail which assumptions must be done on the
observables in order to make this choice consistent. Second we want to give the abstract
theoretical reasons why the phases of the classical KvN waves cannot be felt by those operators 
which depend only on $\widehat{\varphi}=(\widehat{q},\widehat{p})$ when we choose the $(q,p)$ representation. 
In performing this analysis
we shall make use of the notion of superselection rules. For a review about this subject we refer the reader to
\cite{mex}\cite{gal}.

Can a superselection mechanism appear in the Hilbert space of KvN and which are its consequences?
To answer this question we have to analyse in detail the issue of which are the observables in KvN theory.
Usually physicists identify the observables of CM with the functions 
of the phase space variables $\varphi$. In the operatorial formulation of CM this is equivalent 
to postulate that the observables are {\it all} and {\it only} the functions of the operators
$\widehat{\varphi}=(\widehat{q},\widehat{p})$. Let us accept for a while this postulate and see which 
are its consequences for the KvN formulation of CM. 
First of all the algebra of the classical observables turns out to be Abelian and the operators
$\widehat{\varphi}=(\widehat{q},\widehat{p})$ turn out to commute with {\it all} the observables 
of the theory. Therefore the operators $\widehat{\varphi}=(\widehat{q},\widehat{p})$
are {\it superselection operators} and the associated superselection rules have to be taken into account. According to
these rules if
we consider two states corresponding to different eigenvalues of the superselection operators, i.e. 
$|\varphi_{\scriptscriptstyle 1}\rangle$ and $|\varphi_{\scriptscriptstyle 2}\rangle$ satisfying:
%%%
\begin{equation}
\left\{
\begin{array}{l}
\widehat{\varphi}|\varphi_{\scriptscriptstyle 1}\rangle=\varphi_{\scriptscriptstyle 1}|\varphi_{\scriptscriptstyle 1}
\rangle \smallskip \\
\widehat{\varphi}|\varphi_{\scriptscriptstyle 2}\rangle=\varphi_{\scriptscriptstyle 2}|\varphi_{\scriptscriptstyle 2}
\rangle,
\end{array}
\right.
\end{equation} 
%%%
we have that there is no
observable connecting them since:
%%%
\begin{equation}
\langle\varphi_{\scriptscriptstyle 1}|O(\widehat{\varphi})|\varphi_{\scriptscriptstyle 2}\rangle=0.
\end{equation}
%%%
Furthermore if we consider a linear superposition of eigenstates corresponding to different eigenvalues:
%%%
\begin{equation}
|\psi_{\scriptscriptstyle 1}\rangle=\alpha_{\scriptscriptstyle 1}|\varphi_{\scriptscriptstyle 1}\rangle
+\alpha_{\scriptscriptstyle 2}|\varphi_{\scriptscriptstyle 2}\rangle \label{alpha}
\end{equation}
%%%
with $\alpha_{\scriptscriptstyle 1}\alpha_{\scriptscriptstyle 2}\neq 0$ we have that the vector 
$|\psi_{\scriptscriptstyle 1}\rangle$ cannot be considered a pure state. In fact, if 
the observables of the theory
are only the functions $O(\widehat{\varphi})$, then it is impossible to find an 
observable having $|\psi_{\scriptscriptstyle 1}\rangle$
as an eigenstate. So the state $|\psi_{\scriptscriptstyle 1}\rangle$ cannot
be prepared diagonalizing a complete set of observables, like it happens 
for all the pure states of QM.
Besides this when we compute the expectation values of the observables $O(\widehat{\varphi})$ 
on $|\psi_{\scriptscriptstyle 1}\rangle$ we have that:
%%%
\begin{eqnarray}
\displaystyle \overline{O}&\hspace{-0.2cm}=&\hspace{-0.2cm}
\frac{\langle\psi_{\scriptscriptstyle 1}|O(\widehat{\varphi})|\psi_{\scriptscriptstyle 1}\rangle}
{\langle \psi_{\scriptscriptstyle 1}|\psi_{\scriptscriptstyle 1}\rangle}=
\frac{\Bigl(\alpha_{\scriptscriptstyle 1}^*\langle\varphi_{\scriptscriptstyle 1}|+\alpha_{\scriptscriptstyle 2}^*
\langle\varphi_{\scriptscriptstyle 2}|\Bigr)O(\widehat{\varphi})\Bigl(\alpha_{\scriptscriptstyle 1}|\varphi
_{\scriptscriptstyle 1}\rangle+\alpha_{\scriptscriptstyle 2}|\varphi_{\scriptscriptstyle 2}\rangle\Bigr)}
{\Bigl(\alpha_{\scriptscriptstyle 1}^*\langle\varphi_{\scriptscriptstyle 1}|+\alpha_{\scriptscriptstyle 2}^*
\langle\varphi_{\scriptscriptstyle 2}|\Bigr)\Bigl(\alpha_{\scriptscriptstyle 1}|\varphi
_{\scriptscriptstyle 1}\rangle+\alpha_{\scriptscriptstyle 2}|\varphi_{\scriptscriptstyle 2}\rangle\Bigr)}=\nonumber\\
&\hspace{-0.2cm}=&\hspace{-0.2cm}\frac{|\alpha_{\scriptscriptstyle 1}|^2\langle\varphi_{\scriptscriptstyle 1}|
O(\widehat{\varphi})|\varphi_{\scriptscriptstyle 1}\rangle+|\alpha_{\scriptscriptstyle 2}|^2
\langle\varphi_{\scriptscriptstyle 2}|O(\widehat{\varphi})|\varphi_{\scriptscriptstyle 2}\rangle}
{|\alpha_{\scriptscriptstyle 1}|^2\langle\varphi_{\scriptscriptstyle 1}|\varphi_{\scriptscriptstyle 1}\rangle
+|\alpha_{\scriptscriptstyle 2}|^2\langle\varphi_{\scriptscriptstyle 2}|\varphi_{\scriptscriptstyle 2}\rangle}. 
\label{gamma}
\end{eqnarray}
%%%
Therefore the expectation values that can be calculated using the vector $|\psi_{\scriptscriptstyle 1}\rangle$
are just the same as those calculated starting from the mixed density matrix
%%%
\begin{equation}
\widehat{\rho}=|\alpha_{\scriptscriptstyle 1}|^2|\varphi_{\scriptscriptstyle 1}\rangle\langle
\varphi_{\scriptscriptstyle 1}|
+|\alpha_{\scriptscriptstyle 2}|^2|\varphi_{\scriptscriptstyle 2}\rangle\langle\varphi_{\scriptscriptstyle 2}|
\label{beta}
\end{equation}
%%%
via the rule $\displaystyle
\overline{O}=\textrm{Tr}[\widehat{\rho}O(\widehat{\varphi})]/\textrm{Tr}[\widehat{\rho}]$.
So from a physical point of view the state (\ref{alpha}) cannot be distinguished from 
the mixed density matrix (\ref{beta}). This means that ``{\it coherent superpositions of pure states are impossible,
one automatically gets mixed states when attempting to form them}" \cite{lan}. 

All this can be
rephrased also in the following way: the relative phase between $|\varphi_{\scriptscriptstyle 1}\rangle$
and $|\varphi_{\scriptscriptstyle 2}\rangle$ cannot be measured using only observables like $O(\widehat{\varphi})$.
In fact to obtain the mean values of these observables the only important thing 
is the modulus square of $\alpha_{\scriptscriptstyle
1}$ and $\alpha_{\scriptscriptstyle 2}$ as it is clear from (\ref{gamma}) and (\ref{beta}).
These considerations can be extended very easily to the case of a continuous superposition of states 
$|\varphi\rangle$: the two vectors 
$|\psi\rangle=\int d\varphi\,\psi(\varphi)|\varphi\rangle$ and 
$|\widetilde{\psi}\rangle=\int d\varphi\,\psi(\varphi)e^{iA(\varphi)}|\varphi\rangle$
are physically indistinguishable because they give the same expectation values for all the observables 
of the theory. ``{\it But they are also completely different vectors in Hilbert space!$\,$}" \cite{mex}. 
If we want to avoid 
this redundancy we have to forbid the superposition of eigenstates of the superselection operators
and consider only the statistical mixtures (\ref{beta}) which, in the continuous case, become:
%%%
\begin{equation}
\displaystyle \widehat{\rho}=\int d\varphi\,\rho(\varphi)|\varphi\rangle\langle \varphi|. \label{delta}
\end{equation}
%%%
Therefore the Hilbert space must be considered as a direct sum (or, better to say, a direct integral)
%%%
\begin{equation}
\displaystyle {\mathbf{H}}=\oplus_{\scriptscriptstyle
\{\varphi_i\}}{\mathbf{H}}\bigl(\{\varphi_i\}\bigr)
\end{equation}
%%%
of the different eigenspaces ${\mathbf{H}}\bigl(\{\varphi_i\}\bigr)$ corresponding to the different 
eigenvalues $\varphi_i$ 
for the superselection observables
$\widehat{\varphi}$. These eigenspaces are incoherent, i.e. the relative phases between vectors
belonging to different eigenspaces cannot be measured at all; not only, but it is impossible to move from one
eigenspace to the other by means of an observable. All this has a very unpleasant 
consequence: in QM the superselection observables, like the parity or the charge operator, commute with
all the observables of the system and in particular with the Hamiltonian $\widehat{H}$, which is the generator 
of the time evolution. This implies that the eigenvalues of the superselection operators
are constants of the motion. So when we prepare the system in one particular eigenspace the time evolution
cannot bring the system outside it. This is fine if the superposition operator is the parity or the
charge because it only implies that all the states reached by the time evolution have the same parity
and the same charge. But this is catastrophic in the KvN formulation of CM. In fact there the superselection
operators are the $\widehat{\varphi}$ and so the eigenspaces of the superselection operators are in 1-1
correspondence with the points of the phase space ${\cal M}$; when we consider the time evolution of the system 
we pass from one point of the phase space to the other and, therefore, from one eigenspace of the superselection 
operators to the other. Therefore if the observables of CM are only the functions of 
$\widehat{\varphi}$ and the superselection mechanism is automatically triggered, then we have to admit 
that the time evolution of the system cannot be performed by an operator belonging to the observables of
the system.
We note that this is perfectly consistent with the fact that the generator of the evolution is
the Liouvillian which depends also on the variables $\widehat{\lambda}$ and not only on $\widehat{\varphi}$.

The possible ways out are basically two. We can insist on having as observables all and only the functions
$O(\widehat{\varphi})$ and on considering as physically significant only the statistical mixtures
(\ref{delta}). From the probability density $\rho(\varphi)$ we can construct its {\it real} square root
$\psi(\varphi)\equiv |\sqrt{\rho(\varphi)}|$ and use it to build in any case the following 
linear superposition of the eigenstates $|\varphi\rangle$:
%%%
\begin{equation}
\displaystyle |\psi\rangle=\int d\varphi\, \psi(\varphi)|\varphi\rangle. \label{epsilon}
\end{equation}
%%%
Now the coefficients of $|\varphi\rangle$ in the superposition (\ref{epsilon}) are just real functions of $\varphi$.
As a consequence there is a 1-1 correspondence between the statistical mixtures (\ref{delta}) and the 
vectors $|\psi\rangle$ of (\ref{epsilon}). Not only, let us construct the pure density matrix 
%%%
\begin{equation}
\widehat{\rho}^{\,\prime}=|\psi\rangle\langle \psi|=\int d\varphi d\varphi^{\prime}\,\psi(\varphi)\psi(\varphi^{\prime})
|\varphi\rangle\langle \varphi^{\prime}|. \label{delta2}
\end{equation}
%%%
Since the algebra of the observables $O(\widehat{\varphi})$ is Abelian we have that only the diagonal terms 
of $\widehat{\rho}^{\,\prime}$ contribute to the mean values of $O$. As a consequence it is easy to realize 
that the mean values of the observables calculated from the $\widehat{\rho}$ 
of (\ref{delta}) and the $\widehat{\rho}^{\,\prime}$ of (\ref{delta2}) are just the same:
%%%
\begin{equation}
\displaystyle \frac{\textrm{Tr}[\widehat{\rho}O(\widehat{\varphi})]}{\textrm{Tr}[\widehat{\rho}]}=
\frac{\textrm{Tr}[\widehat{\rho}^{\,\prime}
O(\widehat{\varphi})]}{\textrm{Tr}[\widehat{\rho}^{\,\prime}]}.
\end{equation}
%%%
So we can say that via the linear superposition (\ref{epsilon}) with real
coefficients we can perform the same physics as with the statistical mixture $\widehat{\rho}$ of
(\ref{delta}). Furthermore there is also a 1-1 correspondence between the real vectors $|\psi\rangle$
and the statistical mixtures $\widehat{\rho}$. So there is no redundancy in the description and we can 
look at the states $|\psi\rangle$ just as useful mathematical tools to perform all the calculations
and make all the predictions of statistical classical mechanics. Since the coefficients of the superposition
(\ref{epsilon}) are real we have that the operators depending on $\widehat{\varphi}$ cannot feel the phases  
just because there is no phase at all.

Another possible way out is to identify the observables with all the Hermitian operators of the theory.
With this assumption also
the operators $\widehat{\lambda}$ and their Hermitian functions become observables.
Now we cannot say anymore that $\widehat{\varphi}$ commute with all the observables
since $[\widehat{\varphi}^a,\widehat{\lambda}_b]=i\delta_b^a$. As a consequence the superselection mechanism is not triggered
and a coherent superposition, even with complex coefficients, 
of the eigenstates $|\varphi\rangle$ can be prepared by diagonalizing 
a complete set of commuting operators. This is the solution that also Koopman and von Neumann must have had in mind
in their original papers when they
postulated that the elements of their Hilbert space were square integrable and complex functions, like
in QM\footnote[5]{We note that the first paper on superselection rules made its appearance
only in 1952 \cite{wick}, more or less twenty years after KvN's papers.}.
This is also the solution that we have implicitly adopted in the first sections of this thesis. 
Such a theory is a {\it generalization}
of standard CM since it contains a much larger set of observables. It would be interesting
to understand the physical meaning of the extra observables depending on $\widehat{\lambda}$. 
One of these observables is the Liouvillian $\widehat{L}=\widehat{\lambda}_a\omega^{ab}\partial_bH$
and we know that its spectrum
(which, being $\widehat{L}$ an observable, could be measured) gives us information
on such properties as the {\it ergodicity} of the system. It is well-known in fact \cite{Arnold}
that if the zero eigenvalue of $\widehat{L}$ is non degenerate then the system is ergodic. Anyhow 
we could get the same information on the ergodicity of the system using only the Hamiltonian 
$H(\varphi)$ and the correlations of $\widehat{\varphi}$ at different times,
with no need of $\widehat{L}$. Other quantities involving the variables
$\widehat{\lambda}$ were studied in Ref. \cite{deker} where the correlation functions of $\widehat{\lambda}$
and $\widehat{\varphi}$ at different times were related with the so-called response functions 
\cite{Forster} of the system.
Also in this case, anyhow, the correlations between $\widehat{\lambda}$ and $\widehat{\varphi}$ could be calculated 
(via the fluctuation-dissipation theorem) using only correlations among the variables $\widehat{\varphi}$ 
at different times. This implies that the operators $\widehat{\varphi}$, being at different times, superselect 
different Hilbert eigenspaces. Nevertheless all these phenomena of CM can be explained via the incoherent 
superposition of states given by the mixed density 
matrix $\widehat{\rho}$ of (\ref{delta}). Anyhow 
in principle there could be phenomena which need coherent superpositions of states like the one given 
in (\ref{alpha}). These phenomena would measure the relative phases of such states by means
of Hermitian operators of $\widehat{\lambda}$
and $\widehat{\varphi}$. For sure these operators are not among those which have been studied in the literature
(like the Liouville operators or the $\widehat{\lambda}$-$\widehat{\varphi}$ correlations of the response functions). 
Of course there are many other operators which have not been considered and 
they may be of some importance when we explore that tricky 
region which is at the interface between classical and quantum mechanics. This region is described neither by CM nor by QM
and it may be the region where ``classical" coherence phenomena described by the state (\ref{alpha}) and by
observables of $\widehat{\lambda}$ and $\widehat{\varphi}$ emerge. That regime would then be the realm of the KvN theory.
This is just a hypothesis. To conclude this section 
we can say that the KvN theory is a generalization of CM once we admit
among the observables also the Hermitian operators containing $\widehat{\lambda}$.
It reduces to CM if we restrict the observables to be only the Hermitian operators made out of $\widehat{\varphi}$.
Using the arguments explained above via the superselection principle these last observables could never 
detect the relative phases contained in the state (\ref{alpha}) as we explained above. 

If the reader is not convinced of this 
we will work out in detail in the next section the classical analog of the two-slit experiment. 
In this case the observables that we will measure are only functions of $\widehat{\varphi}$ and we will observe
no interference at all.

\bigskip

\section{Two-Slit Experiment}

In QM phases play a crucial role in producing the interference effects which characterize
the two-slit experiment. The mystery of this kind of experiments is best summarized in these words of
Feynman \cite{Feynman}: {\it ``The question is, how does [the two-slit experiment] really work? What
machinery is actually producing this thing? Nobody knows any machinery. The mathematics can be made more precise;
you can mention that they are {\it complex} numbers, and a couple of other minor points which have nothing to 
do with the main idea. But the deep mystery is what I have described, and no one can go any deeper today"}.
As Feynman mentions in the lines above one could think that the interference effects are there because of the {\it complex}
nature of the wave functions. Then it is natural to check what happens in the classical KvN
case where we deal with a Hilbert space of complex, square integrable functions. We will actually show
that, despite the complex nature of these KvN waves, interference effects do not appear. This confirms,
as Feynman suspected, that the mystery of QM is deeper than that. 

If we want to describe a classical two-slit experiment we have to
deal with a problem in two dimensions. Let us call
$y$ the axis along which our beam propagates and
$x$ the orthogonal axis. We suppose that $y=0$ is the starting coordinate of our beam. 
The centers of the two slits $\Delta_1$ and $\Delta_2$ are placed respectively at $x_{\scriptscriptstyle A}$ and
$-x_{\scriptscriptstyle_A}$ on a first plate which has coordinate $y_{\scriptscriptstyle F}$ along the $y$ axis. The
final screen is placed at
$y_{\scriptscriptstyle S}$ like in the figure below: 
\bigskip

\begin{center}
\begin{picture}(248,80)
\put(0,20){\makebox(0,0){$\scriptstyle{-x_A}$}}
\put(0,60){\makebox(0,0){$\scriptstyle{x_A}$}}
\put(16,80){\line(0,-1){80}}
\put(16,20){\makebox(0,0){$\bullet$}}
\put(16,60){\makebox(0,0){$\bullet$}}
\put(24,0){\makebox(0,0){$\scriptstyle{0}$}}
\put(230,80){\line(0,-1){80}}
\put(123,80){\line(0,-1){18}}
\put(120,62){\line(1,0){6}}
\put(120,58){\line(1,0){6}}
\put(123,58){\line(0,-1){36}}
\put(120,22){\line(1,0){6}}
\put(120,18){\line(1,0){6}}
\put(123,18){\line(0,-1){18}}
\put(135,20){\makebox(0,0){$\scriptstyle{\Delta_2}$}}
\put(135,60){\makebox(0,0){$\scriptstyle{\Delta_1}$}}
\put(135,0){\makebox(0,0){$\scriptstyle{y_F}$}}
\put(242,0){\makebox(0,0){$\scriptstyle{y_S}$}}
\end{picture}
\end{center}

\bigskip
To simplify the problem we will make the assumption that the motion of the particles along the $y$
direction is known precisely. This means that at the initial time we know with absolute precision the position and the
momentum of the particles along that axis, for example $y(0)=0,\;p_y(0)=p_y^0$. With this prescription we are sure that the beam will 
arrive at 
the two slits after a time $t_{\scriptscriptstyle F}=y_{\scriptscriptstyle F}m/p_y^0$ and at the
final screen after a time 
$t_{\scriptscriptstyle S}=y_{\scriptscriptstyle S}m/p_y^0$. In this way we can concentrate
ourselves only on the behaviour of the particles along the $x$-axis and the two-slit experiment becomes a problem
in one dimension.
Let us consider, along $x$, a double Gaussian
with an arbitrary phase factor $G(x,p_x)$:
%%%
\begin{equation}
\displaystyle
\psi(x,p_x,t=0)=\frac{1}{\sqrt{\pi ab}}\textrm{exp}\biggl[-\frac{x^2}{2a^2}-\frac{p_x^2}{2b^2}+iG(x,p_x)\biggr]. \label{ijmpa.dunia}
\end{equation}
%%%
We assume $a$ and $b$ sufficiently large, i.e. the initial classical KvN wave sufficiently spread, 
in order to allow the beam to arrive at both slits. The evolution of the wave
function is via the free kernel of propagation (\ref{ijmpa.kernel}) up to the time 
$t_{\scriptscriptstyle F} =y_{\scriptscriptstyle F}m/p_y^0$, when the beam arrives at the first plate. 
The KvN wave at the time $t_{\scriptscriptstyle F}$ is given by:
%%%
\begin{equation}
\displaystyle
\psi(x,p_x,t_{\scriptscriptstyle F})=\frac{1}{\sqrt{\pi
ab}}\textrm{exp}\biggl[-\frac{1}{2a^2}\biggl(x-\frac{p_xy_{\scriptscriptstyle F}}{p_y^0}\biggr)^2-\frac{p_x^2}{2b^2}
+iG\biggl(x-\frac{p_xy_{\scriptscriptstyle F}}{p_y^0},p_x\biggr)\biggl]. \label{ijmpa.dunia2} 
\end{equation}
%%%
If the width of the two slits is $2\delta$ then 
the particles which at time $t_{\scriptscriptstyle F}$ are outside the
two intervals $\Delta_1=(x_{\scriptscriptstyle A}-\delta, x_{\scriptscriptstyle A}+\delta)$ and
$\Delta_2=(-x_{\scriptscriptstyle A}-\delta, -x_{\scriptscriptstyle A}+\delta)$ are absorbed by the first plate and they 
do not arrive at the final screen at all. 
Using Feynman's words again: ``{\it All particles which miss the slit[s] are captured and
removed from the experiment} \cite{Hibbs}". Therefore the KvN wave just after
$t_{\scriptscriptstyle F}$ can be rewritten, using a series of $\theta$-Heavyside functions, in the 
following compact way:
%%%
\begin{equation}
\displaystyle 
\psi(x,p_x,t_{\scriptscriptstyle F}+\epsilon)=N\,
\psi(x,p_x,t_{\scriptscriptstyle F})\,[C_1(x)+C_2(x)] \label{ijmpa.mar}
\end{equation}
%%%
where 
$C_1(x)=\theta(x-x_{\scriptscriptstyle A}+\delta)-\theta(x-x_{\scriptscriptstyle
A}-\delta)$ is the function that parameterizes the slit $\Delta_1$,
$C_2(x)=\theta(x+x_{\scriptscriptstyle A}+\delta)-\theta(x+x_{\scriptscriptstyle
A}-\delta)$ is the one that parameterizes the slit $\Delta_2$ and $N$ is a suitable normalization factor
chosen in such a way that:
$\displaystyle \int dxdp_x \,|\psi(x,p_x,t_{\scriptscriptstyle F}+\epsilon)|^2=1$. 
Beyond the slits we propagate 
the $\psi$ of (\ref{ijmpa.mar}). With our choice of the cut-off functions $C_1$ and $C_2$, 
at $t_{\scriptscriptstyle F}+\epsilon$ the KvN wave
$\psi$ is different from $0$ only  if $x\in\Delta_1$ or $x\in\Delta_2$.
Since there is no limitation in the momentum along the $x$-axis we expect that the $\psi$ will spread 
along $x$ and it will become different from zero also outside the intervals $\Delta_1$ and $\Delta_2$. 

Using the kernel of evolution for free particles (\ref{ijmpa.kernel}) 
we can obtain from (\ref{ijmpa.mar}) the KvN wave at time $t_{\scriptscriptstyle S}
=y_{\scriptscriptstyle S}m/p_y^0$, when the beam arrives at the final screen:
%%%
\begin{eqnarray}
\displaystyle
\psi(x,p_x,t_{\scriptscriptstyle S})&\hspace{-0.2cm}=&\hspace{-0.2cm}N \cdot
\textrm{exp}\biggl[-\frac{1}{2a^2}\biggl(x-\frac{p_xy_{\scriptscriptstyle
S}}{p_y^0}\biggr)^2-\frac{p_x^2}{2b^2}\biggl]\cdot \textrm{exp}\biggl[iG\biggl(x-\frac{p_xy_{\scriptscriptstyle S}}
{p_y^0},p_x\biggr)\biggr]\nonumber\\ \hspace{-0.2cm}&&\hspace{-0.2cm}\cdot
\{C_1(x-\bar{a}p_x)
+C_2(x-\bar{a}p_x)\} \label{ijmpa.marfin}
\end{eqnarray}
%%%
where $ \bar{a}=(y_{\scriptscriptstyle S}-y_{\scriptscriptstyle F})/p_y^0$.
The probability density of finding a particle in a certain point $x$ on the last screen is:
%%%
\begin{equation}
P(x)=\int_{-\infty}^{\infty}dp_x|\psi(x,p_x,t_{\scriptscriptstyle S})|^2. \label{ijmpa.pi}
\end{equation}
%%% 
We have to integrate over $p_x$ because we are interested in the number of
particles arriving at the final plate, independently of their momentum.
At this point we notice a first important property: even starting from an initial KvN wave
with an arbitrary phase factor $G(x,p_x)$, at time $t_{\scriptscriptstyle S}$ we have for the entire KvN wave
the following common phase factor:
$\displaystyle G\biggl(x-\frac{p_xy_{\scriptscriptstyle S}}{p_y^0},p_x\biggr)$, see (\ref{ijmpa.marfin}). 
So $G$ will disappear completely from the modulus square of the KvN wave and, consequently,
from the probability density $P(x)$ of (\ref{ijmpa.pi}). Therefore the phase $G$ of the initial wave 
function (\ref{ijmpa.dunia})
do not have any observable consequence for the figure on the final screen.

The second important thing to notice is that, because of the properties of the $\theta$-functions,
we have that the cut-off term $C_1+C_2$ in (\ref{ijmpa.marfin}) is idempotent:
%%%
\begin{equation}
\displaystyle
(C_1+C_2)^2=C_1+C_2. 
\end{equation}
%%%
Therefore we can rewrite (\ref{ijmpa.pi}) as:
%%%
\begin{eqnarray}
\label{ijmpa.rui}
P(x)&=&\int_{-\infty}^{\infty}dp_x|\psi(x,p_x,t_{\scriptscriptstyle S})|^2=\\
&=&N\int_{-\infty}^{\infty}dp_x\biggl[F^2(x,p_x,t_{\scriptscriptstyle S})C_1(x-\bar{a}p_x)
+F^2(x,p_x,t_{\scriptscriptstyle S})C_2(x-\bar{a}p_x)\biggr] \nonumber
\end{eqnarray}
%%%
where $N$ is a normalization factor and $F$ is given by:
%%%
\begin{equation}
\displaystyle
F(x,p_x, t_{\scriptscriptstyle S})\equiv \textrm{exp}\biggl[-\frac{1}{2a^2}\biggl(x-\frac{p_xy_{\scriptscriptstyle
S}}{p_y^0}\biggr)^2-\frac{p_x^2}{2b^2}\biggr]. \label{ijmpa.intcla}
\end{equation}
%%%

Let us now re-arrange the arguments inside the $\theta$-functions appearing in $C_1$ and  
$C_2$ as follows:
%%%
\begin{eqnarray}
\displaystyle
&&C_1(x-\bar{a}p_x)=\theta\biggl(-p_x+\frac{x-x_{\scriptscriptstyle
A}+\delta}{\bar{a}}\biggr)-\theta\biggl(-p_x+\frac{x-x_{\scriptscriptstyle
A}-\delta}{\bar{a}}\biggr)\nonumber\\
&&C_2(x-\bar{a}p_x)=\theta\biggl(-p_x+\frac{x+x_{\scriptscriptstyle
A}+\delta}{\bar{a}}\biggr)-\theta\biggl(-p_x+\frac{x+x_{\scriptscriptstyle
A}-\delta}{\bar{a}}\biggr).
\end{eqnarray} 
%%%
For the properties of the $\theta$-Heavyside functions it is easy to realize that when $p_x$ is not in 
the interval:
$\displaystyle D_1=\biggl[\frac{x-x_{\scriptscriptstyle A}-\delta}{\bar{a}}, \frac{x-x_{\scriptscriptstyle
A}+\delta}{\bar{a}}\biggr]$ or in
$\displaystyle D_2=\biggl[\frac{x+x_{\scriptscriptstyle A}-\delta}{\bar{a}}, \frac{x+x_{\scriptscriptstyle
A}+\delta}{\bar{a}}\biggr]$ there is no contribution to the modulus square. 
Therefore the final plot $P(x)$ given by (\ref{ijmpa.rui}) can be written as:
%%%
\begin{equation}
\displaystyle
P(x)=N\cdot\biggl[\int_{D_1} dp_x\,F^2(x,p_x,t_{\scriptscriptstyle S})+\int_{D_2}
dp_x\,F^2(x,p_x,t_{\scriptscriptstyle S})\biggr]
\label{ijmpa.imp}
\end{equation}
%%%
where $F$ is the function of (\ref{ijmpa.intcla}).
\vspace{-1cm}
\begin{center}
\begin{figure}
\includegraphics{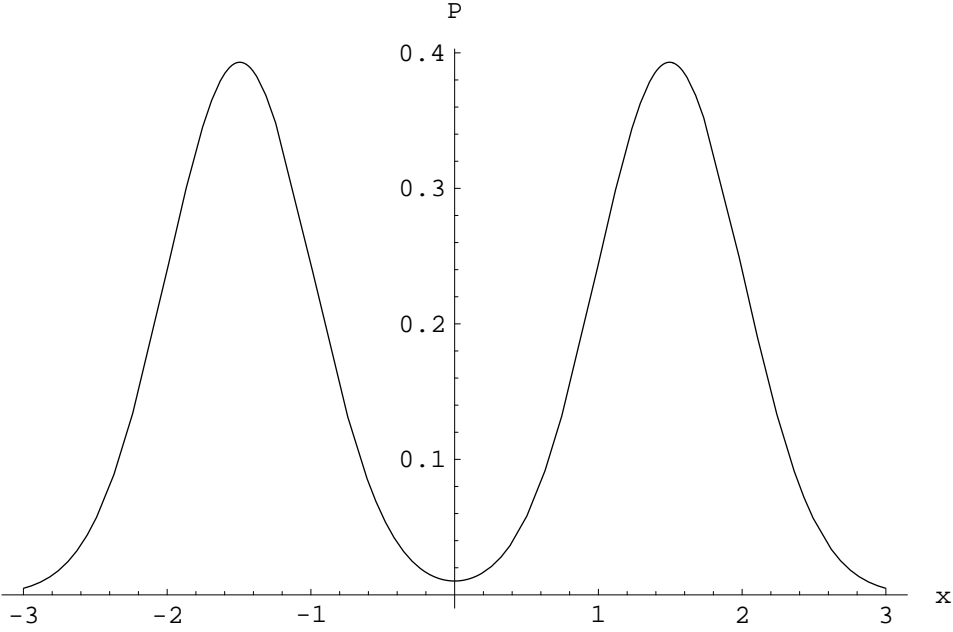}
\caption{\rm{Classical Two-Slit Experiment.}}
{\bf{\label{ijmpa.classical}}}
\end{figure}
\end{center}

Now let us keep open only the first slit $\Delta_1$ and repeat the previous steps. We can propagate
the initial KvN wave (\ref{ijmpa.dunia}) up to the time $t_{\scriptscriptstyle F}$ when the system is again described
by the $\psi(x,p_x,t_{\scriptscriptstyle F})$ of (\ref{ijmpa.dunia2}). The difference is that now only
the first slit $\Delta_1$ is open and so the second cut-off function $C_2$ is identically zero. $C_1$ itself 
is an idempotent function and therefore we can repeat the steps (\ref{ijmpa.mar})-(\ref{ijmpa.imp}),
as before, freezing $C_2$ to zero everywhere. The final result for the probability on the last screen is:
%%%
\begin{equation}
P(x)=K\int_{D_1}dp_xF^2(x,p_x,t_{\scriptscriptstyle S}) \label{ijmpa.imp1}
\end{equation}
%%%
where $F$ is given, as usual, by (\ref{ijmpa.intcla}) and $K$ is the normalization factor. 
In the same manner, keeping open only the slit $\Delta_2$,
we will obtain that:
%%%
\begin{equation}
P(x)=K\int_{D_2}dp_xF^2(x,p_x,t_{\scriptscriptstyle S}). \label{ijmpa.imp2}
\end{equation}
%%%
So, comparing (\ref{ijmpa.imp}) with (\ref{ijmpa.imp1})-(\ref{ijmpa.imp2}), it is clear that when we keep
open both slits $\Delta_1+\Delta_2$ {\it the total probability is the sum of the probabilities} we have
when we keep open first the slit $\Delta_1$ and then the slit $\Delta_2$. The first integral in (\ref{ijmpa.imp})
is then the probability for the particle to pass through the slit $\Delta_1$, while the second integral
is the probability to pass through the slit $\Delta_2$. 
So, even if we start from complex functions in 
the classical Hilbert space,
every interference effect disappears. This is very clear from
Fig. {\bf \ref{ijmpa.classical}} 
which shows the plot of the $P(x)$ of (\ref{ijmpa.imp}) with the particular numerical values
$\displaystyle y_{\scriptscriptstyle S}/p_y^0=2,\;a=b=1, x_{\scriptscriptstyle A}=1,\;\delta=0.1$.

Now we want to perform the same exercise at the quantum level and compare it with the previous 
classical experiment. In order to get an analytic result we will
assume that the motion along $y$ is the same classical motion analysed before. The reason for this
assumption is that otherwise we would not be able to determine the times $t_{\scriptscriptstyle F}$
and $t_{\scriptscriptstyle S}$
when the wave function arrives on the two plates.
Along $x$ instead we will assume that the motion is fully quantum mechanical. This will be sufficient to see 
the difference with the purely classical case we have analysed before and to create interference phenomena. 

In our approach we assume that at beginning the system 
along the $y$-axis is described by a double Dirac delta
$\delta(y)\delta(p_y-p_y^0)$ evolving in time with the Liouvillian. In this way we know 
that the time the particles arrive at the two slits is exactly the same as before. 
Along the other axis, $x$, we consider instead an initial wave function given by:
%%%
\begin{equation}
\displaystyle
\psi(x)=\sqrt{\frac{1}{\sqrt{\pi}a}}\textrm{exp}\biggl(-\frac{x^2}{2a^2}\biggr). \label{ijmpa.binn}
\end{equation}
%%%
With this choice at the beginning the mean value of both $x$ and $p_x$ is zero as in the classical case described by
(\ref{ijmpa.dunia}). Making the above wave function evolve in time via the quantum Schr\"odinger operator, at time
$t_{\scriptscriptstyle F}$ we obtain:
%%%
\begin{equation}
\displaystyle
\psi(x,t_{\scriptscriptstyle F})=\sqrt{\frac{ma}{\sqrt{\pi}(ma^2+i\hbar t_{\scriptscriptstyle F})}}
\textrm{exp}\biggl[-\frac{1}{2}\frac{mx^2}{ma^2+i\hbar t_{\scriptscriptstyle F}}\biggr].
\end{equation}
%%%
Let us parameterize the two slits via the same $\theta$-Heavyside functions 
we have used in the classical case:
%%%
\begin{equation}
C_1(x)=\theta(x-x_{\scriptscriptstyle A}+\delta)-\theta(x-x_{\scriptscriptstyle A}-\delta),\;\;\;\;
C_2(x)=\theta(x+x_{\scriptscriptstyle A}+\delta)-\theta(x+x_{\scriptscriptstyle A}-\delta).
\end{equation}
%%%
Just after the two slits we have that:
%%%
\begin{equation}
\displaystyle
\psi(x,t_{\scriptscriptstyle F}+\epsilon)=\overline{N}\cdot \textrm{exp}\biggl(-\frac{1}{2}\frac{mx^2}{ma^2+i\hbar t_{\scriptscriptstyle
F}}\biggr)\bigl[C_1(x)+C_2(x)\bigr].
\end{equation}
%%%
Using now the kernel of propagation for a quantum free particle \cite{Hibbs} which is given by:
%%%
\begin{equation}
\displaystyle
K(x_b,t_b|x_a,t_a)=\biggl[\frac{2\pi i\hbar(t_b-t_a)}{m}\biggr]^{-1/2}\textrm{exp}\;\frac{im(x_b-x_a)^2}{2\hbar(t_b-t_a)}
\end{equation}
%%%
we get that at time $t_{\scriptscriptstyle S}$ the wave function is:
%%%
\begin{eqnarray}
\displaystyle
&&\psi(x,t_{\scriptscriptstyle S})=\overline{N}_1\int_{-\infty}^{+\infty}dx_{\scriptscriptstyle F}
\,\textrm{exp}\biggl[\frac{im(x-x_{\scriptscriptstyle
F})^2}{2\hbar(t_ {\scriptscriptstyle S}-t_{\scriptscriptstyle F})}-\frac{mx_{\scriptscriptstyle F}^2}{2(ma^2+
i\hbar t_{\scriptscriptstyle F})}\biggr][C_1(x_{\scriptscriptstyle F})+C_2(x_{\scriptscriptstyle F})]\nonumber\\
\label{ijmpa.psifin}
\end{eqnarray}
%%%
where $\overline{N}_1$ is a new normalization constant.
Differently from the classical case, the quantum kernel of propagation is not a simple Dirac delta and
the previous integral cannot be done explicitly. Anyway we can employ the properties
of the $\theta$-functions in order to rewrite (\ref{ijmpa.psifin}) as:
%%%
\begin{eqnarray}
\displaystyle
\psi(x,t_{\scriptscriptstyle S})&=&\overline{N}_1\biggl\{\int_{x_{\scriptscriptstyle A}-\delta}^{x_{\scriptscriptstyle A}+\delta}
dx_{\scriptscriptstyle F}\,\textrm{exp} \biggl[\frac{im(x-x_{\scriptscriptstyle F})^2}
{2\hbar(t_{\scriptscriptstyle S}-t_{\scriptscriptstyle
F})} -\frac{mx_{\scriptscriptstyle F}^2}{2(ma^2+
i\hbar t_{\scriptscriptstyle F})}\biggr]+\nonumber\\
&&+\int_{-x_{\scriptscriptstyle A}-\delta}^{-x_{\scriptscriptstyle A}+\delta}
dx_{\scriptscriptstyle F}\,\textrm{exp}\biggl[\frac{im(x-x_{\scriptscriptstyle F})^2}
{2\hbar(t_{\scriptscriptstyle S}-t_{\scriptscriptstyle
F})} -\frac{mx_{\scriptscriptstyle F}^2}{2(ma^2+
i\hbar t_{\scriptscriptstyle F})}\biggr]\biggr\}. \label{ijmpa.previous}
\end{eqnarray}

\begin{figure}
\centering
\includegraphics[width=14cm]{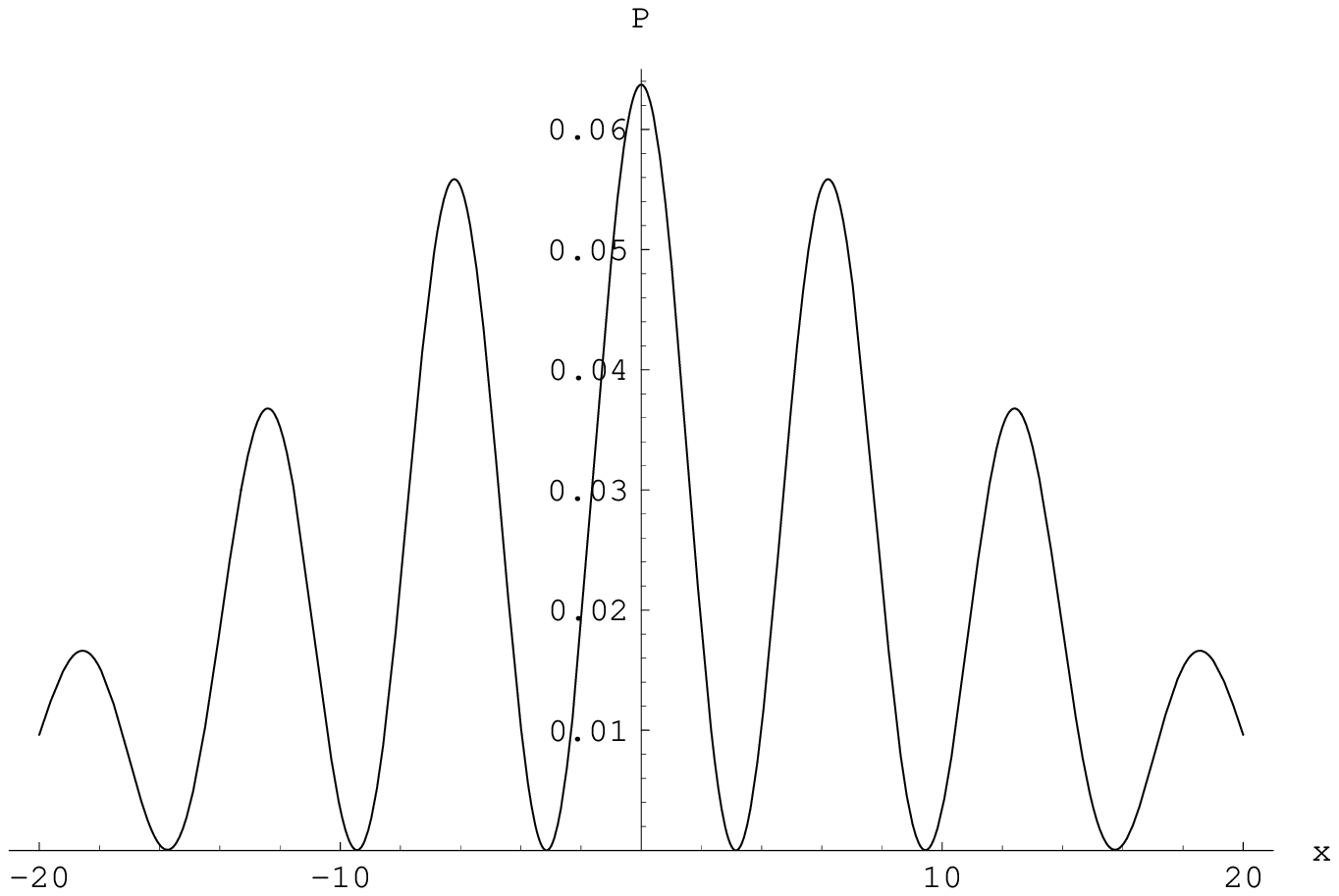}

\includegraphics[width=14cm]{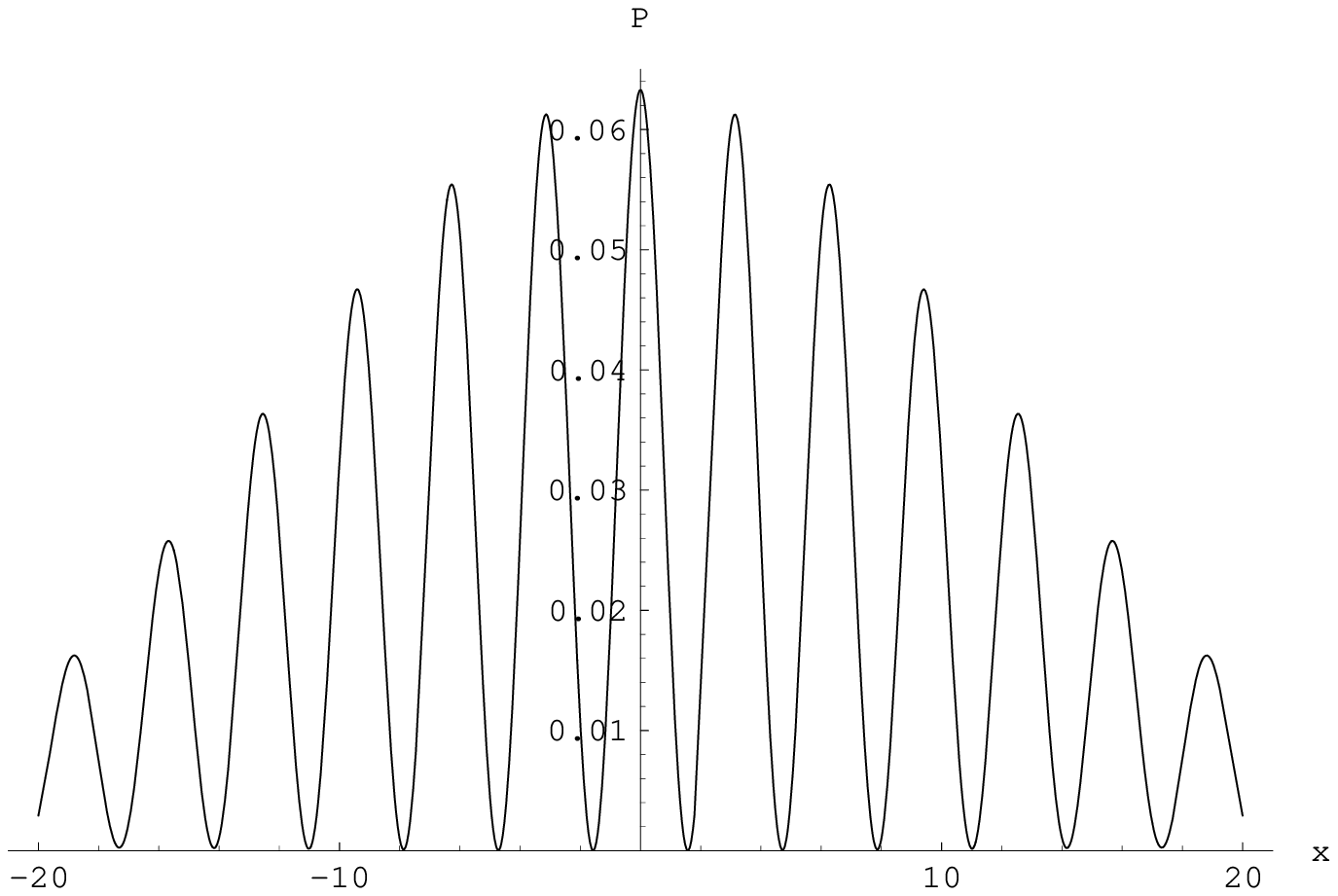}
\caption{\rm{Quantum Two-Slit Experiment.}} 
\label{ijmpa.quantum2}
\end{figure}

\noindent In (\ref{ijmpa.previous}) we have two integrals of the {\it same} function over two {\it different} intervals
$\Delta_1=(x_{\scriptscriptstyle A}-\delta,x_{\scriptscriptstyle A}+\delta)$ and 
$\Delta_2=(-x_{\scriptscriptstyle A}-\delta,-x_{\scriptscriptstyle A}+\delta)$. 
The results will be two complex numbers $\psi_1$ and $\psi_2$ with 
{\it different phases}. So, differently from the classical case (\ref{ijmpa.marfin}), the quantum wave function
on the final screen $\psi(x,t_{\scriptscriptstyle S})$ has not a common phase factor and so the 
relative phases of $\psi_1$ and $\psi_2$ will play a crucial role in giving interference effects. In fact if we
re-write the final wave function as:  
%%%
\begin{equation}
\psi(x, t_{\scriptscriptstyle S})=\overline{N}_1\,
[\psi_1(x, t_{\scriptscriptstyle S})+\psi_2(x, t_{\scriptscriptstyle S})]
\end{equation}
%%%. 
the probability on the last screen 
is given by the modulus square of $\psi(x, t_{\scriptscriptstyle S})$:
%%%
\begin{equation}
P(x, t_{\scriptscriptstyle S})=|\psi_1(x, t_{\scriptscriptstyle S})|^2+
|\psi_2(x, t_{\scriptscriptstyle S})|^2+\psi_1^*(x, t_{\scriptscriptstyle S})\psi_2(x, t_{\scriptscriptstyle S})
+\psi_1(x, t_{\scriptscriptstyle S})\psi_2^*(x, t_{\scriptscriptstyle S}).
\end{equation}
%%%
Note that the last two terms in the previous formula are not identically zero.
If we make a plot of $P(x, t_{\scriptscriptstyle S})$ as a function of $x$ we can
see the typical quantum figure of interference, with a central
maximum and a series of secondary maxima. This can be seen from 
Fig. {\bf \ref{ijmpa.quantum2}} 
which is the plot of $P(x,t_{\scriptscriptstyle S})$ in the case $t_{\scriptscriptstyle S}=2,
t_{\scriptscriptstyle F}=1$, $
m=a=1,\hbar=1,\delta=0.1$ for two different distances of the slits: 
$2x_{\scriptscriptstyle A}=1$ and $2x_{\scriptscriptstyle A}=2$
respectively. 
We can count six minima in the first case and twelve minima in the second one. This in perfect
agreement with the well-known relation that the distance $\Delta x$ between two successive maxima or minima in an
interference figure is inversely proportional to the distance $2x_{\scriptscriptstyle A}$ between the slits.
Therefore, even considering a quantum evolution only along the
$x$-axis, the quantum wave functions create interference effects and the final result reproduces the real experiment. 

Summarizing the results of this section we can say that if we make the evolution along the $x$ axis with 
the Schr\"odinger Hamiltonian $\widehat{H}$, even starting from a real wave function, 
like (\ref{ijmpa.binn}), phases will appear
during the evolution in a non trivial way and they will contribute to create interference effects.
Instead if we make the evolution along $x$ with the Liouvillian $\widehat{L}$, even starting from a complex KvN wave 
like (\ref{ijmpa.dunia}),
the phase will appear as a common factor for the entire $\psi$ on the final screen, so it will not 
contribute to $|\psi|^2$ and it will not have observable consequences.
So we can say that the two different behaviours in the classical and in the quantum case are basically due to 
the different forms of the evolution operators. Nevertheless, in order to get a better understanding of the role 
of the evolution and of the crucial differences between classical and QM in the two-slit experiment,
we want to further simplify our model. In particular we want to strip down the previous calculation of all the
mathematical details which are not necessary in explaining the presence or not of an interference
figure on the final screen.

\begin{figure}
\centering
\includegraphics[width=14cm]{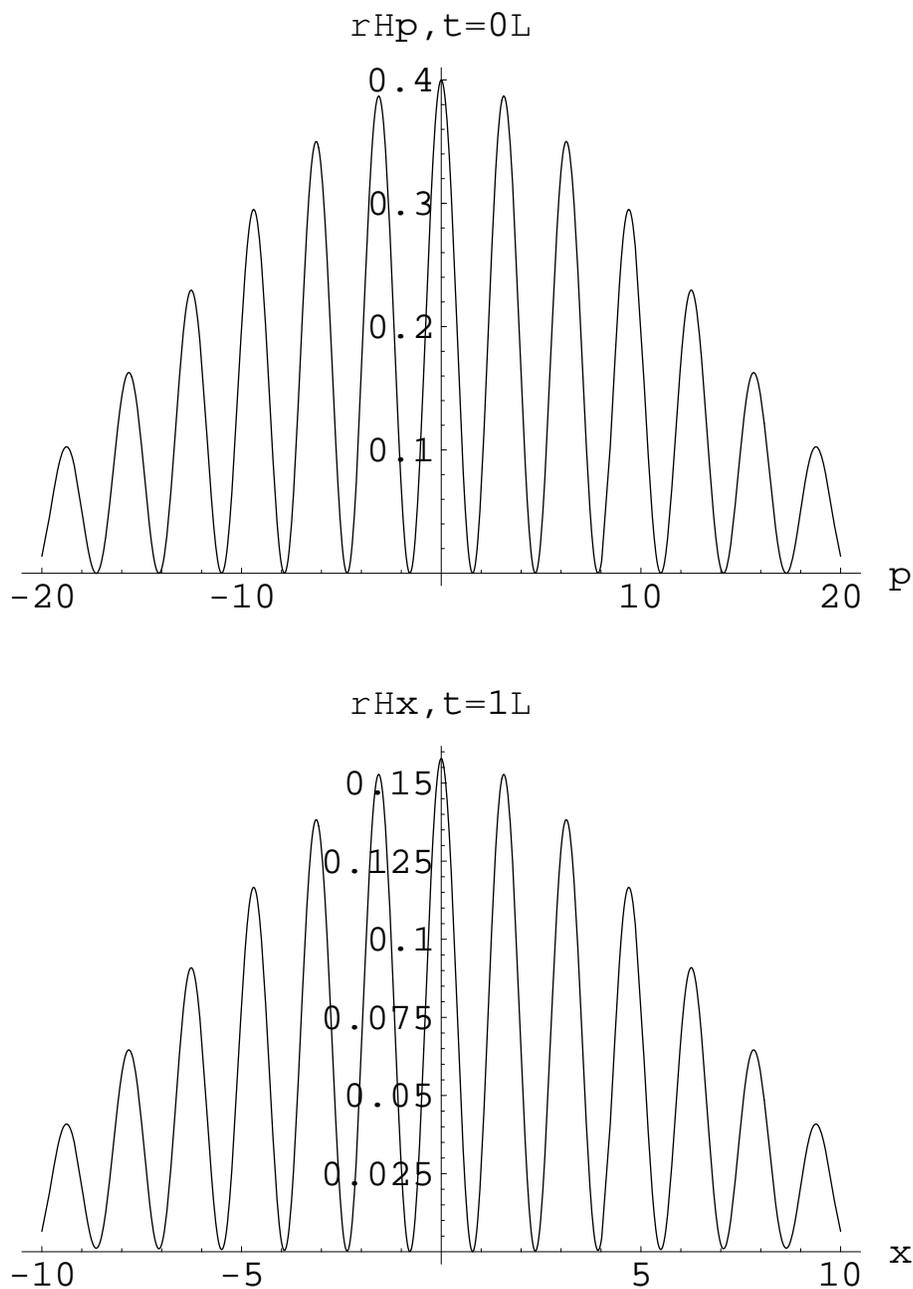}
\caption{{\rm{A simplified version of the two-slit experiment: the figure with maxima and minima
in $p$ at $t=0$ propagates at time $t=1$ to the distribution in $x$ creating an interference figure.}}}
\label{simple}
\end{figure}

So let us prepare a real initial quantum wave function reproducing the probability distribution 
of the particles near the two slits. For example let us assume a uniform distribution within the two slits:
%%%
\begin{equation}
\psi(x,t=0)=\left\{ 
\begin{array}{l}
\sqrt{\frac{5}{2}}\qquad -1.1\le x\le -0.9; \; 0.9\le x\le 1.1\medskip\\
0 \qquad\qquad \textrm{otherwise}
\end{array} \label{initialdis}
\right.  
\end{equation}
%%%
which implies
%%%
\begin{equation}
\rho(x,t=0)=\left\{
\begin{array}{l}
\frac{5}{2}\qquad -1.1\le x\le -0.9; \; 0.9\le x\le
1.1\medskip\\
0 \qquad\quad \textrm{otherwise}.
\end{array}
\right. 
\end{equation}
%%%
So at the beginning the distribution of particles along $x$ does not show any interference figure.
According to the rules of QM to obtain the probability distribution in the momentum space 
$p$ we have first to perform 
the Fourier transform of $\psi(x)$:
%%%
\begin{eqnarray}
\displaystyle \overline{\psi}(p,t=0)&\hspace{-0.2mm}=&\hspace{-0.2mm}
\frac{1}{\sqrt{2\pi\hbar}}\int_{-\infty}^{\infty}dx\,\psi(x,t=0) e^{-ipx/\hbar}=\nonumber\\
&\hspace{-0.2mm}=&\hspace{-0.2mm}\frac{1}{\sqrt{2\pi\hbar}}\int_{-1.1}^{-0.9}dx\,e^{-ipx/\hbar}
+\frac{1}{\sqrt{2\pi\hbar}}\int_{0.9}^{1.1}dx \,e^{-ipx/\hbar}. \label{gea}
\end{eqnarray}
%%%
In the RHS of (\ref{gea}) we have the same complex function, $\displaystyle e^{-ipx/\hbar}$, 
integrated over two different intervals.
This gives two complex functions with two different phases and, consequently, it will give 
a figure with maxima and minima in the
momenta distribution $\rho(p,t=0)=|\overline{\psi}(p,t=0)|^2$. 
Note that the momenta $p$ enter explicitly into the equations of motion of $x$ since
$\dot{x}=p/m$ and this implies that during the time evolution the figure with maxima and minima in $p$ is
inherited by the coordinates $x$ creating the well-known interference effects. 
This emerges very clearly from Fig.
{\bf \ref{simple}}, where we have put $\hbar=m=1$ and plot the probability distribution in $x$ at time $t=1$.
%%%
%%%

The situation is completely different in CM: in fact if
the initial KvN wave $\widetilde{\psi}(x,p)$, for the part in $x$, 
is given by (\ref{initialdis}), 
then the modulus square of its Fourier transform shows again a series of maxima and minima but 
now this modulus square is the probability density in $\lambda_x$. So it is the distribution 
$\rho(\lambda_x,t=0)$ which is identical to the $\rho(p,t=0)$ of Fig. {\bf
\ref{simple}}.
The crucial difference is that in CM the variables $\lambda_x$ do not enter the equations of $x$. Therefore
the figure with maxima and minima in $\lambda_x$ is not inherited by $x$ during the motion
and so it does not create interference 
effects on the final screen, like it happens in QM.
So we can say the following: the formal structure of KvN theory and of QM is the same. Just after 
the first screen we do not have an interference figure in $x$ but a distribution well-localized
near the two slits. Both in quantum and in CM the presence of the slits produces immediately
a figure with maxima and minima in the distribution of the variables conjugate to $x$ which are $p$ 
for QM 
and $\lambda_x$ for CM. Since only $p$ (and not $\lambda_x$) enters the equation
of motion of $x$ the figure with maxima and minima in the
conjugate variable is inherited by $x$ during the motion 
only in the quantum case creating interference effects. This cannot happen
in the classical case where there is no interference at all on the final plate.

\bigskip

\section{Measurements in Quantum and Classical Mechanics} 

Another aspect which is worth investigating in order to get a better
understanding of the differences and the similarities between CM and QM is 
related with the effect of measurements. We know in fact that in QM a measurement
disturbs the system and modifies the probabilities of the 
outcomes of subsequent measurements. This disturbance is present 
even if the measurement is a {\it non selective} one. 
We call non selective those measurements
which are explicitly performed but whose results are not read out.
As we will see later on in 
this case the system does not ``collapse" on a particular eigenstate 
but on a incoherent superposition of all its possible eigenstates. 
We want to begin this analysis with a very simple but pedagogical exercise \cite{Ghirardi}:

\subsection{Effect of non Selective Measurements in QM}

Let us consider a quantum mechanical system characterized by a Hermitian Hamiltonian
with eigenvalues $E_{\scriptscriptstyle 1}=\hbar\omega$; $E_{\scriptscriptstyle 2}=-\hbar\omega$
and eigenfunctions $|+\rangle$ and $|-\rangle$ respectively. Let us consider also an observable 
$\widehat{\Omega}$ with eigenvectors:
%%%
\begin{equation}
\displaystyle
|a\rangle=\frac{1}{\sqrt{2}}\Bigl[|+\rangle+|-\rangle\Bigr],\qquad\quad
|b\rangle=\frac{1}{\sqrt{2}}\Bigl[|+\rangle-|-\rangle\Bigr].
\end{equation}
%%%
Let us choose the initial state of the system to be the following one:
%%%
\begin{equation}
\displaystyle |\psi,0\rangle=\frac{1}{2}|+\rangle+\sqrt{\frac{3}{4}}|-\rangle.
\end{equation}
%%%

\begin{itemize}
\item[{\bf 1)}] Let us evolve the system up to time $t=2\tau$ and calculate the probability
of obtaining $a$ as result of a measurement of $\widehat{\Omega}$. The wave function at time $t=2\tau$
is given by:
%%%
\begin{equation}
\displaystyle |\psi,2\tau\rangle=e^{-\frac{i}{\hbar}\widehat{H}2\tau}\Biggl[\frac{1}{2}|+\rangle
+\sqrt{\frac{3}{4}}|-\rangle\Biggr]=\frac{1}{2}|+\rangle e^{-2i\omega\tau}+\sqrt{\frac{3}{4}}|-\rangle
e^{2i\omega\tau}. \label{purestate}
\end{equation}
%%%
So, when the system is described by the pure state (\ref{purestate}), 
the probability of finding $a$ as result of a measurement of $\widehat{\Omega}$ at time $t=2\tau$ is given by
\begin{equation}
P_{\scriptscriptstyle P}(\Omega=a|t=2\tau)=\Bigl| \langle \psi,2\tau|a\rangle\Bigr|^2=
\frac{1}{2}\Bigl(1+\sqrt{\frac{3}{4}}\textrm{cos}\,4\omega\tau
\Bigr). \label{vercinge}
\end{equation}
%%%
\item[{\bf 2)}] Let us now perform at time $t=\tau$ a {\it non selective} measurement (NSM)
of the observable $\widehat{\Omega}$.
Are the probabilities at time $t=2\tau$ influenced by the fact that we have measured $\widehat{\Omega}$
at time $t=\tau$? To answer this question let us consider what happens just after the measurement. Because 
of the postulate of the collapse of the wave function and because of the fact that in a NSM 
we do not read out the result of the measurement, 
the system will be described by a statistical mixture of the two
eigenstates $|a\rangle$ and $|b\rangle$ of $\widehat{\Omega}$ where the weights are just the probabilities 
$P_a(\tau)$ and $P_b(\tau)$ that the results $a$ and $b$ are obtained at time $t=\tau$:
%%%
\begin{equation}
\widehat{\rho}_{\scriptscriptstyle M}(\tau)=P_a(\tau)|a\rangle
\langle a|+P_b(\tau)|b\rangle\langle b|.
\end{equation}
%%%
The index $M$ on $\widehat{\rho}$ is for ``mixed" to indicate that $\widehat{\rho}_{\scriptscriptstyle M}$ 
is a mixed density matrix.
Now let the system evolve up to time 
$t=2\tau$. What we obtain is:
%%%
\begin{equation}
\widehat{\rho}_{\scriptscriptstyle M}(2\tau)=P_a(\tau)|\psi_a,2\tau\rangle
\langle \psi_a,2\tau|+P_b(\tau)|\psi_b,2\tau\rangle\langle \psi_b,2\tau|
\end{equation}
%%%
where $|\psi_a,2\tau\rangle$ and $|\psi_b,2\tau\rangle$ are given by the evolution of
the eigenstates $|a\rangle$ and $|b\rangle$ from $t=\tau$ to
$t=2\tau$:
%%%
\begin{eqnarray}
\displaystyle &&|\psi_a,2\tau\rangle=e^{-\frac{i}{\hbar}\widehat{H}\tau}|a\rangle=
\frac{1}{\sqrt{2}}\Bigl(e^{-i\omega\tau}|+\rangle+e^{i\omega\tau}|-\rangle\Bigr),\nonumber\\
\displaystyle &&|\psi_b,2\tau\rangle=e^{-\frac{i}{\hbar}\widehat{H}\tau}|b\rangle=
\frac{1}{\sqrt{2}}\Bigl(e^{-i\omega\tau}|+\rangle-e^{i\omega\tau}|-\rangle\Bigr).
\end{eqnarray}
%%%
Let us now calculate the probability of obtaining $\Omega=a$ as result of a further measurement 
of $\widehat{\Omega}$ at time $t=2\tau$. We have to calculate:
%%%
\begin{equation}
P_{\scriptscriptstyle M}(\Omega=a|t=2\tau)=\textrm{Tr}\Bigl[\widehat{\rho}_{\scriptscriptstyle M}(2\tau)|a\rangle
\langle a|\Bigr]
\end{equation}
%%%
whose result is given by 
%%%
\begin{equation}
\displaystyle P_{\scriptscriptstyle M}(\Omega=a|t=2\tau)=\frac{1}{2}\Bigl(1+\sqrt{\frac{3}{4}}\textrm{cos}^2
2\omega\tau\Bigr).
\end{equation}
%%%
Note that the result that we have obtained is different with respect to (\ref{vercinge}). 
\end{itemize}

So we can conclude this 
exercise by saying that, if a NSM 
of $\widehat{\Omega}$ is performed at time $t=\tau$ like in the case {\bf 2)},
then the probabilities of the outcomes 
of $\widehat{\Omega}$ itself at time $t=2\tau$ are {\it modified} with respect to the case {\bf 1)}
in which such a measurement is not performed. 
This can be also summarized in the following scheme:

\bigskip
\bigskip

\begin{center}
{\Large{$\boxed{\;\textrm{Initial Wave function} \; |\psi,0\rangle}\quad $}} 

{\large{
$\begin{array}{cc}
\quad \Big|  \qquad  & \qquad \qquad  \Big| \\
\quad \Big|  \qquad  & \qquad \qquad  \Big| \\
\quad \textrm{case} \;{\bf 1)} \qquad  & \qquad \qquad {\bf 2)} \; \textrm{NSM \;of\;} \widehat{\Omega} \\
\quad \Big|       \qquad  &  \qquad \qquad  
|\psi,\tau\rangle\rightarrow \widehat{\rho}_{\scriptscriptstyle M}(\tau)\\
\quad \Big|  \qquad  &  \qquad \qquad  \Big|\\
\quad \Big\downarrow \qquad & \qquad \qquad \Big\downarrow \medskip\\
\quad |\psi,2\tau\rangle \qquad & \qquad \qquad \widehat{\rho}_{\scriptscriptstyle M}(2\tau)
\end{array}$}}

\hspace{-0.5cm} {\large Outcome 1 for $\widehat{\Omega}$ \qquad $\neq$ \qquad Outcome 2 for $\widehat{\Omega}$}
\end{center}

\bigskip

\subsection{Non Selective Measurements of $\widehat{x}$ in QM}

In order to make a more direct comparison between classical and quantum mechanics in this subsection
we want to analyse what happens in QM when we perform at a
certain time a non selective measurement of a continuous operator like $\widehat{x}$. To be as clear as possible
let us distinguish the following cases:

\begin{itemize}
\item[{\bf a)}] the case in which we do not perform any measurement of $\widehat{x}$ at time $t=0$, we let the system
evolve and we calculate the probabilities of the outcomes of a measurement of $\widehat{x}$ at time $t=\tau$.

\item[{\bf b)}] the case in which we perform a NSM of $\widehat{x}$ at $t=0$ and then

\begin{itemize}
\item[{\bf b1)}] either we perform another measurement (of $\widehat{x}$ or $\widehat{p}$) 
just after the first one at 
$t=0_+$ (we will use the notation $0_+$ in order to indicate that the second measurement is performed 
again at $t=0$ but after the first one) 

\item[{\bf b2)}] or we let the system evolve from $t=0$ to a finite $t=\tau$ and
we calculate at that time the probabilities of the outcomes of a measurement of $\widehat{x}$.
\end{itemize}
\end{itemize}

\begin{itemize}
\item[{\bf a)}] Let us start with a quantum wave function $|\psi,0\rangle$; performing its evolution
up to time $t=\tau$, we obtain another state $|\psi,\tau\rangle$. Let us then ask ourselves: which is
the probability density of finding a particle between $x^{\prime}$ and $x^{\prime}+dx^{\prime}$ 
as result of a measurement of $\widehat{x}$
at time $t=\tau$? The answer is:
%%%
\begin{equation}
\rho_{\scriptscriptstyle P}(x^{\prime},\tau)=\textrm{Tr}\Bigl[|x^{\prime}\rangle\langle x^{\prime}
|\widehat{\rho}_{\scriptscriptstyle P}(\tau)\Bigr]
=|\psi(x^{\prime},\tau)|^2.
\end{equation}
%%%
The index $P$ stands for ``pure" in the sense that the matrix $\widehat{\rho}_{\scriptscriptstyle P}(\tau)=
|\psi,\tau\rangle\langle \psi,\tau|$ is a pure one.
For example if we consider the Gaussian wave function:
%%%
\begin{equation}
\displaystyle
\psi(x,0)=\frac{1}{\sqrt{\sqrt{\pi}a}}\textrm{exp}\biggl(-\frac{x^2}{2a^2}+\frac{i}{\hbar}p_ix\biggr). \label{init1}
\end{equation}
%%% at time $t=\tau$ we obtain that 
the probability density $\rho_{\scriptscriptstyle P}(x^{\prime},\tau)$ is given by the modulus square of 
$\psi(x^{\prime},\tau)$:
%%%
\begin{equation}
\displaystyle
\rho_{\scriptscriptstyle P}(x^{\prime},\tau)=
\textrm{exp}\biggl[-\frac{m^2a^2}{m^2a^4+\hbar^2
\tau^2}\biggl(x^{\prime}-\frac{p_i\tau}{m}\biggr)^2\biggr]. \label{bip}
\end{equation}
%%% 
i.e. another Gaussian with a well-defined mean value and a finite standard deviation.

\item[{\bf b)}] Let us now perform a {\it non selective} measurement (NSM)
of the quantum position $\widehat{x}$ at time $t=0$.
With the initial wave function $|\psi,0\rangle$ the probability of finding the system 
between $x_0$ and $x_0+dx_0$ as result of a measurement is given by 
%%%
\begin{equation}
P(x_0)dx_0=|\psi(x_0,0)|^2dx_0.
\end{equation}
%%%
Just after the measurement, if its result is not read out, 
the system is described by the statistical mixture 
%%%
\begin{equation}
\widehat{\rho}_{\scriptscriptstyle M}(0_+)=\int dx_0 |\psi(x_0,0)|^2|x_0\rangle\langle x_0| \label{sm}
\end{equation}
%%%
instead of the pure one:
%%%
\begin{equation}
\widehat{\rho}_{\scriptscriptstyle P}(0_-)=|\psi,0\rangle\langle\psi,0|
\end{equation}
%%%
which described the system just before the measurement. Since now on we will use the notation
$0_-$ when we want to indicate that we are at $t=0$ but before the measurement.
The question we want to answer is the following: has the NSM 
of $\widehat{x}$ changed the probability distributions 
of the observables? 

\item[{\bf b1)}] First of all, without taking into account any time evolution, let us suppose we perform 
a further measurement at $t=0_+$. 
If the observable we measure is again the position operator $\widehat{x}$, than  
the probability distributions of the outcomes are left unchanged.
In fact before the measurement done at $t=0$ we had the following 
distribution in $x$:
%%%
\begin{equation}
\rho_{\scriptscriptstyle P}(x,0_-)=
\textrm{Tr}\Bigl[|x\rangle\langle x|\widehat{\rho}_{\scriptscriptstyle P}(0_-)\Bigr]
=\psi^*(x,0)\psi(x,0). \label{measbef}
\end{equation}
%%%
Note that it remains the same also after the measurement at $t=0$:
%%%
\begin{equation}
\displaystyle \rho_{\scriptscriptstyle M}(x,0_+)=\frac{\textrm{Tr}\Bigl[|x\rangle\langle x|
\widehat{\rho}_{\scriptscriptstyle M}(0_+)\Bigr]}{\textrm{Tr}\Bigl[\widehat{\rho}_{\scriptscriptstyle M}
(0_+)\Bigr]}=\psi^*(x,0)\psi(x,0). \label{measaft}
\end{equation}
%%%
Let us now look at other observables different from $\widehat{x}$. 
The NSM of $\widehat{x}$ at $t=0$ modifies immediately the probability density
of finding a particular outcome in a measurement of $\widehat{p}$ for example at time $t=0_+$.
In fact before the NSM of $\widehat{x}$ at time $t=0$ the distribution in the momentum 
space was given by:
%%%
\begin{equation}
\rho_{\scriptscriptstyle P}(p,0_-)=
\textrm{Tr}\Bigl[|p\rangle\langle p|\widehat{\rho}_{\scriptscriptstyle P}(0_-)\Bigr]=
\overline{\psi}^*(p,0)\overline{\psi}(p,0). \label{before}
\end{equation}
%%%
So, for example, if we consider as initial wave function the Gaussian $\psi(x,0)$ of (\ref{init1}),
its Fourier transform $\overline{\psi}(p,0)$ will be another Gaussian with a well-defined mean value $p=p_i$ 
and a finite standard deviation. After the non selective measurement of $\widehat{x}$ at $t=0$ the distribution
probability in $p$ will be given by:
%%%
\begin{equation}
\displaystyle \rho_{\scriptscriptstyle M}(p,0_+)=
\frac{\textrm{Tr}\Bigl[|p\rangle\langle p|\widehat{\rho}_{\scriptscriptstyle M}(0_+)\Bigr]}
{\textrm{Tr} \Bigl[\widehat{\rho}_{\scriptscriptstyle M}(0_+)\Bigr]}\textrm{\;\;independent \,of \;}p \label{166}
\end{equation}
%%%
and, if we perform explicitly the calculation, it is easy to realize that this probability density is completely 
independent of the momentum $p$. This can be easily understood also without any kind of calculation:
in fact after the measurement at $t=0$ the system is described by a superposition of the eigenstates $|x_0\rangle$ 
like in (\ref{sm}) but, according 
to Heisenberg's uncertainty principle, every eigenstate of the position $|x_0\rangle$ corresponds to a 
completely uniform probability distribution in the momenta. So after the measurement at $t=0$ the overall 
distribution $\rho_{\scriptscriptstyle M}(p,0_+)$ of (\ref{166}) will be different than the one in
(\ref{before}): it will be a uniform one, without any mean value and with an infinite 
standard deviation. Therefore we can say that, without taking into account any time evolution, 
the NSM of $\widehat{x}$
at the initial time has instantaneously modified the probability distributions of the 
outcomes\footnote[6]{This phenomenon is the same as the previous quantum
experiment of the two slits which had created immediately after the screen a figure with maxima and minima in $p$.}
of the momenta $p$.
The probability outcomes of the positions, instead, are left completely unchanged by the NSM
of $\widehat{x}$. We can now ask ourselves whether this property is maintained or not during the evolution.

\item[{\bf b2)}] So, after the NSM of $\widehat{x}$ at time $t=0$, let us perform
the evolution of (\ref{sm}). At time $t=\tau$ the system is described by
the following statistical mixture:
%%%
\begin{equation}
\displaystyle \widehat{\rho}_{\scriptscriptstyle M}(\tau)=\int dx_0 |\psi(x_0, 0)|^2
|x_0,\tau\rangle\langle x_0,\tau|
\end{equation}
%%%
where $|x_0,\tau\rangle$ is given by the evolution of the eigenstate $|x_0\rangle$ up to time $t=\tau$.
If we use the Schr\"odinger representation we have that
%%%
\begin{equation}
\displaystyle \langle x|x_0,\tau\rangle=\int dx_i\,
\textrm{exp}\biggl[\frac{im}{\hbar\tau}(x-x_i)^2\biggr]\delta(x_i-x_0)=
\textrm{exp}\biggl[\frac{im}{\hbar\tau}(x-x_0)^2\biggr], \label{purephase}
\end{equation}
%%%
i.e. a pure phase factor.
If we calculate the probability density of obtaining $x^{\prime}$ as result of a measurement 
of the position $\widehat{x}$ at time $\tau$ we obtain:
%%%
\begin{equation}
\displaystyle \rho_{\scriptscriptstyle M}(x^{\prime},\tau)=\frac{\textrm{Tr}\Bigl[|x^{\prime}
\rangle\langle x^{\prime}|\widehat{\rho}_{\scriptscriptstyle M}(\tau)\Bigr]}{\textrm{Tr}\Bigl[
\widehat{\rho}_{\scriptscriptstyle M}(\tau)\Bigr]}=
\frac{1}{\int_{-\infty}^{\infty}dx}.
\end{equation}
%%%
Even from the not so well-defined expression above it is easy to understand that there is an equal probability
of finding the particle in any point of the configuration space. Therefore, differently than 
the $\rho_{\scriptscriptstyle P}(x^{\prime},\tau)$
obtained in the case {\bf a)} in (\ref{bip}),
the probability distribution $\rho_{\scriptscriptstyle M}(x^{\prime},\tau)$ is uniform.
So the NSM of the position $\widehat{x}$ at time $t=0$ 
not only has changed immediately the probability distribution of the momenta $p$, 
as we have analysed in {\bf b1)}, but it has changed also 
the probability distribution of $x$ at any time $t>0$. 
Now the reader may wonder why the probability distributions
$\rho_{\scriptscriptstyle P}(x,t)$ 
and $\rho_{\scriptscriptstyle M}(x,t)$  
were the same at $t=0$, see (\ref{measbef})-(\ref{measaft}),
but they are different at any time $t>0$. The explanation
is that during the evolution, which is given by $\dot{x}=p$ and couples $x$ with $p$,
the distributions in $x$ are influenced by the initial distributions in $p$ which, as shown in (\ref{before})
and (\ref{166}), are different in the two cases in which we perform (case {\bf b)}) or not perform (case {\bf a)})
the NSM of $\widehat{x}$ at $t=0$. Again the situation is similar to the one of the two-slit experiment: 
the NSM of $\widehat{x}$ influences
immediately the distribution of probability of the conjugate variable $p$. Next, since
the momenta $p$ enter explicitly
the equations of the positions $x$, the changes in the distributions of $p$ are inherited by all the 
distributions of the positions at any instant of time $t>0$. 
\end{itemize}

\subsection{Non Selective Measurements of $\widehat{\varphi}$ in CM}

Let us now analyse the case of a non selective measurement of $\widehat{\varphi}$ in KvN approach to CM. 
From a formal point of view the situation is the same as in QM with the
following substitutions: $x\longrightarrow \varphi, \;\; p\longrightarrow \lambda$. 
Therefore we can consider the same three case analysed before:

\begin{itemize}
\item[{\bf a)}] Let us consider an initial pure state 
%%%
\begin{equation}
\widehat{\rho}_{\scriptscriptstyle P}(0_-)=|\psi,0\rangle\langle\psi,0|.
\end{equation}
%%%
Then we let the system evolve up to time $t=\tau$ and at that time we perform a measurement of $\widehat{\varphi}$.
The probability density of the possible outcomes $\varphi^{\prime}$ is given by:
%%%
\begin{equation}
\rho_{\scriptscriptstyle P}(\varphi^{\prime},\tau)=\textrm{Tr}\Bigl[|\varphi^{\prime}\rangle\langle
\varphi^{\prime}|\widehat{\rho}_{\scriptscriptstyle P}(\tau)\Bigr]=|\psi(\varphi^{\prime},\tau)|^2.
\end{equation}
%%%
For example if we consider the Gaussian wave function (\ref{ijmpa.double}) we obtain that at time $t=\tau$
also $\rho_{\scriptscriptstyle P}(\varphi^{\prime},\tau)$ is a Gaussian with the following 
mean values and standard deviations:
%%%
\begin{equation}
\displaystyle
\bar{q}=\frac{p_i\tau}{m},\;\;\;\;\;\;\;\bar{p}=p_i,\;\;\;\;\;\;\;\overline{(\Delta
q(\tau))^2}=\frac{a^2}{2}+\frac{b^2}{2}\frac{\tau^2}{m^2},\;\;\;\;\;\;\; 
\overline{(\Delta p(\tau))^2}=\frac{b^2}{2}.
\end{equation}
%%%

\item[{\bf b)}] Let us now suppose instead that we perform a simultaneous measurement at $t=0$
of the positions and the
momenta, i.e. a measurement of $\widehat{\varphi}$, without reading the result. Like in QM also in the KvN 
theory there is the ``postulate" that after a measurement the system collapses in the eigenstate
associated to the eigenvalue that we get. In fact this is not only a typical feature of QM but it is intrinsic 
to any Hilbert space and operator formulation of a theory and to its probabilistic interpretation.
In particular since the measurement of $\widehat{\varphi}$ we performed at $t=0$
is non selective, just after the measurement
the system will be described by an incoherent superposition of the eigenstates $|\varphi_0\rangle$
where the weights are given by $|\psi(\varphi_0),0|^2$, which is the probability density 
associated to that particular outcome:
%%%
\begin{equation}
\displaystyle \widehat{\rho}_{\scriptscriptstyle M}(0_+)=\int d\varphi_0 |\psi(\varphi_0,0)|^2
|\varphi_0\rangle\langle\varphi_0|. \label{mixin}
\end{equation}

\item[{\bf b1)}] Now, as in QM, it is easy to prove that if, without taking into account the time evolution,
at time $t=0_+$ we perform another measurement of $\widehat{\varphi}$ 
the probability distributions of the outcomes
are left unchanged by the initial NSM since:
%%%
\begin{equation}
\frac{\textrm{Tr}\Bigl[|\varphi\rangle\langle\varphi|\widehat{\rho}_{\scriptscriptstyle M}(0_+)\Bigr]}
{\textrm{Tr}\Bigl[\widehat{\rho}_{\scriptscriptstyle M}(0_+)\Bigr]}
=\frac{\textrm{Tr}\Bigl[|\varphi\rangle\langle\varphi|\widehat{\rho}_{\scriptscriptstyle P}(0_-)\Bigr]}
{\textrm{Tr}\Bigl[\widehat{\rho}_{\scriptscriptstyle P}(0_-)\Bigr]}.
\end{equation}
%%%
The situation changes if, instead, we perform a measurement of the conjugate operator $\widehat{\lambda}$.
In this case, before the NSM of $\widehat{\varphi}$ at time $t=0$,
the probability distribution was:
%%%
\begin{equation}
\rho_{\scriptscriptstyle
P}(\lambda,0_-)=\frac{\textrm{Tr}\Bigl[|\lambda\rangle\langle
\lambda|\widehat{\rho}_{\scriptscriptstyle P}(0_-)\Bigr]}
{\textrm{Tr}\Bigl[\widehat{\rho}_{\scriptscriptstyle P}(0_-)\Bigr]}=
\overline{\psi}^*(\lambda,0)\overline{\psi}(\lambda,0).
\end{equation}
%%%
If $\psi(\varphi,0)$ were given by the double Gaussian of Eq. (\ref{ijmpa.double}) then it is easy to prove
that also $\rho_{\scriptscriptstyle P}(\lambda,0_-)$ 
would be a double Gaussian with well-defined mean values and finite 
standard deviations.
After the NSM of $\widehat{\varphi}$ the probability distribution in $\lambda$ is given 
by:
%%%
\begin{equation}
\displaystyle \rho_{\scriptscriptstyle
M}(\lambda,0_+)=\frac{\textrm{Tr}\Bigl[|\lambda\rangle\langle\lambda|\widehat{\rho}_{\scriptscriptstyle
M}(0_+)\Bigr]}{\textrm{Tr}\Bigl[\widehat{\rho}_{\scriptscriptstyle M}(0_+)\Bigr]}
\end{equation}
%%%
which again is a uniform probability distribution; 
this is consistent with the fact that a superposition of 
the eigenstates $|\varphi_0\rangle$ must correspond to a situation of total
ignorance for what concerns the variable $\lambda$. It is important to underline that
these considerations are completely independent of the kernel of evolution: the measurement of $\widehat{\varphi}$ 
changes immediately the probability distribution in the conjugate variables $\lambda$. 
The same phenomenon happens both in classical and in quantum 
mechanics, even if they have different kernel of evolutions:
we have only to replace $x$ with $\varphi$ and $p$ with $\lambda$ in going from QM to CM. 

\item[{\bf b2)}] What we want to prove now is that, differently from what happens in QM, the probability
distributions in $\varphi$ do not change (with respect to the case {\bf a)} 
in which no measurement was done at $t=0$) not only at $t=0_+$ but also at any later time $t>0$. 
Performing the evolution of the statistical mixture (\ref{mixin}), we get at time $t=\tau$:
%%%
\begin{equation}
\displaystyle \widehat{\rho}_{\scriptscriptstyle M}(\tau)=\int d\varphi_0|\psi(\varphi_0, 0)|^2
|\varphi_0,\tau\rangle\langle \varphi_0,\tau|.
\end{equation}
%%%
If the system we are considering is a free point particle then we have:
%%%
\begin{equation}
\displaystyle |\varphi_0,\tau\rangle=|q_0+p_0\tau/m,p_0\rangle.
\end{equation}
%%%
Once we represent the previous equation on $\langle\varphi|$ we obtain that in CM a Dirac delta in $\varphi$ remains 
a Dirac delta in $\varphi$ and it does not become a pure phase factor as in the quantum case (\ref{purephase}).
Consequently the probability density of finding $\varphi^{\prime}$ as result of a further measurement of
$\widehat{\varphi}$ at time $t=\tau$ is given by:
%%%
\begin{eqnarray}
\displaystyle \rho_{\scriptscriptstyle M}(\varphi^{\prime},\tau)&\hspace{-0.2cm}=&\hspace{-0.2cm}
\textrm{Tr}\Bigl[|\varphi^{\prime}\rangle\langle\varphi^{\prime}|\widehat{\rho}_{\scriptscriptstyle M}
(\tau)\Bigr]\Big/\textrm{Tr}\Bigl[\widehat{\rho}_{\scriptscriptstyle M}(\tau)\Bigr]=\nonumber\\
&\hspace{-0.2cm}=&\hspace{-0.2cm}\int d\varphi_0|\psi(\varphi_0,0)|^2\delta(q^{\prime}-q_0-
p_0\tau/m,p^{\prime}-p_0)=\nonumber\\
&\hspace{-0.2cm}=&\hspace{-0.2cm}|\psi(q^{\prime}-p^{\prime}\tau/m,p^{\prime},0)|^2.\label{1633}
\end{eqnarray}
%%%
The previous expression is just the modulus square of the evolution of the initial 
$\psi(\varphi^{\prime},0)$ up to time 
$t=\tau$. This tells us that, differently than in QM, the probability densities of the outcomes 
of $\widehat{\varphi}$ are left unchanged by the NSM of $\widehat{\varphi}$ 
not only at time $t=0_+$ but also at any later time $t>0$. 
In fact (\ref{1633}) is the same distribution as if the measurement at $t=0$ were not done. 
The reason for this is that in CM the equations of motion 
of $\varphi$ depend only on $\varphi$ and not on $\lambda$:
%%%
\begin{equation}
\displaystyle \dot{\varphi}^a=\omega^{ab}\frac{\partial H}{\partial \varphi^b}.
\end{equation}
%%%
Therefore at every time $t>0$ the
probability distributions in $\varphi$ 
will not feel $\lambda$ and, as a consequence, will not feel that the 
probability distributions in $\lambda$ were affected by the measurement at $t=0$. So 
we can say that the effect on $\lambda$ of the NSM of $\widehat{\varphi}$ is not
inherited by $\varphi$ both at $t=0$ and at $t>0$. 
\end{itemize}

What we have done in the last two subsections can be summarized in the following scheme:

\bigskip
\bigskip

\begin{center}

\noindent {\Large{\bf Quantum Non Selective Measurement of $\widehat{x}$:}}

\bigskip
\qquad {\Large{$\boxed{\;\textrm{Initial Quantum Wave Function}}\quad $}}\\

{\large $\begin{array}{cc}
 \big| \quad\quad & \quad\quad\quad \big| \\
\quad\quad \psi(x,0) \quad\quad   \quad\quad  &
\quad \qquad\textrm{NSM} \rightarrow\widehat{\rho}_{\scriptscriptstyle M}(0) \\
 \Big| \quad\quad & \quad\quad\quad  \Big| \\
 \Big| \quad\quad & \quad\quad\quad  \Big| \smallskip \\
 \textrm{case {\bf a)}} \quad\quad & \quad\quad\quad  \textrm{case {\bf b2)}}\\
 \Big| \quad\quad & \quad\quad\quad  \Big| \\
 \Big\downarrow \quad\quad & \quad\quad\quad \Big\downarrow \smallskip\\
 \psi(x,\tau) \quad\quad & \quad\quad\quad \widehat{\rho}_{\scriptscriptstyle M}(\tau)
\end{array}$}

 {\large \textrm{Outcome 1 for} $\widehat{x} \quad\quad\;\;\; \neq$\quad\quad \textrm{Outcome 2 for} $\widehat{x}$
}
\end{center}

\bigskip

\begin{center}

\noindent {\Large{\bf Classical Non Selective Measurement of $\widehat{\varphi}$:}}

\bigskip
\qquad {\Large{$\boxed{\;\textrm{Initial Koopman-von Neumann Function}}\quad $}}\\

{\large $\begin{array}{cc}
 \big| \quad\quad & \quad\quad\quad \big| \\
\quad\quad \psi(\varphi,0) \quad\quad   \quad\quad  &
\quad \qquad\textrm{NSM} \rightarrow\widehat{\rho}_{\scriptscriptstyle M}(0) \\
 \Big| \quad\quad & \quad\quad\quad  \Big| \\
 \Big| \quad\quad & \quad\quad\quad  \Big| \smallskip \\
 \textrm{case {\bf a)}} \quad\quad & \quad\quad\quad  \textrm{case {\bf b2)}}\\
 \Big| \quad\quad & \quad\quad\quad  \Big| \\
 \Big\downarrow \quad\quad & \quad\quad\quad \Big\downarrow \smallskip\\
 \psi(\varphi,\tau) \quad\quad & \quad\quad\quad \widehat{\rho}_{\scriptscriptstyle M}(\tau)
\end{array}$}

 {\large \textrm{Outcome 1 for} $\widehat{\varphi} \quad\quad\;\;\; =$\quad\quad \textrm{Outcome 2 for}
 $\widehat{\varphi}$
}
\end{center}

\bigskip

We can conclude this section by saying that NSM of $\widehat{\varphi}$ in CM {\it do not disturb}
the probability distributions of $\varphi$ either immediately after the measurement or after a long time 
differently than what happens in QM for $\widehat{x}$. All other features instead, like the disturbance on the conjugate
variables, are very similar both in CM and QM.

%% file: chapter2.tex
\def \HT{L}
\def \LT{{\mathcal L}}
\def \HCT{{\widehat{L}}}
\def \ET{{\mathcal E}}
\def \s{\scriptscriptstyle}

\pagestyle{fancy}
\begin{center}
\chapter*{\begin{center}2. Coupling with a Gauge Field
\end{center}}
\end{center}
\addcontentsline{toc}{chapter}{\numberline{2}Coupling with a Gauge Field}
\setcounter{chapter}{2}
\setcounter{section}{0}
\markboth{{\it{2. Coupling with a Gauge Field}}}{}

\begin{quote}
{\it{
In classical electrodynamics the vector and scalar potentials were first introduced as a convenient 
mathematical aid for calculating the fields. It is true that in order to obtain a classical
canonical formalism, the potentials are needed. Nevertheless, the fundamental equations of motion 
can always be expressed directly in terms of the fields alone. In the quantum mechanics, however, 
the canonical formalism is necessary, and as a result, the potentials cannot be eliminated from the
basic equations.}}\medskip\\
-{\bf Y. Aharonov} and {\bf D. Bohm}, 1959.
\end{quote}

\bigskip

\noindent In the previous chapter we have introduced the KvN operatorial approach to CM and studied
the similarities and differences with QM especially at the level of the role 
played by the phases. The phases actually play a role also in another context that is the one
where a gauge field is present. A typical example of the interplay between phases
and gauge fields is the Aharonov-Bohm phenomenon \cite{Bohm}.
In this chapter we will perform this analysis for the KvN formulation of CM. First we will show how to implement 
the analog of the minimal coupling rules at the KvN level, 
and we will analyse in which manner the associated gauge invariance
makes its appearance in the Hilbert space formulation of CM. As an application we will study 
the Landau problem showing that there are some extra degeneracies present in the classical KvN case 
with respect to the quantum one. Next we will construct 
the KvN analog of the Aharonov-Bohm (AB) set-up. While in QM the AB effect manifests itself 
on the spectrum of the Schr\"odinger Hamiltonian, we will prove that nothing similar appears 
in the spectrum of the Liouville operator in the classical KvN case. The work present in these pages is based
on paper \cite{2P} where the reader can find further details of the calculations.
\bigskip

\section{Minimal Coupling Rules for the Liouvillian $L$}

In QM there is a simple rule to couple a gauge field ${\bf A}$ with the point particle degrees
of freedom, the so-called ``{\it minimal coupling"} (MC) rule \cite{Sak}\cite{Coh}:
%%%
\begin{equation}
\displaystyle
{\bf p}\;\longrightarrow\;{\bf p}-\frac{e}{c}{\bf A}. \label{ann.mincoup}
\end{equation}
%%%
According to this rule, in order to get the interaction of the particle with the 
gauge field, it is enough to replace the momentum ${\bf p}$ with $\displaystyle
{\bf p}-\frac{e}{c}{\bf A}$ in the Hamiltonian
$\displaystyle H({\bf q},{\bf p})$  
and then replace ${\bf p}$ and ${\bf A}$ with the associated operators.
In this section we want to find out the
MC rules which transform the free Liouvillian $\HCT$ into the one
with a gauge field interaction $\HCT_{\s A}$.

Let us start by analysing a particle moving
under a constant magnetic field directed along $z$. The gauge field in this case can be chosen as:
%%%
\begin{equation}
\label{ann.Agal}\left\{
	\begin{array}{l}
	\displaystyle A_x=0\\
	\displaystyle A_y=Bx\\
	\displaystyle A_z=0\\
	\end{array}
	\right.
\end{equation}
%%%
where $B$ is the modulus of the magnetic field. 
Using the associated MC rule (\ref{ann.mincoup}):
%%%
\begin{equation}
\displaystyle
p_y\;\longrightarrow\;p_y-\frac{eB}{c}x \label{ann.pipsilon}
\end{equation}
%%%
the Hamiltonian becomes:
%%%
\begin{equation}
\displaystyle
H_{\s A}=\frac{p_x^2}{2m}+\frac{1}{2m}\biggl(p_y-\frac{eB}{c}x\biggr)^2+\frac{p_z^2}{2m}.
\end{equation}
%%%
Inserting this Hamiltonian into the abstract Liouvillian (\ref{ann.hambos}) we obtain:
%%%
\begin{eqnarray}
\displaystyle
\HT_{\s A}&\hspace{-0.2mm}=&\hspace{-0.2mm}
\frac{1}{m}\lambda_xp_x+\frac{1}{m}\lambda_y\biggl(p_y-\frac{eB}{c}x\biggr)+
\frac{1}{m}\lambda_zp_z-\lambda_{p_x}\frac{\partial H}{\partial x}=\nonumber\\
&\hspace{-0.2mm}=&\hspace{-0.2mm}\frac{1}{m}\lambda_xp_x+\frac{1}{m}\biggl(\lambda_y+\frac{eB}{c}\lambda_{p_x}\biggr)
\biggl(p_y-\frac{eB}{c}x\biggr)+\frac{1}{m}\lambda_zp_z.
\label{ann.lioint}
\end{eqnarray}
%%%
If we compare this Liouville operator with the free one which is:
%%%
\begin{equation}
\displaystyle
\HT=\frac{1}{m}\lambda_xp_x+\frac{1}{m}\lambda_yp_y+\frac{1}{m}\lambda_zp_z \label{ann.liofree}
\end{equation}
%%%
then we see that the substitutions to go from (\ref{ann.liofree}) to (\ref{ann.lioint}) are the following ones:
%%%
\begin{equation}
\label{ann.classmc2}
\left\{
	\begin{array}{l}
	\displaystyle p_y\;\longrightarrow\;p_y-\frac{eB}{c}x\smallskip\\
          \displaystyle \lambda_y\;\longrightarrow\;\lambda_y+\frac{eB}{c}\lambda_{p_x}.\\
	\end{array}
	\right.
\end{equation}
%%%
These are the MC rules for the Liouville operator in the case of a constant magnetic field.
They can be put in a compact form using the concept of superfield (\ref{ann.supb1})-(\ref{ann.supb2}). 
In fact let us take the MC rules in the standard phase space ${\cal M}$ for a constant magnetic field (\ref{ann.Agal}) 
and replace $\varphi$ with the superfields. 
This means the following:
%%%
\begin{eqnarray}
\displaystyle
p_y\;&\longrightarrow &\;p_y-\frac{e}{c}Bx \label{ann.seia}\\
\downarrow \;& &\;\;\;\;\;\;\;\;\downarrow\nonumber\\
\Phi^{p_y}\;&\longrightarrow &\Phi^{p_y}-\frac{e}{c}B\Phi^{x}. \label{ann.seib}
\end{eqnarray}
%%%
Expanding (\ref{ann.seib}) in $\theta,\bar{\theta}$ and using (\ref{ann.supb1})-(\ref{ann.supb2}) we get 
%%%
\begin{equation}
\displaystyle
p_y-i\bar{\theta}\theta\lambda_y \;\longrightarrow\; p_y-i\bar{\theta}\theta \lambda_y-\frac{eB}{c}(x+i\bar{\theta}
\theta\lambda_{p_x}). \label{ann.sette}
\end{equation}
%%%
If we compare the terms with an equal number of $\theta$ and $\bar{\theta}$ then  
(\ref{ann.sette}) reproduces exactly the substitution rules (\ref{ann.classmc2}) for the minimal coupling in the enlarged space $\widetilde{\cal M}$. So the
superfield formalism provides a compact way (\ref{ann.seib}) to write the double MC rules (\ref{ann.classmc2}). 

Let us now check if this compact way of writing MC rules via superfields is an accident of the case of a constant magnetic field or
if it holds in general. Via (\ref{ann.mincoup}) the Hamiltonian $H$ of a particle in interaction with a generic magnetic field is:
%%%
\begin{equation}
\displaystyle
H=\frac{1}{2m}\biggl\{\biggl(p_x-\frac{e}{c}A_x\biggr)^2+\biggl(p_y-\frac{e}{c}A_y\biggr)^2+
\biggl(p_z-\frac{e}{c}A_z\biggr)^2\biggr\}. \label{ann.hmagn}
\end{equation}
%%%
The associated abstract Liouvillian of (\ref{ann.hambos}) is then:
%%%
\begin{eqnarray}
\displaystyle
\HT_{\s A}&\hspace{-0.2mm}=&\hspace{-0.2mm}\frac{\lambda_x}{m}
\biggl(p_x\,-\frac{e}{c}A_{x}\biggr)+\frac{\lambda_y}{m}\biggl(p_y-\frac{e}{c}A_{y}\biggr)
+\frac{\lambda_z}{m}\biggl(p_z-\frac{e}{c}A_{z}\biggr)\nonumber\\
\hspace{-0.2mm}&&\hspace{-0.2mm}-\lambda_{p_x}\frac{\partial H}{\partial x}-
\lambda_{p_y}\frac{\partial H}{\partial y}-\lambda_{p_z}\frac{\partial H}{\partial z}=\nonumber\\
&\hspace{-0.2mm}=&\hspace{-0.2mm}\frac{1}{m}
\biggl(\lambda_x+\lambda_{p_x}\frac{e}{c}\frac{\partial A_x}{\partial x}+\lambda_{p_y}\frac{e}{c}\frac{\partial
A_x}{\partial y}+\lambda_{p_z}\frac{e}{c}\frac{\partial A_x}{\partial
z}\biggr)\biggl(p_x-\frac{e}{c}A_x\biggr) \\
\hspace{-0.2mm}&&\hspace{-0.2mm}+\frac{1}{m}
\biggl(\lambda_y+\lambda_{p_x}\frac{e}{c}\frac{\partial A_y}{\partial x}+\lambda_{p_y}\frac{e}{c}\frac{\partial
A_y}{\partial y}+\lambda_{p_z}\frac{e}{c}\frac{\partial A_y}{\partial
z}\biggr)\biggl(p_y-\frac{e}{c}A_y\biggr)\nonumber\\
\hspace{-0.2mm}&&\hspace{-0.2mm}+\frac{1}{m}
\biggl(\lambda_z+\lambda_{p_x}\frac{e}{c}\frac{\partial A_z}{\partial x}+\lambda_{p_y}\frac{e}{c}\frac{\partial
A_z}{\partial y}+\lambda_{p_z}\frac{e}{c}\frac{\partial A_z}{\partial
z}\biggr)\biggl(p_z-\frac{e}{c}A_z\biggr). \nonumber \label{ann.hbosmag}
\end{eqnarray}
%%%
If we define 
%%%
\begin{equation}
\displaystyle
{\mathcal A}_i\equiv-\sum_{\s j=\{x,y,z\}}\lambda_{p_j}\frac{\partial A_i}{\partial j}
\label{ann.tre}
\end{equation}
%%%
we can rewrite the Liouvillian in the following compact way:
%%%
\begin{equation}
\displaystyle
\HT_{\scriptscriptstyle A}=\frac{1}{m}\sum_{\s i=\{x,y,z\}}
\biggl(\lambda_{i}-\frac{e}{c}{\mathcal A}_{i}\biggr)\biggl(p_i-\frac{e}{c}A_{i}\biggr).
\label{ann.abstham}
\end{equation}
%%%
This last expression can be obtained from the Liouvillian $\HT$ of the free particle
%%%
\begin{equation}
\displaystyle
\HT=\frac{1}{m}\sum_{\s i=\{x,y,z\}}\lambda_ip_i
\end{equation}
%%%
via the substitutions:
%%%
\begin{equation}
\label{ann.general}
\displaystyle
\left\{
	\begin{array}{l}
	\displaystyle p_i\;\longrightarrow\;p_i-\frac{e}{c}A_i\smallskip\\
          \displaystyle \lambda_i\;\longrightarrow\;\lambda_i-\frac{e}{c}{\mathcal A}_i\\
	\end{array}
	\qquad i=\{x,y,z\}.
	\right.
\end{equation}
%%%
These are the MC rules for the Liouvillian $\HT$ in a generic magnetic field and 
they reproduce (\ref{ann.classmc2}) in the case of the constant magnetic field (\ref{ann.Agal}). 
We can note that
(\ref{ann.general}) can be derived from the superfield generalization of the 
standard MC rules in the phase space ${\cal M}$, i.e.:
%%%
\begin{eqnarray}
\displaystyle
p_i\;&\longrightarrow &\;p_i-\frac{e}{c}A_i(q) \label{ann.duea}\\
\downarrow \;& &\;\;\;\;\;\;\;\;\downarrow\nonumber\\
\Phi^{p_i}\;&\longrightarrow &\Phi^{p_i}-\frac{e}{c}A_i(\Phi^q).\label{ann.dueb}
\end{eqnarray}
%%%
In fact, 
expanding (\ref{ann.dueb}) in $\theta$ and ${\bar\theta}$, we get:
%%%
\begin{equation}
p_i-i\bar{\theta}\theta\lambda_i
\;\longrightarrow\;p_i-i\bar{\theta}\theta\lambda_i-\frac{e}{c}(A_i-i\bar{\theta}\theta
{\mathcal A}_i)
\end{equation}
%%%
and, comparing the terms with the same number of $\theta,\bar{\theta}$, we obtain
just the relations (\ref{ann.general}). So this proves that (\ref{ann.dueb}) 
is the most compact and general way to write the MC rules for the Liouvillian $\HT$.

\bigskip

\section{Gauge Invariance in KvN Approach}

In the previous section we have seen that in the MC rules for the Liouvillian also the $\lambda_i$ should be changed
when we turn on the magnetic field. To understand the reason of this we have to analyse
the issue of the gauge invariance
of the system. First of all we want to see what happens in the usual phase space $\cal M$.
Let us remember that the Lagrangian associated to the Hamiltonian $H$ of (\ref{ann.hmagn}) is:
%%%
\begin{equation}
{\mathscr L}=\frac{1}{2}m(\dot{x}^2+\dot{y}^2+\dot{z}^2)+\frac{e}{c}(\dot{x}A_x+\dot{y}A_y+\dot{z}A_z) \label{ann.lmagn}
\end{equation}
%%%
where
%%%
\begin{equation}
\displaystyle
\label{ann.velocities}
\left\{
	\begin{array}{l}
	\displaystyle \dot{x}=\frac{1}{m}\bigl(p_x-\frac{e}{c}A_x\bigr)\smallskip\\
          \displaystyle \dot{y}=\frac{1}{m}\bigl(p_y-\frac{e}{c}A_y\bigr)\smallskip\\
          \displaystyle \dot{z}=\frac{1}{m}\bigl(p_z-\frac{e}{c}A_z\bigr).\smallskip\\
	\end{array}
	\right.
\end{equation}
%%%
The velocities appearing above are measurable quantities and so they must be gauge invariant. Since 
$A_i$ transform under a gauge
transformation as $A_i^{\prime}=A_i+\partial_i\alpha(q)$ where $\alpha(q)$ is an arbitrary function
of $q$, the momenta $p_i$
must transform as
%%%
\begin{equation} 
p_i\;\longrightarrow\;p_i^{\prime}=p_i+\frac{e}{c}\partial_i\alpha(q). \label{ann.six}
\end{equation}
%%%
The Hamiltonian $H$ is a combination of gauge invariant quantities like $\displaystyle p_i-\frac{e}{c}A_i$ and so it 
is gauge invariant. Since $H$ is basically the energy of the system
its gauge invariance is crucial. 

Let us now go to the enlarged space $\widetilde{\cal M}$ and let us 
ask ourselves how the variables $\lambda$ should change under a gauge transformation. If we adopt the
same trick (\ref{ann.dueb}) we used to write the MC rules for the Liouvillian $\HT$, we have 
that the superfield analog of the gauge
transformations in ${\cal M}$ should be
%%%
\begin{equation}
\Phi^{p_i}\;\longrightarrow\;\Phi^{p_i}+\frac{e}{c}[\partial_i\alpha](\Phi^{q}) \label{ann.sevena}
\end{equation}
%%%
and
%%%
\begin{equation}
A_i(\Phi^q)\;\longrightarrow\;A_i(\Phi^q)+[\partial_i\alpha](\Phi^q). \label{ann.sevenb}
\end{equation}
%%%
Expanding (\ref{ann.sevena}) in $\theta,\bar{\theta}$ and equating the terms with the same number
of $\theta,\bar{\theta}$ we get (\ref{ann.six}) and the following gauge transformation for $\lambda_i$:
%%%
\begin{equation}
\lambda_i\;\longrightarrow\;\lambda_i^{\prime}=
\lambda_i-\frac{e}{c}\lambda_{p_j}\partial_{j}\partial_i\alpha(q). \label{ann.lamtra}
\end{equation}
%%%
Similarly from (\ref{ann.sevenb}) we get the usual transformation 
$A_i^{\prime}= A_i+\partial_i\alpha$ and the following one:
%%%
\begin{equation}
{\mathcal A}^{\prime}_i={\mathcal A}_i+\partial_i\widetilde{\alpha}(q,\lambda_p) \label{ann.artemisia}
\end{equation}
%%%
where  
%%%
\begin{equation}
\displaystyle \widetilde{\alpha}(q,\lambda_p)=-\sum_{\s j=\{x,y,z\}}\lambda_{p_j}\frac{\partial\alpha}{\partial j}.
\label{ann.aexp}
\end{equation}
%%% 
It is then easy to see that the combinations 
$\displaystyle \lambda_i-\frac{e}{c}{\mathcal A}_{i}$ which enter the Liouvillian (\ref{ann.abstham})
are gauge invariant if we gauge transform $\lambda_{i}$ as in (\ref{ann.lamtra}) and ${\mathcal A}_{i}$ as in 
(\ref{ann.artemisia}). 
Of course all this is very formal and it is a consequence
of the extension of the standard gauge transformations via the superfields. 
We can ask ourselves: while $H$, which is the energy, must be gauge invariant which is the 
reason why the Liouvillian $\HT_{\scriptscriptstyle A}$ should be gauge invariant?
Equivalently, while $p_i$ should change under a gauge transformation like in (\ref{ann.six}) in order to make
the velocities (\ref{ann.velocities}) gauge invariant, which is the reason why
$\lambda_{i}$ should change under a gauge transformation as in (\ref{ann.lamtra})? Actually there is a physical reason 
and it is  the
following. As the velocities (\ref{ann.velocities}) are gauge invariant then their evolution has to be gauge invariant too. 
The evolution can
occur via the Hamiltonian $H$ and the standard Poisson brackets
$\{\varphi^a,\varphi^b\}=\omega^{ab}$
or via the Liouvillian $\HT_{\scriptscriptstyle A}$ and the extended Poisson brackets (\ref{ann.epb}). 
For example the gauge invariant velocity
%%%
\begin{equation}
v_{x}=\frac{1}{m}\biggl(p_x-\frac{e}{c}A_{x}\biggr)
\end{equation}
%%%
evolves via the extended Poisson brackets according to the following equation: 
%%%
\begin{equation}
\displaystyle \dot{v}_{x}=\{v_{x},\HT_{\scriptscriptstyle A}\}_{epb}=\frac{e}{mc}(B_zv_y-B_yv_z).
\end{equation}
%%%
If we use a different gauge but we transform only the gauge fields $A_{i}$ and the momenta $p_i$, and not the variables 
$\lambda_{i}$, we get the following new Liouvillian $\HT_{\scriptscriptstyle A}$:
%%%
\begin{equation}
\displaystyle
\HT_{\scriptscriptstyle A}^{\prime}=\frac{1}{m}\sum_{\s i=\{x,y,z\}}
\biggl(\lambda_{i}-\frac{e}{c}{\mathcal A}_{i}^{\prime}\biggr)\biggl(p_i^{\prime}-
\frac{e}{c}A^{\prime}_{i}\biggr).
\end{equation}
%%%
The evolution
of the velocity $v_x$, via  the gauge transformed $\HT_{\scriptscriptstyle A}^{\prime}$, would turned out to be:
%%%
\begin{equation}
\displaystyle
\dot{v}_x=\{v_x,\HT_{\scriptscriptstyle A}^{\prime}\}_{epb}=\frac{e}{mc}(B_zv_y-B_yv_z)+\frac{e}{mc}[(\partial_x^2\alpha)
v_x+(\partial_y\partial_x\alpha)v_y+(\partial_z\partial_x\alpha)v_z]. \label{ann.dependence}
\end{equation}
%%%
So we notice that the evolution is not anymore gauge invariant because it depends on the gauge parameters $\alpha$ which
appear on the RHS of (\ref{ann.dependence}). This is absurd because the velocities are gauge invariant quantities and 
this gauge invariance has to be maintained by the evolution. 
The lack of gauge invariance in the evolution of the velocities is the price we would pay by  not allowing
$\lambda_{i}$ to change under gauge transformations or, equivalently, by not imposing on 
$\lambda$ the MC rules (\ref{ann.general}). 
  
Up to now in this chapter we have regarded the Liouvillian $\HT_{\scriptscriptstyle A}$ 
as a function which generates the evolution via some
suitable extended Poisson brackets. We want now to analyse the same issue of gauge invariance
when we turn $\HT_{\scriptscriptstyle A}$ into an operator
$\HCT_{\scriptscriptstyle A}$ like in (\ref{ann.operat}). Let us first briefly review what
happens in going from CM in the phase space to ordinary QM \cite{Sak}\cite{Coh}. 
In the standard CM the gauge transformations leave the positions ${\bf q}$ invariant 
but change the momenta ${\bf p}$ as 
%%%
\begin{equation}
p_i^{\prime}=p_i+\frac{e}{c}\partial_{i}\alpha(q). \label{ann.classgau}
\end{equation}
%%%
As a consequence the Poisson brackets are left invariant by these transformations:\break
$\{{\bf q},{\bf p}\}=\{{\bf q},{\bf p}^{\prime}\}=\mathbb{I}$.
%%%
So when we quantize the gauge transformed variables we obtain: 
%%%
\begin{equation}
\{{\bf q},{\bf p}^{\prime}\}=\mathbb{I}\;\longrightarrow\;
[\widehat{{\bf q}},\widehat{{\bf p}}^{\,\prime}]=i\hbar\widehat{\mathbb{I}}. \label{ann.brack}
\end{equation}
%%%
This implies that $\displaystyle \widehat{p}_j^{\,\prime}$ can be realized operatorially 
like the original momenta $\widehat{p}_j$, i.e. $\widehat{p}_j^{\,\prime}=-i\hbar\partial_j$. 
Therefore when we perform a gauge transformation we get that only the gauge fields $\widehat{A}_j$ 
have to be changed
inside the quantum Hamiltonian $\widehat{H}_{\scriptscriptstyle A}$:
%%%
\begin{equation}
\displaystyle 
\widehat{H}_{\scriptscriptstyle A}=\frac{1}{2m}\sum_{\s j=\{x,y,z\}}
\biggl(-i\hbar\frac{\partial}{\partial j}-\frac{e}{c}\widehat{A}_{j}\biggr)^2
\;\longrightarrow\; \widehat{H}_{\scriptscriptstyle A}^{\prime}=\frac{1}{2m}\sum_{\s j=\{x,y,z\}}
\biggl[-i\hbar\frac{\partial}{\partial
j}-\frac{e}{c}(\widehat{A}_{j}+\partial_{j}\widehat{\alpha})\biggr]^2.
\end{equation}
%%%
It is easy to check that one can pass from $\widehat{H}_{\scriptscriptstyle A}$ to 
$\widehat{H}_{\scriptscriptstyle A}^{\prime}$ via the following 
unitary transformation:
%%%
\begin{equation}
\widehat{H}_{\scriptscriptstyle A}^{\prime}=
U\widehat{H}_{\scriptscriptstyle A}U^{-1}, \qquad\qquad U=\textrm{exp}\biggl(i\frac{e}{c\hbar}\alpha(\widehat{q})\biggr). 
\label{ann.unit}
\end{equation}
%%%
So the quantum Hamiltonian $\widehat{H}_{\scriptscriptstyle A}$ is
not gauge invariant, but nevertheless the expectation values of all the measurable
quantities are gauge invariant. This is due to the fact that, if $\widehat{H}_{\scriptscriptstyle A}$
transforms as (\ref{ann.unit}), then also the states have to be changed according to:
%%%
\begin{equation}
|\psi^{\prime}\rangle=U|\psi\rangle\;\Longrightarrow\;
\psi^{\prime}(q)=\textrm{exp}\biggl(i\frac{e}{c\hbar}\alpha(q)\biggr)\psi(q)
\end{equation}
%%%
which is the usual gauge transformation by a phase. We can notice that the expectation values of 
$\langle\psi^{\prime}|\widehat{p}^{\,\prime}_i|\psi^{\prime}\rangle$ and $\langle\psi|\widehat{p}_i|\psi\rangle$ 
are related
exactly as the classical momenta in (\ref{ann.classgau}), i.e.:
%%%
\begin{equation}
\langle\psi^{\prime}|\widehat{p}_i^{\,\prime}|\psi^{\prime}\rangle=
\langle\psi|\widehat{p}_i|\psi\rangle+\frac{e}{c}\partial_{i}\alpha(q).
\end{equation}
%%%

Let us now turn to the KvN operatorial theory and check how the gauge transformations can be implemented. At the
operatorial level we have to construct everything in order to satisfy the gauge invariance of the following 
expectation values:
%%%
\begin{equation}
\langle\psi|\widehat{p}_i-\frac{e}{c}\widehat{A}_{i}|\psi\rangle,\qquad\qquad
\langle\psi|\widehat{\lambda}_{i}-\frac{e}{c}\widehat{{\mathcal A}}_{i}|\psi\rangle. \label{ann.gaugeinv}
\end{equation}
%%%
Let us start by noticing that the commutation
relations $[\widehat{\varphi}^a,\widehat{\lambda}_b]=i\delta^a_b$ are the operatorial counterpart of
the extended Poisson brackets $\{\varphi^a,\lambda_b\}_{epb}=\delta_b^a$ and the  gauge transformed
coordinates  $\varphi^{\prime a},\lambda_b^{\prime}$ given by (\ref{ann.six}) and (\ref{ann.lamtra}) have the same epb
$\{\varphi^{\prime a},\lambda^{\prime}_b\}_{epb}=\delta_b^a$ as the original variables. So we expect that 
also the associated commutators among the gauge transformed operators would be the same as the original ones:
$[\widehat{\varphi}^{\prime a},\widehat{\lambda}^{\prime}_b]=i\delta_b^a$.
This means that we can represent $\widehat{\varphi}^{\prime a}$ and 
$\widehat{\lambda}^{\prime}_b$ in the same manner as
$\widehat{\varphi}^a$ and $\widehat{\lambda}_b$. As a consequence  the gauge transformed expectation
values of (\ref{ann.gaugeinv}) are:
%%%
\begin{eqnarray}
&&\langle\psi^{\prime}|\widehat{p}_i-\frac{e}{c}\widehat{A}_{i}-\frac{e}{c}[\partial_{i}\alpha](\widehat{q})|\psi^{\prime}
\rangle, \quad\langle\psi^{\prime}|\widehat{\lambda}_{i}-\frac{e}{c}\widehat{{\mathcal A}}_{i}-
\frac{e}{c}[\partial_{i}
\widetilde{\alpha}](\widehat{q},\widehat{\lambda}_p)|\psi^{\prime}\rangle. \label{ann.exp}
\end{eqnarray}
%%%
Note that, via the introduction of the operator
%%%
\begin{equation}
\widetilde{U}=\textrm{exp}\bigg\{-i\frac{e}{c}\widehat{\lambda}_{p_j}[\partial_{j}\alpha](\widehat{q})\biggr\} 
\label{ann.uexp}
\end{equation}
%%%
we can write the following transformations:
%%%
\begin{equation}
\left\{
	\begin{array}{l}
\displaystyle  \widehat{p}_i-\frac{e}{c}\widehat{A}_{i}-\frac{e}{c}[\partial_{i}\alpha](\widehat{q})=\widetilde{U}
\biggl[\widehat{p}_i-\frac{e}{c}
\widehat{A}_{i}\biggr]\widetilde{U}^{-1}\\
\displaystyle \widehat{\lambda}_{i}-\frac{e}{c}\widehat{{\mathcal A}}_{i}-\frac{e}{c}[\partial_{i}
\widetilde{\alpha}](\widehat{q},\widehat{\lambda}_p)=
\widetilde{U}\biggl[\widehat{\lambda}_{i}-\frac{e}{c}\widehat{{\mathcal A}}_{i}\biggr]\widetilde{U}^{-1}.
	\end{array}
	\right.
\end{equation}
%%% 
This implies that (\ref{ann.exp}) are gauge invariant provided we transform also the states with the operator 
$\widetilde{U}$:
%%%
\begin{equation}
|\psi^{\prime}\rangle=\widetilde{U}|\psi\rangle=
\textrm{exp}\biggl\{-i\frac{e}{c}\widehat{\lambda}_{p_j}[\partial_{j}\alpha]
(\widehat{q})\biggr\}|\psi\rangle. \label{ann.abstract}
\end{equation}
%%%
Let us now represent this transformation law on the two basis given by (\ref{ann.diagmix0}) and 
(\ref{ijmpa.eq2}). 
In the $\langle q,p|$ basis we obtain:
%%%%
\begin{eqnarray}
\displaystyle
\psi^{\prime}(q,p)&\hspace{-0.2mm}=&\hspace{-0.2mm}\langle q,p|\psi^{\prime}\rangle=\int
dq^{\prime}d\lambda_p^{\prime}\langle
q,p|q^{\prime},\lambda_p^{\prime}\rangle\langle q^{\prime},\lambda_p^{\prime}|\widetilde{U}|\psi\rangle \nonumber\\
\hspace{-0.2mm}&=&\hspace{-0.2mm}\int
\frac{d\lambda_p^{\prime}}{\sqrt{2\pi}}\textrm{exp}\biggl[ip_j\lambda_{p_j}^{\prime}+
i\frac{e}{c}\widetilde{\alpha}
(q,\lambda_p^{\prime})\biggr]\langle
q,\lambda_p^{\prime}|\psi\rangle.
\end{eqnarray}
%%%
where in the last step we have used (\ref{ijmpa.fiftysix}).
Inserting a further completeness relation we get:
%%%
\begin{eqnarray}
\displaystyle
\psi^{\prime}(q,p)\hspace{-0.2mm}&=&\hspace{-0.2mm}\int
\frac{d\lambda_p^{\prime}dq^{\prime}dp^{\prime}}{\sqrt{2\pi}}
\textrm{exp}\biggl[ip_j\lambda_{p_j}^{\prime}+
i\frac{e}{c}\widetilde{\alpha}(q,\lambda_p^{\prime})
\biggr]
\langle q,\lambda_p^{\prime}|q^{\prime},p^{\prime}\rangle\langle q^{\prime},p^{\prime}|\psi\rangle=\nonumber\\
\hspace{-0.2mm}&=&\hspace{-0.2mm}\int
\frac{d\lambda_{p}^{\prime}dp^{\prime}}{2\pi}\textrm{exp}
\biggl[i\lambda_{p_j}^{\prime}\bigl(p_j-p_j^{\prime}-\frac{e}{c}\partial_j\alpha\bigr)
\biggr]\psi(q,p^{\prime})=\nonumber\\ \hspace{-0.2mm}&=&\hspace{-0.2mm}\int
dp^{\prime}\delta\biggl(p-p^{\prime}-\frac{e}{c}\nabla\alpha\biggr)\psi(q,p^{\prime})=
\psi\biggl(q,p-\frac{e}{c}\nabla\alpha\biggr).
\end{eqnarray}
%%%
So in the $(q,p)$ representation of the KvN Hilbert space 
the gauge transformations are implemented by just a shift in the argument $p$ of the wave function. 

Now let us represent (\ref{ann.abstract}) in the $\langle q,\lambda_p|$
basis. We have:
%%%
\begin{eqnarray}
\psi^{\prime}(q,\lambda_p)\hspace{-0.2mm}&\equiv&\hspace{-0.2mm}\langle q,\lambda_p|\psi^{\prime}\rangle=\langle
q,\lambda_p|\textrm{exp}\biggl\{-i\frac{e}{c}\widehat{\lambda}_{p_j}[\partial_{j}\alpha](\widehat{q})\biggr\}|\psi\rangle=
\nonumber\\
\hspace{-0.2mm}&=&\hspace{-0.2mm}\textrm{exp}\biggl\{-i\frac{e}{c}\lambda_{p_j}\partial_{j}\alpha(q)\biggr\}\langle
q,\lambda_p|\psi\rangle=\nonumber\\
\hspace{-0.2mm}&=&\hspace{-0.2mm}\textrm{exp}\biggl(i\frac{e}{c}\widetilde{\alpha}\biggr)\psi(q,\lambda_p).
\label{ann.sixtysix}
\end{eqnarray}
%%%
So in this basis the gauge transformation is just the multiplication by the {\it local phase factor} 
$\widetilde{\alpha}(q,\lambda_p)$ defined in (\ref{ann.aexp}).

To end this section we want to prove that also the Liouville equation
$\displaystyle i\frac{d}{dt}|\psi,t\rangle=\HCT_{\scriptscriptstyle A}|\psi,t\rangle$
is invariant under the gauge transformations. 
As we have seen in (\ref{ann.abstract}) the gauge transformation on the ket is:
%%% 
\begin{equation}
|\psi^{\prime},t\rangle=\widetilde{U}|\psi,t\rangle
\end{equation}
%%%
where $\widetilde{U}$ is given by (\ref{ann.uexp}). To prove the gauge invariance of the Liouville equation 
we have to prove that the state $|\psi^{\prime},t\rangle$ satisfies the following equation:
%%%
\begin{equation}
i\frac{d}{dt}|\psi^{\prime},t\rangle=\widehat{\HT}_{\scriptscriptstyle A}
^{\prime}|\psi^{\prime},t\rangle. \label{ann.btre}
\end{equation}
%%%
$\widehat{\HT}_{\scriptscriptstyle A}^{\prime}$ is the gauge transformed Liouvillian:
%%%
\begin{equation}
\widehat{\HT}_{\scriptscriptstyle A}^{\prime}=\frac{1}{m}\biggl(\widehat{\lambda}_{i}+\frac{e}{c}\widehat{\lambda}_{p_k}
\partial_{k}A^{\prime}_{i}(\widehat{q})\biggr)\biggl(\widehat{p}_i-\frac{e}{c}A^{\prime}_{i}
(\widehat{q})\biggr)-e\widehat{\lambda}_{p_i}\partial_{i}\phi^{\prime}(\widehat{q})
\label{ann.gaugeinvlio}
\end{equation}
%%%
where we have introduced also the gauge transformed
scalar potential $\displaystyle \phi^{\prime}=\phi-\frac{1}{c}\frac{\partial\alpha}{\partial t}$.
Let us evaluate the LHS of (\ref{ann.btre}):
%%%
\begin{eqnarray}
\displaystyle
\label{ann.bcinque}
i\frac{d}{dt}|\psi^{\prime},t\rangle&\hspace{-0.2mm}=&\hspace{-0.2mm}
i\frac{d}{dt}\Bigl[\widetilde{U}(t)|\psi,t\rangle\Bigr]=i\biggl[\frac{d}{dt}\widetilde{U}(t)
\biggr]|\psi,t\rangle+i\widetilde{U}(t)\frac{d}{dt}|\psi,t\rangle\\\
&\hspace{-0.2mm}=&\hspace{-0.2mm}-\frac{e}{c}\frac{\partial\widetilde{\alpha}}{\partial
t}\widetilde{U}|\psi,t\rangle
+\widetilde{U}\HCT_{\scriptscriptstyle A}|\psi,t\rangle =-\frac{e}{c}\frac{\partial\widetilde{\alpha}}{\partial
t}|\psi^{\prime},t\rangle+\widetilde{U}\HCT_{\scriptscriptstyle A}\widetilde{U}^{-1} |\psi^{\prime},t\rangle. \nonumber
\end{eqnarray}
%%%
The explicit expression for $\widetilde{U}\HCT_{\scriptscriptstyle A}\widetilde{U}^{-1}$ is:
%%%
\begin{equation}
\widetilde{U}\HCT_{\scriptscriptstyle A}\widetilde{U}^{-1}=\frac{1}{m}
\biggl(\widehat{\lambda}_{i}^{\prime}+\frac{e}{c}\widehat{\lambda}_{p_k}^{\prime}\partial_{k}A_{i}
(\widehat{q}^{\prime})\biggr)\biggl(\widehat{p}_i^{\prime}-\frac{e}{c}A_{i}(\widehat{q}^{\prime})\biggr)
-e\widehat{\lambda}_{p_i}^{\prime}\partial_{i}
\Phi(\widehat{q}^{\prime})
\label{ann.tildetilde}
\end{equation}
%%%
where
%%%
\begin{equation}
\left\{
	\begin{array}{l}
	\displaystyle 
	\widehat{\lambda}_{i}^{\prime}=\widetilde{U}\widehat{\lambda}_{i}\widetilde{U}^{-1}=
	\widehat{\lambda}_{i}+\frac{e}{c}\widehat{\lambda}_{p_k}
          \partial_{k}\partial_{i}\alpha(\widehat{q})\smallskip\\
          \displaystyle \widehat{p}_i^{\prime}=\widetilde{U}\widehat{p}_i\widetilde{U}^{-1}=
          \widehat{p}_i-\frac{e}{c}\partial_{i}\alpha(\widehat{q})\smallskip\\
          \displaystyle \widehat{q}^{\prime}=\widetilde{U}\widehat{q}\widetilde{U}^{-1}=\widehat{q}\smallskip\\
          \displaystyle \widehat{\lambda}_{p_i}^{\prime}=\widetilde{U}\widehat{\lambda}_{p_i}
          \widetilde{U}^{-1}=\widehat{\lambda}_{p_i}.\\
	\end{array}
	\right.
\end{equation}
%%% 
Remembering that 
%%%
\begin{equation}
\left\{
	\begin{array}{l}
	\displaystyle A^{\prime}_{i}=A_{i}+\partial_{i}\alpha\smallskip\\
          \displaystyle \phi^{\prime}=\phi-\frac{1}{c}\frac{\partial\alpha}{\partial t}\\
          \end{array}
	\right.
\end{equation}
%%%
we can rewrite (\ref{ann.tildetilde}) as
%%%
\begin{eqnarray}
\widetilde{U}\HCT_{\scriptscriptstyle A}\widetilde{U}^{-1}\hspace{-0.2mm}&=&\hspace{-0.2mm}\frac{1}{m}
\Bigl(\widehat{\lambda}_{i}+\frac{e}{c}\widehat{\lambda}_{p_k}\partial_{k}\partial_{i}\alpha(\widehat{q})
+\frac{e}{c}\widehat{\lambda}_{p_k}\partial_{k}A_{i}(\widehat{q})\Bigr)
\cdot\nonumber\\\hspace{-0.2mm}&&\hspace{-0.2mm}\cdot
\Bigl(\widehat{p}_i-\frac{e}{c}\partial_{i}\alpha(\widehat{q})-\frac{e}{c}A_{i}(\widehat{q})\Bigr)
-e\widehat{\lambda}_{p_i}\partial_{i}\phi\nonumber\\
&\hspace{-0.2mm}=&\hspace{-0.2mm}\frac{1}{m}\Bigl(\widehat{\lambda}_{i}+\frac{e}{c}
\widehat{\lambda}_{p_k}\partial_{k}A_i^{\prime}(\widehat{q})\Bigr)\Bigl(\widehat{p}_i-\frac{e}{c}
A_{i}^{\prime}(\widehat{q})
\Bigr)-e\widehat{\lambda}_{p_i}\partial_i\Bigl[\Phi^{\prime}(\widehat{q})+\frac{1}{c}\frac{\partial
\alpha}{\partial t}\Bigl]=\nonumber\\
&\hspace{-0.2mm}=&\hspace{-0.2mm}\widehat{\HT}_{\scriptscriptstyle A}^{\prime}+
\frac{e}{c}\frac{\partial}{\partial t}\widetilde{\alpha}.
\end{eqnarray}
%%%
Inserting this result into (\ref{ann.bcinque}) we obtain:
%%%
\begin{equation}
i\frac{d}{dt}|\psi^{\prime},t\rangle=\widehat{\HT}_{\scriptscriptstyle A}^{\prime}|\psi^{\prime},t\rangle
\end{equation}
%%%
which is just what we wanted to prove.

\bigskip

\section{Landau Problem}

In this section, as a first application of the MC scheme in the KvN formalism, we will analyse
the Landau problem. First of all we want to review it in QM. In the Landau problem 
the main goal is to find the spectrum of a particle under a 
constant magnetic field directed along
$z$. We make the usual choice (\ref{ann.Agal}) for the gauge potential:
%%%
\begin{equation}
A_x=0,\;\;\;\;\;A_y=Bx,\;\;\;\;A_z=0. \label{ann.choice}
\end{equation}
%%%
The Schr\"odinger Hamiltonian is then
%%%
\begin{equation}
\displaystyle \widehat{H}=\frac{1}{2m}\biggl[\widehat{p}_x^{\,2}+\biggl(\widehat{p}_y-\frac{eB}{c}\widehat{x}\biggr)^2+
\widehat{p}_z^{\,2}\biggr].
\end{equation}
%%%
As $\widehat{p}_y$ and $\widehat{p}_z$ commute with $\widehat{H}$ we can diagonalize all these 
three operators simultaneously. 
Then the eigenfunctions will be labeled by the eigenvalues $E$, $p^0_y$ and $p^0_z$ of 
$\widehat{H}$, $\widehat{p}_y$ and $\widehat{p}_z$ respectively. Their form is:
%%%
\begin{equation}
\displaystyle
\psi_{\s E,p^0_y,p^0_z}(x,y,z)=\frac{1}{2\pi\hbar}\textrm{exp}\biggl[\frac{i}{\hbar}(p_y^0y+p_z^0z)\biggr]\psi(x). 
\label{ann.lannine}
\end{equation}
%%%
The stationary eigenvalue problem $\widehat{H}\psi_{\s E,p^0_y,p^0_z}=E\psi_{\s E,p^0_y,p^0_z}$
leads to the following differential equation for $\psi(x)$
%%%
\begin{equation}
\displaystyle
-\frac{\hbar^2}{2m}\psi^{\prime\prime}(x)+\frac{1}{2m}\biggl(p_y^0-\frac{eBx}{c}\biggr)^2\psi(x)=\biggl(E-\frac{p^{0^2}_z}
{2m}\biggr)\psi(x). \label{ann.diffeq}
\end{equation}
%%%
If we introduce the notation:
%%%
\begin{equation}
E_t\equiv E-\frac{p_z^{0^2}}{2m},\qquad\quad x^{\prime}\equiv p_y^0-\frac{eBx}{c} \label{ann.moden}
\end{equation}
%%%
then (\ref{ann.diffeq}) is turned into the following equation:
%%%
\begin{equation}
\displaystyle
-\frac{\hbar^2}{2m}\psi^{\prime\prime}(x^{\prime})+\frac{1}{2m}\biggl[\frac{c}{eB}\biggr]^2x^{\prime^2}\psi(x^{\prime})
=\biggl[\frac{c}{eB}\biggr]^2E_t\;\psi(x^{\prime}).
\end{equation}
%%%
We immediately notice that this  is the equation of the eigenvalue problem for a harmonic oscillator
with the 
frequency replaced by $\displaystyle \omega\equiv\frac{c}{eBm}$ 
and the energy replaced by $\displaystyle
\biggl[\frac{c}{eB}\biggr]^2E_t$.
Therefore this last quantity is discretized like in the harmonic oscillator problem:
%%%
\begin{equation}
\biggl[\frac{c}{eB}\biggr]^2E_t=\hbar\omega\biggl(n+\frac{1}{2}\biggr)=\hbar\frac{c}{eBm}\biggl(n+\frac{1}{2}\biggr). 
\label{ann.modeigen}
\end{equation}
%%%
Inserting the previous expression into (\ref{ann.moden}) we get for the energy of the particle in the gauge potential
(\ref{ann.choice}) the following values:
%%%
\begin{equation}
E_{n,p_z^0}=\frac{e\hbar B}{mc}\biggl(n+\frac{1}{2}\biggr)+\frac{p^{0^2}_z}{2m}.
\end{equation}
%%%
So the eigenfunctions (\ref{ann.lannine}) can be labeled by the quantum numbers $(n,p_y^0,p_z^0)$ but
there is an infinite degeneracy because all the wave functions with different values of $p_y^0$ have the same 
energy $E_{n,p_z^0}$. 

Let us now analyse the same problem at the classical level using the operatorial formalism of KvN.
With the gauge choice (\ref{ann.choice}), the Liouvillian (\ref{ann.abstham}) is:
%%%
\begin{equation}
\HT_{\scriptscriptstyle A}=\frac{1}{m}\lambda_xp_x+\frac{1}{m}\biggl(\lambda_y-\frac{e}{c}{\mathcal A}_y\biggr)\biggl(p_y-\frac{e}{c}A_y
\biggr)+\frac{1}{m}\lambda_zp_z.
\end{equation}
%%%
If we turn $\HT_{\scriptscriptstyle A}$ into an operator $\HCT_{\scriptscriptstyle A}$ 
using the ``mixed" representation (\ref{ann.mixrep}) then what 
we get is: 
%%%
\begin{equation}
\displaystyle
\HCT_{\scriptscriptstyle A}=\frac{1}{m}\frac{\partial}{\partial x}\frac{\partial}{\partial\lambda_{p_x}}+
\frac{1}{m}\biggl(-i\frac{\partial}{\partial y}+\frac{eB}{c}\lambda_{p_x}\biggr)
\biggl(i\frac{\partial}{\partial\lambda_{p_y}}-\frac{eB}{c}x\biggr)+
\frac{1}{m}\frac{\partial}{\partial z}\frac{\partial}{\partial\lambda_{p_z}}. \label{ann.mixham}
\end{equation}
%%%
Let us now diagonalize this operator. The reason to do that is because in the KvN theory the equation to solve is 
(\ref{ijmpa.lio1}):
$i\partial_t\psi=\widehat{\HT}_{\scriptscriptstyle A}\psi$.
So, like for the Schr\"odinger equation, one should first diagonalize the Liouvillian $\HCT_{\scriptscriptstyle A}$
%%%
\begin{equation}
\HCT_{\scriptscriptstyle A}\,\psi_{\s {\mathcal{E}}}={\mathcal{E}}\,\psi_{\s {\mathcal{E}}} \label{ann.mixeigen}
\end{equation}
%%%
and then the evolution of a generic wave function is given by:
%%%
\begin{equation}
\displaystyle
\psi(t)=\sum_{\s {\mathcal{E}}}C_{\s {\mathcal{E}}}e^{-i{\mathcal{E}}t}\psi_{\s {\mathcal{E}}} \label{ann.sevensix}
\end{equation}
%%%
where the $C_{\s {\mathcal{E}}}$ are derived from the expansion of the initial wave function 
$\psi$ on the eigenstates $\psi_{\s {\mathcal{E}}}$.
We want to underline that the eigenvalues ${\mathcal{E}}$ appearing in (\ref{ann.mixeigen})-(\ref{ann.sevensix}) 
have nothing
to do with the physical energy of the system. They are simply the possible eigenvalues of the evolution operator
$\HCT_{\scriptscriptstyle A}$ and, using them and the associated eigenfunctions, we can reconstruct the evolution of the $\psi$ like it is done 
in formula (\ref{ann.sevensix}). 

Going back to the operator (\ref{ann.mixham}) we can notice that the operators:
$ \displaystyle -i\frac{\partial}{\partial y}$, $\displaystyle i\frac{\partial}{\partial\lambda_{p_y}}$,
$\displaystyle-i\frac{\partial}{\partial z}$ and $\displaystyle
i\frac{\partial}{\partial
\lambda_{p_z}}$ commute with
$\HCT_{\scriptscriptstyle A}$ and so they can be diagonalized simultaneously. In the ($q,\lambda_p$)
representation the generic eigenfunction of $\HCT_{\scriptscriptstyle A}$ has the form 
%%%
\begin{equation}
\displaystyle
\psi(q,\lambda_p)=\frac{1}{(2\pi)^2}\textrm{exp}\Bigl[i\lambda_y^0y-i\lambda_{p_y}p_y^0+
i\lambda_z^0z-i\lambda_{p_z}p_z^0\Bigr]
\psi(x,\lambda_{p_x}) \label{ann.sevsev}
\end{equation}
%%%
where $\lambda_y^0, p_y^0, \lambda_z^0$ and $p_z^0$ are the eigenvalues of the operators $\displaystyle
-i\frac{\partial}{\partial y},
i\frac{\partial}{\partial\lambda_{p_y}}, -i\frac{\partial}{\partial z}, i\frac{\partial}{\partial\lambda_{p_z}}$
respectively. We see the similarity with the quantum case except for the fact that the dimension of the space is double.
Inserting (\ref{ann.sevsev}) in (\ref{ann.mixeigen}) we get the following equation for $\psi(x,\lambda_{p_x})$:
%%%
\begin{equation}
\displaystyle
\biggl[\frac{1}{m}\frac{\partial}{\partial x}\frac{\partial}{\lambda_{p_x}}+\frac{1}{m}\biggl(\lambda_y^0+\frac{eB}{c}
\lambda_{p_x}\biggr)\biggl(p_y^0-\frac{eB}{c}x\biggr)+\frac{1}{m}\lambda_z^0p_z^0\biggr]\psi(x,\lambda_{p_x})
={\mathcal{E}}\,\psi(x,\lambda_{p_x}). \label{ann.lanotto}
\end{equation}
%%%
Via the new quantity
%%%
\begin{equation}
{\mathcal{E}}^+\equiv{\mathcal{E}}-\frac{1}{m}\lambda_z^0p_z^0 \label{ann.nq}
\end{equation}
%%%
we can rewrite (\ref{ann.lanotto}) as
%%%
\begin{equation}
\displaystyle
\biggl[\frac{1}{m}\frac{\partial}{\partial x}\frac{\partial}{\partial\lambda_{p_x}}+\frac{1}{m}\biggl(\lambda_y^0+
\frac{eB}{c}\lambda_{p_x}\biggr)\biggl(p_y^0-\frac{eB}{c}x\biggr)\biggl]\psi(x,\lambda_{p_x})
={\mathcal{E}}^+\,\psi(x,\lambda_{p_x}). \label{ann.lannove}
\end{equation}
%%%
Performing the following change of variables
%%%
\begin{equation}
\left\{
	\begin{array}{l}
	\displaystyle x^{\prime}\equiv x-\frac{c}{eB}p_y^0\\
\displaystyle \lambda_{p_x}^{\prime}\equiv\lambda_{p_x}+\frac{c}{eB}\lambda_y^0
\end{array}
\right.
\end{equation}
%%%
(\ref{ann.lannove}) becomes:
%%%
\begin{equation}
\biggl[\frac{1}{m}\frac{\partial}{\partial x^{\prime}}\frac{\partial}{\partial\lambda_{p_x}^{\prime}}
-m\omega^2\lambda_{p_x}^{\prime}x^{\prime}\biggr]\psi(x^{\prime},\lambda_{p_x}^{\prime})
={\mathcal{E}}^+\,\psi(x^{\prime},\lambda_{p_x}^{\prime}) \label{ann.lannovebis}
\end{equation}
%%%
where $\displaystyle \omega\equiv\frac{eB}{mc}$ has the dimensions of an angular velocity and it is related to the
well-known  Larmor frequency of rotation of a particle in a constant magnetic field.
Like it happens in the quantum case also (\ref{ann.lannovebis}) is nothing more than
the KvN eigenvalue equation for a harmonic oscillator. If we
introduce the following new variables
%%%
\begin{equation}
Z_+\equiv\frac{x^{\prime}+\Delta\lambda_{p_x}^{\prime}}{\sqrt{2}},\;\;\;\;\;\;\;\;
Z_-\equiv\frac{x^{\prime}-\Delta\lambda_{p_x}^{\prime}}{\sqrt{2}} \label{ann.paolo}
\end{equation}
%%%
where $\Delta$ is a constant with the dimensions of an action, the operator appearing in (\ref{ann.lannovebis})
can be written as
%%%
\begin{eqnarray}
\displaystyle
\HCT_{\scriptscriptstyle A}&=&\frac{1}{\Delta}\biggl[-\frac{\Delta^2}{2m}\frac{\partial^2}{\partial
Z_-^2}+\frac{m\omega^2}{2}Z_-^2\biggr] -\frac{1}{\Delta}\biggl[-\frac{\Delta^2}{2m}\frac{\partial^2}{\partial
Z_+^2}+\frac{m\omega^2}{2}Z_+^2\biggr]=\nonumber\\ &=&\frac{1}{\Delta}\biggl[H^{osc}\biggl(Z_-,\frac{\partial}{\partial
Z_-}\biggr)-H^{osc}\biggl(Z_+,\frac{\partial}{\partial Z_+}\biggr)\biggr]. \label{aha}
\end{eqnarray}
%%%
As indicated in the second step above, we notice that $\HCT_{\scriptscriptstyle A}$ 
is the difference of two quantum harmonic oscillators
respectively in $Z_-$ and $Z_+$, where the role of $\hbar$ is taken by the constant $\Delta$.
The eigenstates $\psi(Z_+,Z_-)$ of $\HCT_{\scriptscriptstyle A}$
can  be easily obtained. They are:
%%%
\begin{equation}
\psi(Z_+,Z_-)=\psi_n^{osc}(Z_+)\psi_m^{osc}(Z_-). \label{ann.eve}
\end{equation}
%%%
In the previous expression $\psi^{osc}_n(Z_{\pm})$ indicate the eigenfunctions of the quantum harmonic oscillator:
%%%
\begin{equation}
\displaystyle
\psi^{osc}_n(Z_{\pm})=(\sqrt{\pi}2^nn!\sigma_0)^{-1/2}H_n\biggl(\frac{Z_{\pm}}{\sigma_0}\biggr)
\textrm{exp}\biggl(-\frac{Z_{\pm}^2}{2\sigma_0^2}\biggr),\;\;\;\;n=0,+1,+2,\cdots
\end{equation}
%%%
where $H_n$ are the Hermite polynomials and $\displaystyle \sigma_0=\sqrt{\frac{\Delta}{m\omega}}$.
The eigenvalues associated to the eigenfunctions (\ref{ann.eve}) are:
%%%
\begin{equation}
{\mathcal{E}}^+_{n,m}=\frac{1}{\Delta}\biggl[\biggl(m+\frac{1}{2}\biggr)\Delta\omega-\biggl(n+\frac{1}{2}\biggr)
\Delta\omega\biggl]=(m-n)\omega=N\omega \label{ann.dsei}
\end{equation}
%%%
where $N$ can take every positive or negative integer value: $N=0,\pm 1,\pm 2,\cdots$.
Let us notice that the spectrum of the Liouvillian (\ref{aha}) is discretized and unbounded below
and that, by making the difference of the two oscillators, the quantum zero-point energy disappears completely
from (\ref{ann.dsei}). 
Note also that there is
an $\infty$-order degeneracy in the sense that every eigenvalue ${\mathcal{E}}^+=N\omega$ has  
an entire set of eigenfunctions labeled by $n$:
$\psi=\psi^{osc}_n(Z_+)\psi^{osc}_{n+N}(Z_-)$, where $n=-N, -N+1, -N+2,\cdots$ if $N<0$ and $n=0, 1, 2,\cdots$
if $N\ge 0$.

Coming back to the Landau problem and inserting (\ref{ann.dsei}) into (\ref{ann.nq}) 
we have that the final spectrum is  
%%%
\begin{equation}
\displaystyle {\mathcal{E}}=N\omega+\frac{1}{m}\lambda_z^0p_z^0=
N\biggl(\frac{eB}{mc}\biggr)+\frac{1}{m}\lambda_z^0p_z^0
\end{equation}
%%%
while the associated eigenfunctions are:
%%%
\begin{equation}
\displaystyle
\psi_{N,n,\lambda_y^0,p_y^0,\lambda_z^0,p_z^0}=\frac{1}{(2\pi)^2}\textrm{exp}\Bigl[i\lambda_y^0y-i\lambda_{p_y}p_y^0+
i\lambda_z^0z-i\lambda_{p_z}p_z^0\Bigr]\psi_n^{osc}(Z_+)\psi_{n+N}^{osc}(Z_-). \label{ann.liouvham}
\end{equation}
%%%
From the previous eigenfunctions we see that the degeneracy in the KvN case
is much more than in the quantum one. In fact not only 
the eigenfunctions with different 
values of $p_y^0$ have the same eigenvalue ${\mathcal{E}}$, but the same happens also
for all the eigenfunctions with different values
of $\lambda_y^0$, $n$, $\lambda_z^0$ and $p_z^0$ with the only constraint 
that the product $\lambda_z^0p_z^0$ must be the same.
So there is a much more wider degeneracy here
than in the quantum case. This is due to the fact that the ``wave functions" in the KvN formalism have a
number of variables $(q,\lambda_p)$ that is double than the ones in QM.

We want to conclude this analysis of the Landau problem 
finding out which are the constants of motion, i.e. 
the operators that commute with the generator of the time evolution $\widehat{L}_{\scriptscriptstyle A}$. 
These operators will give us some indications
concerning the trajectory of the classical particle in the constant magnetic field. 

Let us remember the form of the Liouvillian in the Landau problem:
%%%
\begin{equation}
\displaystyle
\widehat{L}_{\scriptscriptstyle A}=\frac{1}{m}\widehat{\lambda}_x\widehat{p}_x+\frac{1}{m}\biggl(\widehat{\lambda}_y+\frac{e}{c}B
\widehat{\lambda}_{p_x}\biggr)
\biggl(\widehat{p}_y-\frac{eB}{c}\widehat{x}\biggr)+\frac{1}{m}\widehat{\lambda}_z\widehat{p}_z.
\end{equation}
%%%
The commutator of the gauge invariant velocity $\displaystyle \widehat{v}_y=\frac{1}{m}
\biggl(\widehat{p}_y-\frac{e}{c}\widehat{A}_y\biggr)$ with the Liouvillian $\HCT_{\scriptscriptstyle A}$ is given by:
%%%
\begin{equation}
\displaystyle
[\widehat{v}_y,\HCT_{\scriptscriptstyle
A}]=\frac{1}{m}\biggl[\widehat{p}_y-\frac{e}{c}\widehat{A}_y,\HCT_{\scriptscriptstyle A}\biggr]=
-\frac{e}{m^2c}\Bigl[B\widehat{x},\widehat{\lambda}_x\Bigr]\widehat{p}_x
=-\frac{ieB}{m^2c}\widehat{p}_x. \label{ann.cos1}
\end{equation}
%%%
If we introduce the Larmor frequency: $\displaystyle \omega=\frac{eB}{mc}$ we can then easily prove
that\break $\displaystyle \widehat{x}_0\equiv \widehat{x}+\widehat{v}_y/\omega$ is a constant of motion.
In fact, using (\ref{ann.cos1}) we get:
%%%
\begin{equation}
[\widehat{x}_0,\HCT_{\scriptscriptstyle A}]=\biggl[\widehat{x}+\frac{\widehat{v}_y}{\omega},
\HCT_{\scriptscriptstyle A}\biggr]=\frac{i}{m}\widehat{p}_x-\frac{mc}{eB}\cdot
\frac{ieB}{m^2c}\widehat{p}_x=0.
\end{equation}
%%%
In the same way the commutators of $\widehat{y}$ and $\displaystyle 
\widehat{v}_x=\frac{1}{m}\biggl(\widehat{p}_x-\frac{e}{c}\widehat{A}_x\biggr)$ 
with the Liouvillian are:
%%%
\begin{equation}
\displaystyle
[\widehat{y},\HCT_{\scriptscriptstyle A}]=\frac{i}{m}\biggl(\widehat{p}_y-\frac{eB}{c}\widehat{x}\biggr),\;\;\;\;\;\;\;
[\widehat{v}_x,\HCT_{\scriptscriptstyle A}]=\frac{ieB}{m^2c}\biggl(\widehat{p}_y-\frac{eB}{c}\widehat{x}\biggr) \label{ann.cos2}
\end{equation}
%%%
and so we obtain that also $\displaystyle \widehat{y}_0\equiv \widehat{y}-\widehat{v}_x/\omega$ commutes with 
$\HCT_{\scriptscriptstyle A}$:
%%%
\begin{equation}
[\widehat{y}_0,\HCT_{\scriptscriptstyle A}]=[\widehat{y},\HCT_{\scriptscriptstyle A}]
-\frac{mc}{eB}[\widehat{v}_x,\HCT_{\scriptscriptstyle A}]=0.
\end{equation}
%%%
Now a classical particle in a constant magnetic field
directed along $z$ describes an helicoidal orbit whose projection 
on the $x,y$-plane is a circumference with a radius equal to the Larmor one $\varrho_{Lar}$:
%%%
\begin{equation}
\displaystyle
\varrho_{Lar}^2\equiv \frac{1}{\omega^2}(v_x^2+v_y^2).
\end{equation}
%%%
Using (\ref{ann.cos1})-(\ref{ann.cos2}) it is possible to prove that also the 
Larmor radius is a constant of motion:
%%%
\begin{equation}
\displaystyle
[\widehat{\varrho}_{Lar}^2,\HCT_{\scriptscriptstyle A}]=\frac{1}{\omega^2}[\widehat{v}_x^2,
\HCT_{\scriptscriptstyle A}]+\frac{1}{\omega^2}
[\widehat{v}_y^2,\HCT_{\scriptscriptstyle A}]=0.
\end{equation}
%%%
The Larmor radius can be written also in terms of the operators $\widehat{x}$, $\widehat{x}_0$, $\widehat{y}$, 
$\widehat{y}_0$ in the following way:
%%%
\begin{equation}
\displaystyle
\widehat{\varrho}^2_{Lar}=\frac{1}{\omega^2}(\widehat{v}_x^2+\widehat{v}_y^2)=\frac{1}{\omega^2}\Bigl[\omega(\widehat{y}-
\widehat{y}_0)\Bigr]^2+
\frac{1}{\omega^2}\Bigl[\omega(\widehat{x}_0-\widehat{x})\Bigr]^2=(\widehat{x}-\widehat{x}_0)^2+(\widehat{y}-\widehat{y}_0)^2.
\end{equation}
%%%
Therefore $(x_0,y_0)$ is the center of a circumference which is the projection of the orbit of the particle
onto the plane $(x,y)$ and $\varrho_{Lar}$ is the corresponding radius.
Note that in the KvN operatorial formalism the operators $\widehat{x}_0$ and $\widehat{y}_0$ are suitable combinations
of the operators  $\widehat{\varphi}$ and they commute among themselves. This implies that they can be determined 
with arbitrary precision. In QM, instead, one can prove that the following relation holds:
%%%
\begin{equation}
[\widehat{x}_0,\widehat{y}_0]=\frac{-i\hbar c}{eB}
\end{equation}
%%%
and therefore, differently than in CM, 
there is an uncertainty relation involving the coordinates of 
the center of the circumference.

\bigskip

\section{Aharonov-Bohm Effect}

The second application of the MC rules that we want to study now is the well-known Aharonov-Bohm (AB) effect, \cite{Bohm}. 
According to this phenomenon, while the classical motion of a particle 
feels only the magnetic field and not the gauge potential, 
the quantum wave functions can be changed by the presence of a gauge potential even in
regions where the associated magnetic field is zero. The change in the wave functions can be detected for example 
by an interference experiment.
In this section we will study this 
phenomenon by considering not the wave functions or the interference effects but 
the spectra of the quantum Schr\"odinger operator $\widehat{H}$
and the classical KvN Liouvillian $\HCT$. 
The geometrical set up that we will use for the AB effect is
illustrated in  Fig.  {\bf \ref{ann.ABset}}: 
basically we have a hollow cylindrical shell and a particle completely confined to the interior of the shell
with rigid walls \cite{Sak}.  We will show that 
the spectrum of $\widehat{H}$ {\it changes} once we turn on the magnetic field in the region enclosed by the shell 
and this happens even if the magnetic field is identically zero in the shell. Using
the same geometrical configuration, we will study the spectrum of the KvN-Liouville operator 
$\HCT$ and we will
prove that it {\it does not change} once we turn on the magnetic field differently than what happens in QM.
We feel that this, in the framework of the operatorial formulation of CM, is the best mathematical proof that there is no
AB effect in CM.

\setcounter{figure}{0}
\begin{figure}
\centering
\includegraphics{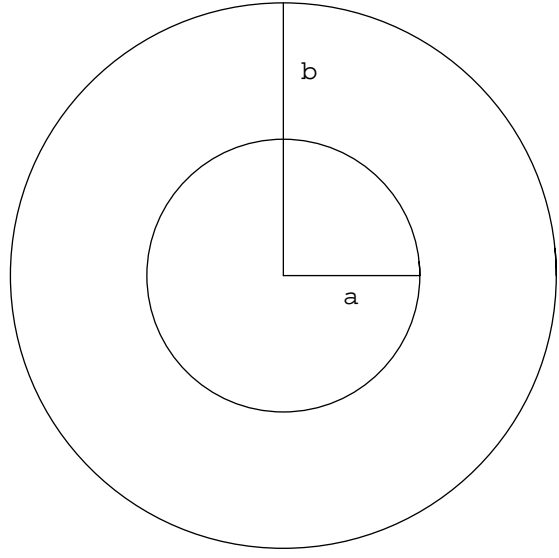}
\caption{\rm{Aharonov-Bohm geometrical set up.}} 
\label{ann.ABset}
\end{figure}

\begin{figure}
\centering
\includegraphics[width=14cm]{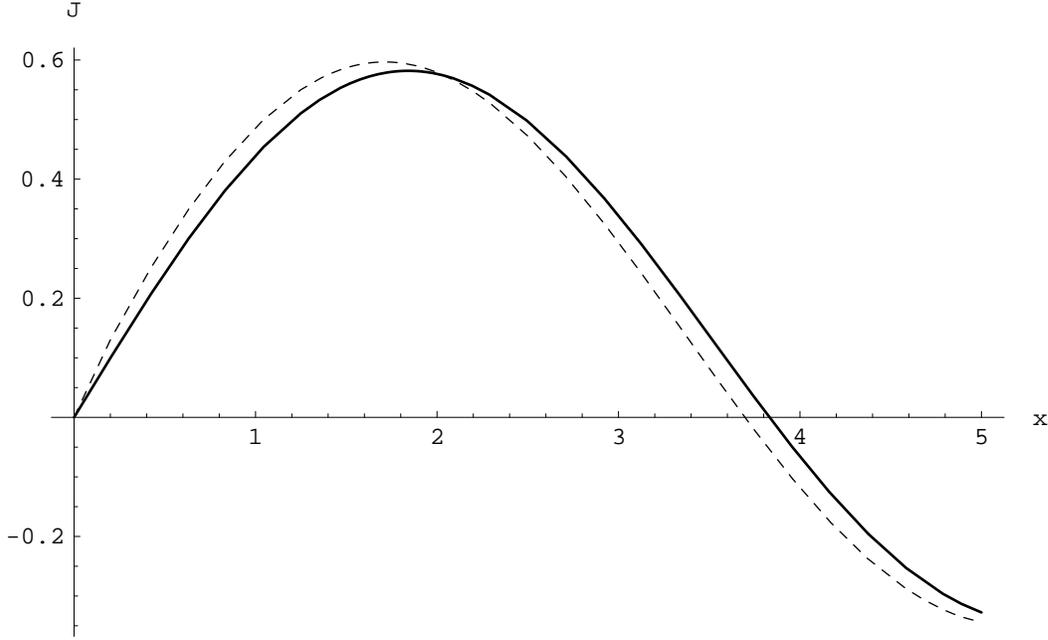}
\caption{\rm{
Zeros of Bessel functions: m=1 (continuous line), m=0.9 (dashed line).}} 
\label{ann.Bes}
\end{figure}

Let us now study the Schr\"odinger operator in the geometrical set up of Fig. {\bf \ref{ann.ABset}} and 
without magnetic field. Using cylindrical coordinates
%%%
\begin{equation}
\label{ann.cylindrical}
\left\{
	\begin{array}{l}
	\displaystyle x=\rho \,\textrm{cos}\theta\smallskip\\
          \displaystyle y=\rho \,\textrm{sin}\theta\smallskip\\
          \displaystyle z=z\\
	\end{array}
	\right.
\end{equation}
%%%
the Schr\"odinger operator for a free particle is:
%%%
\begin{equation}
\displaystyle
\widehat{H}=-\frac{\hbar^2}{2\mu}\biggl(\frac{\partial^2}{\partial\rho^2}+\frac{1}{\rho}\frac{\partial}{\partial\rho}
+\frac{1}{\rho^2}\frac{\partial^2}{\partial\theta^2}+\frac{\partial^2}{\partial z^2}\biggr) \label{ann.freeH}
\end{equation}
%%%
and the equation for the eigenvalues becomes:
%%%
\begin{equation}
\displaystyle
-\frac{\hbar^2}{2\mu}\biggl(\frac{\partial^2}{\partial\rho^2}+\frac{1}{\rho}\frac{\partial}{\partial\rho}
+\frac{1}{\rho^2}\frac{\partial^2}{\partial\theta^2}+\frac{\partial^2}{\partial z^2}\biggr)\psi(\rho,\theta,z)
=E\,\psi(\rho,\theta,z). \label{ann.bessel1}
\end{equation}
%%%
As the operators $\displaystyle 
\frac{\partial}{\partial\theta}$ and $\displaystyle \frac{\partial}{\partial z}$ commute with $\widehat{H}$ we can diagonalize
these three operators simultaneously and search for the eigenfunctions in the form:
%%%
\begin{equation}
\displaystyle
\psi(\rho,\theta,z)=\frac{1}{2\pi}\textrm{exp}\biggl[\frac{ip_z^0z}{\hbar}+im\theta\biggr]R(\rho) \label{ann.decomp1}
\end{equation}
%%%
where $p_z^0$ is a real number and $m$ is an integer. Inserting (\ref{ann.decomp1}) in (\ref{ann.bessel1}) we get the following
equation for $R(\rho)$:
%%%
\begin{equation}
R^{\prime\prime}(\rho)+\frac{R^{\prime}(\rho)}{\rho}+\biggl(\bar{s}-\frac{m^2}{\rho^2}\biggr)R(\rho)=0 \label{ann.besselr}
\end{equation}
%%%
where 
%%%
\begin{equation}
\bar{s}\equiv\frac{2\mu E}{\hbar^2}-\frac{p_z^{0^2}}{\hbar^2}. \label{ann.abar}
\end{equation}
%%%
Using the new variables 
$r\equiv\sqrt{\bar{s}}\rho$ Eq. (\ref{ann.besselr}) becomes the well-known Bessel equation \cite{Watson}:
%%%
\begin{equation}
\displaystyle
\frac{\partial^2R}{\partial r^2}+\frac{1}{r}\frac{\partial R}{\partial r}+\biggl(1-\frac{m^2}{r^2}\biggr)R=0.
\label{ann.besselr2}
\end{equation}
%%%
Since we want to have a wave function confined to the interior of the shell
we will solve (\ref{ann.besselr2}) with the boundary conditions:
%%%
\begin{equation}
R(\sqrt{\bar{s}}a)=R(\sqrt{\bar{s}}b)=0 \label{ann.bouncon}
\end{equation}
%%%
where $a$ and $b$ are respectively the smaller and the larger radius
of the cylindrical shell, see Fig. {\bf \ref{ann.ABset}}. 
The most general solution of (\ref{ann.besselr2}) with m integer is given by the linear 
combination
of the Bessel functions of the first and second kind \cite{Watson}:
%%%
\begin{equation}
R(\sqrt{\bar{s}}\rho)=a_{\scriptscriptstyle 1}
J_m(\sqrt{\bar{s}}\rho)+a_{\scriptscriptstyle 2}Y_m(\sqrt{\bar{s}}\rho) \label{ann.solutions}
\end{equation}
%%%
where
%%%
\begin{eqnarray}
\displaystyle
&&J_m(r)=\biggl(\frac{r}{2}\biggr)^m\sum_{n=0}^{\infty}\frac{(-1)^n\bigl(\frac{r}{2}\bigr)^{2n}}{n!\,\Gamma(n+m+1)}
\nonumber\\
&&Y_m(r)=\lim_{\epsilon\to 0}\frac{1}{\epsilon}[J_{m+\epsilon}(r)-(-1)^mJ_{-m-\epsilon}(r)].
\end{eqnarray}
%%%
The spectrum of the system is completely determined by the boundary conditions (\ref{ann.bouncon}). 
In order to simplify things 
we will
consider the limit case in which the radius of the internal cylinder $a$ goes to zero. In this case the boundary 
conditions
(\ref{ann.bouncon}) become
%%%
\begin{equation}
R(0)=0, \;\;\;\; R(\sqrt{\bar{s}}b)=0. \label{ann.bouncon2}
\end{equation}
%%%
Now the Bessel functions of the second  kind $Y_m$ are singular in the origin \cite{Watson}, so we have to
restrict ourselves to solutions (\ref{ann.solutions}) of the form:
%%%
\begin{equation}
R(\sqrt{\bar{s}}\rho)=a_{\scriptscriptstyle 1}J_m(\sqrt{\bar{s}}\rho).
\end{equation}
%%%
With this restriction the first of the boundary conditions (\ref{ann.bouncon2}) is automatically satisfied because 
$J_m(0)=0$ for $m\ge 1$. So we have to impose only the second of the conditions (\ref{ann.bouncon2}) which implies:
%%%
\begin{equation}
J_m(\sqrt{\bar{s}}b)=0. \label{ann.last}
\end{equation}
%%%
This relation tells us that we have to look for the zeros of the Bessel functions of the first kind $J_m$. 
Let us call them $\alpha_{k,m}$ where $m$ indicates the Bessel
function which we refer to and $k=1,2,\cdots$ labels the various zeros of the $m\textrm{th}$ Bessel function.
Then the solutions of (\ref{ann.last}) can be formally written as 
%%%
\begin{equation}
\sqrt{\bar{s}}b=\alpha_{k,m}.
\end{equation}
%%%
Replacing $\bar{s}$ in the equation above with its expression (\ref{ann.abar}), we get the following energy levels:
%%%
\begin{equation}
\displaystyle
E_{k,m}=\hbar^2\frac{\alpha^2_{k,m}}{2\mu b^2}+\frac{p_z^{0^2}}{2\mu}.
\end{equation}
%%%
To give an example useful for the following discussion, if we choose the second zero ($k=2$) of the first Bessel function
($m=1$), which is $\alpha_{2,1}\approx 3.83$, we obtain the following energy level, see Fig. {\bf \ref{ann.Bes}}:
%%%
\begin{equation}
E_{2,1}=\hbar^2\frac{7.33}{\mu b^2}+\frac{p_z^{0^2}}{2\mu}. \label{ann.edueuno}
\end{equation}
%%%

Let us now turn on a magnetic field \cite{Bohm} which is zero everywhere except for the $B_z$ component on
the line $x^2+y^2=0$ and which has a fixed flux $\Phi_{\s B}$. A choice of the gauge potential is
%%%
\begin{equation}
\left\{
	\begin{array}{l}
	\displaystyle A_x=\frac{-y\Phi_{\s B}}{2\pi(x^2+y^2)}\smallskip\\
          \displaystyle A_y=\frac{x\Phi_{\s B}}{2\pi(x^2+y^2)}\smallskip\\
          \displaystyle A_z=0. \label{ann.Abohm}\\
	\end{array}
	\right.
\end{equation}
%%%
In cylindrical coordinates (\ref{ann.Abohm}) becomes: 
%%%
\begin{equation}
A_{\rho}=0,\;\;\;A_{\theta}=\frac{\Phi_{\s B}}{2\pi\rho},\;\;\; A_z=0.
\end{equation}
%%%
So with the above gauge field the MC rules change only the operator 
$\displaystyle -i\hbar\frac{\partial}{\partial\theta}$ in the following way:
%%%
\begin{equation}
-i\hbar\frac{\partial}{\partial\theta}\;\longrightarrow\;
-i\hbar\frac{\partial}{\partial\theta}-\frac{e}{c}\frac{\Phi_
{\s B}}{2\pi}. \label{ann.val}
\end{equation}
%%%
Inserting (\ref{ann.val}) in (\ref{ann.freeH}) we have that the Schr\"odinger operator becomes: 
%%%
\begin{equation}
\displaystyle
\widehat{H}\;\longrightarrow\;\widehat{H}_{\s A}=\frac{-\hbar^2}{2\mu}\biggl[\frac{\partial^2}{\partial\rho^2}
+\frac{1}{\rho}\frac{\partial}{\partial\rho}+\frac{1}{\rho^2}\biggl(\frac{\partial}{\partial\theta}-\frac{ie}{ch}
\Phi_{\s B}\biggr)^2+\frac{\partial^2}{\partial z^2}\biggr]
\end{equation}
%%%
while the equation for the eigenvalues is:
%%%
\begin{equation}
\displaystyle
\frac{-\hbar^2}{2\mu}\biggl[\frac{\partial^2}{\partial\rho^2}
+\frac{1}{\rho}\frac{\partial}{\partial\rho}+\frac{1}{\rho^2}\frac{\partial^2}{\partial\theta^2}
-\frac{2ie}{ch}\Phi_{\s B}\frac{1}{\rho^2}\frac{\partial}{\partial\theta}-\frac{e^2}{c^2h^2}
\frac{\Phi^2_{\s B}}{\rho^2}+\frac{\partial^2}{\partial z^2}\biggr]\psi=E\psi(\rho,\theta,z). \label{ann.ninenine}
\end{equation}
%%%
Like in the free case for the previous equation we can choose solutions of the form:
%%%
\begin{equation}
\displaystyle
\psi(\rho,\theta,z)=\frac{1}{2\pi}\textrm{exp}\biggl[\frac{ip_z^0z}{\hbar}+im\theta\biggr]R(\rho)
\end{equation}
%%%
which, inserted in (\ref{ann.ninenine}), give the following differential equation for $R(\rho)$
%%%
\begin{equation}
\displaystyle
-\frac{\hbar^2}{2\mu}\biggl[\frac{R^{\prime\prime}}{R}+\frac{1}{\rho}\frac{R^{\prime}}{R}-\frac{1}{\rho^2}
\biggl(m-\frac{e\Phi_{\s B}}{ch}\biggr)^2-\frac{p_z^{0^2}}{\hbar^2}\biggr]=E. \label{ann.eigeneq}
\end{equation}
%%%
If $\displaystyle \alpha\equiv\frac{e\Phi_{\s B}}{ch}$ and 
$\displaystyle \bar{s}\equiv\frac{2\mu E}{\hbar^2}-\frac{p_z^{0^2}}{\hbar^2}$, then 
(\ref{ann.eigeneq}) can be written as 
%%%
\begin{equation}
R^{\prime\prime}(\rho)+\frac{R^{\prime}(\rho)}{\rho}+\biggl[\bar{s}-\frac{(m-\alpha)^2}{\rho^2}\biggr]R(\rho)=0.
\label{ann.besselmag}
\end{equation}
%%%
If we compare the previous equation with (\ref{ann.besselr}) we notice that the magnetic field has only shifted 
$m\rightarrow m_{\s A}\equiv m-\alpha$. Therefore $m_{\s A}$ will no longer be an integer but a real number.
Doing the same change of variables as before, $r\equiv\sqrt{\bar{s}}\rho$, we can transform (\ref{ann.besselmag}) into
%%%
\begin{equation}
\frac{\partial^2 R}{\partial r^2}+\frac{1}{r}
\frac{\partial R}{\partial r}+\biggl(1-\frac{m^2_{\s A}}{r^2}\biggr)R=0.
\label{ann.besselmag2}
\end{equation}
%%%
For this equation with $m_{\s A}$ real there are two linearly independent solutions which are two Bessel functions of the 
first kind with opposite indices:
%%%
\begin{eqnarray}
\displaystyle
&& J_{m_{\s A}}(r)=\biggl(\frac{r}{2}\biggr)^{m_{\s A}}\sum_{n=0}^{\infty}\frac{(-1)^n\bigl(\frac{r}{2}\bigr)^{2n}}
{n!\,\Gamma(n+m_{\s A}+1)}\nonumber\\
&& J_{-m_{\s A}}(r)=\biggl(\frac{r}{2}\biggr)^{-m_{\s A}}\sum_{n=0}^{\infty}\frac{(-1)^n\bigl(\frac{r}{2}\bigr)^{2n}}
{n!\,\Gamma(-n+m_{\s A}+1)}.
\end{eqnarray}
%%%
One immediately notices that $J_{m_{\s A}}(0)=0$ while $J_{-{m_{\s A}}}(0)$ diverges.
As before when we impose the boundary conditions (\ref{ann.bouncon2}) 
the general solution of (\ref{ann.besselmag2}) becomes:
%%%
\begin{equation}
R(\sqrt{\bar{s}}\rho)=a_{\scriptscriptstyle 1}J_{m_{\s A}}(r).
\end{equation}
%%%
From the boundary condition
%%%
\begin{equation}
J_{m_{\s A}}(\sqrt{\bar{s}}b)=0
\end{equation}
%%%
we can derive, as before, the energy levels
%%%
\begin{equation}
\displaystyle
E_{k,m_{\s A}}=\hbar\frac{\alpha^2_{k,m_{\s A}}}{2\mu b^2}+\frac{p_z^{0^2}}{2\mu}. \label{ann.ekappa}
\end{equation}
%%%
If the magnetic flux $\Phi_{\s B}$ gives $\alpha=0.1$, then we have to consider 
the Bessel functions $J_{m-0.1}$. The second zero of $J_{0.9}$, analog
to the second one of $J_1$ that we considered before, is $\alpha_{2, 0.9}=3.70$, 
see Fig. {\bf \ref{ann.Bes}}.
Inserting this value  in (\ref{ann.ekappa}) we get
%%%
\begin{equation}
\displaystyle
E_{2, 0.9}=\hbar^2\frac{6.84}{\mu b^2}+\frac{p_z^{0^2}}{2\mu}<E_{2,1}.
\end{equation}
%%%
According to the last inequality the energy level $E_{2,0.9}$ with the magnetic field 
is smaller than the corresponding level $E_{2,1}$ without the magnetic field
calculated in (\ref{ann.edueuno}). So this is a clear indication that the presence
of a gauge potential modifies the spectrum of the quantum 
Schr\"odinger operator even if the wave function is restricted to an
area with zero magnetic field. 

Let us now perform the same analysis in the classical case, studying what happens to the Liouvillian $\HCT$. 
We have to be careful and write in
cylindrical coordinates all the variables which enter the Liouvillian $\HCT$. We can start from the free 
Lagrangian in cylindrical coordinates which is:
%%%
\begin{equation}
{\mathscr L}=\frac{1}{2}\mu\dot{\rho}^2+\frac{1}{2}\mu\rho^2\dot{\theta}^2+\frac{1}{2}\mu\dot{z}^2. \label{previous}
\end{equation}
%%%
From (\ref{previous}) we can derive the following momenta conjugate to $\rho,\theta$ and $z$:
%%%
\begin{equation}
\left\{
	\begin{array}{l}
	\displaystyle p_{\rho}=\mu\dot{\rho}\smallskip\\
          \displaystyle p_{\theta}=\mu\rho^2\dot{\theta}\smallskip\\
          \displaystyle p_z=\mu\dot{z}.\\
	\end{array}
	\right.
\end{equation}
%%%
Then the relations between the Cartesian momenta $p_x,p_y,p_z$ and the cylindrical ones
$p_{\rho},p_{\theta},p_z$ are: 
%%%
\begin{equation}
\left\{
	\begin{array}{l}
	\displaystyle p_x=\mu\dot{x}=\mu\dot{\rho}\;\textrm{cos}\theta-\mu\rho\dot{\theta}\;\textrm{sin}\theta=
	p_{\rho}\textrm{cos}\theta-\frac{p_{\theta}}{\rho}
	\textrm{sin}\theta\smallskip\\
          \displaystyle p_{y}=\mu\dot{y}=\mu\dot{\rho}\;\textrm{sin}\theta+\mu\rho\dot{\theta}\;\textrm{cos}\theta=p_{\rho}\textrm{sin}\theta+\frac{p_{\theta}}{\rho}
	\textrm{cos}\theta\smallskip\\
          \displaystyle p_z=p_z.\\
	\end{array}
	\right.
\end{equation}
%%%
As the coordinate $z$ and the momentum $p_z$ are the same in the two coordinate systems we can 
summarize the basic transformation rules in the following scheme:
%%%
\begin{equation}
\left\{
	\begin{array}{l}
	\displaystyle x=\rho \;\textrm{cos}\theta\smallskip\\
          \displaystyle y=\rho \;\textrm{sin}\theta\smallskip\\
          \displaystyle p_x=p_{\rho}\textrm{cos}\theta-\frac{p_{\theta}}{\rho}\textrm{sin}\theta\smallskip\\
          \displaystyle p_y=p_{\rho}\textrm{sin}\theta+\frac{p_{\theta}}{\rho}\textrm{cos}\theta. \label{ann.cyl}\\
	\end{array}
	\right.
\end{equation}
%%%
Note that the transformations of the momenta $p_x,p_y$ include not only the new momenta $p_{\rho},p_{
\theta}$ but also the new coordinates $\theta,\rho$. Let us remember that the Liouvillian $\HCT$ 
in Cartesian coordinates contains derivatives with respect to both the coordinates and the momenta 
and so we should check how they 
are related to derivatives in cylindrical coordinates. Using (\ref{ann.cyl}) it is a long but easy calculation to show that:
%%%
\begin{equation}
\displaystyle
\begin{pmatrix}
\partial/\partial x \\
               \partial/\partial y\\
               \partial/\partial p_x\\
               \partial/\partial p_y
               \end{pmatrix}=
                    \begin{pmatrix}
                    \textrm{cos}\theta & -\textrm{sin}\theta/\rho & -p_{\theta}\textrm{sin}\theta/\rho^2 &
                    p_{\theta}/\rho\cdot \textrm{cos}\theta+p_{\rho}\textrm{sin}\theta\\
                    \textrm{sin}\theta & \textrm{cos}\theta/\rho & p_{\theta}\textrm{cos}\theta/\rho^2 &
                    p_{\theta}/\rho\cdot \textrm{sin}\theta-p_{\rho}\textrm{cos}\theta\\
                    0 & 0 & \textrm{cos}\theta & -\rho \textrm{sin}\theta\\
                    0 & 0 & \textrm{sin}\theta & \rho \textrm{cos}\theta\\
                    \end{pmatrix}
                    \cdot\begin{pmatrix}\partial/\partial \rho\\
                    \partial/\partial \theta \\
                    \partial/\partial p_{\rho}\\
                    \partial/\partial p_{\theta}\end{pmatrix}. \label{ann.cyl2}
                    \end{equation}
%%%
Equipped with the transformations (\ref{ann.cyl}) and (\ref{ann.cyl2})
we can then easily transform $\HCT$ from Cartesian coordinates to cylindrical ones. 
In the  case of a free particle we get
%%%
\begin{eqnarray}
\HCT&\hspace{-0.2mm}=&\hspace{-0.2mm}-i\frac{p_x}{\mu}\frac{\partial}{\partial x}-i\frac{p_y}{\mu}\frac{\partial}{\partial y}
-i\frac{p_z}{\mu}\frac{\partial}{\partial z}=\nonumber\\
&\hspace{-0.2mm}=&\hspace{-0.2mm}-\frac{i}{\mu}p_{\rho}
\frac{\partial}{\partial\rho}-\frac{i}{\mu}\frac{p_{\theta}}{\rho^2}\frac{\partial}{\partial\theta}
-\frac{i}{\mu}p_z\frac{\partial}{\partial z}-
\frac{i}{\mu}\frac{p_{\theta}^2}{\rho^3}\frac{\partial}{\partial p_{\rho}}.
\label{ann.freecyl}
\end{eqnarray}
%%%
Next we will turn on the magnetic field whose
gauge potential is given by (\ref{ann.Abohm}). To implement the MC rules (\ref{ann.general}), we need to 
build the variables ${\mathcal A}_{i}$ which, for our potential, are: 
%%%
\begin{equation}
\label{ann.gouno}
\left\{
	\begin{array}{l}
	\displaystyle {\mathcal A}_x=-\lambda_{p_x}\frac{\Phi_{\s B}}{\pi}\frac{xy}{(x^2+y^2)^2}+\lambda_{p_y}\frac{\Phi_{\s B}}{2\pi}
	\frac{x^2-y^2}{(x^2+y^2)^2}\smallskip\\
          \displaystyle 
          {\mathcal A}_y=\lambda_{p_x}\frac{\Phi_{\s B}}{2\pi}\frac{x^2-y^2}{(x^2+y^2)^2}+\lambda_{p_y}\frac{\Phi_{\s B}}{\pi}
	\frac{xy}{(x^2+y^2)^2}\smallskip\\
          \displaystyle {\mathcal A}_z=0.\\
	\end{array}
	\right.
\end{equation}
%%%
In the expression above we have now to turn $\lambda_{p_x}$, and $\lambda_{p_y}$ into operators, 
like in (\ref{ijmpa.operatorial}), and next we have to change everything into 
cylindrical coordinates using (\ref{ann.cyl}) and (\ref{ann.cyl2}). The result is:
%%%
\begin{equation}
\label{ann.godue}
\left\{
	\begin{array}{l}
	\displaystyle \widehat{{\mathcal A}}_x=i\frac{\Phi_{\s B}}{2\pi}\frac{1}{\rho^2}
	\biggl[\textrm{sin}\theta\frac{\partial}{\partial p_{\rho}}
	-\rho\; \textrm{cos}\theta\frac{\partial}{\partial p_{\theta}}\biggr]\smallskip\\
          \displaystyle \widehat{{\mathcal A}}_y=-i\frac{\Phi_{\s
B}}{2\pi}\frac{1}{\rho^2}\biggl[\textrm{cos}\theta\frac{\partial}{\partial p_{\rho}}
	+\rho \;\textrm{sin}\theta\frac{\partial}{\partial p_{\theta}}\biggr]\smallskip\\
          \displaystyle \widehat{{\mathcal A}}_z=0.\\
	\end{array}
	\right.
\end{equation}
%%%
Inserting (\ref{ann.Abohm})-(\ref{ann.godue}) into (\ref{ann.abstham}) and turning all the variables
into operators we get, after a long but simple calculation, that the Liouvillian is:
%%%
\begin{equation}
\displaystyle
\HCT_{\s A}\equiv -\frac{i}{\mu}p_{\rho}\frac{\partial}{\partial \rho}-\frac{i}{\mu\rho^2}\biggl(p_{\theta}
-\frac{e\Phi_{\s B}}{2\pi c}\biggr)\frac{\partial}{\partial\theta}-\frac{i}{\mu}p_z\frac{\partial}{\partial
z}-\frac{i}{\mu\rho^3}\biggl(p_{\theta}-
\frac{e\Phi_{\s B}}{2\pi c}\biggr)^2\frac{\partial}{\partial p_{\rho}}.
\label{ann.oneoneone}
\end{equation}
%%%
Notice that we could obtain this $\HCT_{\s A}$ from the free Liouvillian 
$\HCT$ of (\ref{ann.freecyl}) by just doing the replacement
%%%
\begin{equation}
p_{\theta}\;\longrightarrow\;p_{\theta}-\frac{e\Phi_{\s B}}{2\pi c}. \label{ann.finiu}
\end{equation}
%%%

Let us now turn to the eigenvalue equations: in the free case
%%%
\begin{equation}
\HCT\,\psi={\mathcal{E}}\,\psi
\label{ann.psio}
\end{equation}
%%%
can be solved by noticing that $\HCT$ commutes with 
$\displaystyle -i\frac{\partial}{\partial\theta}$,$\displaystyle -i\frac{\partial}
{\partial z},p_{\theta},p_z$. So these five operators can be 
diagonalized simultaneously and the solution of (\ref{ann.psio}) have the 
form
%%%
\begin{equation}
\displaystyle
\psi=\frac{1}{2\pi}\widetilde{R}(\rho,p_{\rho})\delta(p_{\theta}-p_{\theta}^0)\delta(p_z-p_z^0)\textrm{exp}
(in\theta+i\lambda_z^0 z)
\label{ann.psir}
\end{equation}
%%%
where $n$ is an integer and $\lambda_z^0,p^0_{\theta},p_z^0$ 
are the eigenvalues of $\displaystyle -i\frac{\partial}{\partial z},\widehat{p}_{\theta},
\widehat{p}_z$. Inserting (\ref{ann.psir}) in (\ref{ann.psio}) we get the following 
equation for $\widetilde{R}
(\rho,p_{\rho})$:
%%%
\begin{equation}
\biggl(-\frac{i}{\mu}p_{\rho}\frac{\partial}{\partial\rho}+\frac{p_{\theta}^0n}{\mu\rho^2}-\frac{i}{\mu\rho^3}
p^{0^2}_{\theta}\frac{\partial}{\partial
p_{\rho}}+\frac{\lambda_z^0p_z^0}{\mu}-{\mathcal{E}}\biggr)\widetilde{R}(\rho,p_{\rho})=0.
\label{ann.besselfree}
\end{equation}
%%%
We showed before that we could turn the free Liouvillian 
$\HCT$ into the Liouvillian $\HCT_{\s A}$ of (\ref{ann.oneoneone}) by just doing the 
substitution (\ref{ann.finiu}) $\displaystyle p_{\theta}\rightarrow p_{\theta}-\frac{e\Phi_{\s B}}{2\pi c}$. 
It is then clear
that the solutions of the eigenvalue equation:
%%%
\begin{equation}
\HCT_{\s A}\psi_{\s A}={\mathcal{E}}_{\s A}\psi_{\s A} \label{ann.sedici}
\end{equation}
%%%
can be obtained from the solutions (\ref{ann.psir}) of the free one by means of the 
same replacement (\ref{ann.finiu}). The result is:
%%%
\begin{equation}
\displaystyle
\psi_{\s A}=\frac{1}{2\pi}\widetilde{R}_{\s A}(\rho,p_{\rho})\delta\biggl(p_{\theta}-p_{\theta}^0-\frac{e\Phi_{\s B}}{2\pi
c}\biggr)
\delta(p_z-p_z^0) \textrm{exp}(in\theta+i\lambda_z^0z). \label{ann.psir2}
\end{equation}
%%%
Inserting $\psi_{\s A}$ into (\ref{ann.sedici}) we will get for $\widetilde{R}_{\s A}(\rho,p_{\rho})$ the following
equation:
%%%
\begin{equation}
\displaystyle
\biggl(-\frac{i}{\mu}p_{\rho}\frac{\partial}{\partial \rho}+\frac{p^{0}_{\theta}n}{\mu\rho^2}-\frac{i}{\mu\rho^3}
p_{\theta}^{0^2}\frac{\partial}{\partial p_{\rho}}+\frac{\lambda_z^0p_z^0}{\mu}-{\mathcal{E}}_{\s
A}\biggr)\widetilde{R}_{\s A}(\rho,p_{\rho})=0.
\label{ann.diciotto}
\end{equation}
%%%
(\ref{ann.diciotto}) is the same equation as the free one 
(\ref{ann.besselfree}) and as a consequence the eigenvalues ${\mathcal{E}}_{\s A}$ are the same as the 
eigenvalues ${\mathcal{E}}$ obtained from (\ref{ann.besselfree}). 
Therefore the spectrum of the Liouville operator {\it is not changed} by the presence 
of the gauge potential in the shell. Of course the two eigenfunctions which have the same eigenvalues ${\mathcal{E}}=
{\mathcal{E}}_{\s A}$ are different because they are labeled by different eigenvalues of the operator $\widehat{p}_{\theta}$.
In fact the eigenfunction $\psi$ of (\ref{ann.psir}) has eigenvalue $p_{\theta}^0$ 
while the eigenfunction $\psi_{\s A}$ of (\ref{ann.psir2}) has eigenvalue 
$\displaystyle p_{\theta}^0+\frac{e\Phi_{\s B}}{2\pi c}$. 
So the two eigenfunctions are related by a shift in one of their ``classical"
numbers $p^0_{\theta}$.
The difference with the quantum case is that the corresponding equations 
(\ref{ann.besselr}) and (\ref{ann.besselmag}) cannot
be turned one into the other like in the KvN case because in (\ref{ann.besselr})
$m$ is an integer and not a continuous real eigenvalue like $p^0_{\theta}$ is in the KvN case.
Also in CM there is an integer eigenvalue, $n$ for $\displaystyle 
-i\frac{\partial}{\partial\theta}$, but as a consequence of the MC rules in KvN Hilbert space, the only 
difference between the free and the interacting case is in the eigenvalue $p_{\theta}^0$ and not in $n$.
The reader may object that, even if the spectrum of the classical Liouvillian is the same in the two cases, 
the eigenfunctions  are different and then
the evolution may lead to different results. Actually it is not so because, as we see from (\ref{ann.sevensix}), 
to reconstruct the evolution of the wave functions we have to integrate
over  all the possible eigenvalues  
which label the eigenfunctions. In our case the different eigenfunctions (\ref{ann.psir})-(\ref{ann.psir2}) 
have only the ``classical" number $p_{\theta}^0$ shifted. Since $p_{\theta}^0$
can assume every real number, when in
(\ref{ann.sevensix}) we integrate over all the $p_{\theta}^0$ a shift in them has no effect on the final result
and the evolution of the classical probability densities is left unchanged by the gauge potential.
We feel that this proof that the spectrum of the classical KvN operator is unchanged by 
the presence of the gauge potential in the shell,
while the spectrum of the Schr\"odinger operator is changed, is a very convincing proof
of the AB phenomenon. 

This concludes the first part of the thesis on the original KvN formulation
of CM. In the following chapters we will include and analyse the geometrical and physical role of the 
differential forms appearing in the Classical Path Integral which, besides a functional counterpart of the 
KvN operatorial theory, is also one of its possible generalizations.

%% file: chapter3.tex
\def \HT{{\mathcal H}}
\def \LT{{\mathcal L}}
\def \ET{{\widetilde{\mathcal E}}}
\def \HCT{\widehat{\mathcal H}}
\def \s{\scriptscriptstyle}
\def\stackrel#1#2{\mathrel{\mathop{#2}\limits^{#1}}}
\newcommand{\quattrova}{($\varphi^{a},c^{a},\lambda_{a},{\bar c}_{a}$)}
\newcommand{\treva}{($\varphi^{a},c^{a},{\bar c}_{a}$)}

\pagestyle{fancy}
\pagestyle{fancy}
\chapter*{\begin{center}
3. Geometrical Aspects of the Classical Path Integral
\end{center}}
\addcontentsline{toc}{chapter}{\numberline{3}Geometrical Aspects of the Classical Path Integral}
\setcounter{chapter}{3}
\setcounter{section}{0}
\markboth{{\it{3. Geometrical Aspects of the Classical Path Integral}}}{}

\begin{quote}
{\it{
I am coming more and more to the conviction that the necessity of our geometry cannot be 
demonstrated, at least neither by, nor for, the human intellect... geometry should be
ranked, not with arithmetic, which is purely aprioristic, but with mechanics.}}\medskip\\
-{\bf Carl Friedrich Gauss}, 1817.
\end{quote}

\bigskip

\noindent In the first part of this thesis we have already shown that in the path integral formulation of CM
some auxiliary variables made their appearance besides the standard phase space ones. 
Both the geometrical \cite{Goz89}\cite{1P} and the physical meaning \cite{Liapunov}
of these variables have already been studied in detail. 
In particular it is possible to rewrite the entire Cartan calculus of symplectic
differential geometry by using the auxiliary variables of
the CPI and the extended Poisson
brackets introduced in (\ref{ann.epb}). 
In this chapter we will show
how it is possible to reproduce via the CPI also some generalizations of the Lie brackets
which were lacking in the original papers. These are the Schouten-Nijenhuis, the Fr\"olicher-Nijenhuis and the
Nijenhuis-Richardson brackets \cite{Kolar}. In all this geometrical construction a crucial role is 
played by the identification of the Grassmann variables present in the CPI with the differential 
forms on phase space. Grassmann variables are more or less known objects to the physics 
community but, as Coleman said in his Erice's lectures, \cite{Coleman} ``{\it 
anticommuting c-numbers are notoriously objects that make strong men quail}".
Therefore in this chapter we will show that it is possible to 
replace the anticommuting variables with less mysterious
objects such as matrices. In performing this operation all the geometrical richness of the original CPI is not lost.
In fact it is possible to map the differential forms of a 
$2n$-dimensional phase space into $2^{2n}$-dimensional vectors. Correspondently all the standard operations
of differential geometry, such as exterior derivatives, Lie derivatives
and so on, can be translated in terms of suitable combinations of Pauli and identity matrices. More details
about the work contained in this chapter can be found in \cite{1P} and \cite{8P}.

\section{Classical Path Integral and Cartan Calculus}

As we have already seen in Chapter {\bf 1} from the Lagrangian of the CPI 
(\ref{ann.suplag}) one can derive the following equations of motion for the Grassmann variables:
%%%
\begin{eqnarray}
&&\dot{c}^b=\omega^{bc}\partial_c\partial_aHc^ a\nonumber\\
&&\dot{\bar{c}}_b=-\bar{c}_a\omega^{ac}\partial_c\partial_bH.
\end{eqnarray}
%%%
Therefore the infinitesimal transformation generated by ${\cal H}$ on the phase space variables 
$\varphi^a$ is:
%%%
\begin{equation}
\displaystyle 
\varphi^{a^{\prime}}=\varphi^a+\epsilon\omega^{ab}\partial_bH
\end{equation}
%%%
and the corresponding transformations on $c$ and $\bar{c}$ are given by:
%%%
\begin{eqnarray}
&&c^{a^{\prime}}=c^a+\epsilon\omega^{ac}\partial_c\partial_bHc^b=
\frac{\partial\varphi^{a^{\prime}}}{\partial\varphi^b}c^b\nonumber\\
&&\bar{c}_a^{\prime}=\bar{c}_a
-\epsilon\bar{c}_b\omega^{bc}\partial_c\partial_aH=\frac{\partial\varphi^b}
{\partial\varphi^{a^{\prime}}}\bar{c}_b.
\end{eqnarray}
%%%
From these equations we notice that, under the diffeomorphism generated by $\HT$, $c^a$ transforms as a basis for 
the differential forms 
$d\varphi^a$, while $\bar{c}_a$ transforms as a basis for the vector fields 
$\displaystyle \frac{\partial}{\partial \varphi^a}$, see \cite{Goz89}\cite{Regini}. This is 
a first clear indication of the geometrical richness of the CPI and we will use it in a little while.

Before going on we want to underline that in \cite{Schw} the 
space whose coordinates are $(\varphi^{a}, c^{a})$ is called  
{\it reversed parity tangent bundle} and it is indicated with $\Pi T{\cal M}$. The specification
``{\it reversed parity}" 
is due to the fact that the $c^{a}$ are Grassmann variables. According to (\ref{ijmpa.comm}) $(\lambda_{a},
{\bar c}_{a})$ are the ``momenta" associated to ($\varphi^a,c^a$). 
Therefore $\varphi^a$, $\lambda_a$, $c^a$ and $\bar{c}_a$ span the cotangent bundle to the
reversed parity tangent bundle $T^{\star}(\Pi T{\cal M})$.
For more details about this we refer the interested reader to \cite{Regini}.
Since the superspace of the CPI is a cotangent bundle it has the Poisson
structure we found in (\ref{ann.epb}).
In the remaining part of this section we will show how to reproduce all the abstract Cartan calculus
via these Poisson structures and the Grassmann variables. 
First of all we think that, even if the auxiliary variables $\lambda_a,c^a, \bar{c}_a$ 
have a well-defined geometrical
meaning \cite{Goz89}\cite{Regini}, still the reader could claim that they are somehow
redundant because we can do classical mechanics by using only the standard phase space variables $\varphi^a$. 
This redundancy is actually signaled by the presence of some {\it universal} symmetries \cite{Goz89} whose charges
are:
%%%
\begin{eqnarray}
&&\displaystyle \label{jmp.eq:ventuno} Q\equiv ic^a\lambda_a, \;\;\;\;\;\;\;\;\;
\overline{Q}\equiv  i\bar{c}_a\omega^{ab}\lambda_b,\nonumber\\
&&Q_f\equiv c^a\bar{c}_a,\;\;\;\;K\equiv \frac{1}{2}\omega_{ab}c^ac^b,\;\;\;\; 
\overline{K}\equiv \frac{1}{2}
\omega^{ab}\bar{c}_a\bar{c}_b. 
\end{eqnarray}
where $\omega_{ab}$ are the elements of the inverse matrix of $\omega^{ab}$.
%%%
There are also the following {\it supersymmetry} charges \cite{Goz89}\cite{Deotto}:
%%%
\begin{eqnarray}
&&Q_{\scriptscriptstyle H}=Q-\beta N=ic^a\lambda_a-\beta 
c^a\partial_aH\nonumber\\
&&\overline{Q}_{\scriptscriptstyle H}=\overline{Q}+
\beta\overline{N}=i\bar{c}_a\omega^{ab}\lambda_b+\beta\bar{c}_a\omega^{ab}
\partial_bH
\end{eqnarray}
%%%
whose anticommutator gives $\HT$:
%%%
\begin{equation}
[Q_{\scriptscriptstyle H},\overline{Q}_{\scriptscriptstyle H}]_{\scriptscriptstyle +}
=2i\beta\HT.
\end{equation}
%%%

Since the $c^{a}$ transform
as a basis for forms $d\varphi^{a}$ and $\bar{c}_a$ 
as a basis for vector fields $\displaystyle \frac{\partial}
{\partial \varphi^{a}}$, one can start building the following map, called ``hat"
map $\wedge$:
%%%
\begin{eqnarray}
\label{jmp.eq:ventisei}
\alpha=\alpha_{a}d\varphi ^{a} & \hat{\longrightarrow} &  {\widehat\alpha}\equiv
\alpha_{a}c^{a}\\
\label{jmp.eq:ventisette}
V=V^{a}\partial_{a}  & \hat{\longrightarrow} & {\widehat V}\equiv V^{a}{\bar
c}_{a}.
\end{eqnarray}  
%%%
\noindent It is  actually a much more general map between forms $\alpha$, antisymmetric
tensors $V$ 
and functions of $\varphi, c, \bar{c}$:
%%%
\begin{eqnarray}
\label{jmp.eq:ventotto}
F^{\scriptscriptstyle (p)}={\frac{1}{p!}}F_{a_{1}\cdots a_{p}}d\varphi ^{a_{1}}\wedge\cdots\wedge
d\varphi ^{a_{p}}  & \hat{\longrightarrow} &{\widehat F}^{\scriptscriptstyle (p)}\equiv {\frac{1}{p!}}
F_{a_{1}\cdots a_{p}}c^{a_{1}}\cdots c^{a_{p}}\\
\label{jmp.eq:ventinove}
V^{\scriptscriptstyle (p)}={\frac{1}{p!}}V^{a_{1}\cdots a_{p}}\partial_{a_{1}}\wedge\cdots\wedge 
\partial_{a_{p}} & \hat{\longrightarrow} & {\widehat V}^{\scriptscriptstyle (p)}\equiv {\frac{1}{p!}}V^{a_{1}
\cdots a_{p}}{\bar c}_{a_{1}}\cdots {\bar c}_{a_{p}}.
\end{eqnarray}  
%%%
\noindent Once the correspondence (\ref{jmp.eq:ventisei})-(\ref{jmp.eq:ventinove}) is established 
we can easily find out how to implement the various Cartan operations 
such as the exterior derivative {\bf d}
of a form, or the interior contraction between a vector field $V$ and a form $F$. It
is easy to check that \cite{Goz89}:
%%%
\begin{eqnarray}
\label{jmp.eq:trenta}
{\bf d}F^{\scriptscriptstyle (p)} & \hat{\longrightarrow} & 
i\{Q,{\widehat F}^{\scriptscriptstyle (p)}\}_{\scriptscriptstyle epb} \\
\label{jmp.eq:trentuno}
\iota_{{\scriptscriptstyle V}}F^{\scriptscriptstyle (p)} & \hat{\longrightarrow} & i\{{\widehat V},
{\widehat F}^{\scriptscriptstyle (p)}\}_{\scriptscriptstyle epb}
\\
\label{jmp.eq:trentadue}
pF^{\scriptscriptstyle (p)} & \hat{\longrightarrow} & i\{Q_{f}, 
{\widehat F}^{\scriptscriptstyle (p)}\}_{\scriptscriptstyle epb}
\end{eqnarray}  
%%%
\noindent where $Q$ and $Q_f$ are the charges of (\ref{jmp.eq:ventuno}).
We can translate in our language also the usual mapping~\cite{Marsd}
between vector fields $V$
and forms $V^{\flat}$ realized by the symplectic two-form $\omega(V,0)\equiv V^{\flat}$:
%%%
\begin{equation}
V^{\flat} \;\hat{\longrightarrow}\; i\{K,{\widehat V}\}_{\scriptscriptstyle epb}
\end{equation}
%%%
or the inverse operation of building a vector field $\alpha^{\sharp}$ out of a form
$\alpha=(\alpha^{\sharp})^{\flat}$:
%%%
\begin{equation}
\label{jmp.eq:trentaquattro}
\alpha^{\sharp}\; \hat{\longrightarrow}\; i\{{\overline K},{\widehat\alpha}\}_{\scriptscriptstyle epb}
\end{equation}  
%%%
\noindent where again $K$ and $\overline{K}$ are the charges of (\ref{jmp.eq:ventuno}).
We can also translate the standard operation of building a vector field out of a function 
$f(\varphi)$:
%%%
\begin{equation}
\label{jmp.eq:trantacinque}
({\bf d}f)^{\sharp} \; \hat{\longrightarrow} \;i\{{\overline Q},f\}_{\scriptscriptstyle epb}
\end{equation}
%%%
and the Poisson brackets between two functions $f$ and $g$:
%%%
\begin{equation}
\label{jmp.eq:trentasei}
\{f,g\}_{\scriptscriptstyle pb} \;\hat{\longrightarrow}\;
i\Bigl\{\{f,Q\},\{\overline{Q},g\}\Bigr\}_{\scriptscriptstyle epb}.
\end{equation}  
%%%
\noindent The next thing to do is to translate the concept of Lie derivative along a vector field
$V$ which is defined as:  
~${\cal L}_{\scriptscriptstyle V}={\bf d}\iota_{\scriptscriptstyle V}
+\iota_{\scriptscriptstyle V}{\bf d}$. It is easy to prove that:
%%%
\begin{equation}
\label{jmp.eq:trentasette}
{\cal L}_{\scriptscriptstyle V}F^{\scriptscriptstyle (p)} \;\; \hat{\longrightarrow} \;\; 
\{-{\HT}_{\scriptscriptstyle V},{\widehat F}
^{\scriptscriptstyle (p)}\}_{\scriptscriptstyle epb}
\end{equation}  
%%%
\noindent where ${\HT}_{\scriptscriptstyle V}=\lambda_aV^{a}+i\bar{c}_a
\partial_bV^{a}c^{b}$; note that,
for $V^{a}=\omega^{ab}\partial_bH$,  ${\HT}_{\scriptscriptstyle V}$ becomes just the Hamiltonian ${\HT}$
of (\ref{ann.supham}), which appears in the weight of the CPI. 
This tells us that ${\HT}$ 
is just the Lie derivative of the Hamiltonian flow. Since the exterior derivative ${\bf d}$ commutes with every 
Lie derivative ${\cal L}_{\scriptscriptstyle V}$ now it is also clear the reason why 
the symmetry charge $Q$ commutes with the Hamiltonian $\HT$. Finally
the Lie brackets between two vector fields $V,\;W$ is another vector field given by:
%%%
\begin{equation}
\label{jmp.eq:trentotto}
[V,W]_{Lie\,brack} \; \hat{\longrightarrow} \; \{-\HT_{\scriptscriptstyle 
V},{\widehat W}\}_{\scriptscriptstyle epb}.
\end{equation} 
%%%
In the next section we will show that at least
three generalizations of the Lie brackets, which are well-known in the literature on differential geometry,
can be reproduced and unified within the formalism of the CPI.
%%%%%%%%%%%%%%%%%%%%%%%%%%%%%%%%%%%%%%%%%%%%%%%%%%%%%%%
%%%%%%%%%%%%%%%%%%%%%%%%%%%%%%%

\bigskip

\section{Generalizations of the Lie Brackets}

\subsection{Schouten-Nijenhuis (SN) Brackets}

The first generalization of the Lie brackets we will analyse in this section is given by
the so-called Schouten-Nijenhuis (SN) brackets: they involve two {\it multivector fields}  and they reduce to the usual Lie 
brackets in the case of vector fields.
As the Lie brackets map two vector fields $X$ and $Y$ into another vector field $[X,Y]$, so the SN 
brackets map two multivector fields of rank $p$ ($P=X_{\scriptscriptstyle (1)}\wedge\cdots\wedge
X_{\scriptscriptstyle (p)}$)
and $r$ ($R=Y_{\scriptscriptstyle (1)}\wedge\cdots\wedge Y_{\scriptscriptstyle (r)}$) 
into a multivector field $[\,\cdot\,,\,\cdot\,]_{\scriptscriptstyle SN}$ of rank $p+r-1$ via the
following rule~\cite{vaisman}:
%%%
\begin{eqnarray}
&&\qquad\qquad [\,\cdot\,,\,\cdot\,]_{\scriptscriptstyle SN}: \; {\cal V}^p({{\cal M}})
\times{\cal V}^r({{\cal M}}) \; 
\longrightarrow \; {\cal V}^{p+r-1}({{\cal M}})
\nonumber\\
&&[P,R]_{\scriptscriptstyle SN}
\equiv \sum_{i=1}^{p}(-1)^{i+1}X_{\scriptscriptstyle (1)}\wedge\cdots
\wedge{\widehat {\widehat  X}} _{\scriptscriptstyle (i)}\cdots\wedge X_{\scriptscriptstyle (p)}\wedge[X_{\scriptscriptstyle
(i)},R] 
\end{eqnarray}  
%%%
\noindent where ${\cal V}^{s}$ indicates the space of multivector fields of rank
$s$, the double hat ${\widehat {\widehat  X}}_{\scriptscriptstyle (i)}$ indicates that we have removed
$X_{\scriptscriptstyle (i)}$ and
$[X_{\scriptscriptstyle (i)},R]={\cal L}_{\scriptscriptstyle X_{\scriptscriptstyle (i)}}R$ is the 
Lie derivative of a multivector which can be defined in terms of the Lie brackets $[X_{\scriptscriptstyle
(i)},Y_{\scriptscriptstyle (j)}]$ between usual vector fields as:
%%%
\begin{equation}
{\cal L}_{{\scriptscriptstyle X}_{\scriptscriptstyle (i)}}R=
\sum_{j=1}^rY_{\scriptscriptstyle (1)}\wedge\cdots\wedge[X_{\scriptscriptstyle (i)},Y_{\scriptscriptstyle (j)}]
\wedge\cdots\wedge Y_{\scriptscriptstyle (r)}. \label{jmp.liederivative}
\end{equation}  
%%%
Now the Lie derivative along a vector field of a 
multivector (\ref{jmp.liederivative}) can be translated in our language
as\footnote[1]{Since now on we will omit the ``epb"
acronym for our brackets $\{\,,\,\}$.}:
%%%
\begin{equation}
{\cal L}_{\scriptscriptstyle X_{\scriptscriptstyle (i)}}R \; \hat{\longrightarrow} \; \{-\HT_{
\scriptscriptstyle X_{\scriptscriptstyle (i)}},
\widehat{R}\}\equiv
-\Bigl\{\{\widehat{X}_{\scriptscriptstyle (i)}, Q\},
\widehat{R}\Bigr\}
\end{equation}  
%%%
where $\widehat{R}=Y^{j_1}_{\scriptscriptstyle (1)}\bar{c}_{j_1}\cdots Y^{j_r}_{\scriptscriptstyle (r)}\bar{c}_{j_r}=
\widehat{Y}_{\scriptscriptstyle (1)}\widehat{Y}_{\scriptscriptstyle (2)}\cdots\widehat{Y}_{\scriptscriptstyle (r)}$.
\noindent In fact:
%%%
\begin{eqnarray}
\{-\HT_{\scriptscriptstyle X_{\scriptscriptstyle (i)}},\widehat{R}\}&\hspace{-0.2mm}=&\hspace{-0.2mm}
\{-\HT_{\scriptscriptstyle X_{\scriptscriptstyle (i)}},\widehat{Y}_{\scriptscriptstyle (1)}
\widehat{Y}_{\scriptscriptstyle (2)}\cdots
\widehat{Y}_{\scriptscriptstyle (r)}\}=
\sum_{j=1}^r\widehat{Y}_{\scriptscriptstyle (1)}\widehat{Y}_{\scriptscriptstyle (2)}
\cdots\{-\HT_{\scriptscriptstyle X_{\scriptscriptstyle (i)}},\widehat{Y}_{\scriptscriptstyle (j)}\}\cdots
\widehat{Y}_{\scriptscriptstyle (r)}=\nonumber\\
&\hspace{-0.2mm}=&\hspace{-0.2mm}\sum_{j=1}^r\widehat{Y}_{\scriptscriptstyle (1)}\widehat{Y}_{\scriptscriptstyle (2)}
\cdots\Bigl([X_{\scriptscriptstyle (i)},Y_{\scriptscriptstyle (j)}]\Bigr)^{\wedge}\cdots
\widehat{Y}_{\scriptscriptstyle (r)}=({\cal L}_{\scriptscriptstyle X_{\scriptscriptstyle (i)}}R)^{\wedge}.
\end{eqnarray}  
%%%
\noindent We note that the extended Poisson brackets between $-\HT_{\scriptscriptstyle X_{\scriptscriptstyle (i)}}$
and $\widehat{R}$ take automatically into account the sum over $j$ which appears 
in the definition of the Lie derivative of a multivector.

Now we can consider the SN brackets. According to their definition we have
%%%
\begin{equation}
[P,R]_{\scriptscriptstyle SN} \; \hat{\longrightarrow} \; 
\sum_{i=1}^p(-1)^{i+1}\widehat{X}_{\scriptscriptstyle (1)}\cdots{\widehat {\widehat  X}}_{\scriptscriptstyle (i)}
\cdots\widehat{X}_{\scriptscriptstyle (p)}\Bigl\{-\{\widehat{X}_{\scriptscriptstyle (i)}, Q\},\widehat{R}\Bigr\}. 
\label{jmp.SNlong}
\end{equation}  
%%%
\noindent The previous formula can be written in a very compact way as:
%%%
\begin{equation}
\label{jmp.compact}
[P,R]_{\scriptscriptstyle SN} \;\; \hat{\longrightarrow} \;\; -\Bigl\{\{Q,\widehat {P}\},\widehat {R}\Bigr\}.
\end{equation}   
%%%
\noindent In fact:
%%%
\begin{eqnarray}
-\Bigl\{\{Q,\widehat {P}\},\widehat {R}\Bigr\}&\hspace{-0.2mm}=&\hspace{-0.2mm}
-\Bigl\{\{Q,\widehat{X}_{\scriptscriptstyle (1)}
\cdots\widehat{X}_{\scriptscriptstyle (p)}\},\widehat{R}\Bigr\}=\\
&\hspace{-0.2mm}=&\hspace{-0.2mm}
\sum_{i=1}^p(-)^{i+1}\widehat{X}_{\scriptscriptstyle (1)}\cdots{\widehat {\widehat  X}}_{\scriptscriptstyle (i)}
\cdots\widehat{X}_{\scriptscriptstyle (p)}\Bigl\{-\{Q,\widehat{X}_{\scriptscriptstyle (i)}\},\widehat{R}\Bigr\}
=[P,R]^{\wedge}_{\scriptscriptstyle SN}. \nonumber
\end{eqnarray}  
%%%
\noindent We note that on the RHS of (\ref{jmp.compact}) we have the images, via the $\wedge$-map,
of the multivectors $P$ and $R$, which appear on the LHS of the same equation, 
and the usual BRS charge $Q$ which appears naturally also in this context. 

Like in the case of vector fields, where 
$\HT_{\scriptscriptstyle X}=
\{Q,\widehat{X}\}$,
we can define a Hamiltonian associated with a multivector field $P$
in the following way:
%%%
\begin{equation}
\HT_{\scriptscriptstyle P}=\{Q,\widehat{X}_{\scriptscriptstyle (1)}\cdots\widehat{X}_{\scriptscriptstyle (p)}\}
=\sum_{i=1}^p(-1)^{i+1}\widehat{X}_{\scriptscriptstyle (1)}\cdots{\widehat{\widehat X}}_{\scriptscriptstyle (i)}
\cdots\widehat{X}_{\scriptscriptstyle (p)}
\HT_{\scriptscriptstyle X_{\scriptscriptstyle (i)}}. \label{jmp.generelazing1}
\end{equation}  
%%%
and write the hat map of SN brackets as:
%%%
\begin{equation}
[P,R]_{\scriptscriptstyle SN} \;\; \hat{\longrightarrow} \;\; -\{\HT_
{\scriptscriptstyle P},\widehat {R}\}. \label{jmp.generelazing2}
\end{equation}  
%%%
From (\ref{jmp.generelazing1}) and (\ref{jmp.generelazing2}) we can 
notice  how 
the SN brackets become the usual Lie brackets in the case of vector fields. 

\subsection{Fr\"olicher-Nijenhuis (FN) Brackets}

The second kind of brackets we want to analyse in this section are the Fr\"olicher-Nijenhuis 
(FN) ones. These brackets associate to two {\it vector-valued
forms}\footnote[2]{Regarding the manner to
indicate the space of vector-valued forms with $\Omega^{k}({\cal M},T{{\cal M}})$ 
we follow the notation of Ref.~\cite{Kolar}.}
$K\in\Omega^{k+1}({{\cal M}};T{{\cal M}})$ 
of degree $k+1$ and $L\in\Omega^{l+1}({{\cal M}};T{{\cal M}})$ of degree $l+1$ 
a vector-valued form of degree $k+l+2$:
%%%
\begin{equation}
[\,\cdot\,,\,\cdot\,]_{\scriptscriptstyle FN}: \; \Omega^{k+1}({{\cal M}};T{{\cal M}})
\times\Omega^{l+1}({{\cal M}};T{{\cal M}})
\; \longrightarrow \; \Omega^{k+l+2}({{\cal M}};T{{\cal M}}).
\end{equation}  
%%%
\noindent They are defined in the following manner~\cite{Kolar}:

\noindent {\bf a)} let us first define the interior contraction $\iota_{\scriptscriptstyle J}$ 
with a vector-valued form $J$ of degree $j+1$, and its action on a form $\omega$ of degree $l$. As $J$
is a $(j+1)$-form, $\iota_{\scriptscriptstyle J}\omega$ is a ($j+l$)-form and
when we apply it on $j+l$ vectors, 
we obtain the following number:
%%%
\begin{eqnarray}
&&(\iota_{{\scriptscriptstyle J}}\omega)(X_{\scriptscriptstyle (1)},\cdots,X_{\scriptscriptstyle (j+l)}) \equiv \nonumber\\
&&\equiv \frac{1}{ (j+1)!(l-1)!} \sum_{\{\sigma\in S_{j+l}\}} 
(\textrm{sign}~\sigma)\;
\omega \Bigl[J(X_{\sigma(1)},\cdots,X_{\sigma(j+1)}), X_{\sigma(j+2)},\cdots, X_{\sigma(j+l)}\Bigr] \nonumber\\
&&
\end{eqnarray}  
%%%
\noindent where $S_{j+l}$ is the set of permutations of the vector fields $X_{\scriptscriptstyle (1)}
\cdots X_{\scriptscriptstyle (j+l)}$. 

\noindent {\bf b)} having a generalized interior contraction $\iota_{\scriptscriptstyle J}$, we can also define a 
generalized Lie derivative as: 
%%%
\begin{equation}
{\cal L}_{\scriptscriptstyle J} \; = \; [\iota_{\scriptscriptstyle J}, {\bf d}]
\end{equation}  
%%%
\noindent where $[\;\cdot\; , \;\cdot\;]$ is the usual graded commutator and 
$J\in \Omega^{j+1}({{\cal M}};T{\cal M})$.

\noindent {\bf c)} Now the FN brackets are defined in the 
following implicit way:
%%%
\begin{equation}
[{\cal L}_{\scriptscriptstyle J},{\cal L}_{\scriptscriptstyle L}] \; \equiv \;
{\cal L}_{\scriptscriptstyle [J,L]_{\scriptscriptstyle FN}}
\end{equation}  
%%%
\noindent where $[{\cal L}_{\scriptscriptstyle J}, {\cal L}_{\scriptscriptstyle L}]$  
is again the graded commutator among the Lie derivatives.

How can we translate all this in our language?

\noindent {\bf a)} First of all via our $\wedge$-map the vector-valued forms 
$J\in\Omega^{j+1}({{\cal M}};T{{\cal M}})$ become:
%%%
\begin{equation}
\displaystyle
J \; \hat{\longrightarrow} \; \frac{1}{(j+1)!}\;J^{i}_{i_1i_2\cdots i_{j+1}}
\;[c^{i_1}c^{i_2}\cdots c^{i_{j+1}}][\bar{c}_i]. \label{jmp.Kmap}
\end{equation}  
%%%
Which is the $\wedge$-map of the interior contraction $\iota_{\scriptscriptstyle J}\omega$? As in the case
of the interior contraction with a vector field, we expect that:
%%%
\begin{equation}
\iota_{\scriptscriptstyle J}\omega \; \hat{\longrightarrow} \; i\{\widehat{J},\widehat{\omega}\}.
\label{jmp.Kcontr}
\end{equation}  
%%%
If we rewrite $J=\alpha\otimes X$, where $\alpha\in\Omega^{j+1}({{\cal M}})$ and $X$ is a vector field,
then $\iota_{\scriptscriptstyle X}\omega\in\Omega^{l-1}({{\cal M}})$ and we can translate the interior contraction
$\iota_{\scriptscriptstyle J}\omega$ as the exterior product between two forms:
%%%
\begin{equation}
\iota_{\scriptscriptstyle J}\omega \;\; = \;\; \alpha\wedge\iota_{\scriptscriptstyle X}\omega.
\end{equation}  
%%%
\noindent Now,
if $\alpha\in\Omega^{j+1}({{\cal M}})$ and $\beta\in\Omega^{l-1}({{\cal M}})$ 
are two differential 
forms then the hat-map of their exterior product is simply the product of the hat-map of the two forms:
%%%
\begin{equation}
(\alpha\wedge\beta) \;\; \hat{\longrightarrow} \;\; \widehat{\alpha}\widehat{\beta}. \label{jmp.alphabeta}
\end{equation}  
%%%
In the case of our interior contraction we have that:
%%%
\begin{equation}
\label{chapman1}
\iota_{\scriptscriptstyle J}\omega \;\; = \;\; \alpha\wedge\iota_{\scriptscriptstyle X}\omega \;\; 
\hat{\longrightarrow} \;\; \widehat{\alpha}
(\iota_{\scriptscriptstyle X}\omega)^{\wedge}.
\end{equation}  
%%%
\noindent Using (\ref{jmp.eq:trentuno}) and $\{\widehat{\alpha},\widehat{\omega}\}=0$ we can go on 
writing:
%%%
\begin{equation}
\label{chapman2}
\widehat{\alpha}(\iota_{\scriptscriptstyle 
X}\omega)^{\wedge}=i\widehat{\alpha}\{\widehat{X},\widehat{\omega}\}=
i\{\widehat{\alpha}\widehat{X},\widehat{\omega}\}=i\{\widehat{J},\widehat{\omega}\}.
\end{equation}  
%%%
\noindent Inserting (\ref{chapman2}) into the RHS of (\ref{chapman1}) we finally obtain:
%%%
\begin{equation}
\iota_{\scriptscriptstyle J}\omega \;\;\; \hat{\longrightarrow} \;\;\; 
i\{\widehat{J},\widehat{\omega}\}
\end{equation}  
%%%
which is what we wanted to prove.

\noindent {\bf b)} At this point, having defined the concept of interior contraction with a
vector-valued form, we can go on finding out which is the mapping of the Lie derivative associated with a vector-valued 
form $J$:
%%%
\begin{equation}
{\cal L}_{\scriptscriptstyle J}\;\;=\;\;[\iota_{\scriptscriptstyle J},{\bf d}]
\;\;=\;\;\iota_{\scriptscriptstyle J}{\bf d}
+(-1)^{j+1}{\bf d}\iota_{\scriptscriptstyle J}.
\end{equation}  
%%%
\noindent Since we know how to translate  in our language both the interior contraction with a vector-valued 
form and the exterior
derivative, we have the following mapping:
%%%
\begin{eqnarray}
{\cal L}_{\scriptscriptstyle J}\omega \;& \hat{\longrightarrow} &\; 
i\{\widehat{J},({\bf d}\omega)^{\wedge}
\}+(-1)^{j+1}i\{Q,(\iota_{\scriptscriptstyle J}\omega)^{\wedge}\}=\nonumber\\
&&=-\Bigl\{\widehat{J},\{Q,\widehat{\omega}\}\Bigr\}+(-1)^{j}\Bigl\{Q,\{\widehat{J},\widehat{\omega}\}\Bigr\}
=-\Bigl\{\{\widehat{J},Q\},\widehat{\omega}\Bigr\}
\end{eqnarray}  
%%%
\noindent where, in the last step, we have used the Jacobi identity.
So we have:
%%%
\begin{equation}
{\cal L}_{\scriptscriptstyle J}\omega \; \hat{\longrightarrow} \; -\{\HT_
{\scriptscriptstyle J},\widehat{\omega}\} \label{jmp.ellacca}
\end{equation}  
%%%
\noindent where we have  defined, as usual, $\HT_{\scriptscriptstyle J}=
\{\widehat{J},Q\}$. From this definition and from (\ref{jmp.Kmap}) it follows that the explicit expression
of $\HT_{\scriptscriptstyle J}$ is:
%%%
\begin{equation}
\displaystyle
\HT_{\scriptscriptstyle J}=\frac{1}{(j+1)!}\Biggl(\lambda_iJ^i_{i_1i_2\cdots i_{j+1}}
+i\bar{c}_i(\partial_dJ^{i}_{i_1i_2\cdots i_{j+1}}c^d)\Biggr)c^{i_1}\cdots c^{i_{j+1}}.
\end{equation}  
%%%
We note that if $j$ is Grassmannian even then $\HT_
{\scriptscriptstyle J}$ is odd and if $j$ is odd then $\HT_
{\scriptscriptstyle J}$ is even. Moreover, from (\ref{jmp.ellacca}), the Grassmannian parity 
of $\HT_{\scriptscriptstyle J}$ coincides 
with that of the correspondent Lie derivative ${\cal L}_{\scriptscriptstyle J}$.

\noindent {\bf c)} Finally we have all the elements to translate in our language the FN brackets.
They are defined~\cite{Kolar} in implicit way by the equation:
%%%
\begin{equation}
[{\cal L}_{\scriptscriptstyle J},{\cal L}_{\scriptscriptstyle L}] \;\; = \;\;
{\cal L}_{\scriptscriptstyle
[J,L]_{\scriptscriptstyle FN}}. \label{jmp.FNdef}
\end{equation}  
%%%
\noindent Now if we apply the LHS of (\ref{jmp.FNdef}) on a generic 
form $\omega$ we have:
%%%
\begin{equation}
[{\cal L}_{\scriptscriptstyle J},{\cal L}_{\scriptscriptstyle L}]\omega=
({\cal L}_{\scriptscriptstyle J}{\cal L}_{\scriptscriptstyle L})\omega
-(-1)^{[\HT_{\scriptscriptstyle L}][\HT_{\scriptscriptstyle J}]}
({\cal L}_{\scriptscriptstyle L}{\cal L}_{\scriptscriptstyle J})\omega
\end{equation}  
%%%
where $[\HT]$ is the Grassmannian parity of $\HT$.
Via the hat-map we obtain:
%%%
\begin{eqnarray}
\label{jmp.comparison1}
[{\cal L}_{\scriptscriptstyle J},{\cal L}_{\scriptscriptstyle L}]\omega & \hat{\longrightarrow} &
\Bigl\{\HT_{\scriptscriptstyle J},\{\HT_{\scriptscriptstyle L},
\widehat{\omega}\}\Bigr\}-(-1)^{[\HT_{\scriptscriptstyle L}][\HT_
{\scriptscriptstyle J}]}\Bigl\{\HT_{\scriptscriptstyle L},
\{\HT_{\scriptscriptstyle J},\widehat{\omega}\}\Bigr\}=\nonumber\\
&&=\Bigl\{\{\HT_
{\scriptscriptstyle J},\HT_{\scriptscriptstyle L}\},\widehat{\omega}\Bigr\}
\end{eqnarray}  
%%%
\noindent where in the last step we have used, as usual, the Jacobi identity.
From (\ref{jmp.ellacca}) we have that the RHS of (\ref{jmp.FNdef}) can be translated as:
%%%
\begin{equation}
\label{jmp.comparison2}
{\cal L}_{\scriptscriptstyle [J,L]_{\scriptscriptstyle FN}}\omega \; \hat{\longrightarrow} \;
-\{\HT_{\scriptscriptstyle [J,L]_{\scriptscriptstyle FN}},\widehat{\omega}\}.
\end{equation}  
%%%
\noindent Comparing (\ref{jmp.comparison1}) and (\ref{jmp.comparison2}) we obtain:
%%%
\begin{equation}
\label{jmp.FNimportant}
\Bigl\{([J,L]_{\scriptscriptstyle FN})^{\wedge},Q\Bigr\}=\HT_{\scriptscriptstyle [J,L]_{\scriptscriptstyle FN}}
=-\{\HT_
{\scriptscriptstyle J},\HT_{\scriptscriptstyle L}\}.
\end{equation}  
%%%
\noindent Therefore if we want the correct representation of the FN brackets, we have to write
$\{\HT_{\scriptscriptstyle J},\HT_{\scriptscriptstyle L}\}$ as 
$\{\,\cdot\, , Q\}$.
This is not difficult to do. In fact, using the nilpotency of $Q$ and the Jacobi identities one obtains:
%%%
\begin{equation}
\{\HT_{\scriptscriptstyle J},\HT_{\scriptscriptstyle L}\}=
\Bigl\{\{\widehat{J},Q\},\{\widehat{L},Q\}\Bigr\}=\Bigl\{\{\{\widehat{J},Q\},\widehat{L}\},Q\Bigr\}. \label{jmp.Qform}
\end{equation}  
%%%
Substituting (\ref{jmp.Qform}) into the RHS of (\ref{jmp.FNimportant}) we finally obtain:
%%%
\begin{equation}
[J,L]_{\scriptscriptstyle FN} \; \hat{\longrightarrow} \; 
=-\{ \HT_{\scriptscriptstyle J},\widehat{L} \}.
\end{equation}  
%%%
\noindent We notice that, if $J$ and $L$ are vector-valued zero-forms, i.e.
if they are usual vector fields, then the FN brackets reduce to the usual Lie ones. So
we can say that, while the SN brackets generalize Lie brackets in the case of 
multivector fields, 
the FN ones generalize the Lie brackets in the case of vector-valued forms. 

\subsection{Nijenhuis-Richardson (NR) Brackets}

The last brackets we will analyse in this section are the Nijenhuis-Richardson (NR) ones:
they map two vector-valued forms
$J\in\Omega^{j+1}({{\cal M}};T{{\cal M}})$ and $L\in\Omega^{l+1}({{\cal M}};T{{\cal M}})$ to another vector-valued form of 
degree $j+l+1$ defined 
in an implicit way as \cite{Kolar}:
%%%
\begin{eqnarray}
&&[\,\cdot\, ,\,\cdot\,]_{\scriptscriptstyle NR}: \; \Omega^{j+1}({{\cal M}};T{{\cal M}})
\times\Omega^{l+1}({{\cal M}};T{{\cal M}}) \;
\longrightarrow \; \Omega^{j+l+1}({{\cal M}};T{{\cal M}})\nonumber\\
&&\qquad\qquad\qquad\qquad\iota_{\scriptscriptstyle [J,L]_{\scriptscriptstyle NR}}\; \equiv 
\;[\iota_{\scriptscriptstyle J},
\iota_{\scriptscriptstyle L}]. \label{jmp.NRdef}
\end{eqnarray}  
%%%
\noindent Let us now apply on a generic form $\omega\in\Omega^m({{\cal M}})$ the LHS of (\ref{jmp.NRdef})
and let us use the hat-map. We obtain:
%%%
\begin{equation}
\iota_{\scriptscriptstyle [J,L]_{\scriptscriptstyle NR}}\omega \;\;\; \hat{\longrightarrow} \;\;\; i\Bigl\{([J,L]_
{\scriptscriptstyle NR})^{\wedge},\widehat{\omega}\Bigr\} \label{jmp.alpha}.
\end{equation}  
%%%
\noindent Instead if we apply on $\omega$ the RHS of (\ref{jmp.NRdef}) we get:
%%%
\begin{eqnarray}
[\iota_{\scriptscriptstyle J},\iota_{\scriptscriptstyle L}]\omega&=&
\iota_{\scriptscriptstyle J}(\iota_{\scriptscriptstyle L}\omega)-(-1)^{jl}
\iota_{\scriptscriptstyle L}(\iota_{\scriptscriptstyle J}\omega) 
\;\hat{\longrightarrow}\; i\{\widehat{J},(\iota_{\scriptscriptstyle L}\omega)^{\wedge}\}
-(-1)^{jl}i\{\widehat{L},(\iota_{\scriptscriptstyle J}\omega)^{\wedge}\}=\nonumber\\
&=&-\{\widehat{J},\{\widehat{L},\widehat{\omega}\}\}+(-1)^{jl}\{\widehat{L},\{\widehat{J},
\widehat{\omega}\}\}.
\end{eqnarray}  
%%%
\noindent Using the Jacobi identity we can rewrite the previous equation as:
%%%
\begin{equation}
[\iota_{\scriptscriptstyle J},\iota_{\scriptscriptstyle L}]\omega \;\;\; \hat{\longrightarrow} \;\;\;
-\Bigl\{\{\widehat{J},\widehat{L}\},\widehat{\omega}\Bigr\}. \label{jmp.beta}
\end{equation}  
%%%
\noindent and, from the definition (\ref{jmp.NRdef}) and the comparison of (\ref{jmp.alpha}) with (\ref{jmp.beta}),
we obtain the following hat-map for the NR brackets:
%%%
\begin{equation}
[J,L]_{\scriptscriptstyle NR} \;\;\; \hat{\longrightarrow} \;\;\; i\{\widehat{J},\widehat{L}\}.
\label{jmp.gamma}
\end{equation}  
%%%
\noindent So the NR brackets between two vector-valued forms are just proportional to the extended Poisson
brackets of the vector-valued forms themselves.

We can now summarize all SN, FN, NR brackets in the following very compact way:
%%%
\begin{eqnarray}
\label{centrale}
&&[P,R]_{\scriptscriptstyle SN} \;\; \hat{\longrightarrow} \;\; -\{
{\cal H}_{\scriptscriptstyle P},\widehat {R}\}_{\scriptscriptstyle epb}\nonumber\\
&&[J,L]_{\scriptscriptstyle FN} \;\; \hat{\longrightarrow} \;\; 
-\{{\cal H}_{\scriptscriptstyle J}, \widehat{L}\}_{\scriptscriptstyle epb}\\
&&[J,L]_{\scriptscriptstyle NR} \;\; \hat{\longrightarrow} \;\;\;\;\; i\{\widehat{J},\widehat{L}\}_{\scriptscriptstyle
epb} \nonumber
\end{eqnarray}
%%%
\noindent where:
%%%
\begin{eqnarray}
&&P=Y_{\scriptscriptstyle (1)}\wedge\cdots\wedge Y_{\scriptscriptstyle (r)} \;\;\;\;\,
\hat{\longrightarrow} \;\;\;
Y_{\scriptscriptstyle (1)}^{j_1}\bar{c}_{j_1}\cdots
Y_{\scriptscriptstyle (r)}^{j_r}\bar{c}_{j_r} \nonumber \\
&&R=Y_{\scriptscriptstyle (1)}\wedge\cdots\wedge Y_{\scriptscriptstyle (r)} \;\;\;\;\, \hat{\longrightarrow} \;\;\;
Y_{\scriptscriptstyle (1)}^{j_1}\bar{c}_{j_1}\cdots
Y_{\scriptscriptstyle (r)}^{j_r}\bar{c}_{j_r} \nonumber \\
&&{\cal H}_{\scriptscriptstyle P}=\{Q,\widehat{X}_{\scriptscriptstyle (1)}\cdots
\widehat{X}_{\scriptscriptstyle (p)}\}
=\sum_{i=1}^p(-1)^{i+1}\widehat{X}_{\scriptscriptstyle (1)}\cdots{\widehat{\widehat X}}_{\scriptscriptstyle 
(i)}\cdots\widehat{X}_{\scriptscriptstyle (p)}
{\cal H}_{\scriptscriptstyle X_{\scriptscriptstyle (i)}}\nonumber\\
&&\displaystyle
{\cal H}_{\scriptscriptstyle J}=\frac{1}{(j+1)!}\Biggl(\lambda_iJ^i_{i_1i_2\cdots i_{j+1}}
+i\bar{c}_i(\partial_dJ^{i}_{i_1i_2\cdots i_{j+1}}c^d)\Biggr)c^{i_1}\cdots c^{i_{j+1}}\\
&&J\;\in\Omega^{j+1}({{\cal M}};T{{\cal M}}) \;\;\;\;\; \hat{\longrightarrow} \;\;  
\frac{1}{(j+1)!}J^{i}_{i_1i_2\cdots i_{j+1}}[c^{i_1}c^{i_2}\cdots c^{i_{j+1}}]
[\bar{c}_i] \nonumber\\
&&L\;\;\in\Omega^{l+1}({{\cal M}};T{{\cal M}}) \;\;\;\;\;\, \hat{\longrightarrow} \;\;
\frac{1}{(l+1)!}\;\,L^{j}_{j_1j_2\cdots j_{l+1}}[c^{j_1}c^{j_2}\cdots c^{j_{l+1}}]
[\bar{c}_j]. \nonumber
\end{eqnarray} 
%%%
By looking
at (\ref{centrale}) one immediately realizes that we  have
reduced {\it three} different generalizations of the Lie brackets, like the SN, FN
and NR ones, to the
epb of (\ref{ann.epb}). Basically to change the brackets we have only to consider, as
entries of the epb, different functions of the CPI variables. In this sense we have provided a unifying structure 
for all the possible generalizations of the standard Lie brackets between vector fields. Further calculation
details about the content of this section can be found in \cite{1P}.

\bigskip

\section{Grassmann Variables and Matrices in $n=1$ Symplectic Manifolds}
In the previous sections we have implemented and analysed all the geometrical aspects of the CPI
and, in doing this, a very important role was played by the variables $c^a$, $\bar{c}_a$ which 
are the generators of a Grassmann algebra. It is well-known
that every Grassmann algebra can be realized in terms of suitable square matrices, 
see for example \cite{Henneaux}.
This matrix realization was already used in supersymmetric
quantum mechanics \cite{Witten}\cite{Salomonson}. What we want to see now is
whether the same matrix realization can be used
also in the case of the CPI.
The content of the remaining part of this chapter can be found also in \cite{8P}.

First of all let us notice that the CPI can be used to compute transition probabilities 
and classical amplitudes \cite{Goz89}. In fact let us consider:
%%%
\begin{equation}
\displaystyle K(\varphi_f,c_f,t_f|\varphi_i,c_i,t_i)=\int {\cal D}^{\prime\prime}\varphi{\cal D}\lambda
{\cal D}^{\prime\prime}c{\cal D}\bar{c}\,\textrm{exp}\Bigl[i\int_{t_i}^{t_f}dt\,{\cal L}\Bigr] \label{zoe}
\end{equation}
%%%
with the boundary conditions
%%%
\begin{equation}
\varphi^a(t_i)=\varphi^a_i,\qquad\varphi^a(t_f)=\varphi^a_f,\qquad c^a(t_i)=c^a_i,\qquad c^a(t_f)=c^a_f.
\end{equation}
%%%
It is easy to realize that such a path integral makes the evolution of the generalized wave functions
%%%
\begin{equation}
\psi(\varphi,c)=\psi(\varphi)+\psi_a(\varphi)c^a+\psi_{ab}(\varphi)c^ac^b+\ldots
\end{equation}
%%%
where $\psi(\varphi)$, $\psi_a(\varphi)$ and $\psi_{ab}(\varphi)$ are complex functions of $\varphi$. 
In the case 
$n=1$, i.e. $\varphi^a=(q,p)$, $c^a=(c^q,c^p)$, these generalized wave functions become:
%%%
\begin{equation}
\displaystyle 
\psi(\varphi,c)=\psi_{\scriptscriptstyle 0}(\varphi)+\psi_q(\varphi)c^q+\psi_p(\varphi)c^p+
\psi_{\scriptscriptstyle 2}(\varphi)c^pc^q.  \label{matrix.genw}
\end{equation}
%%%
The 4 arbitrary functions of the phase space variables 
$\psi_{\scriptscriptstyle 0}, \psi_q,\psi_p$ and $\psi_{\scriptscriptstyle 2}$
can be thought of as the components of a 4-vector:
%%%
\begin{equation}
\displaystyle 
\psi(\varphi,c)=\psi_{\scriptscriptstyle 0}(\varphi)+\psi_q(\varphi)c^q+\psi_p(\varphi)c^p+
\psi_{\scriptscriptstyle 2}(\varphi)c^pc^q\equiv
\left( \begin{array}{c}
\psi_{\scriptscriptstyle 0}\\ \psi_q\\ \psi_p\\ \psi_{\scriptscriptstyle 2} \end{array}\right).  \label{matrix.genw2}
\end{equation}
%%%
With this choice it is then possible to represent every operator of the theory as a 
$4\times 4$ matrix. For example
if we apply the operator of multiplication by $c^q$ on $\psi$ we get
$\psi^{\prime}=c^q\psi=c^q\psi_{\scriptscriptstyle 0}+c^qc^p\psi_p$.
This wave function $\psi^{\prime}$, in the 4-vector notation (\ref{matrix.genw2}), has the form:
%%%
\begin{equation}
\psi^{\prime}= \left( \begin{array}{c} 0\\ 
\psi_{\scriptscriptstyle 0}\\ 0 \\ 
-\psi_p \end{array}\right)
\end{equation}
%%%
and it could be obtained from the 4-vector representation of $\psi$ as
%%%
\begin{equation}
\label{adelgisa}
\psi^{\prime}=\begin{pmatrix}0 & 0 & 0 & 0\\
1 & 0 & 0 & 0\\
0 & 0 & 0 & 0\\
0 & 0 & -1 & 0\end{pmatrix} \left( \begin{array}{c}
\psi_{\scriptscriptstyle 0}\\ \psi_q\\ \psi_p\\ \psi_{\scriptscriptstyle 2} \end{array}\right).
\end{equation}
%%%
From (\ref{adelgisa}) we have that the matrix realization of the operator $c^q$
is given by\footnote[3]{Since now on
in this chapter we will put the hat symbol only to indicate the matrix realizations of the Grassmann operators.}:
%%%
\begin{equation}
\widehat{c}^q=\begin{pmatrix}0 & 0 & 0 & 0\\
1 & 0 & 0 & 0\\
0 & 0 & 0 & 0\\
0 & 0 & -1 & 0\end{pmatrix}. \label{matrix.operator1}
\end{equation}
%%%
Likewise the operator of multiplication by $c^p$ 
is represented by the following $4\times 4$ matrix:
%%%
\begin{equation}
\widehat{c}^p=\begin{pmatrix}0 & 0 & 0 & 0\\
0 & 0 & 0 & 0\\
1 & 0 & 0 & 0\\
0 & 1 & 0 & 0\end{pmatrix}. \label{matrix.operator2}
\end{equation}
%%%
$\bar{c}_q$ and $\bar{c}_p$ are instead the derivative operators 
$\displaystyle \frac{\partial}{\partial c^q}$ and 
$\displaystyle \frac{\partial}{\partial c^p}$. If we apply them on the wave function 
$\psi$ written in (\ref{matrix.genw})
and we perform steps similar to those which lead to (\ref{matrix.operator1}), then we can obtain the following matrix 
representations:
%%%
\begin{equation}
\widehat{\bar{c}}_q=\begin{pmatrix}0 & 1 & 0 & 0\\
0 & 0 & 0 & 0\\
0 & 0 & 0 & -1\\
0 & 0 & 0 & 0\end{pmatrix}, \qquad\quad \widehat{\bar{c}}_p=
\begin{pmatrix}0 & 0 & 1 & 0\\
0 & 0 & 0 & 1\\
0 & 0 & 0 & 0\\
0 & 0 & 0 & 0\end{pmatrix}. \label{matrix.operator3}
\end{equation}
%%%
Note that the matrices $\widehat{\bar{c}}$ are just the transpose 
of the associated matrices $\widehat{c}$.
It is also easy to verify that the matrices (\ref{matrix.operator1})-(\ref{matrix.operator3}) 
satisfy the correct anticommutation relations given by the Grassmann algebra:
$[\widehat{c},\widehat{c}]_+=0$, $[\widehat{\bar{c}},\widehat{\bar{c}}]_+=0$, 
$[\widehat{c}^a,\widehat{\bar{c}}_b]_+=\delta_b^a$.

The $4\times 4$ matrices we have obtained so far can be written in a more compact form via tensor products
of Pauli matrices. Let us introduce: 
%%%
\begin{equation}
\displaystyle
\frac{\sigma^{\scriptscriptstyle (+)}}{2}=\frac{\sigma_x+i\sigma_y}{2}=
\begin{pmatrix}0 & 1\\ 0 & 0\end{pmatrix},
\qquad\quad \frac{\sigma^{\scriptscriptstyle (-)}}{2}
=\frac{\sigma_x-i\sigma_y}{2}=
\begin{pmatrix}0 & 0\\ 1 & 0
\end{pmatrix}.
\end{equation}
%%%
It is then easy to prove that (\ref{matrix.operator1})-(\ref{matrix.operator3})
can be written as tensor products $\otimes$ of Pauli matrices as:
%%%
\begin{eqnarray}
\displaystyle
&&\widehat{c}^p=\frac{\sigma^{\scriptscriptstyle(-)}}{2}\otimes {\bf 1},\qquad\quad 
\widehat{c}^q=\sigma_z\otimes 
\frac{\sigma^{\scriptscriptstyle(-)}}{2}\nonumber\\
&&\widehat{\bar{c}}_p=\frac{\sigma^{\scriptscriptstyle(+)}}{2}\otimes {\bf 1},
\qquad\quad \widehat{\bar{c}}_q=\sigma_z\otimes 
\frac{\sigma^{\scriptscriptstyle(+)}}{2}.
\end{eqnarray}
%%%
These formulae are very useful because, as we will see in Sec. {\bf 3.5}, 
they can be easily generalized to the case
of systems with an arbitrary great number of degrees of freedom.

Via the representation (\ref{matrix.operator1})-(\ref{matrix.operator3}) 
for the Grassmann operators of the theory we can build also the
matrix representation of the symmetry charges of the CPI:
%%%
\begin{displaymath}
\displaystyle 
\widehat{Q}=i\widehat{c}^q\lambda_q+i\widehat{c}^p\lambda_p=\widehat{c}^q\partial_q+
\widehat{c}^p\partial_p=\begin{pmatrix}0 & 0 & 0 & 0\\ \partial_q & 0 & 0 & 0\\ 
\partial_p & 0 & 0 & 0\\ 0 & \partial_p
& -\partial_q  & 0\end{pmatrix}
\end{displaymath}
%%%
\begin{equation}
\widehat{\overline{Q}}=\begin{pmatrix}0 & \partial_p & -\partial_q & 0\\
0 & 0 & 0 & -\partial_q\\ 0 & 0 & 0 & -\partial_p\\ 0 & 0 & 0 & 0\end{pmatrix},
\quad\qquad \widehat{Q}_f=\begin{pmatrix}0 & 0 & 0 & 0\\ 0 & 1 & 0 & 0\\
 0 & 0 & 1 & 0\\ 0 & 0
& 0  & 2\end{pmatrix} \label{matrix.opmat}
\end{equation}
%%%
\begin{displaymath}
\widehat{K}=\begin{pmatrix}0 & 0 & 0 & 0\\ 0 & 0 & 0 & 0\\ 0 & 0 & 0 & 0\\ 1 & 0
& 0 & 0\end{pmatrix}, \qquad \quad \widehat{\overline{K}}=\begin{pmatrix}0 & 0 & 0 & 1\\
0 & 0 & 0 & 0\\ 0 & 0 & 0 & 0\\ 0 & 0 & 0 & 0\end{pmatrix}. 
\end{displaymath}
%%%
One can easily check that the algebra of these charges is the one of Ref. \cite{Goz89}:
%%%
\begin{eqnarray}
&&[\widehat{Q},\widehat{Q}]_{\scriptscriptstyle +}=[\widehat{\overline{Q}},
\widehat{\overline{Q}}]_{\scriptscriptstyle +}=[\widehat{Q},\widehat{\overline{Q}}]_
{\scriptscriptstyle +}=0\nonumber\\
&&[\widehat{Q}_f,\widehat{K}]_{\scriptscriptstyle -}=2\widehat{K},\;\;[\widehat{Q}_f,
\widehat{\overline{K}}]_{\scriptscriptstyle -}=-2\widehat{\overline{K}},\;\;
[\widehat{K},\widehat{\overline{K}}]_{\scriptscriptstyle -}=\widehat{Q}_f-{\bf 1}\nonumber\\
&&[\widehat{Q}_f,\widehat{Q}]_{\scriptscriptstyle -}=\widehat{Q},\;\;
[\widehat{Q}_f,\widehat{\overline{Q}}]_{\scriptscriptstyle -}=
-\widehat{\overline{Q}},\;\; [\widehat{K},\widehat{Q}]_{\scriptscriptstyle -}=0\nonumber\\
&&[\widehat{K},\widehat{\overline{Q}}]_{\scriptscriptstyle -}=\widehat{Q},\;\;
[\widehat{\overline{K}},\widehat{Q}]_{\scriptscriptstyle -}=\widehat{\overline{Q}},\;\;
[\widehat{\overline{K}}, \widehat{\overline{Q}}]_{\scriptscriptstyle -}=0. \label{matrix.algcpi}
\end{eqnarray}
%%%
The representation of the supersymmetry charges is given by:
%%%
\begin{eqnarray}
\bigskip
&&\widehat{Q}_{\scriptscriptstyle H}=\begin{pmatrix}0 & 0 & 0 & 0\\ \partial_q-\beta
\partial_qH & 0 & 0 & 0\\ 
\partial_p-\beta\partial_pH & 0 & 0
& 0\\ 0 & \partial_p-\beta\partial_pH & -\partial_q+\beta\partial_qH & 0\end{pmatrix}
\nonumber\\
\bigskip
&&\widehat{\overline{Q}}_{\scriptscriptstyle H}=\begin{pmatrix}0 &
\partial_p+\beta\partial_pH & -\partial_q-\beta\partial_qH & 0\\ 0 & 0 & 0 & 
-\partial_q-\beta\partial_qH\\ 0 & 0 & 0 &
-\partial_p-\beta\partial_pH \\ 0 & 0 & 0 & 0\end{pmatrix}. \label{matrix.matrixn1}
\end{eqnarray}
%%%
Finally the matrix representing the operator of evolution:
%%%
\begin{equation}
{\cal
H}=\widehat{L}+i\bar{c}_q\partial_p\partial_pHc^p+i\bar{c}_q\partial_p
\partial_qHc^q-i\bar{c}_p\partial_q\partial_pHc^p
-i\bar{c}_p\partial_q\partial_qHc^q,
\end{equation}
%%%
where $\widehat{L}=\lambda_a\omega^{ab}\partial_bH$ is the Liouville operator,
is given by:
%%%
\begin{equation}
\displaystyle 
\widehat{\HT}=\begin{pmatrix}\widehat{L} & 0 & 0 & 0\\ 0 & 
\widehat{L}-i\partial_q\partial_p H & i\partial_q\partial_qH & 0\\
0 & -i\partial_p\partial_pH & \widehat{L}+i\partial_p\partial_qH & 0\\ 0 & 0 & 0 & 
\widehat{L}\end{pmatrix}.
\label{matrix.accatilde}
\end{equation}
%%%
From the above expression of $\widehat{\HT}$ it is clear that the zero- and the
two-forms,
which in the 4-vector representation (\ref{matrix.genw2}) have respectively only the 
first and the last components different from zero, evolve only
with  the Liouvillian $\widehat{L}$. The one-forms instead evolve with the central $2\times 2$ submatrix 
of (\ref{matrix.accatilde}) which contains all the possible second derivatives of the 
Hamiltonian $H(\varphi)$ and mixes the two central 
components
of the 4-vector $\psi$. 
It is easy to check that also at the matrix level the usual supersymmetry algebra holds,
i.e.: $[\widehat{Q}_{\scriptscriptstyle H},\widehat{\overline{Q}}_{\scriptscriptstyle H}]_{\scriptscriptstyle +}=
2i\beta\widehat{\HT}$, $[\widehat{Q}_{\scriptscriptstyle H},\widehat{\HT}]_{\scriptscriptstyle -}=
[\widehat{\overline{Q}}_{\scriptscriptstyle H},\widehat{\HT}]_{\scriptscriptstyle -}=0$. 
So, using the matrix realization of the Grassmann operators in the case $n=1$,
we automatically represent the superalgebra of the symmetry charges of the CPI
in terms of $4\times 4$ matrices of operators. Now 
irreducible representations of superalgebras are well-known in literature. 
In Sec. {\bf 3.7} we shall use the results of Ref. \cite{Rittenberg} to
show how it is possible to build the irreducible representations for the superalgebra of the CPI
in terms of $4\times 4$ matrices whose entries are real numbers.

All the symmetries of the CPI turn the states into each other within the 
eigenspaces of the operator of evolution $\widehat{\HT}$. For
example we can start considering an eigenstate of the Liouvillian $\widehat{L}$ 
with eigenvalue $l$: 
$\widehat{L}\psi_{\scriptscriptstyle 0}^l=l\psi_{\scriptscriptstyle 0}^l$. 
Since $\psi_{\scriptscriptstyle 0}^l$ is a zero-form we can represent 
it by means of the following 4-vector:
%%%
\begin{equation}
\psi_{\scriptscriptstyle 0}^l=\left( \begin{array}{c} \psi_{\scriptscriptstyle 0}^l \\ 0 \\ 0 \\ 0\end{array}\right)
\end{equation}
%%%
which is an eigenstate for $\widehat{\HT}$ with eigenvalue $l$: 
$\widehat{\HT}\psi_{\scriptscriptstyle 0}^l=l\psi_{\scriptscriptstyle 0}^l$.
Since $\widehat{Q}$ commutes with $\widehat{\HT}$ also $\widehat{Q}\psi_{\scriptscriptstyle 0}^l$ 
is an eigenstate for $\widehat{\HT}$ 
with the same eigenvalue:
%%%
\begin{equation}
[\widehat{\HT}, \widehat{Q}]=0\;\Longrightarrow\; \widehat{\HT}
(\widehat{Q}\psi_{\scriptscriptstyle 0}^l)=l(\widehat{Q}\psi_{\scriptscriptstyle 0}^l).
\end{equation}
%%%
The explicit form of $\widehat{Q}\psi_{\scriptscriptstyle 0}^l$ is given by the
following one-form:
%%%
\begin{equation}
\displaystyle 
\widehat{Q}\psi_{\scriptscriptstyle 0}^l=\begin{pmatrix}0 & 0 & 0 & 0\\ \partial_q & 0 & 0 & 0\\ \partial_p & 0 
& 0 & 0\\ 0 & \partial_p & -\partial_q & 0\end{pmatrix}
\left( \begin{array}{c} \psi_{\scriptscriptstyle 0}^l\\ 0\\ 0\\ 0 \end{array} \right)=
\left( \begin{array}{c} 0\\ \partial_q\psi_{\scriptscriptstyle 0}^l\\ \partial_p\psi_{\scriptscriptstyle 0}^l\\ 0 
\end{array} \right). \label{matrix.BRS}
\end{equation}
%%%
In the same way $\widehat{K}\psi_{\scriptscriptstyle 0}^l$ is another eigenstate for 
$\widehat{\HT}$ with eigenvalue $l$. Its explicit form is given by:
%%%
\begin{equation}
\widehat{K}\psi_{\scriptscriptstyle 0}^l=\begin{pmatrix}0 & 0 & 0 & 0\\ 0 & 0 & 0 & 0\\ 0 & 0 & 0 & 0\\ 1 & 0 & 0 & 0\end{pmatrix}
 \left( \begin{array}{c}\psi_{\scriptscriptstyle 0}^l\\ 0\\ 0\\ 0\end{array}\right)=
\left( \begin{array}{c}0\\ 0\\ 0\\ \psi_{\scriptscriptstyle 0}^l\end{array}\right)
\end{equation}
%%%
and it is a two-form. So
by means of the symmetry charges we can move within the eigenspaces of $\widehat{\HT}$:
from the zero-forms to the one-forms (by the charge $\widehat{Q}$) and from the
zero-forms to the two-forms (by the charge 
$\widehat{K}$). 

A similar role is played also by the supersymmetry charges. 
First of all, following Ref. \cite{Ioffe2}, we can rewrite the matrix realization of 
$Q_{\scriptscriptstyle H}$ and 
$\overline{Q}_{\scriptscriptstyle H}$, that we derived in (\ref{matrix.matrixn1}), as:
%%%
\begin{equation}
\widehat{Q}_{\scriptscriptstyle H}=\begin{pmatrix}0 & 0 & 0 & 0\\ 
Q_1^- & 0 & 0 & 0\\ Q_2^- & 0 & 0 & 0\\ 
0 & Q_2^- & -Q_1^- & 0\end{pmatrix}, \qquad\quad \widehat{\overline{Q}}_{\scriptscriptstyle H}
=\begin{pmatrix}0 & Q_1^+ & Q_2^+ & 0 \\
0 & 0 & 0 & Q_2^+\\ 0 & 0 & 0 & -Q_1^+\\ 0 & 0 & 0 & 0\end{pmatrix} \label{matrix.susy}
\end{equation}
%%%
where 
%%%
\begin{eqnarray}
&&Q_1^-=\partial_q-\beta\partial_qH, \qquad\quad Q_1^+=\partial_p+
\beta \partial_pH\nonumber\\
&&Q_2^-=\partial_p-\beta\partial_pH, \qquad\quad Q_2^+=-\partial_q-
\beta\partial_qH.
\end{eqnarray}
%%%
Since $\widehat{Q}_{\scriptscriptstyle H}$ commutes with the Hamiltonian $\widehat{\HT}$, we have that
if $\psi_{\scriptscriptstyle 0}^l$ is an eigenstate for the Liouvillian with eigenvalue $l$ then 
%%%
\begin{equation}
\widehat{Q}_{\scriptscriptstyle H} \left( \begin{array}{c}\psi_{\scriptscriptstyle 0}^l\\ 0\\ 0\\ 0\end{array}\right)
=\left( \begin{array}{c}0\\ Q_1^-\psi_{\scriptscriptstyle 0}^l\\ Q_2^-\psi_{\scriptscriptstyle 0}^l\\ 0\end{array}\right)
\label{matrix.3-19}
\end{equation} 
%%%
is also an eigenstate for $\widehat{\HT}$ with the same eigenvalue. Not only, but let us rewrite 
the operator of evolution (\ref{matrix.accatilde}) as
%%%
\begin{equation}
\widehat{\HT}=\begin{pmatrix}\widehat{L} & 0 & 0\\ 0 & 
\widehat{\HT}_{kj}^{\scriptscriptstyle (1)} & 0\\ 0 & 0 & \widehat{L}\end{pmatrix}
\end{equation}
%%%
where $\widehat{\HT}$ is the following $2\times 2$ matrix:
%%%
\begin{equation}
\widehat{\HT}^{\scriptscriptstyle (1)}=\begin{pmatrix}\widehat{L}-i\partial_q\partial_pH & 
i\partial_q\partial_qH\\ -i\partial_p\partial_p H & \widehat{L}
+i\partial_p\partial_qH\end{pmatrix}.
\end{equation}
%%%
It is easy to show that the 2-vector $Q_j^-\psi_{\scriptscriptstyle 0}^l$ is an
eigenstate for $\widehat{\HT}^{\scriptscriptstyle (1)}$ with eigenvalue
$l$ $\widehat{\HT}^{\scriptscriptstyle (1)}_{kj}(Q_j^-\psi_{\scriptscriptstyle 0}^l)=
l(Q_k^-\psi_{\scriptscriptstyle 0}^l)$, i.e. the following relation holds:
%%%
\begin{equation}
\widehat{\HT}^{\scriptscriptstyle (1)}\left(\begin{array}{c}Q_1^-
\psi_{\scriptscriptstyle 0}^l\\ Q_2^-\psi_{\scriptscriptstyle 0}^l
\end{array}\right)=
l\left(\begin{array}{c} Q_1^-\psi_{\scriptscriptstyle 0}^l \\ Q_2^-
\psi_{\scriptscriptstyle 0}^l\end{array}
\right). \label{matrix.3-22}
\end{equation}
%%%
Vice versa if a state $\psi_{k}^{\scriptscriptstyle (1)}$ is an eigenstate for the operator 
$\widehat{\HT}^{\scriptscriptstyle (1)}$ with eigenvalue $l$, then 
the associated 4-vector $\displaystyle \left(\begin{array}{c} 0 \\ \psi_{k}^{\scriptscriptstyle (1)} 
\\ 0\end{array}\right)$ is
an eigenstate for $\widehat{\HT}$ with the same eigenvalue. 
As $\widehat{\overline{Q}}_{\scriptscriptstyle H}$ commutes with $\widehat{\HT}$,
the 4-vectors $\displaystyle \left(\begin{array}{c} 0 \\ \psi_{k}^{\scriptscriptstyle (1)} 
\\ 0\end{array}\right)$ and
%%%
\begin{equation}
\widehat{\overline{Q}}_{\scriptscriptstyle H}\left(\begin{array}{c} 0 \\
\psi_{k}^{\scriptscriptstyle (1)} \\ 0\end{array}\right)=
\begin{pmatrix}0 & Q_1^+ & Q_2^+ & 0\\
0 & 0 & 0 & Q_2^+ \\ 0 & 0 & 0 & -Q_1^+\\ 0 & 0 & 0 & 0\end{pmatrix}\cdot\left(\begin{array}{c}
0 \\ \psi_{1}^{\scriptscriptstyle (1)}\\ \psi_{2}^{\scriptscriptstyle (1)} \\ 0\end{array}\right)=\left(\begin{array}{c} 
Q_k^+\psi_{k}^{\scriptscriptstyle(1)}\\ 0 \\ 0\\ 0
\end{array}\right) \label{matrix.3.24}
\end{equation}
%%%
are degenerate. We can also phrase this degeneracy
by saying that, if $\psi_k^{\scriptscriptstyle (1)}$ is an eigenstate of $\widehat{\HT}^{\scriptscriptstyle (1)}$,
then the operators $Q_k^+$ map the eigenstates of 
$\widehat{\HT}^{\scriptscriptstyle (1)}$ into eigenstates of the Liouvillian
$\widehat{L}$, according to the following relation:
%%%
\begin{equation}
\widehat{L}(Q_k^+\psi_{k}^{\scriptscriptstyle (1)})=
l(Q_k^+\psi_{k}^{\scriptscriptstyle (1)})
\end{equation}
%%%
where an implicit sum over $k$ is understood.
A more complete and refined analysis can be performed on the basis of what has been done for supersymmetric
quantum mechanics in Ref. \cite{Ioffe2}. The final result is that 
two quite different operators like $\widehat{L}$ and $\widehat{\HT}^{\scriptscriptstyle (1)}$
have equivalent spectra and the only difference might be in the handling of the zero eigenvalue.
We should notice that in our case one of the two operators, the Liouvillian $\widehat{L}$, has a deep 
physical meaning and its spectrum gives us information on important properties like the ergodicity, 
the mixing of the system, etc. \cite{Arnold}. So the fact that its spectrum
is equivalent to the one of $\widehat{\HT}^{\scriptscriptstyle (1)}$ may help in discovering further 
things on dynamical systems. 

Up to now we have specified only which is the space of the wave functions whose evolution
is given by the CPI, i.e. the space of 4-vectors $\psi$ of (\ref{matrix.genw2}). 
To build a true Hilbert space we have to introduce also a
suitable scalar product between two different wave functions $\psi$ and $\Phi$. Following Ref. \cite{Salomonson}
one of the most natural choices is:
%%%
\begin{equation}
\displaystyle \langle \psi|\Phi\rangle =\int d\varphi\,[\psi_{\scriptscriptstyle 0}^*\Phi_{\scriptscriptstyle 0}
+\psi_q^*\Phi_q+\psi_p^*\Phi_p+\psi^*_{\scriptscriptstyle 2}\Phi_{\scriptscriptstyle 2}]. \label{matrix.scp}
\end{equation} 
%%%
With this scalar product all the states have positive definite norms and the only state with zero norm is the null state.
It is also easy to prove that:
%%%
\begin{equation}
\langle \psi|c^a\Phi\rangle =\langle\bar{c}_a\psi|\Phi\rangle,\;\;\;\;\; \langle \psi|\bar{c}_a\Phi\rangle
=\langle c^a\psi|\Phi\rangle \label{matrix.3-31}
\end{equation}
%%%
i.e. the operators $c$ and $\bar{c}$ are one the Hermitian conjugate of the other:
$\bar{c}=c^{\dagger}=(c^{\scriptscriptstyle T})^*$.
Therefore the two number operators $N_q=c^q\bar{c}_q$ and $N_p=c^p\bar{c}_p$ are Hermitian and commute. 
Since $N_q^2=N_q$ and $N_p^2=N_p$ the only possible eigenvalues of the number operators are 0 and 1 as it is 
particularly clear using their matrix representation derived from the matrix representation 
of $c$ and $\bar{c}$:
%%%
\begin{eqnarray}
&&\widehat{N}_q=\widehat{c}^q\widehat{\bar{c}}_q=
\begin{pmatrix}0 & 0 & 0 & 0\\ 1 & 0 & 0 & 0\\ 0 & 0 & 0 & 0\\ 0 & 0 & 1 &
0\end{pmatrix}\cdot
\begin{pmatrix}0 & 1 & 0 & 0\\ 0 & 0 & 0 & 0\\ 0 & 0 & 0 & 1\\ 0 & 0 & 0 & 0\end{pmatrix}=
\begin{pmatrix}0 & 0 & 0 & 0\\ 0 & 1 & 0 & 0\\ 0 & 0 & 0 & 0\\ 0 & 0 & 0 & 1\end{pmatrix}\\
&&\widehat{N}_p=\widehat{c}^p\widehat{\bar{c}}_p=
\begin{pmatrix}0 & 0 & 0 & 0\\ 0 & 0 & 0 & 0\\ 1 & 0 & 0 & 0\\ 0 & -1 & 0 & 0\end{pmatrix}\cdot
\begin{pmatrix}0 & 0 & 1 & 0\\ 0 & 0 & 0 & -1\\ 0 & 0 & 0 & 0\\ 0 & 0 & 0 & 0\end{pmatrix}=
\begin{pmatrix}0 & 0 & 0 & 0\\ 0 & 0 & 0 & 0\\ 0 & 0 & 1 & 0\\ 0 & 0 & 0 & 1\end{pmatrix}.\nonumber
\end{eqnarray}
%%%
$\widehat{N}_q$ and $\widehat{N}_p$ are a complete set of commuting and Hermitian operators 
for what concerns the Grassmannian part of
the theory. This means that the knowledge of the simultaneous eigenvalues of $\widehat{N}_q$ and 
$\widehat{N}_p$ allows us to select in a
unique way one of the 4 basis vectors. 
The correspondence between the couple of eigenvalues $(n_q,n_p)$ of the operators
$(\widehat{N}_q,\widehat{N}_p)$ and the basis state vectors is given by the following table:
%%%
\begin{eqnarray}
&& (0,0)\;\Longleftrightarrow\;\left( \begin{array}{c} 1\\ 0\\
0\\ 0\end{array}\right),\;\;\;\;\;\;\;\;\;\; (1,0)\;\Longleftrightarrow\;\left(
\begin{array}{c} 0\\ 1 \\ 0\\ 0\end{array}\right)\nonumber\\
&& (0,1)\;\Longleftrightarrow\; \left( \begin{array}{c} 0\\ 0\\ 1\\
0\end{array}\right),\;\;\;\;\;\;\;\;\;\; (1,1)\;\Longleftrightarrow\;
\left( \begin{array}{c} 0\\ 0\\ 0\\ 1\end{array}\right).
\end{eqnarray}
%%%
Therefore every wave function $\psi$ of the
generalized Hilbert space can be expanded on the basis of the common eigenstates of 
$\widehat{N}_q$ and $\widehat{N}_p$ and it is possible to construct a resolution of the identity involving only 
these eigenstates. These considerations will be useful in Chapter {\bf 4} when we will analyse
more in detail the Hilbert space structure underlying the path integral formulation of CM.

\bigskip

\section{Cartan Calculus in $n=1$ Symplectic Manifolds}

In the previous sections we have seen that a lot of operations
of the Cartan calculus can be performed via the symmetry charges of the CPI \cite{Goz89} and that
all these symmetry charges can be represented
via $4\times 4$ matrices, in the case of $n=1$ symplectic manifolds labeled by two variables
$(q,p)$. Therefore in this case we expect that also the operations of
differential geometry can be performed by means of $4\times 4$ matrices. We 
start remembering that if $n=1$ then
a basis for the cotangent bundle of the phase space is given by $(dq\equiv c^q, dp\equiv c^p)$ and, as we have seen 
in (\ref{matrix.genw2}), the most general non-homogeneous differential form 
$\psi=\psi_{\scriptscriptstyle 0}+\psi_qc^q+\psi_pc^p+\psi_{\scriptscriptstyle 2}c^pc^q$ 
can be represented by the 4-vector $\displaystyle \psi=\left( \begin{array}{c} 
\psi_{\scriptscriptstyle 0}\\ \psi_q\\ \psi_p\\ \psi_{\scriptscriptstyle 2} \end{array} \right)$. 
In this section we will translate into matrices all the operations one can do on forms in ordinary differential 
geometry.

\medskip

\noindent $\bullet$ {\bf Exterior Derivative}. 
With the identifications $dq\equiv c^q$ and $dp\equiv c^p$ 
the action of the exterior derivative {\bf d} on a homogeneous form increases by 1 the degree 
of the form itself according to the following equations:
%%%
\begin{equation}
\left\{
	\begin{array}{l}
	{\bf d}\psi_{\scriptscriptstyle 0}=\partial_q\psi_{\scriptscriptstyle 0}c^q
	+\partial_p\psi_{\scriptscriptstyle 0}c^p\\
	{\bf d}(\psi_qc^q+\psi_pc^p)=(\partial_p\psi_q-\partial_q\psi_p)c^pc^q\\
	{\bf d}(\psi_{\scriptscriptstyle 2}c^pc^q)=0.
	\end{array}
	\right.
\end{equation}
%%%
Then it is easy to prove that the symmetry charge
$\widehat{Q}$ can be interpreted as the exterior derivative also at the matrix level. 
In fact when we apply $\widehat{Q}$ over the 4-vector
$\psi$ we produce a new 4-vector whose components are just the 
4 components of the differential form ${\bf d}\psi$ obtained acting with the exterior derivative
${\bf d}$ over $\psi$:
%%%
\begin{equation}
\widehat{Q}\psi=\begin{pmatrix}0 & 0 & 0 & 0\\ \partial_q & 0 & 0 & 0\\ \partial_p & 
0 & 0 & 0\\ 0 & \partial_p & -\partial_q & 0\end{pmatrix}
\left( \begin{array}{c} \psi_{\scriptscriptstyle 0}\\ \psi_q\\ \psi_p\\ \psi_{\scriptscriptstyle 2} 
\end{array} \right)=
\left( \begin{array}{c} 0 \\ \partial_q \psi_{\scriptscriptstyle 0} \\ \partial_p \psi_
{\scriptscriptstyle 0} \\ \partial_p
\psi_q-\partial_q\psi_p \end{array} \right)
\equiv {\bf d}\psi. \label{matrix.extder}
\end{equation}
%%%

\medskip

\noindent $\bullet$ {\bf Form Number}. The symmetry charge $\widehat{Q}_f$ 
provides the form number of $\psi$. In fact:
%%%
\begin{equation}
\widehat{Q}_f\psi^{\scriptscriptstyle (p)}=\begin{pmatrix}0 & 0 & 0 & 0\\ 0 & 1 & 0 & 0\\ 0 & 0 & 1 & 0\\ 
0 & 0 & 0 & 2\end{pmatrix}\left( \begin{array}{c}
\psi^{\scriptscriptstyle (0)}_{\scriptscriptstyle 0} \\ \psi_q^{\scriptscriptstyle (1)} \\ 
\psi_p^{\scriptscriptstyle (1)} \\ \psi_{\scriptscriptstyle 2}^{\scriptscriptstyle (2)}\end{array}\right)
=\left( \begin{array}{c}
0\cdot \psi^{\scriptscriptstyle (0)}_{\scriptscriptstyle 0} \\ 1\cdot\psi_q^{\scriptscriptstyle (1)} \\ 
1\cdot\psi_p^{\scriptscriptstyle (1)} \\ 2\cdot\psi_{\scriptscriptstyle 2}^{\scriptscriptstyle (2)}\end{array}\right)
=p\psi^{\scriptscriptstyle (p)}
\end{equation}
%%%
where the previous relation means that all the homogeneous forms are eigenstates
for $\widehat{Q}_f$. In particular the zero-forms $\psi^{\scriptscriptstyle (0)}$
are eigenstates for $\widehat{Q}_f$ 
with eigenvalue 0, the one-forms $\psi^{\scriptscriptstyle (1)}$ are eigenstates
with eigenvalue 1 and the two-forms $\psi^{\scriptscriptstyle (2)}$ are eigenstates with eigenvalue 2. 

\medskip

\noindent $\bullet$ {\bf Interior Contraction}.
The interior contraction of a homogeneous form with the 
vector field $V=V^a\bar{c}_a=V^q\bar{c}_q+V^p\bar{c}_p$ is given by
%%%
\begin{equation}
\left\{
	\begin{array}{l}
	\iota_{\scriptscriptstyle V}\psi^{\scriptscriptstyle (0)}=0\\
	\iota_{\scriptscriptstyle V}\psi^{\scriptscriptstyle (1)}=
	V^q\psi^{\scriptscriptstyle (1)}_q+V^p\psi^{\scriptscriptstyle (1)}_p\\
	\iota_{\scriptscriptstyle V}\psi^{\scriptscriptstyle (2)}=V^p\psi^{\scriptscriptstyle (2)}
	_{\scriptscriptstyle 2}c^q-V^q
	\psi^{\scriptscriptstyle (2)}_{\scriptscriptstyle 2}c^p.
	\end{array}
	\right.
\end{equation}
%%%
So the interior contraction of a form with the vector field $V$
maps $\displaystyle \left( \begin{array}{c} \psi^{\scriptscriptstyle (0)}_{\scriptscriptstyle 0}\\ 
\psi_q^{\scriptscriptstyle (1)}\\ \psi_p^{\scriptscriptstyle (1)}\\ 
\psi_{\scriptscriptstyle 2}^{\scriptscriptstyle (2)} \end{array}\right)$ 
into a new 4-vector $\displaystyle \left( \begin{array}{c} 
V^q\psi_q^{\scriptscriptstyle (1)}+V^p\psi_p^{\scriptscriptstyle (1)}\\ 
V^p\psi_{\scriptscriptstyle 2}^{\scriptscriptstyle (2)} \\ -V^q\psi_{\scriptscriptstyle 2}^{\scriptscriptstyle (2)}\\ 0
\end{array}\right)$.  It is easy to see that the $4\times 4$ matrix realizing the previous mapping 
is given by:
%%%
\begin{equation}
\displaystyle 
\iota_{\scriptscriptstyle V}=\begin{pmatrix}0 & V^q & V^p & 0\\ 0 & 0 & 0 &
V^p\\ 0 & 0 & 0 & -V^q\\ 0 & 0 & 0 & 0\end{pmatrix}. \label{matrix.intcon}
\end{equation}
%%%
The matrix (\ref{matrix.intcon}) is just equal to the matrix representation of the vector field\break
$\widehat{V}=V^q\widehat{\bar{c}}_q+V^p
\widehat{\bar{c}}_p$ where $\widehat{\bar{c}}_q$ and 
$\widehat{\bar{c}}_p$
are the matrix representations of the Grassmann operators, see (\ref{matrix.operator3}). 

\medskip

\noindent $\bullet$ {\bf Lie Derivative along the Hamiltonian Flow}. 
As a particular case of the previous analysis, if we take
a Hamiltonian vector field $h^a=\omega^{ab}\partial_bH$, then the components of $V$ are: 
$V^q=\partial_pH$, $V^p=-\partial_qH$
and the interior contraction (\ref{matrix.intcon}) becomes:
%%%
\begin{equation}
\iota_{h}=\begin{pmatrix}0 & \partial_pH & -\partial_qH & 0\\ 0 & 0 & 0 & 
-\partial_qH\\ 0 & 0 & 0 & -\partial_pH\\
0 & 0 & 0 & 0\end{pmatrix}. \label{matrix.iota}
\end{equation}
%%%
From the matrix representation of the exterior derivative (\ref{matrix.extder}) 
and of the interior contraction (\ref{matrix.iota}) 
we can easily derive the matrix representation
for the Lie derivative along the Hamiltonian vector field $h$:
%%%
\begin{equation}
\displaystyle 
{\cal L}_h={\bf d}\iota_h+\iota_h{\bf d}=\begin{pmatrix}i\widehat{L} & 0 & 0 & 0\\ 0 & i\widehat{L}+
\partial_q\partial_pH & -\partial_q\partial_qH & 0\\
0 & \partial_p\partial_pH & i\widehat{L}-\partial_p\partial_qH & 0\\ 0 & 0 & 
0 & i\widehat{L}\end{pmatrix}. \label{matrix.lieder}
\end{equation}
%%%
By comparing (\ref{matrix.accatilde}) and (\ref{matrix.lieder}) we have that 
also at the matrix level the Hamiltonian $\HT$ is nothing more 
than the Lie derivative along the Hamiltonian vector field:
$\widehat{\HT}=-i{\cal L}_h$. This means that, when we apply the CPI operator of 
evolution $\widehat{\HT}$ on a generic 
$\psi$ we obtain another form whose components are given, modulus a factor 
$-i$, by the Lie 
derivative of the Hamiltonian flow: $\widehat{\HT}\psi=-i{\cal L}_h\psi$.

\medskip

\noindent $\bullet$ {\bf Hodge Star}. The Hodge $*$ transformation 
is defined as \cite{Eguchi}:
%%%
\begin{equation}
\displaystyle 
*(dx^{i_1}\wedge dx^{i_2}\wedge \ldots \wedge dx^{i_p})=\frac{1}{(n-p)!}
\epsilon_{i_1i_2\ldots i_pi_{p+1}\ldots i_N}
dx^{i_{p+1}}\wedge dx^{i_{p+2}}\wedge\cdots \wedge dx^{i_N}.
\end{equation}
%%%
In the language of the CPI we have that in the case of one degree of freedom ($N=2$): 
%%%
\begin{eqnarray}
&&*(1)=\epsilon_{qp} dq\wedge dp=dq\wedge dp\;\;\longrightarrow\;\; *(1)=c^qc^p\nonumber\\
&&*(dx^i)=\epsilon_{ij}dx^j\;\;\longrightarrow\;\; *(c^q)=c^p,\;\,*(c^p)=-c^q\\
&&*(dp\wedge dq)=\epsilon_{pq}=-1\;\;\longrightarrow\;\; *(c^pc^q)=-1.\nonumber
\end{eqnarray}
%%%
With the convention (\ref{matrix.genw2}), the action of $*$ on the basis state vectors is given by:
%%%
\begin{eqnarray}
&&*\left( \begin{array}{c} 1\\0\\0\\0\end{array} \right)=
\left( \begin{array}{c} 0\\0\\0\\-1\end{array} \right),\qquad\quad
*\left( \begin{array}{c} 0\\1\\0\\0\end{array} \right)=\left( \begin{array}{c} 
0\\0\\1\\0\end{array} \right),\nonumber\\
&&*\left( \begin{array}{c} 0\\0\\1\\0\end{array} \right)=\left( \begin{array}{c} 
0\\-1\\0\\0\end{array} \right),\qquad\quad
*\left( \begin{array}{c} 0\\0\\0\\1\end{array} \right)=\left( 
\begin{array}{c} -1\\0\\0\\0\end{array} \right).
\end{eqnarray}
%%%
Therefore the matrix representation for the Hodge $*$ transformation is:
%%%
\begin{equation}
\displaystyle
*=\begin{pmatrix}0 & 0 & 0 & -1\\ 0 & 0 & -1 & 0\\ 0 & 1 & 0 & 0\\ -1 & 0 & 0 & 0\end{pmatrix}. \label{matrix.hodge}
\end{equation}
%%%

\medskip

\noindent $\bullet$ {\bf The Adjoint of {\bf d}}. The previous $*$ transformation can be used
to define the following ``inner product" between two $l$-forms:
%%%
\begin{equation}
\displaystyle
(\alpha_l,\beta_l)=\int_{\cal M} \alpha_l\wedge *\beta_l. \label{matrix.innpro}
\end{equation}
%%%
Using it the adjoint ${\bf \delta}$ of the exterior derivative {\bf d} can be defined
as:
%%%
\begin{equation}
(\alpha_l,{\bf d}\beta_{l-1})\equiv({\bf \delta}\alpha_l,\beta_{l-1}). \label{matrix.4-11}
\end{equation}
%%%
In particular it is possible to prove \cite{Eguchi} that, in the case of manifolds with 
even dimension like the symplectic ones, ${\bf \delta}$ can be expressed in terms of ${\bf d}$ 
and $*$ as: ${\bf \delta}=-*{\bf d}*$.
Using (\ref{matrix.extder}) for the exterior derivative, and (\ref{matrix.hodge}) for the Hodge $*$ 
transformation, 
we obtain for ${\bf \delta}$ the following matrix representation:
%%%
\begin{equation}
{\bf \delta}=\begin{pmatrix}0 & -\partial_q & -\partial_p & 0\\ 0 & 0 & 0 & 
-\partial_p\\ 0 & 0 & 0 &\partial_q\\ 
0 & 0 & 0 & 0\end{pmatrix}.
\end{equation}
%%%
${\bf \delta}$ is a nilpotent matrix and
it lowers the degree of the forms by one. So it acts like 
the symmetry charge $\widehat{\overline{Q}}$
but, nevertheless, it does not coincide with it. In fact, using
the language of the CPI, we have that $\partial_a=i\lambda_a$ and, 
from the explicit form (\ref{matrix.operator3}) of the matrices $\widehat{\bar{c}}$ we easily obtain that:
%%%
\begin{equation}
{\bf \delta}=\begin{pmatrix}0 & -i\lambda_q & -i\lambda_p & 0\\ 0 & 0 & 0 & -i\lambda_p\\ 
0 & 0 & 0 & i\lambda_q\\ 0 & 0 & 0 & 0\end{pmatrix}\equiv -i\lambda_q\widehat{\bar{c}}_q-i
\lambda_p\widehat{\bar{c}}_p \label{matrix.delta}
\end{equation}
%%%
which is different from $\widehat{\overline{Q}}=-i\lambda_q\widehat{\bar{c}}_p
+i\lambda_p\widehat{\bar{c}}_q$. 
It is also possible to prove that the ``inner product" (\ref{matrix.innpro}) used in 
differential geometry
coincides with the positive definite inner product defined in (\ref{matrix.scp}). 
For example from the associated hermiticity relations (\ref{matrix.3-31}) among the Grassmann operators
$c^{q^{\dagger}}=\bar{c}_q,\;c^{p^{\dagger}}=\bar{c}_p$ we have that:
%%%
\begin{equation} 
{\bf \delta}=(ic^q\lambda_q+ic^p\lambda_p)^{\dagger}\;\Longrightarrow\;{\bf \delta}={\bf d}^{\dagger}
\end{equation}
%%%
which is nothing more than (\ref{matrix.4-11}).

\medskip

\noindent $\bullet$ {\bf The Laplacian}. The Laplacian operator can be defined 
starting from ${\bf d}$ and ${\bf \delta}$
as:
%%%
\begin{equation}
{\bf \Delta}=({\bf d}+{\bf \delta})^2={\bf d}{\bf \delta}+{\bf \delta}{\bf d}
\end{equation} 
%%%
where in the last step we have used the property that ${\bf d}$ and ${\bf \delta}$ are nilpotent.
From (\ref{matrix.extder}) and (\ref{matrix.delta}) we have that the matrix representation of 
the Laplacian is given by:
%%%
\begin{equation}
{\bf d}{\bf \delta}+{\bf \delta} {\bf d}=(-\partial_q^2-
\partial_p^2){\bf 1}_{\scriptscriptstyle 4\times 4}=
{\bf \Delta}=(\lambda_q^2+\lambda_p^2){\bf 1}_{\scriptscriptstyle 4\times 4}
\end{equation}
from which it is particularly clear that ${\bf \Delta}$ is a positive definite operator.

\bigskip

\section{Grassmann Algebras and Pauli Matrices}

In order to generalize the results of the previous sections to the case 
of a system with an arbitrary great number of degrees of freedom, it is particularly useful to find a
representation of the Grassmann operators in terms of tensor products of Pauli 
matrices like we did
in the case of $n=1$ where we identified $\displaystyle \widehat{c}^p=\frac{\sigma^
{\scriptscriptstyle (-)}}{2}\otimes {\bf 1}$ and $\displaystyle \widehat{c}^q=\sigma_z\otimes
\frac{\sigma^{\scriptscriptstyle(-)}}{2}$. In the case of a generic symplectic manifold  
we will label the indices $1,2,\ldots, k$ on $\varphi^k$ and $c^k$ in such a way that: 
$\varphi^{\scriptscriptstyle 1}=p_{\scriptscriptstyle 1},\,\varphi^{\scriptscriptstyle 2}=
q_{\scriptscriptstyle 1},\,\varphi^{\scriptscriptstyle 3}=p_{\scriptscriptstyle 2},\,
\varphi^{\scriptscriptstyle 4}=q_{\scriptscriptstyle 2}$ and so on. 
The correspondence between Grassmann operators and Pauli
matrices becomes:
%%%
\begin{equation}
\left\{
	\begin{array}{l}
	\displaystyle
	\widehat{c}^k=(\sigma_z)^{\otimes k-1}\otimes \frac{\sigma^{
        \scriptscriptstyle (-)}}{2}
        \otimes ({\bf 1})^{\otimes 2n-k}, \;\;\;\;\; k=1,\cdots, 2n \smallskip\\
	\displaystyle \widehat{\bar{c}}_j=(\sigma_z)^{\otimes j-1}\otimes\frac{\sigma^{
        \scriptscriptstyle (+)}}{2}\otimes({\bf 1})^{\otimes 2n-j}, \;\;\;\;\; j=1,\cdots,2n\\
	\end{array} \label{matrix.5-1}
	\right.
\end{equation}
%%%
where, for example, $(\sigma_z)^{\otimes k-1}=\underbrace{\sigma_z\otimes \sigma_z \otimes
\cdots \otimes \sigma_z}_{k-1 \;\textrm{times}}$
indicates the tensor product of $k-1$ Pauli matrices $\sigma_z$.
The matrices $\widehat{c}^k$ and $\widehat{\bar{c}}_j$ 
built in (\ref{matrix.5-1}) satisfy the usual Grassmann algebra, i.e.:
%%%
\begin{equation}
[\widehat{c}^a,\widehat{\bar{c}}_b]_+=\delta^a_b, \;\;\;\;\;\; 
[\widehat{c}^a,\widehat{c}^b]_+=[\widehat{\bar{c}}_a,\widehat{\bar{c}}_b]_+=0,\;\;\;\; a,b=1,\dots, 2n
\label{matrix.5-3}.
\end{equation}
%%%
So we can say that the construction (\ref{matrix.5-1}) allowed us to realize the Grassmann
operators of the CPI in terms of tensor
products of suitable Pauli or identity matrices. Since this property is crucial in order to derive all the other
formulae of this section we will prove it in detail in Appendix
{\bf \ref{app:pm}}. Here we want only to underline that 
the Grassmann algebra of $\widehat{c}$ and $\widehat{\bar{c}}$ 
becomes a direct consequence of the algebra of 
Pauli matrices. Using the construction (\ref{matrix.5-1}) we will now generalize all 
the results obtained in the case $n=1$
to the case of an arbitrary great number of degrees of freedom, without 
losing a certain compactness in the
appearance of the formulae. 

If we want to represent the Grassmann operators of the CPI as tensor 
products of Pauli matrices we have to 
represent also the states of the associated Hilbert space 
as tensor products of 
2-dimensional vectors in a consistent way. 
For example in the
$n=1$ case, where we represented
$\displaystyle \widehat{c}^p=\frac{\sigma^{\scriptscriptstyle (-)}}{2}\otimes 
{\bf 1}$ and $\displaystyle \widehat{c}^q=\sigma_z\otimes 
\frac{\sigma^{\scriptscriptstyle (-)}}{2}$,
the correspondent Hilbert space could be constructed from all the possible tensor products  
of the 2-dimensional vectors on which the Pauli matrices act:
%%%
\begin{equation}
\left\{
\begin{array}{l}
\left( \begin{array}{c} 1\\ 0 \end{array}\right)\otimes \left( 
\begin{array}{c} 1 \\ 0 \end{array}\right)=
\left( \begin{array}{c} 1 \\ 0 \\ 0 \\ 0 \end{array}\right)\Leftrightarrow 1,\qquad \quad
\left( \begin{array}{c} 1\\ 0 \end{array}\right)\otimes \left( 
\begin{array}{c} 0 \\ 1 \end{array}\right)=
\left( \begin{array}{c} 0 \\ 1 \\ 0 \\ 0 \end{array}\right)\Leftrightarrow c^q,\smallskip \\
\left( \begin{array}{c} 0\\ 1 \end{array}\right)\otimes \left( 
\begin{array}{c} 1 \\ 0 \end{array}\right)=
\left( \begin{array}{c} 0 \\ 0 \\ 1 \\ 0 \end{array}\right)\Leftrightarrow c^p,\qquad\;\;
\left( \begin{array}{c} 0\\ 1 \end{array}\right)\otimes \left( 
\begin{array}{c} 0 \\ 1\end{array} \right)=
\left( \begin{array}{c} 0 \\ 0 \\ 0 \\ 1 \end{array}\right)\Leftrightarrow c^pc^q.
\end{array}
\right.
\label{matrix.basis}
\end{equation}
%%%
These four states and the identifications we have indicated on their RHS 
are of course consistent with the expression 
(\ref{matrix.genw2}) since it is from there 
that we began our analysis. What we mean is that from (\ref{matrix.basis}) we obtain 
that the generic form
%%%
\begin{equation}
\psi(\varphi,c)=\psi_{\scriptscriptstyle 0}(\varphi)\cdot 1
+\psi_q(\varphi)\cdot c^q+\psi_p(\varphi)\cdot c^p+
\psi_{\scriptscriptstyle 2}(\varphi)\cdot c^pc^q \label{matrix.5-7}
\end{equation}
%%%
can be identified with the 4-vector:
%%%
\begin{equation}
\psi(\varphi,c)=\left(\begin{array}{c} \psi_{\scriptscriptstyle 0}(\varphi) \\ \psi_q(\varphi) 
\\ \psi_p(\varphi) \\ \psi_{\scriptscriptstyle 2}(\varphi)
\end{array} \right) 
\end{equation}
%%%
that is just the RHS of (\ref{matrix.genw2}).
An important thing to underline is that, once we have fixed the matrix 
representation of the Grassmann operators, the representation of the 
states must be derived by consistency.
Therefore the choice (\ref{matrix.5-1}) implies that we have to order the components of the generic form $\psi$ 
in a very peculiar way. 
For example to make the identifications (\ref{matrix.basis}) we have used the following empirical rule:
we have indicated the {\it lacking} of a Grassmann variable with respect to the reference string $c^pc^q$ 
with the vector $\left( \begin{array}{c}
1\\ 0 \end{array}\right)$ and its {\it presence} with $\left( \begin{array}{c}
0\\ 1 \end{array}\right)$.
For example $c^p$ has the first Grassmann variable {\it present} and the second {\it lacking}, so it has to be identified
with $c^p\Leftrightarrow \left( \begin{array}{c} 0\\ 1 \end{array}\right)\otimes 
\left( \begin{array}{c} 1\\ 0 \end{array}\right)$.
Let us now apply the same rule in the case of 2 degrees of freedom ($n=2$), i.e.: 
$\displaystyle \varphi^a=(p^{\scriptscriptstyle 1},q^{\scriptscriptstyle 1},p^{\scriptscriptstyle 2},
q^{\scriptscriptstyle 2})$, $\displaystyle 
c^a=(c^{\scriptscriptstyle p_1},c^{\scriptscriptstyle q_1},c^{\scriptscriptstyle p_2},
c^{\scriptscriptstyle q_2})$, $\displaystyle \bar{c}_a=(\bar{c}_{\scriptscriptstyle p_1},
\bar{c}_{\scriptscriptstyle q_1},\bar{c}_{\scriptscriptstyle p_2},
\bar{c}_{\scriptscriptstyle q_2})$. The basis of the zero-forms
is given by the following vector with 16 components:
%%%
\begin{equation}
1=\left(\begin{array}{c} 1\\0\end{array}\right)\otimes 
\left(\begin{array}{c} 1\\0 \end{array} \right)\otimes 
\left(\begin{array}{c} 1\\0 \end{array} \right) \otimes 
\left(\begin{array}{c} 1\\0 \end{array} \right)=\delta_{i,1}.
\end{equation}
%%%
For reasons of space we have not written down explicitly the 16-components vector. We have indicated it
with $\delta_{i,1}$ which means that it has an element $1$ in the first position and all the other 15
elements equal to $0$. For the one-forms we have instead:
%%%
\begin{eqnarray}
&& c^{q_2}\equiv dq_{\scriptscriptstyle 2}\Leftrightarrow\left(\begin{array}{c} 1\\0\end{array}\right)\otimes 
\left(\begin{array}{c} 1\\0\end{array}\right)\otimes
\left(\begin{array}{c} 1\\0\end{array}\right)\otimes\left(\begin{array}{c} 
0\\1\end{array}\right)
=\delta_{i,2}\nonumber\\
&& c^{p_2}\equiv dp_{\scriptscriptstyle 2}\Leftrightarrow\left(\begin{array}{c} 1\\0\end{array}\right)\otimes 
\left(\begin{array}{c} 1\\0\end{array}\right)\otimes
\left(\begin{array}{c} 0\\1\end{array}\right)\otimes\left(\begin{array}{c} 
1\\0\end{array}\right)
=\delta_{i,3}\nonumber\\
&& c^{q_1}\equiv dq_{\scriptscriptstyle 1}\Leftrightarrow\left(\begin{array}{c} 1\\0\end{array}\right)\otimes 
\left(\begin{array}{c} 0\\1\end{array}\right)\otimes
\left(\begin{array}{c} 1\\0\end{array}\right)\otimes\left(\begin{array}{c} 
1\\0\end{array}\right)
=\delta_{i,5}\nonumber\\
&& c^{p_1}\equiv dp_{\scriptscriptstyle 1}\Leftrightarrow\left(\begin{array}{c} 0\\1\end{array}\right)\otimes 
\left(\begin{array}{c} 1\\0\end{array}\right)\otimes
\left(\begin{array}{c} 1\\0\end{array}\right)\otimes\left(\begin{array}{c} 
1\\0\end{array}\right)
=\delta_{i,9}. \label{matrix.5-10}
\end{eqnarray}
%%%
If we perform explicitly the previous tensor products we obtain four 
16-dimensional vectors with all the elements 
equal to 0 except for one element equal to 1 placed respectively in the second, 
the third, the fifth and the ninth
position. Here we note one defect of the representation we have introduced: 
the components of the one-form are scattered 
inside the 16-dimensional vector and they do not form a unique block of 
adjacent components, from the second to the 
fifth one, like it happens for $n=1$.

Now, having represented the states as tensor products, it is quite evident 
the reason why  all the Grassmann
algebra can be reconstructed starting from $\displaystyle 
\frac{\sigma^{\scriptscriptstyle (-)}}{2},
\frac{\sigma^{\scriptscriptstyle (+)}}{2}$ and $\sigma_z$. In fact
$\displaystyle \frac{\sigma^{\scriptscriptstyle (-)}}{2}$ plays the role 
of the operator of multiplication by $c$:
%%%
\begin{equation}
\begin{array}{l}
\begin{pmatrix}0 & 0\\ 1 & 0\end{pmatrix}\cdot\left(\begin{array}{c} 1\\0\end{array}\right)=
\left(\begin{array}{c} 0\\1\end{array}\right)
\;\;\Longleftrightarrow\;\;\widehat{c}\cdot 1=c\medskip\\
\begin{pmatrix}0 & 0\\ 1 & 0\end{pmatrix}\cdot\left(\begin{array}{c} 0\\1\end{array}\right)=
\left(\begin{array}{c} 0\\0\end{array}\right)
\;\;\Longleftrightarrow\;\;\widehat{c}\cdot c=0.
\end{array}
\end{equation}
%%%
while $\displaystyle \frac{\sigma^{\scriptscriptstyle (+)}}{2}$ plays the role 
of $\bar{c}$ which is the derivative operator with respect to $c$:
%%%
\begin{equation}
\begin{array}{l}
\displaystyle
\begin{pmatrix}0 & 1\\ 0 & 0\end{pmatrix}\left(\begin{array}{c} 1\\0\end{array}\right)=
\left(\begin{array}{c} 0\\0\end{array}\right)
\;\;\Longleftrightarrow\;\;\frac{\partial}{\partial c}1=0\medskip\\
\displaystyle \begin{pmatrix}0 & 1\\ 0 & 0\end{pmatrix}\left(\begin{array}{c} 0\\1\end{array}\right)=
\left(\begin{array}{c} 1\\0\end{array}\right)
\;\;\Longleftrightarrow\;\;\frac{\partial}{\partial c}c=1.
\end{array}
\end{equation}
%%%
In (\ref{matrix.5-1}) also the matrix $\sigma_z$ made its appearance and the reader 
may wonder on which is its role.
Actually $\sigma_z$ allows us to give to the states
$\left(\begin{array}{c} 1\\0\end{array}\right)$ and $\left(\begin{array}{c} 0\\1\end{array}\right)$
the grading factors that 
we have to introduce for the anticommutativity
of the Grassmann variables. In fact the state 
$\left(\begin{array}{c} 1\\0\end{array}\right)$ must be Grassmannian even in order to represent ``1", 
while
$\left(\begin{array}{c} 0\\1\end{array}\right)$ must be Grassmannian odd in order
to represent ``$c$". 
The matrix $\sigma_z$ gives exactly the correct grading factor to the vectors $\left(\begin{array}{c}
1\\0\end{array}\right)$ and $\left(\begin{array}{c} 0\\1\end{array}\right)$:
%%%
\begin{equation}
\displaystyle \sigma_z\left(\begin{array}{c} 1\\0\end{array}\right)=
\left(\begin{array}{c} 1\\0\end{array}\right), \;\;\;\;\;\;\;\sigma_z\left
(\begin{array}{c} 0\\1\end{array}\right)=-
\left(\begin{array}{c} 0\\1\end{array}\right).
\end{equation}
%%% 
Just to give an example the representation of the equation 
$\displaystyle \widehat{\bar{c}}_q[c^pc^q]=\frac{\partial}{\partial c^q}(c^pc^q)=-c^p$ 
in terms of Pauli matrices in the case $n=1$ is:
%%%
\begin{equation}
\displaystyle 
\sigma_z\otimes\frac{\sigma^{\scriptscriptstyle (+)}}{2}\Biggl[\left
(\begin{array}{c} 0\\1\end{array}\right)\otimes
\left(\begin{array}{c} 0\\1\end{array}\right)\Biggr]=\sigma_z\left(
\begin{array}{c} 0\\1\end{array}\right)
\otimes \frac{\sigma^{\scriptscriptstyle (+)}}{2}\left(\begin{array}{c} 
0\\1\end{array}\right)=-
\left(\begin{array}{c} 0\\1\end{array}\right)\otimes\left(\begin{array}
{c} 1\\0\end{array}\right). \label{matrix.5-14}
\end{equation}
%%%
Note that the matrix $\sigma_z$ is crucial in order to reproduce the minus sign 
on the RHS of (\ref{matrix.5-14}). That minus sign was there in the original equations written
in terms of $c$ because the derivative
$\displaystyle \frac{\partial}{\partial c^q}$ had to go through a Grassmannian odd variable
$c^p$ before acting on $c^q$. 

\bigskip

\section{Cartan Calculus and Pauli Matrices}

With the tools developed in the previous section, we can now generalize to more than one degree of freedom
what we did in Sec. {\bf 3.4}, i.e. we can write down all the operations of the Cartan calculus via
Pauli matrices. 

\medskip

\noindent $\bullet$ {\bf Exterior Derivative}. The exterior
derivative ${\bf d}=\widehat{c}^a\partial_a$ is a linear operator
in the variables $c$. From (\ref{matrix.5-1}) its matrix representation is given by:
%%%
\begin{equation}
\displaystyle {\bf d}=\sum_{a=1}^{2n}=\widehat{c}^a\partial_a=
\sum_{j=1}^{2n}(\sigma_z)^{\otimes j-1}\otimes \frac{\sigma^{
\scriptscriptstyle (-)}}{2}\partial_j\otimes ({\bf 1})^{\otimes 
2n-j} \label{matrix.exteriorder}.
\end{equation} 
%%%

\medskip

\noindent $\bullet$ {\bf Form Number}. From the tensor representation (\ref{matrix.5-1})
of $\widehat{c}$ and $\widehat{\bar{c}}$ it is very easy to find the 
expression of $\widehat{Q}_f$ in the general case:
%%%
\begin{equation}
\displaystyle
\widehat{Q}_f=\sum_{j=1}^{2n}\widehat{c}^j\widehat{\bar{c}}_j=
\sum_{j=1}^{2n}(\sigma_z\cdot\sigma_z)
^{\otimes j-1}\otimes 
\frac{\sigma^{\scriptscriptstyle (-)}}{2}\cdot
\frac{\sigma^{\scriptscriptstyle (+)}}{2}\otimes ({\bf 1})^{\otimes 2n-j}.
\end{equation}
%%%
Since $\sigma_z\cdot\sigma_z={\bf 1}$ and $\displaystyle 
\frac{\sigma^{
\scriptscriptstyle (-)}}{2}\cdot
\frac{\sigma^{\scriptscriptstyle (+)}}{2}=\begin{pmatrix}0 & 0\\ 0 & 1\end{pmatrix}=
\frac{1}{2}({\bf 1}-\sigma_z)$
we can rewrite $\widehat{Q}_f$ as:
%%%
\begin{equation}
\displaystyle
\widehat{Q}_f=\sum_{j=1}^{2n}\widehat{c}^j\widehat{\bar{c}}_j=
\sum_{j=1}^{2n}({\bf 1})^{\otimes j-1}
\otimes \frac{1}{2}
({\bf 1}-\sigma_z)\otimes ({\bf 1})^{\otimes 2n-j}.
\end{equation}
%%%
We call $\widehat{Q}_f$ the form number because 
every homogeneous form of degree $p$ 
is an eigenstate for the matrix $\widehat{Q}_f$ with eigenvalue $p$. 

\medskip

\noindent $\bullet$ {\bf Interior Contraction}. In the case $n=1$ 
the interior contraction with a vector
field was given by $\iota_{\scriptscriptstyle V}=V^q\widehat{\bar{c}}_q+V^p
\widehat{\bar{c}}_p$. In general we have:
$\iota_{\scriptscriptstyle V}=V^j\widehat{\bar{c}}_j$ whose matrix 
representation is given by:
%%%
\begin{equation}
\displaystyle \iota_{\scriptscriptstyle V}=\sum_{j=1}^{2n}(\sigma_z)^
{\otimes j-1}\otimes \frac{\sigma^{
\scriptscriptstyle (+)}}{2}V^j
\otimes ({\bf 1})^{\otimes 2n-j}.
\end{equation}
%%%
In the particular case of a Hamiltonian vector field $h^j=\omega^{jk}
\partial_kH$ we obtain:
%%%
\begin{equation}
\displaystyle \iota_{\scriptscriptstyle h}=\sum_{j,k=1}^{2n}(\sigma_z)^
{\otimes j-1}\otimes \frac{\sigma^{
\scriptscriptstyle (+)}}{2}\omega^{jk}
\partial_kH\otimes ({\bf 1})^{\otimes 2n-j}. \label{matrix.inthamvec}
\end{equation}
%%%

\medskip

\noindent $\bullet$ {\bf Lie Derivative along the Hamiltonian Flow}. 
It is easy to represent the Lie derivative along the Hamiltonian flow
as a matrix starting from the matrix representation of the exterior derivative 
{\bf d},  (\ref{matrix.exteriorder}), and of the interior contraction with a 
Hamiltonian vector field, (\ref{matrix.inthamvec}). In fact
such a Lie derivative is given by the anticommutator of ${\bf d}$ and $\iota_h$ \cite{Marsd}: 
%%%
\begin{eqnarray}
\displaystyle {\cal L}_h={\bf d}\iota_h+\iota_h{\bf d}&=&\sum_{j<k}({\bf 1})^{\otimes j-1}
\otimes \biggl[\sigma_z,\frac{\sigma^{\scriptscriptstyle (+)}}{2}\biggr]\otimes (\sigma_z)^{\otimes k-j-1}
\otimes \frac{\sigma^{\scriptscriptstyle (-)}}{2}\otimes ({\bf 1})^{\otimes 2n-k}\omega^{jl}\partial_lH
\partial_k\nonumber\\
& &+\sum_{j<k}({\bf 1})^{\otimes j-1}
\otimes \sigma_z\cdot\frac{\sigma^{\scriptscriptstyle (+)}}{2}\otimes (\sigma_z)^{\otimes k-j-1}
\otimes \frac{\sigma^{\scriptscriptstyle (-)}}{2}\otimes ({\bf 1})^{\otimes 2n-k}\omega^{jl}\partial_k\partial_lH
\nonumber\\
& &+\sum_j({\bf 1})^{\otimes j-1}
\otimes {\bf 1} \otimes ({\bf 1})^{\otimes 2n-j}\omega^{jl}\partial_lH\partial_j+ \label{matrix.6-6}
\\
& &\sum_{j>k}({\bf 1})^{\otimes k-1}
\otimes \biggl[\frac{\sigma^{\scriptscriptstyle (-)}}{2},\sigma_z\biggr]\otimes (\sigma_z)^{\otimes j-k-1}
\otimes \frac{\sigma^{\scriptscriptstyle (+)}}{2}\otimes ({\bf 1})^{\otimes 2n-j}\omega^{jl}\partial_lH\partial_k
\nonumber\\
& &+\sum_{j>k}({\bf 1})^{\otimes k-1}
\otimes \frac{\sigma^{\scriptscriptstyle (-)}}{2}\cdot\sigma_z\otimes (\sigma_z)^{\otimes j-k-1}
\otimes \frac{\sigma^{\scriptscriptstyle (+)}}{2}\otimes ({\bf 1})^{\otimes 2n-j}\omega^{jl}\partial_k\partial_lH
\nonumber
\end{eqnarray}
%%%
Using the anticommutation relations $\displaystyle \biggl[\sigma_z,\frac{\sigma^{\scriptscriptstyle (+)}}{2}
\biggr]_{\scriptscriptstyle +}=\biggl[\frac{\sigma^{\scriptscriptstyle (-)}}{2},\sigma_z
\biggr]_{\scriptscriptstyle +}=0$ we can write (\ref{matrix.6-6}) in the more compact form:
%%%
\begin{eqnarray}
\label{matrix.seiotto} 
{\cal L}_h&=&(\omega^{ab}\partial_bH\partial_a)({\bf 1})
^{\otimes 2n}+\\
&&-\sum_{j<k}({\bf 1})^{\otimes j-1}\otimes \frac{\sigma^
{\scriptscriptstyle (+)}}{2}
\sigma_z\otimes (\sigma_z)^{\otimes k-1-j}\otimes 
\frac{\sigma^{\scriptscriptstyle (-)}}{2}\otimes 
({\bf 1})^{\otimes 2n-k}\cdot \omega^{jl}\partial_l\partial_kH+\nonumber\\
&&-\sum_{j>k}({\bf 1})^{\otimes k-1}\otimes \sigma_z\frac{\sigma^{\scriptscriptstyle
(-)}}{2}
\otimes (\sigma_z)^{\otimes j-1-k}\otimes \frac{\sigma^{\scriptscriptstyle 
(+)}}{2}
\otimes ({\bf 1})^{\otimes 2n-j}\cdot \omega^{jl}\partial_l\partial_kH. \nonumber 
\end{eqnarray}
%%%
It is easy to realize that (\ref{matrix.seiotto}) is just $i$ times the matrix representation of
$\HT$:
%%%
\begin{equation}
{\cal L}_h=i\widehat{\HT}=\omega^{ab}\partial_bH\partial_a{\bf 1}^{\otimes 2n}-
\widehat{\bar{c}}_j\omega^{jl}\partial_l\partial_kH\widehat{c}^k. \label{88}
\end{equation}
%%%
To prove (\ref{88}) it is sufficient to insert in $\widehat{\HT}$ the usual matrix representation 
(\ref{matrix.5-1}) for the Grassmann operators $\widehat{c},\widehat{\bar{c}}$.
This confirms that the operator $\HT$ of (\ref{ann.supham})
which appears in the weight of the CPI is nothing more than 
the Lie derivative along the Hamiltonian flow.
\medskip

\noindent $\bullet$ {\bf The Adjoint of {\bf d} and the Laplacian}. 
The expression of ${\bf \delta}$ in terms of Grassmann variables:
${\bf \delta}=-\bar{c}_j\partial_j$ can be translated into Pauli matrices 
using (\ref{matrix.5-1}):
%%%
\begin{equation}
\displaystyle
{\bf \delta}=-\widehat{\bar{c}}_j\partial_j=-\sum_{j=1}^{2n}(\sigma_z)^
{\otimes j-1}\otimes \frac{\sigma^{
\scriptscriptstyle (+)}}{2}\partial_j
\otimes ({\bf 1})^{\otimes 2n-j}.
\end{equation}
%%%
This is the expression of ${\bf \delta}$ for an arbitrary number of degrees of freedom $n$. It 
is possible to prove that ${\bf d}$ and ${\bf \delta}$
are nilpotent. Let us check that for the exterior derivative ${\bf d}$:
%%%
\begin{eqnarray}
\displaystyle
{\bf d}^2 &=&\sum_{jk}\biggl[(\sigma_z)^{\otimes j-1}\otimes \frac{\sigma^
{\scriptscriptstyle 
(-)}}{2}\partial_j\otimes ({\bf 1})^{\otimes 2n-j}\biggr]
\biggl[(\sigma_z)^{\otimes k-1}\otimes \frac{\sigma^{\scriptscriptstyle 
(-)}}{2}\partial_k\otimes ({\bf 1})^{\otimes 2n-k}\biggr]=\nonumber\\
&=&\sum_{j<k}({\bf 1})^{\otimes j-1}\otimes\frac{\sigma^{\scriptscriptstyle 
(-)}}{2}\sigma_z\partial_j\otimes (\sigma_z)^{\otimes k-1-j}
\otimes \frac{\sigma^{\scriptscriptstyle (-)}}{2}\partial_k\otimes 
({\bf 1})^{\otimes 2n-k}+\nonumber\\
& &+\sum_{k<j}({\bf 1})^{\otimes k-1}\otimes \sigma_z\frac{\sigma^
{\scriptscriptstyle 
(-)}}{2}\partial_k\otimes (\sigma_z)^{\otimes j-1-k}
\otimes \frac{\sigma^{\scriptscriptstyle (-)}}{2}\partial_j\otimes 
({\bf 1})^{\otimes 2n-j}=\nonumber\\
&=&\sum_{j<k}({\bf 1})^{\otimes j-1}\otimes \biggl[\frac{\sigma^
{\scriptscriptstyle 
(-)}}{2},\sigma_z\biggr]_+\partial_j\otimes 
(\sigma_z)^{\otimes k-1-j}\otimes \frac{\sigma^{\scriptscriptstyle (-)}}
{2}\partial_k\otimes ({\bf 1})^{\otimes 2n-k}=\nonumber\\
&=&0
\end{eqnarray}
where we have used respectively the fact that:\newline\smallskip
{\bf 1)} the terms with $j=k$ do not contribute to the sum since 
$\displaystyle \frac{\sigma^{\scriptscriptstyle (-)}}{2}$ is nilpotent;\newline \smallskip
{\bf 2)} $j$ and $k$ are dummy indices and so they can be interchanged; \newline \smallskip
{\bf 3)} $\displaystyle\biggl[\frac{\sigma^{\scriptscriptstyle (-)}}{2},\sigma_z\biggr]_+=0$. \newline \smallskip
With an analogous 
calculation one can prove that ${\bf \delta}^2=0$.
Because of this the Laplacian turns out to be just the anticommutator of ${\bf d}$ and ${\bf \delta}$:
${\bf \Delta}=({\bf d}+{\bf \delta})^2=[{\bf d},{\bf \delta}]_+$. Therefore, using the matrix representations 
of ${\bf d}$ and ${\bf \delta}$ we have: 
%for the Laplacian the following matrix expression:
%%%
\begin{eqnarray}
\displaystyle {\bf \Delta} &=&-\biggl[\sum_j(\sigma_z)^{\otimes j-1}\otimes \frac{\sigma^{
\scriptscriptstyle (-)}}
{2}\partial_j\otimes ({\bf 1})^{\otimes 2n-j},
\sum_k(\sigma_z)^{\otimes k-1}\otimes \frac{\sigma^{
\scriptscriptstyle (+)}}{2}\partial_k
\otimes ({\bf 1})^{\otimes 2n-k}\biggr]_+=\nonumber\\
&=&-\sum_{j<k}({\bf 1})^{\otimes j-1}\otimes \biggl[\frac{\sigma^{\scriptscriptstyle (-)}}{2},
\sigma_z\biggr]_+\partial_j
\otimes (\sigma_z)^{\otimes k-j-1}\otimes \frac{\sigma^{\scriptscriptstyle (+)}}{2}
\partial_k\otimes({\bf 1})^{\otimes 2n-k}\nonumber\\
& &-\sum_j({\bf 1})^{\otimes j-1}\otimes\biggl[\frac{\sigma^{\scriptscriptstyle (-)}}{2},
\frac{\sigma^{\scriptscriptstyle (+)}}{2}\biggr]_+
\partial_j^2\otimes ({\bf 1})^{\otimes 2n-j}\\
& &-\sum_{j>k}({\bf 1})^{\otimes k-1}\otimes\biggl[\sigma_z,
\frac{\sigma^{\scriptscriptstyle (+)}}{2}\biggr]_+\partial_k
\otimes (\sigma_z)^{\otimes j-k-1}\otimes\frac{\sigma^{\scriptscriptstyle (-)}}{2}
\partial_j\otimes ({\bf 1})^{\otimes 2n-j}=\nonumber\\
&=&-\sum_j({\bf 1})^{\otimes j-1}\otimes {\bf 1}\partial_j^2
\otimes ({\bf 1})^{\otimes 2n-j}.\nonumber
\end{eqnarray}
%%%
In the notations of the CPI ${\bf \Delta}$ becomes the following positive definite operator:
%%%
\begin{equation}
\displaystyle
{\bf \Delta}=\sum_{j=1}^{2n}({\bf 1})^{\otimes j-1}\otimes {\bf 1}
\lambda_j^2\otimes ({\bf 1})^{\otimes 2n-j}.
\end{equation}
%%%

\medskip

\noindent $\bullet$ {\bf Other Symmetry Charges of the CPI}. 
In the CPI some other charges were found which had also a clear geometrical meaning \cite{Goz89}.
For completeness we will write down here their expressions in terms of Pauli matrices:
%%%
\begin{eqnarray}
\displaystyle
&&\widehat{\overline{Q}}=\widehat{\bar{c}}_a\omega^{ab}\partial_b=
\sum_{j,l=1}^{2n}(\sigma_z)^{\otimes j-1}
\otimes \frac{\sigma^{\scriptscriptstyle (+)}}{2}\omega^{jl}\partial_l
\otimes ({\bf 1})^{\otimes 2n-j}\nonumber\\
&&\widehat{K}=\frac{1}{2}\omega_{ab}\widehat{c}^a\widehat{c}^b=\sum_{i=1}^n
\widehat{c}^{p_i}\widehat{c}^{q_i}=\nonumber\\
&&\qquad\;=\sum_{i=1}^n\biggl[(\sigma_z)^{\otimes 2(i-1)}\otimes
\frac{\sigma^{\scriptscriptstyle (-)}}{2}
\otimes ({\bf 1})^{\otimes 2(n-i)+1}\biggr]
\biggl[(\sigma_z)^{\otimes 2i-1}\otimes\frac{\sigma^{\scriptscriptstyle 
(-)}}{2}\otimes ({\bf 1})^{\otimes 2(n-i)}\biggr]=\nonumber\\
&&\qquad\;=\sum_{i=1}^n({\bf 1})^{\otimes 2(i-1)}\otimes 
\frac{\sigma^{\scriptscriptstyle 
(-)}}{2}\sigma_z\otimes\frac{\sigma^{\scriptscriptstyle (-)}}{2}\otimes
({\bf 1})^{\otimes 2(n-i)}=\\
&&\qquad\;=\sum_{i=1}^n({\bf 1})^{\otimes 2i-2}\otimes 
\biggl(\frac{\sigma^{\scriptscriptstyle 
(-)}}{2}\biggr)^{\otimes 2}\otimes({\bf 1})^{\otimes 2n-2i}\nonumber\\
&&\widehat{\overline{K}}=\sum_{i=1}^{n}\widehat{\bar{c}}_{q_i}\widehat{\bar{c}}_{p_i}
=\sum_{i=1}^n({\bf 1})^{\otimes 2i-2}
\otimes\biggl(\frac{\sigma^{\scriptscriptstyle 
(+)}}{2}\biggr)^{\otimes 2}\otimes({\bf 1})^{\otimes
2n-2i} \nonumber\\
&&\widehat{Q}_{\scriptscriptstyle H}=\sum_{j=1}^{2n}\widehat{c}^j(\partial_j-\beta\partial_jH)=
\sum_{j=1}^{2n}(\sigma_z)
^{\otimes j-1}\otimes\frac{\sigma^{\scriptscriptstyle
(-)}}{2}(\partial_j-\beta\partial_jH)
\otimes({\bf 1})^{\otimes 2n-j}\nonumber\\
&&\widehat{\overline{Q}}_{\scriptscriptstyle H}=\sum_{j,l=1}^{2n}\widehat{\bar{c}}_j\omega^{jl}
(\partial_l+\beta\partial_lH)=\sum_{j,l=1}^{2n}
(\sigma_z)^{\otimes j-1}\otimes\frac{\sigma^{\scriptscriptstyle
(+)}}{2}\omega^{jl}
(\partial_l+\beta\partial_lH)\otimes ({\bf 1})^{\otimes 2n-j}. \nonumber
\end{eqnarray}
%%%

\bigskip

\section{Irreducible Representations of the Algebra of the Symmetry Charges}

It is well-known, see \cite{Goz89}, that the charges $Q_f$, $K$ and $\overline{K}$ 
form an algebra which is, modulus a central extension, an Sp(2) algebra:
%%%
\begin{eqnarray}
&&\label{matrix.spp2} [Q_f,K]_{\scriptscriptstyle -}=2K\nonumber\\
&&[Q_f,\overline{K}]_{\scriptscriptstyle -}=-2\overline{K}\\
&&[K,\overline{K}]_{\scriptscriptstyle -}=Q_f-1. \nonumber 
\end{eqnarray}
%%%
It is easy to check that the $4\times 4$ matrices $\widehat{K},\widehat{\overline{K}}, \widehat{Q}_f$ 
defined in (\ref{matrix.opmat}) satisfy (\ref{matrix.spp2}).
Since the matrix ${\bf 1}_{\scriptscriptstyle 4\times 4}$ commutes with 
$\widehat{K}$ and $\widehat{\overline{K}}$ 
it is also possible to throw away 
the central extension of the algebra by replacing $\widehat{Q}_f\to 
\widehat{Q}_f-{\bf 1}_{\scriptscriptstyle 4\times 4}$.
What we want to prove now is that the Sp(2) algebra can be reproduced via the 
$2\times 2$ matrices $\displaystyle \frac{\sigma^{\scriptscriptstyle (+)}}{2},
\frac{\sigma^{\scriptscriptstyle (-)}}{2},\sigma_z$ 
used several times in the previous sections.
In fact:
%%%
\begin{eqnarray}
\displaystyle \label{matrix.spp3}
&&\biggl[\sigma_z,\frac{\sigma^{\scriptscriptstyle (+)}}{2}\biggr]_{\scriptscriptstyle -}=
\begin{pmatrix}0 & 2\\ 0 & 0\end{pmatrix}=2\biggl(\frac{\sigma^{\scriptscriptstyle (+)}}{2}\biggr)\nonumber\\
&&\biggl[\sigma_z,\frac{\sigma^{\scriptscriptstyle (-)}}{2}\biggr]_{\scriptscriptstyle -}=
\begin{pmatrix}0 & 0\\ -2 & 0\end{pmatrix}=-2\biggl(\frac{\sigma^{\scriptscriptstyle (-)}}{2}\biggr)\\
&&\biggl[\frac{\sigma^{\scriptscriptstyle (+)}}{2},\frac{\sigma^{\scriptscriptstyle (-)}}{2}\biggr]_{\scriptscriptstyle -}=
\begin{pmatrix}1 & 0\\ 0 & -1\end{pmatrix}=\sigma_z. \nonumber
\end{eqnarray} 
%%%
From (\ref{matrix.spp2}) and (\ref{matrix.spp3}) we have that it is possible to 
reproduce the Sp(2) algebra, without any central extension, by identifying
$\displaystyle Q_f=\sigma_z,\; K
=\frac{\sigma^{\scriptscriptstyle (+)}}{2}, \; 
\overline{K}=\frac{\sigma^{\scriptscriptstyle (-)}}{2}$.
Furthermore the only $2\times 2$ matrices commuting with the 3 generators of 
Sp(2) are those proportional to the identity.
This confirms that the representation we found is irreducible.
We note however that, while we can look at the 
generators of Sp(2) as $2\times 2$ matrices,
the same cannot happen for the charges $Q$ and $\overline{Q}$. 
For example it is impossible to find a $2\times 2$
matrix $Q$ which satisfies the relation 
$[Q_f,Q]_{\scriptscriptstyle -}=Q$. 
Therefore if we want to extend Sp(2) including all the other symmetries
of the CPI and finding non trivial representations of the algebra then
we have to consider $4\times 4$ instead of $2\times 2$ matrices. 
For sure the matrices (\ref{matrix.opmat})-(\ref{matrix.matrixn1}) whose entries 
are operators satisfy the correct algebra
of the symmetry charges of the CPI. What we want to prove now is that it is possible 
to find an irreducible representation 
in terms of $4\times 4$ matrices whose entries are just {\it real numbers}.
Let us consider, as independent charges, $Q=ic^a\lambda_a,\; 
\overline{Q}=i\bar{c}_a\omega^{ab}\lambda_b,
\; -i\overline{N}=-i\bar{c}_a\omega^{ab}\partial_bH,\; iN=ic^a\partial_aH$. 
They are conserved charges and among them the 
only anticommutators different from zero are: $[Q,-i\overline{N}]_{\scriptscriptstyle +}=
[\overline{Q},iN]_{\scriptscriptstyle +}=\HT$. Since $\HT$ is a Casimir for
the whole algebra the irreducible representations will be labeled by its 
eigenvalues $h$. Using
the results of the Appendix of \cite{Rittenberg} we can start by considering
two basis vectors $e_1,e_2$ to represent the subalgebra $[Q,-i\overline{N}]_{\scriptscriptstyle +}=\HT$ 
and two basis vectors $f_1,f_2$
to represent the subalgebra $[\overline{Q},iN]_{\scriptscriptstyle +}=\HT$:
%%%
\begin{eqnarray}
&&Qe_1=\sqrt{h}e_2,\;\;\; -i\overline{N}e_1=0,\;\;\;\qquad\quad iNf_1=
\sqrt{h}f_2,\;\;\;\overline{Q}f_1=0,\nonumber\\
&&Qe_2=0,\;\;\; -i\overline{N}e_2=\sqrt{h}e_1,\;\;\;\qquad \quad iNf_2=0,
\;\;\;\overline{Q}f_2=\sqrt{h}f_1.
\end{eqnarray}
%%%
A basis to represent the
algebra of $Q,\overline{Q},iN,-i\overline{N}, \HT$ is given by:
$F_1=e_1f_1$, $F_2=e_1f_2$, $F_3=e_2f_1$, $F_4=e_2f_2$.
According to our conventions $e_1$ and  $f_1$ are Grassmannian even while  
$e_2$ and $f_2$ are Grassmannian odd. 
Consequently $F_1$ and $F_4$ are Grassmannian even while $F_2$ and $F_3$ 
are Grassmannian odd. The matrix
representation of $Q$ is given by:
%%%
\begin{eqnarray}
\label{matrix.qmat}
&&QF_1=(Qe_1)f_1=\sqrt{h}e_2f_1=\sqrt{h}F_3\nonumber\\
&&QF_2=(Qe_1)f_2=\sqrt{h}e_2f_2=\sqrt{h}F_4\nonumber\\
&&\qquad\qquad\qquad\qquad\quad\Downarrow\\
&&\qquad\quad Q=\begin{pmatrix}0 & 0 & 0 & 0\\ 0 & 0 & 0 & 0\\ \sqrt{h} 
& 0 & 0 & 0\\ 0 & \sqrt{h} & 0 & 0\end{pmatrix}. \nonumber 
\end{eqnarray}
%%%
In the same way we can find the matrices associated to the other 
symmetry charges:
%%%
\begin{eqnarray}
&&-i\overline{N}=\begin{pmatrix}0 & 0 & \sqrt{h} & 0\\ 0 & 0 & 0 & \sqrt{h}\\ 
0 & 0 & 0 & 0\\ 0 & 0 & 0 & 0\end{pmatrix},\;\;\;\;
\overline{Q}=\begin{pmatrix}0 & \sqrt{h} & 0 & 0\\ 0 & 0 & 0 & 0\\ 0 & 0 & 
0 & -\sqrt{h}\\ 0 & 0 & 0 & 0\end{pmatrix}\nonumber\\
&&iN=\begin{pmatrix}0 & 0 & 0 & 0\\ \sqrt{h} & 0 & 0 & 0\\ 0 & 0 & 0 & 0\\ 
0 & 0 & -\sqrt{h} & 0\end{pmatrix},\;\;\;\;
\HT=\begin{pmatrix}h & 0 & 0 & 0\\ 0 & h & 0 & 0\\ 0 & 0 & h & 0\\ 0 & 0 & 
0 & h\end{pmatrix}. \label{matrix.nummat}
\end{eqnarray}
%%%
\setcounter{figure}{0}
\begin{figure}
\centering
\includegraphics[width=13cm, height=8cm]{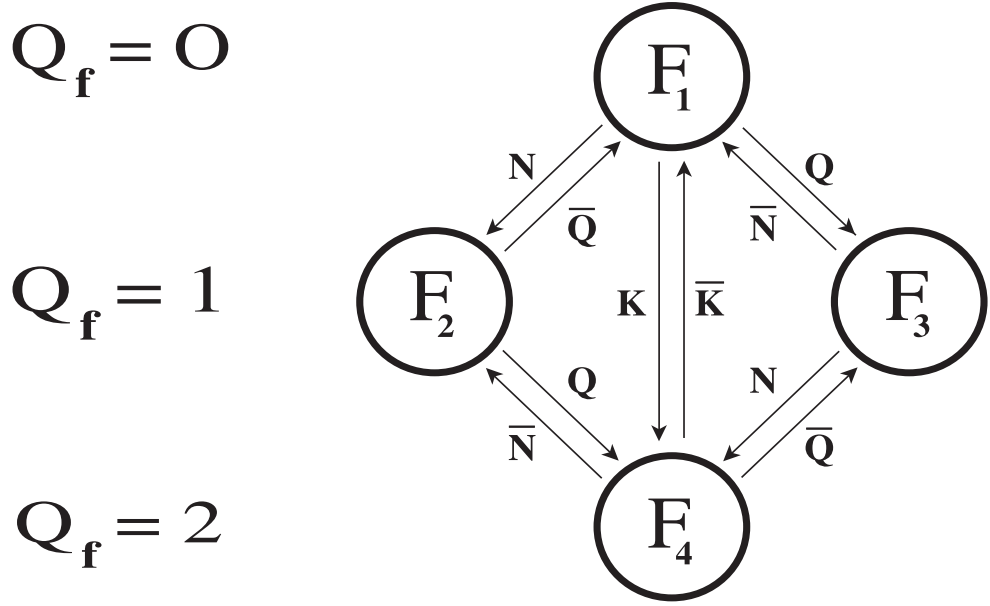}
\caption{\rm{Representation of the CPI charges.}} 
\label{matrix.rep}
\end{figure}
%%%
Considering the algebra of (\ref{matrix.qmat})-(\ref{matrix.nummat}) the only anticommutators
different from zero are: $[Q,-i\overline{N}]_{\scriptscriptstyle +}=
[\overline{Q},iN]_{\scriptscriptstyle +}=\HT$.
Among these $4\times 4$ matrices there is enough space also for the 
charges $Q_f=c^a\bar{c}_a$, $\displaystyle K=\frac{1}{2}\omega_{ab}c^ac^b$ 
and $\displaystyle \overline{K}=\frac{1}{2}\omega^{ab}\bar{c}_a\bar{c}_b$:
%%%
\begin{eqnarray}
&&\qquad \qquad Q_f=\begin{pmatrix}0 & 0 & 0 & 0\\ 0 & 1 & 0 & 0\\ 0 & 0 & 1 & 0\\ 0 & 
0 & 0 & 2\end{pmatrix},\nonumber\\
&& K=\begin{pmatrix}0 & 0 & 0 & 0\\ 0 & 0 & 0 & 0\\ 0 & 0 & 0 & 0\\ 1 & 0 & 
0 & 0\end{pmatrix},\;\;\;\;\;\;
\overline{K}=\begin{pmatrix}0 & 0 & 0 & 1\\ 0 & 0 & 0 & 0\\ 0 & 0 & 0 & 
0\\ 0 & 0 & 0 & 0\end{pmatrix}. \label{matrix.sp2}
\end{eqnarray}
%%%
In (\ref{matrix.sp2}) we have found again the matrices of (\ref{matrix.opmat}).
The novelty is entirely contained in Eqs. (\ref{matrix.qmat})-(\ref{matrix.nummat})
since there we have matrices whose entries are real numbers instead of operators.
All the matrices we have defined here satisfy the correct algebra 
of the symmetry charges of the CPI, which 
can be found in the original papers \cite{Goz89} or in (\ref{matrix.algcpi}).
It is possible to prove that the only $4\times 4$ matrix that commutes
with all the symmetry charges of the theory is given, modulus a proportionality 
factor, by the identity matrix. Therefore the representation we have found
is irreducible.  
So, in order to construct a non trivial irreducible representation of the symmetry charges of the CPI,
we just need 2 Grassmannian even states ($F_1$ and $F_4$) and 2 Grassmannian odd states ($F_2$ and $F_3$). 
All these 4 states can be connected each other by means of the symmetry charges as it emerges from Fig.
{\bf \ref{matrix.rep}}.
For example the charges $Q$ and $N$ which increase the form number by one allow us to go from $F_1$ 
to $(F_2,F_3)$ and from $(F_2,F_3)$ to $F_4$. The charge $K$ which increases the form number by 2 allows us to go from 
$F_1$ to $F_4$ while in the opposite direction we can go via $\overline{K}$. 

\newenvironment{appendices}{
\chapter*{Appendix to Chapter 3}
%\addcontentsline{toc}{chapter}{\numberline{}Appendix to Chapter 3}
\let\appendix\section
\setcounter{section}{0}
\def\thesection{\thechapter.\Alph{section}}
}{\clearpage}

\begin{appendices}
\appendix{Proof of (\ref{matrix.5-3})}
\label{app:pm}
\markboth{{\it{Appendix to Chapter 3}}}{}

\noindent In this Appendix we want to check explicitly that the Grassmann algebra can be reproduced 
in terms of tensor products of Pauli matrices, i.e. we want to prove in detail
Eq. (\ref{matrix.5-3}). First of all we want to prove that all the
$\widehat{c}$ anticommute.
If we take two indices $k$ and $l$ with $k<l$ we have that:
%%%
\begin{equation}
\displaystyle \widehat{c}^k\widehat{c}^l=
{\bf 1}\otimes {\bf 1}\otimes\ldots
\otimes {\bf 1}\otimes \underbrace{\frac{\sigma^{\scriptscriptstyle
(-)}}{2}\sigma_z}_{k}
\otimes \sigma_z\otimes\sigma_z\otimes
\ldots\otimes \underbrace{\frac{\sigma^{\scriptscriptstyle (-)}}{2}}_{l}\otimes
{\bf 1}\otimes\dots\otimes {\bf 1}
\end{equation}
%%%
while
%%%
\begin{equation}
\displaystyle \widehat{c}^l\widehat{c}^k=
{\bf 1}\otimes {\bf 1}\otimes\ldots
\otimes {\bf 1}\otimes \underbrace{\sigma_z\frac{\sigma^{\scriptscriptstyle
(-)}}{2}}_{k}
\otimes \sigma_z\otimes\sigma_z\otimes
\ldots\otimes \underbrace{\frac{\sigma^{\scriptscriptstyle (-)}}{2}}_{l}\otimes
{\bf 1}\otimes\dots\otimes {\bf 1}.
\end{equation}
%%%
Therefore the anticommutator is given by:
%%%
\begin{equation}
\displaystyle [\widehat{c}^k,\widehat{c}^l]_{\scriptscriptstyle +}=
{\bf 1}\otimes {\bf 1}\otimes\ldots
\otimes {\bf 1}\otimes \underbrace{\biggl[\frac{\sigma^{\scriptscriptstyle
(-)}}{2},\sigma_z
\biggr]_{\scriptscriptstyle +}}_{k}
\otimes \sigma_z\otimes\sigma_z\otimes
\ldots\otimes \underbrace{\frac{\sigma^{\scriptscriptstyle (-)}}{2}}_{l}\otimes
{\bf 1}\otimes\dots\otimes {\bf 1}
\end{equation}
%%%
and $\widehat{c}^k$ and $\widehat{c}^l$ anticommute because $\displaystyle 
\biggl[\frac{\sigma^{\scriptscriptstyle
(-)}}{2},\sigma_z\biggr]_{\scriptscriptstyle +}=0$. If instead we take $k=l$ 
we have:
%%%
\begin{equation}
\displaystyle \widehat{c}^k\widehat{c}^k={\bf 1}\otimes {\bf 1}\otimes
\ldots\otimes{\bf 1}\otimes\biggl(\frac{\sigma^{\scriptscriptstyle
(-)}}{2}\biggr)^2\otimes {\bf 1}\otimes
\ldots \otimes {\bf 1}
\end{equation}
%%%
and $\widehat{c}^k \widehat{c}^k=0$ because $\displaystyle 
\biggl(\frac{\sigma^{\scriptscriptstyle (-)}}{2}\biggr)^2=0$.
The proof that all the $\widehat{\bar{c}}$ anticommute is the same as the
previous one 
with $\sigma^{\scriptscriptstyle (-)}$
replaced everywhere by $\sigma^{\scriptscriptstyle (+)}$. 

The only thing that remains to be proved is the result of the anticommutator of
$\widehat{c}$
with $\widehat{\bar{c}}$.
If $k<l$ we have
%%%
\begin{equation}
\displaystyle \widehat{c}^k\widehat{\bar{c}}_l=
{\bf 1}\otimes {\bf 1}\otimes\ldots
\otimes {\bf 1}\otimes \underbrace{\frac{\sigma^{\scriptscriptstyle
(-)}}{2}\sigma_z}_{k}
\otimes \sigma_z\otimes\sigma_z\otimes
\ldots\otimes \underbrace{\frac{\sigma^{\scriptscriptstyle (+)}}{2}}_{l}\otimes
{\bf 1}\otimes\dots\otimes {\bf 1}
\end{equation}
%%%
while
%%%
\begin{equation}
\displaystyle \widehat{\bar{c}}_l\widehat{c}^k=
{\bf 1}\otimes {\bf 1}\otimes\ldots
\otimes {\bf 1}\otimes \underbrace{\sigma_z\frac{\sigma^{\scriptscriptstyle
(-)}}{2}}_{k}
\otimes \sigma_z\otimes\sigma_z\otimes
\ldots\otimes \underbrace{\frac{\sigma^{\scriptscriptstyle (+)}}{2}}_{l}\otimes
{\bf 1}\otimes\dots\otimes {\bf 1}.
\end{equation}
%%%
Therefore from $\displaystyle \biggl[\sigma_z,\frac{\sigma^{\scriptscriptstyle
(-)}}{2}\biggr]_{\scriptscriptstyle +}=0$
we get that $[\widehat{c}^k,\widehat{\bar{c}}_l]_{\scriptscriptstyle +}=0$.
If $k>l$ we have 
%%%
\begin{equation}
\displaystyle \widehat{c}^k\widehat{\bar{c}}_l=
{\bf 1}\otimes {\bf 1}\otimes\ldots
\otimes {\bf 1}\otimes \underbrace{\sigma_z\frac{\sigma^{\scriptscriptstyle
(+)}}{2}}_{l}
\otimes \sigma_z\otimes\sigma_z\otimes
\ldots\otimes \underbrace{\frac{\sigma^{\scriptscriptstyle (-)}}{2}}_{k}\otimes
{\bf 1}\otimes\dots\otimes {\bf 1}
\end{equation}
%%%
and 
%%%
\begin{equation}
\displaystyle \widehat{\bar{c}}_l\widehat{c}^k=
{\bf 1}\otimes {\bf 1}\otimes\ldots
\otimes {\bf 1}\otimes \underbrace{\frac{\sigma^{\scriptscriptstyle
(+)}}{2}\sigma_z}_{l}
\otimes \sigma_z\otimes\sigma_z\otimes
\ldots\otimes \underbrace{\frac{\sigma^{\scriptscriptstyle (-)}}{2}}_{k}\otimes
{\bf 1}\otimes\dots\otimes {\bf 1}.
\end{equation}
%%%
So from $\displaystyle \biggl[\sigma_z,\frac{\sigma^{\scriptscriptstyle
(+)}}{2}\biggr]_{\scriptscriptstyle +}=0$
we have that $[\widehat{c}^k,\widehat{\bar{c}}_l]_{\scriptscriptstyle
+}=0$.
%%%
Finally if we take the same index $k=l$ we obtain
%%%
\begin{eqnarray}
&&\displaystyle \widehat{c}^k\widehat{\bar{c}}_k={\bf 1}\otimes {\bf
1}\otimes \ldots \otimes
{\bf 1}\otimes \underbrace{\frac{\sigma^{\scriptscriptstyle
(-)}}{2}\frac{\sigma^{\scriptscriptstyle (+)}}{2}}_{k}\otimes
{\bf 1}\otimes \ldots \otimes {\bf 1}, \nonumber\\
&&\displaystyle \widehat{\bar{c}}_k\widehat{c}^k={\bf 1}\otimes {\bf
1}\otimes \ldots \otimes
{\bf 1}\otimes \underbrace{\frac{\sigma^{\scriptscriptstyle
(+)}}{2}\frac{\sigma^{\scriptscriptstyle (-)}}{2}}_{k}\otimes
{\bf 1}\otimes \ldots \otimes {\bf 1}
\end{eqnarray}
%%%
from which we can derive
%%%
\begin{eqnarray}
\displaystyle [\widehat{c}^k,\widehat{\bar{c}}_k]_{\scriptscriptstyle +}
&=&{\bf 1}\otimes {\bf 1}\otimes \ldots \otimes
{\bf 1}\otimes \underbrace{\biggl[\frac{\sigma^{\scriptscriptstyle (-)}}{2},
\frac{\sigma^{\scriptscriptstyle (+)}}{2}\biggr]_{\scriptscriptstyle
+}}_{k}\otimes
{\bf 1}\otimes \ldots \otimes {\bf 1}\nonumber\\
&=&{\bf 1}^{\otimes 2n}
\end{eqnarray}
%%%
where we have used the fact that:
$\displaystyle \biggl[\frac{\sigma^{\scriptscriptstyle (-)}}{2},
\frac{\sigma^{\scriptscriptstyle (+)}}{2}\biggr]_{\scriptscriptstyle +}={\bf
1}.$
%%%
\end{appendices}

%% file: chapter4.tex
\def \HT{{\mathcal H}}
\def \LT{{\mathcal L}}
\def \ET{{\widetilde{\mathcal E}}}
\def \HCT{\widehat{\mathcal H}}
\def \s{\scriptscriptstyle}

\def \I{1{\hspace*{-.32cm} |}}

\pagestyle{fancy}
\chapter*{\begin{center}
4. Hilbert Space Structure with Forms: I
\end{center}}
\addcontentsline{toc}{chapter}{\numberline{4}Hilbert Space Structure with Forms:
I}
\setcounter{chapter}{4}
\setcounter{section}{0}
\markboth{{\it{4. Hilbert Space Structure with Forms: I}}}{}

\begin{quote}
{\it{
Gentlemen: there's lots of room left in Hilbert space.}}\medskip\\
-{\bf Saunders MacLane}.
\end{quote}

\bigskip

\noindent In the previous chapter we have shown that the Classical Path Integral, with all its auxiliary variables,
provides a generalization of the original KvN theory and it provides a lot of interesting
geometrical structures. In this chapter we want to study more in detail 
the Hilbert space underlying the CPI. In particular we shall show that, while 
the Hilbert space of zero-forms can be endowed with a positive definite scalar
product and a unitary evolution, the same cannot be done when we include higher
forms. In this last case we explore all possible scalar products and prove that for those which are positive
definite the evolution is not unitary. Vice versa in all the scalar products for which the evolution is unitary
there are some states with negative norms. This feature is due to the Grassmannian nature of the forms
but it appears only in {\it classical mechanics}. It is known in fact that in a similar structure, which is
supersymmetric {\it quantum mechanics}, there can be scalar products compatible with both positive norms 
and unitary evolutions. The content of this chapter with many further calculational details can be found in
\cite{6P}.

\bigskip

\section{The Salomonson-van Holten Scalar Product}
As we have already seen in the first chapter, in order to endow their space with a true Hilbert 
structure, KvN used the following scalar product:
%%%
\begin{equation}
\displaystyle \langle \psi|\Phi\rangle =\int d\varphi \;\psi^*(\varphi)\Phi(\varphi). \label{grass.uno-quattro}
\end{equation}
%%%
Under this scalar product all the states have positive definite norms and the Liouville operator $\widehat{L}$
is Hermitian. This is necessary in order to guarantee the unitarity of
the evolution operator which is given by 
$\displaystyle U(\Delta t)=e^{-i\widehat{L}\Delta t}$.
As it is well-known the unitarity of the evolution is crucial in order to guarantee
the conservation of the total probability. 
The question we want to address now is whether the space of higher forms 
$\psi(\varphi, c)$ can also be endowed with a positive definite scalar product under which 
the operator of evolution $\widehat{\HT}$ is Hermitian. 

The first scalar product we shall explore in this section is the one proposed by Salomonson and van Holten in 
\cite{Salomonson} for supersymmetric quantum mechanics (susy QM). 
Their Hamiltonian is similar to our $\widehat{\HT}$ because both have ``bosonic" degrees
of freedom ($\varphi^a$) and Grassmannian ones ($c^a$), which are turned into one another by a
supersymmetry invariance \cite{Witten}. The differences instead are mainly in
the fact that the Hamiltonian of supersymmetric QM, like every 
good quantum Schr\"odinger-like operator, contains second order derivatives in the kinetic term, while
ours, like every good classical Liouville-like operator, contains only first order derivatives.

The basic operators of the CPI can be divided in the following two sets:
%%%
\begin{equation}
(\widehat{q},\widehat{\lambda}_q, \widehat{c}^q,\widehat{\bar{c}}_q)\;\;\; ; \;\;\;(\widehat{p},     \widehat{\lambda}_p,\widehat{c}^p,\widehat{\bar{c}}_p).
\label{grass.due-zero}
\end{equation}
%%%
The operators in the first set commute with those in the second one
according to the graded commutators 
(\ref{ijmpa.comm}). Therefore we can concentrate for a while on the first set. 
The strategy we shall 
follow is {\it to choose} some hermiticity conditions for these operators and
the normalization of one of the basic
states \cite{Salomonson}. These conditions will be sufficient to build a complete basis and
the scalar product which reproduces the chosen hermiticity
conditions. For the bosonic variables
($\widehat{\varphi}^a,\widehat{\lambda}_a$) we choose, once and for
all, the following hermiticity conditions:
%%%
\begin{equation}
\label{grass.due-uno}
\left\{
	\begin{array}{l}
	\displaystyle
	\widehat{\varphi}^{a\dagger}=\widehat{\varphi}^a\smallskip\\
	\displaystyle \widehat{\lambda}_a^{\dagger}=\widehat{\lambda}_a \\
	\end{array}
	\right. 
\end{equation}
%%%
which are consistent with the first of the commutators  (\ref{ijmpa.comm}). For the Grassmann variables
$\widehat{c}^q$ and $\widehat{\bar{c}}_q$, because of the
anticommutator present in (\ref{ijmpa.comm}), we could choose various hermiticity properties. Here we impose:
%%%
\begin{equation}
\label{grass.due-due}
	\left\{
		\begin{array}{l}
		\displaystyle
		\widehat{c}^{q\dagger}=\widehat{\bar{c}}_q\smallskip\\
		\displaystyle \widehat{\bar{c}}_q^{\dagger}=\widehat{c}^q\\
		\end{array}
		\right.
\end{equation}
%%%
which are the analog 
of the hermiticity conditions used by SvH in \cite{Salomonson}. Next let us  start building a basis for our Hilbert space.
First we define the state $|0-\rangle$ as 
%%%
\begin{equation}
\widehat{c}^q|0-\rangle=0 . \label{grass.due-tre}
\end{equation}
%%%
Such a state must exist because if we 
take a generic state $|s\rangle$ such that $\widehat{c}^q|s\rangle\neq 0$, then the following state
$|0-\rangle \equiv \widehat{c}^q|s\rangle$ has precisely the property (\ref{grass.due-tre}).
From $|0 -\rangle$ it is easy to build the complete set of eigenvectors of $\widehat{c}^q$
which are:
%%%
\begin{equation}
\displaystyle |\alpha -\rangle\equiv e^{-\alpha\widehat{\bar{c}}_q}|0-\rangle=
(1-\alpha\widehat{\bar{c}}_q)|0-\rangle= |0-\rangle -\alpha|0+\rangle. \label{grass.due-quattro}
\end{equation}
%%%
Using the anticommutator $[\widehat{c}^q,\widehat{\bar{c}}_q]=1$, it is in fact easy to show that  
%%%
\begin{equation}
\widehat{c}^q|\alpha -\rangle=\alpha|\alpha -\rangle
\end{equation}
%%%
where $\alpha$, due to the ``Grassmannian" nature of $\widehat{c}^q$, is a 
Grassmannian odd parameter. The state 
$|0+\rangle$ which
appears in (\ref{grass.due-quattro}) is defined as: 
%%%
\begin{equation}
|0+\rangle\equiv\widehat{\bar{c}}_q|0 -\rangle. \label{grass.due-cinque}
\end{equation}  
%%%
Note that $|0+\rangle$ cannot be zero, otherwise we would end up in a
contradiction; in fact, if $|0 +\rangle$ were zero, we
would get $\widehat{c}^q|0+\rangle=0$, and this cannot be true
because\break 
$\widehat{c}^q|0+\rangle=\widehat{c}^q\widehat{\bar{c}}_q|0-\rangle=[\widehat{c}^q,\widehat{\bar{c}}_q]|0-\rangle=|0-\rangle$
which is a non-zero state. Via $|0+\rangle$ it is easy to build the
eigenstates of $\widehat{\bar{c}}_q$. In fact, using 
the anticommutation relation (\ref{ijmpa.comm}), it is easy to prove
that the states 
%%%
\begin{equation}
|\beta +\rangle\equiv e^{-\beta \widehat{c}^q}|0+\rangle=(1-\beta\widehat{c}^q)
|0+\rangle=|0+\rangle -\beta|0-\rangle \label{grass.due-sei}
\end{equation}
%%% 
are eigenvectors of $\widehat{\bar{c}}_q$ with eigenvalues $\beta$:
%%%
\begin{equation}
\widehat{\bar{c}}_q|\beta +\rangle=\beta|\beta +\rangle. \label{grass.due-sette}
\end{equation}
%%%
Among the plethora of states that we built we have to choose a basis for our vector space 
by diagonalizing a Hermitian operator. 
Thanks to the hermiticity
conditions (\ref{grass.due-due}), it is easy to see that the operator 
$\widehat{N}_q\equiv\widehat{c}^q\widehat{\bar{c}}_q$ is Hermitian. Moreover it is a 
projection operator which implies that
it has only 1 and 0 as eigenvalues. The corresponding eigenstates are
just $|0-\rangle$ and $|0+\rangle$. In fact:
%%%
\begin{equation}
\label{grass.due-otto}
	\left\{
		\begin{array}{l}
		\displaystyle
		\widehat{N}_q|0-\rangle=\widehat{c}^q\widehat{\bar{c}}_q|0-\rangle=
		[\widehat{c}^q,\widehat{\bar{c}}_q]|0-\rangle=|0-\rangle\smallskip\\
		\displaystyle \widehat{N}_q|0+\rangle=\widehat{c}^q\widehat{\bar{c}}_q|0+\rangle=0.\\
		\end{array}
		\right.
\end{equation}
%%%
So $|0-\rangle$ and $|0+\rangle$ make a basis for our space. The careful reader may object that the eigenstates
$\bigl\{|0+\rangle,|0-\rangle\bigr\}$ of a Hermitian operator 
are a basis of the Hilbert space only if the scalar product is a positive
definite one. We shall see in (\ref{grass.due-venticinque}) that this is just the case of the SvH scalar product 
and so we can
proceed. 

Up to now we have not given any scalar product but only some hermiticity conditions like
(\ref{grass.due-due}). To obtain the  scalar product, which would reproduce those hermiticity rules, the only extra condition we
have to give is the normalization of $|0 -\rangle$, which we {\it choose} to be
%%%
\begin{equation}
\langle-0|0-\rangle=1 .\label{grass.due-nove}
\end{equation}
%%%
As a consequence of this we get 
%%%
\begin{equation}
\langle+0|0+\rangle=1,\qquad\qquad  \langle+0|0-\rangle=0.\label{grass.due-dieci}
\end{equation}
%%%
Using the following von Neumann notation for the scalar product:
$\Bigl(|\alpha \pm\rangle, |\beta\pm\rangle\Bigr)\equiv \langle\pm\alpha|\beta\pm\rangle$
we have that the relations (\ref{grass.due-dieci}) can be proven in the following manner:
%%%
\begin{eqnarray}
&&\Bigl(|0+\rangle,|0+\rangle\Bigr)=
\Bigl(\widehat{\bar{c}}_q|0-\rangle,\widehat{\bar{c}}_q|0-\rangle\Bigr)=
\Bigl(|0-\rangle,\widehat{c}^q\widehat{\bar{c}}_q|0-\rangle\Bigr)
\nonumber=\\
&&\qquad\qquad\quad\;\;\; =\Bigl(|0-\rangle,[\widehat{c}^q,\widehat{\bar{c}}_q]_+
|0-\rangle\Bigr)=\Bigl(|0-\rangle,|0-\rangle\Bigr)=1;\medskip\\
&&\Bigl(|0+\rangle,|0-\rangle\Bigr)=\Bigl(\widehat{\bar{c}}_q|0-\rangle,|0-\rangle\Bigr)=
\Bigl(|0-\rangle, \widehat{c}^q|0-\rangle\Bigr)=\Bigl(|0-\rangle,0\Bigr)=0. \nonumber
\end{eqnarray}
%%%
So $\bigl\{|0+\rangle$, $|0-\rangle\bigr\}$ form an orthonormal basis. As, according to (\ref{grass.due-due}),
$\widehat{c}^q$ and $\widehat{\bar{c}}_q$ are not Hermitian operators we cannot say that their eigenstates 
$|\alpha-\rangle$  
and $|\beta +\rangle$ make up an
orthonormal basis for our Hilbert space. Nevertheless it is interesting to study the relations among these states. For
example it is easy to prove that
%%%
\begin{equation}
     \left\{
		\begin{array}{l}
		\displaystyle
		|\beta +\rangle=-\int d\alpha \,e^{\alpha\beta}|\alpha-\rangle\smallskip \\
		\displaystyle |\gamma -\rangle=-\int d\beta \, e^{\beta\gamma}|\beta+\rangle \label{grass.due-tredici-b}
		\end{array}
		\right.
\end{equation}
%%%
while the scalar products among them are given by:
%%%
\begin{equation}
\left\{
		\begin{array}{l}
		\displaystyle
		\Bigl(|\alpha \pm\rangle,|\beta\pm\rangle\Bigr)=e^{\alpha^*\beta}\smallskip \label{grass.due-quattordici-a}\\
		\displaystyle \Bigl(|\alpha\pm\rangle, |\beta\mp\rangle\Bigr)=\pm\delta(\alpha^*-\beta). 
		\end{array}
		\right.
\end{equation}
%%%
With $|\alpha -\rangle$ and $|\beta +\rangle$ it is possible to build the following 
``twisted" resolutions of the identity:
%%%
\begin{equation}
-\int d\alpha|\alpha\pm\rangle\langle\mp \alpha^*|={\bf 1}. \label{grass.due-quindici}
\end{equation}
%%%
We call them ``twisted" because the bras and kets have the ``$+$" and ``$-$" signs interchanged. 

Since $|0+\rangle$ and $|0-\rangle$ form a basis of our Hilbert space, the completeness relation 
can also be written as:
%%%
\begin{equation}
|0+\rangle\langle+0|+|0-\rangle\langle-0|={\bf 1}. \label{grass.due-sedici}
\end{equation}
%%%
It is easy to prove that (\ref{grass.due-quindici}) and  (\ref{grass.due-sedici}) are
equivalent. In fact let us remember the relations:
%%%
\begin{equation}
	\left\{
		\begin{array}{l}
		\displaystyle \label{grass.due-diciassette}
		|\alpha -\rangle=|0-\rangle-\alpha|0+\rangle\smallskip \\
		\displaystyle |\beta +\rangle=|0+\rangle-\beta|0-\rangle 
		\end{array}
		\right.
\end{equation}
%%%
%%%
\begin{equation}
	\left\{
		\begin{array}{l}
		\displaystyle \label{grass.due-diciotto}
		\langle -\alpha|=\langle  -0|-\alpha^*\langle+0| \smallskip \\
		\displaystyle \langle +\beta|=\langle  +0|+\beta^*\langle-0|
		\end{array}
		\right.
\end{equation}
%%%
and let us insert them into the LHS of (\ref{grass.due-quindici}). What we get is
%%%
\begin{eqnarray}
{\bf 1}&\hspace{-0.2cm}=&\hspace{-0.2cm}-\int d\alpha |\alpha +\rangle\langle -\alpha^*|
=-\int d\alpha\Bigl\{\Bigl(|0+\rangle-\alpha|0-\rangle\Bigr)
\Bigl(\langle -0|-\alpha\langle+0|\Bigr)\Bigr\}=\nonumber\\
&\hspace{-0.2cm}=&\hspace{-0.2cm}-\int d\alpha\Bigl\{|0+\rangle\langle-0|-\alpha |0-\rangle\langle-0|-\alpha |0+\rangle\langle+0|\Bigr\}=
|0-\rangle\langle -0|+|0+\rangle\langle +0| \nonumber\\
\end{eqnarray}
%%%
which is exactly (\ref{grass.due-sedici}). 
For our resolutions of the identity we can either use (\ref{grass.due-sedici}) or
(\ref{grass.due-quindici}). For example we could use the last one to express a generic state
$|\psi\rangle$ in the following manner:
%%%
\begin{eqnarray}
|\psi\rangle&=&-\int d\alpha |\alpha-\rangle\langle+\alpha^*|\psi\rangle=-\int
d\alpha|\alpha-\rangle(\psi_0+\psi_1\alpha)=\nonumber\\
&=&-\int      d\alpha\Bigl(|0-\rangle-\alpha|0+\rangle\Bigr)(\psi_0+\alpha\psi_1)=\psi_0|0+\rangle+\psi_1|0-\rangle.
\label{grass.aggiunta-1}
\end{eqnarray}
%%%
Similarly to what happens in ordinary QM, where $\langle
q|\psi\rangle=\psi(q)$ is a function of $q$, the projection of 
the state $|\psi\rangle$ onto the basis of the eigenstates $\langle+\alpha^*|$
is given by a function of 
the Grassmannian odd number $\alpha$:
%%%
\begin{equation}
\langle+\alpha^*|\psi\rangle=\psi(\alpha)=\psi_0+\psi_1\alpha .\label{grass.aggiunta-2}
\end{equation}
%%%
As we want to reproduce the forms whose coefficients $\psi_0(\varphi)$,
$\psi_a(\varphi),\ldots$ are complex functions we will make the choice that $\psi_0$ and $\psi_1$ in (\ref{grass.aggiunta-2}) are
complex numbers. 

Making use of  (\ref{grass.aggiunta-1}) and of the scalar products
(\ref{grass.due-nove})-(\ref{grass.due-dieci}) among the states $\bigl\{|0+\rangle, |0-\rangle\bigr\}$,
 it is very easy to derive the expression of the scalar product between two generic
states $|\psi\rangle, \;|\Phi\rangle$ of the theory:
%%%
\begin{eqnarray}
\langle\Phi|\psi\rangle&\hspace{-0.2cm}=&\hspace{-0.2cm}\Bigl(|\Phi\rangle,|\psi\rangle\Bigr)=
\Bigl(\Phi_0|0+\rangle+\Phi_1|0-\rangle,\psi_0|0+\rangle+\psi_1|0-
\rangle\Bigr)=\nonumber\\
&\hspace{-0.2cm}=&\hspace{-0.2cm}\Bigl(|0+\rangle,\Phi_0^*\psi_0|0+\rangle\Bigr)+\Bigl(|0-\rangle,\Phi^*_1\psi_0|0+\rangle\Bigr)+\Bigl(|0+\rangle,
\Phi_0^*\psi_1|0-\rangle\Bigr)+\nonumber\\
&&+\Bigl(|0-\rangle,\Phi_1^*\psi_1|0-\rangle\Bigr)=
\Phi_0^*\psi_0+\Phi_1^*\psi_1 .\label{grass.scprod}
\end{eqnarray}
%%%
We can also express (\ref{grass.scprod}) in terms of integrations
over Grassmann variables:
%%%
\begin{equation}
\displaystyle \langle\Phi|\psi\rangle=\int d\eta d\alpha^* e^{\alpha^*\eta} \Phi^*(\alpha)\psi(\eta).
\end{equation}
%%%
From the previous relation it is also easy to obtain the expression of the norm of a generic state $|\psi\rangle$:
%%%
\begin{equation}
\langle\psi|\psi\rangle=|\psi_0|^2+|\psi_1|^2\ge 0 .\label{grass.due-venticinque}
\end{equation}
%%%

Up to now, in all our expressions, we have explicitly indicated the dependence only on the Grassmann
variable $c^q$, but we know that the basic operators of the CPI
(\ref{grass.due-zero}) were many more. Considering for example those
contained in the first set of (\ref{grass.due-zero}), i.e. 
$(\widehat{q},\widehat{\lambda}_q,\widehat{c}^q,\widehat{\bar{c}}_q)$, we see that our basic states should
include also a dependence on $q$ (if we choose the $q$ representation):
%%%
\begin{eqnarray}
&& |q,\alpha +\rangle\equiv |q\rangle\otimes |\alpha +\rangle\nonumber\\
&& |q,\beta -\rangle\equiv |q\rangle\otimes|\beta -\rangle .
\end{eqnarray}
%%%
The variables $\lambda_q$ and $\bar{c}_q$ are the momenta conjugated to $q$ and $c^q$ 
and so they would appear in the wave functions only in the momentum
representation that for the moment we do not consider.
If we include $q$ in the completeness relations (\ref{grass.due-quindici}) then we obtain:
%%%
\begin{equation}
-\int d\alpha dq\,|q,\alpha\pm\rangle\langle\mp\alpha^*,q|={\bf 1} .\label{grass.due-ventisette}
\end{equation}
%%%
Moreover the expansion (\ref{grass.aggiunta-1}) becomes:
%%%
\begin{eqnarray}
|\psi\rangle&=&-\int d\alpha dq\Bigl(|q, 0-\rangle-\alpha|q, 0+\rangle\Bigr)
\Bigl(\psi_0(q)+\alpha\psi_1(q)\Bigr)=\nonumber\\
&=&\int dq\Bigl[\psi_0(q)|q, 0+\rangle+ \psi_1(q)|q, 0-\rangle\Bigr]
\end{eqnarray}
%%%
and as a consequence the scalar product (\ref{grass.due-venticinque}) turns out to be
%%%
\begin{equation}
\langle\psi|\psi\rangle=\int dq\Bigl[|\psi_0(q)|^2+|\psi_1(q)|^2\Bigr]\ge 0. \label{grass.due-ventinove}
\end{equation}
%%%
Next we should start including the variables of the second set in (\ref{grass.due-zero}), i.e. $(\widehat{p}$,
$\widehat{\lambda}_p$, $\widehat{c}^p$, $\widehat{\bar{c}}_p)$.
As these operators commute with those of the first set things are not difficult.
Regarding the Grassmann
variables we should impose on
$\widehat{c}^p$ the following hermiticity conditions:
%%%
\begin{equation}
\label{grass.hermiticityp}
	\left\{
		\begin{array}{l}
		\displaystyle
		\widehat{c}^{p\dagger}=\widehat{\bar{c}}_p\smallskip\\
		\displaystyle \widehat{\bar{c}}_p^{\dagger}=\widehat{c}^p.\\
		\end{array}
		\right.
\end{equation}
%%%
Then we define a new state $|0-,0-\rangle$ as
%%%
\begin{equation}
	\left\{
		\begin{array}{l}
		\displaystyle 
		\widehat{c}^q|0-,0-\rangle=0 \smallskip \label{grass.due-trenta-a}\\
		\displaystyle \widehat{c}^p|0-,0-\rangle=0 
		\end{array}
		\right.
\end{equation}
%%%
where the first ``$0-$" in $|0-,0-\rangle$ is associated to $c^q$ and the second one to $c^p$. Analogously to
(\ref{grass.due-cinque}) we can define also the following other states 
%%%
\begin{equation}
	\left\{
		\begin{array}{l}
		\displaystyle 
		|0+,0-\rangle=\widehat{\bar{c}}_q|0-,0-\rangle \smallskip \\
		\displaystyle |0-,0+\rangle=-\widehat{\bar{c}}_p|0-,0-\rangle \smallskip\\
		|0+,0+\rangle=-\widehat{\bar{c}}_q\widehat{\bar{c}}_p|0-,0-\rangle. \label{grass.due-trentuno-b}
		\end{array}
		\right.
\end{equation}
%%%
The reason for the ``$-$" sign in front of the $\widehat{\bar{c}}_p$ on the RHS of
(\ref{grass.due-trentuno-b}) is because of the Grassmannian odd
nature of the first ``$0-$" in the state $|0-,0-\rangle$. These signs are the only things we should be careful about
when we deal with a great number of Grassmann variables.
Analogously to (\ref{grass.due-quattro}) we can construct the simultaneous eigenstates 
of $\widehat{c}^q$ and $\widehat{c}^p$: 
%%%
\begin{equation}
\displaystyle |\alpha_q-,\alpha_p-\rangle\equiv e^{-\alpha_q\widehat{\bar{c}}_q-\alpha_p\widehat{\bar{c}}_p}|
0-,0-\rangle, \label{grass.due-trentadue}
\end{equation}
%%%
the simultaneous eigenstates of $\widehat{\bar{c}}_q$ and $\widehat{\bar{c}}_p$:
%%%
\begin{equation}
\displaystyle |\beta_q+,\beta_p+\rangle\equiv e^{-\beta_q\widehat{c}^q-\beta_p\widehat{c}^p}|0+,0+\rangle
\label{grass.due-trentaduebis}
\end{equation}
%%%
but we can build also {\it mixed} states like:
%%%
\begin{equation}
\displaystyle
|\alpha_q-,\beta_p+\rangle\equiv e^{-\alpha_q\widehat{\bar{c}}_q-\beta_p\widehat{c}^p}|0-,0+\rangle.
\label{grass.due-trentanove}
\end{equation}
%%%
Like in the case of one Grassmann variable, imposing the normalization condition
%%%
\begin{equation}
\langle -0,-0|0-,0-\rangle=1 \label{grass.due-trentasette-a}
\end{equation}
%%%
and using the hermiticity rules (\ref{grass.due-due})-(\ref{grass.hermiticityp}) 
we get the scalar products among all the states. 
So for example we get that the other scalar products different from zero are:
%%%
\begin{equation}
\Bigl(|0+, 0-\rangle, |0+, 0-\rangle\Bigr)=1 
\end{equation}
%%%
\begin{equation}
\Bigl(|0-, 0+\rangle, |0-, 0+\rangle\Bigr)=1 \label{grass.due-trentasette-b}
\end{equation}
%%%
\begin{equation}
\Bigl(|0+, 0+\rangle, |0+, 0+\rangle\Bigr)=1. \label{grass.due-trentasette-c}
\end{equation}
%%%
We can also easily
derive the scalar products among the states (\ref{grass.due-trentadue}) and
(\ref{grass.due-trentaduebis}). For example:
%%%
\begin{equation}
\label{grass.due-trentotto-c}
	\left\{
		\begin{array}{l}
		\displaystyle 
		\Bigl(|\alpha_q -,\alpha_p-\rangle,|\beta_q-,\beta_p-\rangle\Bigr)=	
        \textrm{exp}[\alpha_q^*\beta_q+\alpha_p^*\beta_p]
        \smallskip\\
        \displaystyle 
		\Bigl(|\alpha_q^*+,\alpha_p^*+\rangle,|\beta_q-,\beta_p-\rangle\Bigr)=
		\delta(\alpha_q-\beta_q)\delta(\alpha_p-\beta_p) 
		\smallskip\\
		\displaystyle
		\Bigl(|\alpha_q^*-,\alpha_p^*-\rangle,|\beta_q+,\beta_p+\rangle\Bigr)=
		-\delta(\alpha_q-\beta_q)\delta(\alpha_p-\beta_p).
		\end{array}
		\right.
\end{equation}
%%%
In the case of more than one Grassmann variable, the round brackets
$\Bigl(|\;\;\rangle,|\;\;\rangle\Bigr)$, which indicate the scalar product,
are turned into the Dirac notation as follows:
%%%
\begin{equation}
\Bigl(|\alpha_q,\alpha_p\rangle,|\beta_q,\beta_p\rangle\Bigr)\equiv\langle\alpha_p,\alpha_q|\beta_q,\beta_p\rangle .\label{grass.due-quaranta}
\end{equation}
%%%
This indicates that to pass from the bra to the ket it is necessary to invert the order of the entries for the $q$ and the $p$. Otherwise if we interpreted for example the scalar product $\displaystyle \Bigl(|0_q-,0_p+\rangle,|0_q-,0_p+\rangle\Bigr)$ as 
$\langle-0_q,+0_p|0_q-,0_p+\rangle$ and
not as in (\ref{grass.due-quaranta}), then we would end up in the following contradiction: 
%%%
\begin{eqnarray}
&& \langle -0_q,+0_p|0_q-,0_p+\rangle=\langle -0_q,-0_p|c^p(-\bar{c}_p)|0_q-,0_p-\rangle=\nonumber\\
&&=-\langle -0_q,-0_p|0_q-,0_p-\rangle=-1=-\Bigl(|0_q-,0_p+\rangle,|0_q-,0_p+\rangle\Bigr).
\end{eqnarray}
%%%

We have now all the ingredients to write down the ``twisted" resolutions of the identity analog to  (\ref{grass.due-quindici}). They are
%%%
\begin{eqnarray}
&&\int d\alpha_qd\alpha_p|\alpha_q+,\alpha_p+\rangle\langle -\alpha_p^*,-\alpha_q^*|={\bf 1}\nonumber\\
&&\int d\alpha_pd\alpha_q|\alpha_q-,\alpha_p-\rangle\langle+\alpha_p^*,+\alpha_q^*|={\bf 1}.
\label{grass.due-quarantuno}
\end{eqnarray}
%%%
It is easy to prove that the LHS of (\ref{grass.due-quarantuno}) turns out to be equal to
%%%
\begin{equation}
|0+,0+\rangle\langle+0,+0|+|0-,0+\rangle\langle+0,-0|+
|0+,0-\rangle\langle-0,+0|+|0-,0-\rangle\langle-0,-0| \label{grass.due-quarantadue}
\end{equation}
%%%
and this is clearly equal to ${\bf 1}$ because it is made of the projectors on the 4 states:
%%%
\begin{equation}
|0+,0+\rangle, \;|0-,0+\rangle,\; |0+,0-\rangle,\; |0-,0-\rangle \label{grass.due-quarantatre}
\end{equation}
%%%
which are a complete basis in the case of 2 Grassmann variables $c^q$ and $c^p$. The
proof that they are a basis is analogous to the one presented in (\ref{grass.due-otto}): as we have
seen also in Sec. {\bf 3.3},
in the case of a 2-dimensional phase space we
can build two commuting Hermitian operators:
%%%
\begin{equation}
\label{grass.due-quarantaquattro}
	\left\{
		\begin{array}{l}
		\displaystyle 
		\widehat{N}_q=\widehat{c}^q\widehat{\bar{c}}_q
		\smallskip\\
		\displaystyle
		\widehat{N}_p=\widehat{c}^p\widehat{\bar{c}}_p 
		\end{array}
		\right.
\end{equation}
%%%
and the states (\ref{grass.due-quarantatre}) are just the 
eigenstates of $\widehat{N}_q$ and $\widehat{N}_p$ with eigenvalues $(0,0)$, $(1,0)$, $(0,1)$, $(1,1)$ respectively:
\begin{equation}
\label{grass.due-quarantacinque}
	\left\{
		\begin{array}{l}
		\displaystyle 
	 \widehat{N}_q|0+,0+\rangle=0,\;\;\;\;\;\qquad\qquad\quad\; \widehat{N}_p|0+,0+\rangle=0
		\smallskip\\
		\displaystyle
		\widehat{N}_q|0-,0+\rangle=|0-,0+\rangle,\;\;\;\;\;\qquad\widehat{N}_p|0-,0+\rangle=0
	 \smallskip\\
	 \displaystyle 
	 \widehat{N}_q|0+,0-\rangle=0,\;\;\;\;\;\qquad\qquad\;\;\;\;\widehat{N}_p|0+,0-\rangle=|0+,0-\rangle
	 \smallskip\\
	 \widehat{N}_q|0-,0-\rangle=|0-,0-\rangle,\;\;\;\;\;\qquad\widehat{N}_p|0-,0-\rangle=|0-,0-\rangle .
		\end{array}
		\right.
\end{equation}
%%%
If we had diagonalized only $\widehat{N}_q$ we would not have had a complete set of operators because 
the states $|0-,0-\rangle$ and
$|0-,0+\rangle$ have the same eigenvalue with respect to $\widehat{N}_q$. The degeneracy is completely 
removed by diagonalizing another Hermitian
and commuting operator like $\widehat{N}_p$ and so $\widehat{N}_q$ and $\widehat{N}_p$ make a complete
set of operators\footnote[1]{$\widehat{N}_q$ and $\widehat{N}_p$ have a
Grassmannian nature and so the reader could doubt that the usual theorems on Hermitian
operators hold for them and that the four states (\ref{grass.due-quarantatre}) really make a complete
basis. To convince himself of that the reader can perform the
long, but boring, calculation of checking  (\ref{grass.due-quarantuno}) on all possible states of the theory.}.

The next thing we are going to prove is that the SvH scalar
product is positive definite in the general case of a system with $2n$ Grassmann
variables. For $n=1$ let us write the second of the
completeness relations (\ref{grass.due-quarantuno}) as follows:
%%%
\begin{equation}
\int dpdqdc^pdc^q|q,p,c^q-,c^p-\rangle\langle+c^{p*},+c^{q*},p,q|={\bf 1} \label{grass.due-quarantasette}
\end{equation}
%%%
where we have denoted with $c^q$ and $c^p$, rather than $\alpha^q$ and $\alpha^p$, the eigenvalues of $\widehat{c}^q$ and $\widehat{c}^p$ and we have included the dependence of 
the wave functions also on $p$ and
$q$. Inserting this completeness relation into the scalar product $\langle\Phi|\psi\rangle$ between the states, we get
%%%
\begin{eqnarray}
\langle\Phi|\psi\rangle &=& \int dp dq dc^pdc^q\langle\Phi|q,p,c^q-,c^p-\rangle\langle+c^{p*},
+c^{q*},p,q|\psi\rangle=\nonumber\\
&=& \int dp dq dc^pdc^q\,\Phi_+^*(q,p,c^q,c^p)\psi_-(q,p,c^q,c^p) \label{grass.due-quarantotto}
\end{eqnarray}
%%%
where
%%%
\begin{equation}
\label{grass.due-quarantanove}
	 \left\{
		\begin{array}{l}
		\displaystyle 
		\Phi_+(q,p,c^q,c^p)\equiv\langle -c^p,-c^q,p,q|\Phi\rangle
		\smallskip \\
		\displaystyle
		\psi_-(q,p,c^q,c^p)\equiv\langle +c^{p*},+c^{q*},p,q|\psi\rangle .
		\end{array}
		\right.
\end{equation}
%%%
The function $\psi_-$ depends on $(q,p,c^q,c^p)$ and, because
$(c^q,c^p)$ are Grassmannian, it can only have the following form
%%%
\begin{eqnarray}
\label{grass.due-cinquanta-a}
\psi_-(q,p,c^q,c^p)&\hspace{-0.2cm}=&\hspace{-0.2cm}\langle +c^{p*},+c^{q*},p,q|\psi\rangle=\\
&\hspace{-0.2cm}=&\hspace{-0.cm}\psi_0(q,p)+\psi_q(q,p)c^q+\psi_p(q,p)c^p+\psi_2(q,p)c^qc^p\nonumber
\end{eqnarray}
%%%
where $\psi_0$ is the zero-form, $\psi_q$ the coefficient of $c^q$, $\psi_p$ the coefficient of $c^p$
and $\psi_2$ is the coefficient of the two-form.
$\Phi_+(q,p,c^q,c^p)$ instead is
%%%
\begin{eqnarray}
&&\Phi_+(q,p,c^q,c^p)=\langle -c^p,-c^q,p,q|\Phi\rangle=\nonumber\\
&&=\int dp^{\prime}dq^{\prime}dc^{p\prime}dc^{q\prime}\langle -c^p,-c^q,p,q|q^{\prime},p^{\prime},c^{q\prime}-,c^{p\prime}-\rangle
\langle +c^{p\prime *},+c^{q\prime *},p^{\prime},q^{\prime}|\Phi\rangle=\nonumber\\
&&=\int dc^{p\prime}dc^{q\prime}(1+c^{q*}c^{q\prime}+c^{p*}c^{p\prime}+c^{q*}c^{q\prime}c^{p*}c^{p\prime})
\cdot(\Phi_0+\Phi_qc^{q\prime}+\Phi_pc^{p\prime}+\Phi_2c^{q\prime}c^{p\prime})=\nonumber\\
&&=\Phi_2(q,p)+\Phi_p(q,p)c^{q*}-\Phi_q(q,p)c^{p*}-\Phi_0(q,p)c^{q*}c^{p*} \label{grass.due-cinquanta-b}
\end{eqnarray}
%%%
where we have made use of the resolution of the identity
(\ref{grass.due-quarantasette}) in the second step above and of the 
scalar product (\ref{grass.due-trentotto-c}) in the third step.
Inserting now (\ref{grass.due-cinquanta-a}) and (\ref{grass.due-cinquanta-b}) into (\ref{grass.due-quarantotto}) we get
%%%
\begin{eqnarray}
\langle\Phi|\psi\rangle &\hspace{-0.2cm}=&\hspace{-0.2cm}
\int dpdq dc^pdc^q(\Phi_2^*+\Phi_p^*c^q 
-\Phi_q^*c^p-\Phi_0^*c^pc^q)\cdot (\psi_0+\psi_qc^q+\psi_pc^p+\psi_2c^qc^p)=\nonumber\\
&\hspace{-0.2cm}=&\hspace{-0.cm}\int dp dq \,[\Phi_0^*\psi_0+\Phi^*_q\psi_q+\Phi^*_p\psi_p+\Phi_2^*\psi_2] .\label{grass.svhprod}
\end{eqnarray}
%%%
Using this equation we see that the norm of a generic state $|\psi\rangle$ 
is:
%%%
\begin{equation}
\langle\psi|\psi\rangle=\int dp dq\biggl[|\psi_0|^2+|\psi_q|^2+|\psi_p|^2+|\psi_2|^2\biggr]. \label{grass.due-cinquantuno}
\end{equation}
%%%
This confirms that the SvH scalar product is {\it positive definite}. 
Note that for the zero-forms the SvH scalar product is reduced to the
KvN one of  (\ref{grass.uno-quattro}). 

The derivation that we have presented here can be repeated for any number of degrees of freedom. 
The state becomes
%%%
\begin{equation}
\displaystyle \psi=\frac{1}{m!}\sum_{m=0}^{2n}\psi_{a_1\ldots a_m}(\varphi)c^{a_1}c^{a_2}\ldots c^{a_m}
\end{equation}
%%%
and the SvH norm turns out to be 
%%%
\begin{equation}
\displaystyle \langle\psi|\psi\rangle=K\sum_{\{a_i\}}\sum_{m=0}^{2n}\int d\varphi\,|\psi_{a_1\ldots a_m}(\varphi)|^2
\end{equation}
%%%
where $K$ is a positive number. The derivation is long but straightforward.

All the construction we have done here is very similar to the one of SvH \cite{Salomonson} but there is a {\it crucial} difference.
The model examined by SvH is supersymmetric QM whose
Hamiltonian is Hermitian under the
SvH scalar product. The Hamiltonian of our model instead is not
Hermitian under the same scalar product as we shall show below. In fact the operator which corresponds to the
$\HT$ of (\ref{ann.supham}) can be written as:
%%%
\begin{equation}
\widehat{\HT}=\widehat{\HT}_{bos}+\widehat{\HT}_{ferm} \label{grass.due-cinquantuno-x}
\end{equation}
%%%
where 
%%%
\begin{equation}
\label{grass.due-cinquantuno-xx}
	 \left\{
		\begin{array}{l}
		\displaystyle 
		\widehat{\HT}_{bos}=\widehat{\lambda}_a\omega^{ab}\partial_bH
		\smallskip \\
		\displaystyle
		\widehat{\HT}_{ferm}=i\widehat{\bar{c}}_a\omega^{ab}\partial_b\partial_dH\widehat{c}^d.
		\end{array}
		\right.
\end{equation}
%%%
Let us check the Hermitian nature of each piece: the bosonic part is nothing more than 
the Liouvillian and it is Hermitian also under the SvH scalar product. In fact:
%%%
\begin{equation}
\widehat{\HT}_{bos}^{\dagger}=(\widehat{\lambda}_a\omega^{ab}\partial_bH)^{\dagger}=
\partial_bH\omega^{ab}\widehat{\lambda}_a=\widehat{\lambda}_a\omega^{ab}\partial_bH=
\widehat{\HT}_{bos} \label{grass.due-cinquantuno-b}
\end{equation}
%%% 
where we have used the fact that, according to (\ref{grass.due-uno}), $\widehat{\lambda}_a^{\dagger}=
\widehat{\lambda}_a$.
Now let us analyse the fermionic part $\widehat{\HT}_{ferm}$ of the
Hamiltonian. The SvH hermiticity conditions (\ref{grass.due-due})-(\ref{grass.hermiticityp}) 
for the Grassmann variables are the following ones:
%%%
\begin{equation}
     \left\{
		\begin{array}{l}
		\displaystyle 
		(\widehat{c}^a)^{\dagger}=\widehat{\bar{c}}_a
		\smallskip \\
		\displaystyle
		(\widehat{\bar{c}}_a)^{\dagger}=\widehat{c}^a .\label{grass.due-cinquantadue}
		\end{array}
		\right.
\end{equation}
%%%
Next let us write $\widehat{\HT}_{ferm}$ as
%%%
\begin{equation}
\widehat{\HT}_{ferm}=i\widehat{\bar{c}}_a\omega^{ab}\partial_b\partial_dH\widehat{c}^d=
i\widehat{\bar{c}}_a{\cal F}^a_d\widehat{c}^d
\end{equation}
%%%
where ${\cal F}_d^a\equiv\omega^{ab}\partial_b\partial_dH$. Then
%%%
\begin{equation}
(\widehat{\HT}_{ferm})^{\dagger}=(i\widehat{\bar{c}}_a{\cal F}^a_d\widehat{c}^d)^{\dagger}=
(-i)(\widehat{c}^d)^{\dagger}({\cal F}^{\dagger})^d_a
(\widehat{\bar{c}}_a)^{\dagger}= -i\widehat{\bar{c}}_d({\cal F}^{\dagger})^d_a\widehat{c}^a.
\end{equation}
%%%
So $\widehat{\HT}_{ferm}$ would be Hermitian if ${\cal F}^{\dagger}=-{\cal F}$. As ${\cal F}$ is real,
the relation ${\cal F}^{\dagger}=-{\cal F}$ implies that ${\cal F}^{T}=-{\cal F}$. 
Let us see if this happens by taking
%%%
\begin{equation}
{\cal F}^q_p=\omega^{qb}\partial_b\partial_pH=\omega^{qp}\partial_p\partial_pH=
\partial_p\partial_pH \label{grass.due-cinquantatre}
\end{equation}
%%%
and comparing it with its transposed element
%%%
\begin{equation}
{\cal F}^p_q=\omega^{pb}\partial_b\partial_qH=\omega^{pq}\partial_q\partial_qH=-\partial_q\partial_qH .
\label{grass.due-cinquantaquattro}
\end{equation}
%%%
We see that (\ref{grass.due-cinquantatre}) and (\ref{grass.due-cinquantaquattro}) are not the opposite 
of each other like it should
be for $\widehat{\HT}_{ferm}$ to be Hermitian. Note anyhow that if we consider a harmonic oscillator with 
$\displaystyle H=\frac{1}{2}p^2+\frac{1}{2}q^2$, then 
${\cal F}^q_p=-{\cal F}^p_q$ and so $\widehat{\HT}$ is Hermitian. 
This of course would not happen for a generic potential and 
this concludes the proof that $\widehat{\HT}$ is {\it not always Hermitian} under the SvH scalar product. 

Originally the CPI model was formulated directly via path integrals without deriving it explicitly from the operatorial 
formalism.
In QM instead the path integral can be derived \cite{Hibbs} by assembling infinitesimal time evolutions in operatorial
form and inserting between them suitable resolutions of the identity. We shall now do the same for the CPI and as
resolutions of the identity we shall use the ones associated to the SvH scalar
product. Before proceeding we should remember that, besides
the Schr\"odinger representation (\ref{ijmpa.operatorial})
in which $\widehat{\varphi}^a$ is a multiplicative operator while 
$\displaystyle \widehat{\lambda}_a=-i\frac{\partial}{\partial\varphi^a}$ is a
derivative one, 
we can also have a sort of ``momentum" representation
in which $\widehat{\lambda}_a$ is a multiplicative operator and $\widehat{\varphi}^a$
is a derivative one; the same happens for the Grassmann variables. 
So in this momentum representation we have:
%%%
\begin{equation}
\label{grass.due-cinquantacinque}
	 \left\{
		\begin{array}{l}
		\displaystyle 
		\widehat{\varphi}^a=i\frac{\partial}{\partial\lambda_a}
		\smallskip \\
		\displaystyle
		\widehat{c}^a=\frac{\partial}{\partial\bar{c}_a}.
		\end{array}
		\right.
\end{equation}
%%%
The ``wave functions" in this representation would depend on 
$\lambda,\bar{c}$ and the resolution of the identity involving $\lambda$ and $\bar{c}$ would be:
%%%
\begin{equation}
\displaystyle
\int d\lambda_qd\lambda_pd\bar{c}_qd\bar{c}_p|\lambda_q,\lambda_p,\bar{c}_q+,\bar{c}_p+\rangle
\langle -\bar{c}^*_p, -\bar{c}^*_q,\lambda_p,\lambda_q|={\bf 1} \label{grass.due-cinquantasei-a}
\end{equation}
%%%
to be contrasted with the one involving $\varphi$ and $\bar{c}$ which is:
%%%
\begin{equation}
\int dqdpdc^pdc^q|q,p,c^q-,c^p-\rangle\langle +c^{p*},+c^{q*},p,q|={\bf 1}. \label{grass.due-cinquantasei-b}
\end{equation}
%%%
Using the two resolutions of the identity above and the SvH scalar product we will prove that
the following kernel:
%%%
\begin{equation}
\displaystyle K(f|i)=\langle+c_f^{p*},+c^{q*}_f,\varphi_f|e^{-i\widehat{\HT}
(t_f-t_i)}|\varphi_i,c^q_i-,c^p_i-\rangle, \label{grass.due-cinquantasette}
\end{equation}
%%%
which is the transition amplitude to go from $(\varphi_i,c^q_i,c^p_i)$ to 
$(\varphi_f,c^q_f,c^p_f)$, 
has the same path integral structure which led to the CPI \cite{Goz89}. 
So we can say that the Hilbert space structure of SvH
leads just to the classical path integral. Later on we will prove the same also for the other 
types of scalar product which we will introduce. The reader may wonder why (\ref{grass.due-cinquantasette}) is really the
transition amplitude to go from $(\varphi_i,c^q_i,c^p_i)$ to $(\varphi_f,c^q_f,c^p_f)$ since the final bra in 
(\ref{grass.due-cinquantasette}) is $\langle +c_f^{p*},+c_f^{q*},\varphi_f|$. The reason is
that this bra is the eigenstate of $\widehat{c}_q$ and $\widehat{c}_p$ with eigenvalues $c^q_f,c^p_f$. The proof goes
as follows. Let us start from the following relations:
\begin{equation}
\label{grass.due-cinquantotto-a}
	 \left\{
		\begin{array}{l}
		\displaystyle 
		\widehat{\bar{c}}_q|\varphi_f,c_f^{q*}+,c_f^{p*}+\rangle=c_f^{q*}|
		\varphi_f,c_f^{q*}+,c_f^{p*}+\rangle
		\smallskip \\
		\displaystyle
		\widehat{\bar{c}}_p|\varphi_f,c_f^{q*}+,c_f^{p*}+\rangle=c_f^{p*}|
		\varphi_f,c_f^{q*}+,c_f^{p*}+\rangle
		\end{array}
		\right.
\end{equation}
%%%
and let us perform their Hermitian conjugation with the
rules of the SvH scalar product. We get
%%%
\begin{equation}
\label{grass.due-cinquantotto-b}
	 \left\{
		\begin{array}{l}
		\displaystyle 
		\langle+c_f^{p*},+c_f^{q*},\varphi_f|\widehat{c}^q=c^q_f\langle+c_f^{p*},+c_f^{q*},\varphi_f|
		\smallskip \\
		\displaystyle
		\langle+c_f^{p*},+c_f^{q*},\varphi_f|\widehat{c}^p=c^p_f\langle+c_f^{p*},+c_f^{q*},\varphi_f|
		\end{array}
		\right.
\end{equation}
%%%
that is what we wanted to prove. 

Now, as it is usually done in QM, let us divide the interval $t_f-t_i$ in (\ref{grass.due-cinquantasette}) 
into $N$ intervals of
length $\epsilon$, so that $N\epsilon=t_f-t_i$. The amplitude $K(f|i)$ of (\ref{grass.due-cinquantasette}) 
can then be written as
%%%
\begin{equation}
K(f|i)=\langle +c_f^{p*},+c_f^{q*},\varphi_f\biggl|\underbrace{\textrm{exp}\bigl[-i\epsilon\widehat{\HT}\bigr]
\ldots \textrm{exp}\bigl[-i\epsilon\widehat{\HT}\bigr]}_{N\;\textrm{terms}}
\biggr|\varphi_i,c^q_i-,c^p_i-\rangle .\label{grass.due-cinquantotto}
\end{equation}
%%%
Let us then insert a resolution of the identity (\ref{grass.due-cinquantasei-a}) in front of each exponential in 
(\ref{grass.due-cinquantotto}) and a resolution of the identity
(\ref{grass.due-cinquantasei-b}) behind each exponential. We get the following expression:
%%%
\begin{eqnarray}
\displaystyle
K(f|i)&\hspace{-0.2cm}=&\hspace{-0.2cm}\langle+c_f^{p*},+c_f^{q*},p_f,q_f|
\biggl\{\prod_{j=1}^{N}\int d\lambda_{q_j} d\lambda_{p_j} 
d\bar{c}_{q_j} d\bar{c}_{p_j}dq_jdp_jdc^{p}_{j}dc^{q}_{j}\nonumber\\
&&|\lambda_{q_j},\lambda_{p_j},\bar{c}_{q_j}+,\bar{c}_{p_j}+\rangle\cdot
\langle -\bar{c}_{p_j}^*,-\bar{c}_{q_j}^*,\lambda_{p_j},\lambda_{q_j}\Bigl|\textrm{exp}[-i\epsilon\widehat{\HT}]
\Bigr|q_{\scriptscriptstyle j},p_{\scriptscriptstyle j},
c^q_{\scriptscriptstyle j}-,c^p_{\scriptscriptstyle j}-\rangle\nonumber\\
&&\cdot\langle +c_j^{p*},+c_j^{q*},p_{\scriptscriptstyle j},q_{\scriptscriptstyle j}|\biggr\}|q_i,p_i,c^q_i-,
c^p_i-\rangle .\label{grass.due-cinquantotto-bis}
\end{eqnarray}
%%%
The subindex $j$ on $(q,p,\lambda_q,\lambda_p,c^q,c^p,\bar{c}_q,\bar{c}_p)$ is the time
label in the subdivision of the interval $(t_f-t_i)$ in $N$ subintervals. 
There are various elements to evaluate in the expression above. We can start
from the last one which is easy to evaluate:
%%%
\begin{equation}
\langle +c_{\scriptscriptstyle 1}^{p*},+c_{\scriptscriptstyle 1}^{q*},p_{\scriptscriptstyle 1},q_{\scriptscriptstyle 1}
|q_i,p_i,c^q_i-,c^p_i-\rangle=\delta(p_{\scriptscriptstyle 1}-p_i)\delta(q_{\scriptscriptstyle 1}-q_i)
\delta(c^q_{\scriptscriptstyle 1}-c^q_i)\delta(c^p_{\scriptscriptstyle 1}-c^p_i) .\label{grass.due-cinquantanove}
\end{equation}
%%%
Another element is 
%%%
\begin{eqnarray}
&&\label{grass.due-sessanta} \langle
+c_j^{p*},+c_j^{q*},p_j,q_j|\lambda_{q_{j-1}},\lambda_{p_{j-1}},\bar{c}_{q_{j-1}}+,\bar{c}_{p_{j-1}}+\rangle=\nonumber\\
&&=\textrm{exp}\Bigl[ip_j\lambda_{p_{j-1}}+iq_j\lambda_{q_{j-1}}\Bigr]\langle
+c^{p*}_j,+c_j^{q*}|\bar{c}_{q_{j-1}}+,\bar{c}_{p_{j-1}}+\rangle=\\
&&=\textrm{exp}\Bigl[ip_j\lambda_{p_{j-1}}+iq_j\lambda_{q_{j-1}}\Bigr]\,\textrm{exp}
\Bigl[c^q_j\bar{c}_{q_{j-1}}+c^p_j\bar{c}_{p_{j-1}}\Bigr]\nonumber.
\end{eqnarray}
%%%
These expressions are the analog of the QM ones which link the momentum and space eigenstates and can be derived in
the same manner using the operatorial expression (\ref{ijmpa.operatorial}) for $\lambda_a$ and $\bar{c}_a$.
The last element that we need in (\ref{grass.due-cinquantotto-bis}) is
%%%
\begin{eqnarray}
\label{grass.due-sessantuno}
&& \langle -\bar{c}_{p_j}^*,-\bar{c}_{q_j}^*,\lambda_{p_j},\lambda_{q_j}|\textrm{exp}[-i\epsilon\widehat{\HT}
]|q_j,p_j,c^q_j-,c^p_j-\rangle=\\
&&=\textrm{exp}\Bigl[-i\epsilon\HT(\varphi_j,\lambda_j,c_j,\bar{c}_j)\Bigr]\textrm{exp}
\Bigl[\bar{c}_{q_j}c^q_j+\bar{c}_{p_j}c^p_j\Bigr]
\textrm{exp}\Bigl[-i\lambda_{q_j}q_j-i\lambda_{p_j}p_j\Bigr]\nonumber. 
\end{eqnarray}
%%%
Inserting (\ref{grass.due-cinquantanove})-(\ref{grass.due-sessanta})-(\ref{grass.due-sessantuno})
into (\ref{grass.due-cinquantotto-bis}) we get\footnote[2]{In the formula below we have suppressed the index ``$a$" on
$(\varphi^a,\lambda_a,c^a,\bar{c}_a)$.} 
%%%
\begin{equation}
\displaystyle 
K(f|i)
=\int {\mathcal D}\mu \,\textrm{exp}\biggl[i\epsilon
\sum_{j=1}^N\lambda_j\frac{(\varphi_{j+1}-\varphi_j)}{\epsilon}
-\epsilon\sum_{j=1}^N\bar{c}_j\frac{(c_{j+1}-c_j)}{\epsilon}-i\epsilon \sum_{j=1}^N\HT(j)\biggr]
\label{grass.due-sessantadue}
\end{equation}
%%%
where the boundary conditions are:
%%%
\begin{equation}
\varphi_0=\varphi_i,\;\;\;\;\;\;\varphi_{\scriptscriptstyle N+1}=\varphi_f,\;\;\;\;\;\;c_0=c_i,\;\;\;\;\;\;
c_{\scriptscriptstyle{N+1}}=c_f
\end{equation}
%%%
and where the measure is:
%%%
\begin{equation}
\displaystyle {\mathcal D}\mu= \biggl(\prod_{j=2}^Nd\varphi_jd\lambda_jd\bar{c}_jdc_j\biggr)
d\lambda_1d\bar{c}_1. \label{grass.due-sessantatre}
\end{equation}
%%%
This measure indicates that the initial and the final $(\varphi,c)$ are not integrated over. The continuum limit of (\ref{grass.due-sessantadue}) can be easily derived:
%%%
\begin{equation}
\displaystyle K(f|i)=\int_{\varphi_ic_i}^{\varphi_fc_f}{\mathcal D}\mu \;\textrm{exp}\Bigl[i\int dt\LT \Bigr]
\end{equation}
%%%
and $\LT$ turns out to be the Lagrangian in (\ref{ann.suplag}). This confirms that, via the scalar products and
the resolutions of the identity of SvH (\ref{grass.due-cinquantasei-a})-(\ref{grass.due-cinquantasei-b}), one gets just the CPI.

We can summarize this long section by saying that with the SvH scalar product the Hilbert space of CM is a true Hilbert space in
the sense that the scalar product is positive definite but unfortunately the Hamiltonian is not Hermitian even if the standard
path integral for CM can be reproduced. 

%%%%%%%%%%%%%%%%%%%%%%%%%%%%%%%%%%%%%%%%%%%%%%%%%%%%%%%%%%%%%%%%%%%%%%%%%%%%%%%%%%%%%%%%%%%%%%%%%%%%%%%%%%%%%%%%%%%%%%%%

\bigskip

\section{The Gauge Scalar Product}
\noindent
In this section we will study another scalar product which is the one typically used in gauge 
theories \cite{Henneaux}; this is the reason why we will call 
it ``gauge scalar product". Proceeding as in (\ref{grass.due-uno})-(\ref{grass.due-due}) 
for the SvH scalar product, we shall first ``postulate" some hermiticity conditions for the 
operators of the theory and then derive the scalar product and the resolutions of the identity. 
For the bosonic variables we choose the same hermiticity conditions as in the SvH case
%%%
\begin{equation}
\label{grass.tre-uno}
	 \left\{
		\begin{array}{l}
		\displaystyle 
		\widehat{\varphi}^{a\dagger}=\widehat{\varphi}^a
		\smallskip\\
		\displaystyle
		\widehat{\lambda}_a^{\dagger}=\widehat{\lambda}_a
		\end{array}
		\right.
\end{equation}
%%%
while for the Grassmann variables we choose:
%%%
\begin{equation}
\label{grass.tre-due}
	 \left\{
		\begin{array}{l}
		\displaystyle 
		\widehat{c}^{a\dagger}=\widehat{c}^a
		\smallskip\\
		\displaystyle
		\widehat{\bar{c}}_a^{\dagger}=\widehat{\bar{c}}_a.
		\end{array}
		\right.
\end{equation}
%%%
This means that the Grassmann variables are Hermitian, differently 
from the SvH case (\ref{grass.due-cinquantadue}). With this choice the Hamiltonian of the
CPI turns out to be Hermitian. In fact from
%%%
\begin{equation}
\widehat{\HT}=\widehat{\HT}_{bos}+\widehat{\HT}_{ferm} \label{grass.tre-due-b}
\end{equation}
%%%
we have that $\widehat{\HT}_{bos}$ is Hermitian as it was in the SvH case (\ref{grass.due-cinquantuno-b}) because it
involves only the bosonic variables which are Hermitian both in the SvH (\ref{grass.due-uno}) and in
the gauge case (\ref{grass.tre-uno}). On the other hand
it is easy to prove that also the $\widehat{\HT}_{ferm}$ of (\ref{grass.due-cinquantuno-xx}) is Hermitian. In fact:
%%%
\begin{equation}
(\widehat{\HT}_{ferm})^{\dagger}=-i\widehat{c}^d\omega^{ab}\partial_b\partial_dH\widehat{\bar{c}}_a=
i\widehat{\bar{c}}_a\omega^{ab}\partial_b\partial_dH\widehat{c}^d=\widehat{\HT}_{ferm} \label{grass.tre-tre}
\end{equation}
%%%
where in the second step of (\ref{grass.tre-tre}) we have used the anticommutation relations
%%%
\begin{equation}
[\widehat{c}^d,\widehat{\bar{c}}_a]=\delta_a^d \label{grass.tre-tre-a}
\end{equation}
%%%
together with the fact that $\omega^{ab}\partial_b\partial_aH=0$. 
Therefore the overall Hamiltonian $\widehat{\HT}$ 
is Hermitian under the gauge scalar product and this is a first important difference from the SvH case where 
$\widehat{\HT}$ was not Hermitian. We
will next check whether the scalar product underlying the gauge hermiticity conditions 
(\ref{grass.tre-uno})-(\ref{grass.tre-due}) is
positive definite, as in the SvH case, or not. 

Let us now perform all the steps necessary to implement the gauge scalar product
in the case of a single Grassmann variable $c$ and its conjugate $\bar{c}$. Their algebra and hermiticity
relations can be summarized in the following table:
%%%
\begin{equation}
\label{grass.tre-quattro}
	 \left\{
		\begin{array}{l}
		\displaystyle 
		[\widehat{c},\widehat{\bar{c}}]_+=1
		\smallskip \\
		\displaystyle
		\widehat{c}^2=\widehat{\bar{c}}^2=0
		\smallskip \\
		\widehat{c}^{\dagger}=\widehat{c}
		\smallskip \\
		\widehat{\bar{c}}^{\dagger}=\widehat{\bar{c}}.
		\end{array}
		\right.
\end{equation}
%%%
As in the SvH case, let us define a state $|0-\rangle$ by
%%%
\begin{equation}
\widehat{c}|0-\rangle=0 \label{grass.tre-cinque}
\end{equation}
%%%
and a state $|0+\rangle$ by
%%%
\begin{equation}
|0+\rangle\equiv \widehat{\bar{c}}|0-\rangle. \label{grass.tre-sei}
\end{equation}
%%%
If we calculate the norm of $|0+\rangle$, we get:
%%%
\begin{eqnarray}
\Bigl(|0+\rangle,|0+\rangle\Bigr)&=&\Bigl(\widehat{\bar{c}}|0-\rangle,\;\widehat{\bar{c}}|0-\rangle\Bigr)=
\Bigl(|0-\rangle,\;\widehat{\bar{c}}^{\dagger}\;\widehat{\bar{c}}|0-\rangle\Bigr)=\nonumber\\ &=&
\Bigl(|0-\rangle,\;\widehat{\bar{c}}\;\widehat{\bar{c}}|0-\rangle\Bigr)=\Bigl(|0-\rangle,0\cdot |0-\rangle\Bigr)=0. \label{grass.tre-sette}
\end{eqnarray}
%%%
The same happens also for $|0-\rangle$:
%%%
\begin{eqnarray}
\Bigl(|0-\rangle,|0-\rangle\Bigr)&=&\Bigl(\widehat{c}|0+\rangle,\widehat{c}|0+\rangle\Bigr)=
\Bigl(|0+\rangle,\widehat{c}^2|0+\rangle\Bigr)=\nonumber\\
&=& \Bigl(|0+\rangle,0\cdot |0+\rangle\Bigr)=0 .\label{grass.tre-otto}
\end{eqnarray}
%%%
First we notice that, differently from the SvH case, neither of these norms
can be chosen as we like; they are completely determined by the algebra given in
(\ref{grass.tre-quattro}). Second we notice that these states, which are
different from the null state, turn out to be of zero norm: 
this is the first sign that the gauge scalar product is not positive definite.
Then if we evaluate the following scalar product:
%%%
\begin{eqnarray}	
\langle +0|0-\rangle &=& \Bigl(|0+\rangle,|0-\rangle\Bigr)=\Bigl(\widehat{\bar{c}}|0-\rangle,|0-\rangle\Bigr)=
\Bigl(|0-\rangle,\widehat{\bar{c}}^{\dagger}|0-\rangle\Bigr)=\nonumber\\
&=& \Bigl(|0-\rangle,\widehat{\bar{c}}|0-\rangle\Bigr)=\Bigl(|0-\rangle,|0+\rangle\Bigr)=\langle -0|0+\rangle \label{grass.tre-nove}
\end{eqnarray}
%%%
we discover that $\langle +0|0-\rangle$ and $\langle -0|0+\rangle$ are not determined by the algebra 
(\ref{grass.tre-quattro}) and so we could choose them to be 1:
%%%
\begin{equation}
\langle +0|0-\rangle=\langle -0|0+\rangle=1. \label{grass.tre-dieci}
\end{equation}
%%%
In this way the complete set of scalar products is the following one:
\begin{equation}
\label{grass.tre-quindici}
	 \left\{
		\begin{array}{l}
		\displaystyle 
		\Bigl(|0+\rangle,|0+\rangle\Bigr)=0
		\smallskip \\
		\displaystyle
		\Bigl(|0-\rangle,|0-\rangle\Bigr)=0
		\smallskip \\
		\displaystyle 
		\Bigl(|0+\rangle,|0-\rangle\Bigr)=1
		\smallskip \\
		\Bigl(|0-\rangle,|0+\rangle\Bigr)=1.
		\end{array}
		\right.
\end{equation}
%%%
Analogously to the SvH case let us now build the eigenstates 
of $\widehat{c}$ and $\widehat{\bar{c}}$ which are respectively:
%%%
\begin{equation}
\label{grass.tre-sedici}
	 \left\{
		\begin{array}{l}
		\displaystyle 
		|\alpha -\rangle=e^{-\alpha\widehat{\bar{c}}}|0-\rangle=|0-\rangle-\alpha|0+\rangle
		\smallskip \\
		\displaystyle
		|\beta +\rangle=e^{-\beta \widehat{c}}|0+\rangle=|0+\rangle-\beta|0-\rangle.
		\end{array}
		\right.
\end{equation}
%%% 
The gauge scalar products among these states can be easily worked out using
 (\ref{grass.tre-quindici}):
%%%
\begin{equation}
\label{grass.tre-diciassette}
	 \left\{
		\begin{array}{l}
		\displaystyle 
		\Bigl(|\alpha\pm\rangle,|\beta\pm\rangle\Bigr)=\mp\delta(\alpha^*-\beta)
		\smallskip \\
		\displaystyle
		\Bigl(|\alpha\pm\rangle,|\beta\mp\rangle\Bigr)=e^{\alpha^*\beta}.
		\end{array}
		\right.
\end{equation}
%%% 
These relations should be compared with the SvH ones which 
are given in (\ref{grass.due-quattordici-a}).

\par 
Next we have to build the resolutions of the identity analogous to the SvH ones 
(\ref{grass.due-quindici}). In the gauge case they become:
%%%
\begin{equation}
-\int d\alpha|\alpha\pm\rangle\langle\pm\alpha^*|={\bf 1}. \label{grass.tre-diciotto}
\end{equation}
%%%
Note that the signs in the bra and ket are no longer ``twisted" as they were
in the SvH case.
To prove (\ref{grass.tre-diciotto}) we have to test it on all the states we are interested in, i.e. $|\beta+\rangle$
and $|\beta-\rangle$. First let us start with $|\beta +\rangle$ and the relation
%%%
\begin{equation}
-\int d\alpha|\alpha +\rangle\langle +\alpha^*|={\bf 1} . \label{grass.tre-venti}
\end{equation}
%%%
Applying the LHS of (\ref{grass.tre-venti}) on $|\beta +\rangle$ we obtain:
%%%
\begin{eqnarray}
&&\displaystyle -\int d\alpha|\alpha+\rangle\langle +\alpha^*|\beta+\rangle=-\int d\alpha|\alpha
+\rangle(-)\delta(\alpha-\beta)=\nonumber\\
&&=\int d\alpha \Bigl(|0+\rangle-\alpha|0-\rangle\Bigr)(\alpha-\beta)=|0+\rangle-\beta|0-\rangle=|\beta +\rangle
\end{eqnarray}
%%%
where in the first step above we used the scalar product (\ref{grass.tre-diciassette}).
If we do the same for $|\beta-\rangle$ we obtain:
%%%
\begin{eqnarray}
&&\displaystyle -\int d\alpha|\alpha+\rangle\langle+\alpha^*|\beta-\rangle=-\int d\alpha|\alpha+\rangle \textrm{exp}(\alpha
\beta)=\nonumber\\
&&=-\int d\alpha\Bigl(|0+\rangle-\alpha|0-\rangle\Bigr)(1+\alpha\beta)=|0-\rangle-\beta|0+\rangle=|\beta-\rangle
\end{eqnarray}
%%%
which proves (\ref{grass.tre-venti}). The proof is the same for 
%%%
\begin{equation}
-\int d\alpha|\alpha-\rangle\langle-\alpha^*|={\bf 1}. \label{grass.tre-venti-b}
\end{equation}
%%%

Let us now find out which is the expression
of the gauge scalar product between two generic states $|\psi\rangle$ and $|\Phi\rangle$. 
Using the resolution of the identity (\ref{grass.tre-venti-b}) we have 
%%%
\begin{equation}
|\psi\rangle=-\int d\alpha|\alpha-\rangle\langle-\alpha^*|\psi\rangle. \label{grass.tre-ventuno}
\end{equation}
%%%
As $\langle -\alpha^*|\psi\rangle$ is a function of $\alpha$, we can write it as
$\langle -\alpha^*|\psi\rangle=\psi_1+\alpha\psi_2$. Inserting in (\ref{grass.tre-ventuno}) 
the expression (\ref{grass.tre-sedici}) for $|\alpha-\rangle$, we get:
%%%
\begin{equation}
|\psi\rangle=
-\int d\alpha\Bigl(|0-\rangle-\alpha|0+\rangle\Bigr)(\psi_1+\alpha\psi_2)=\psi_1|0+\rangle+\psi_2|0-\rangle.
\end{equation}
%%%
Doing the same for $\langle\Phi|$ we get that the scalar product between two states $|\Phi\rangle$ and $|\psi\rangle$ 
is
%%%
\begin{eqnarray}
\langle\Phi|\psi\rangle&=&\Bigl(|\Phi\rangle,|\psi\rangle\Bigr)=\Bigl(\Phi_1|0+\rangle+\Phi_2|0-\rangle,
\psi_1|0+\rangle+\psi_2|0-\rangle\Bigr)=\Phi_1^*\psi_1\Bigl(|0+\rangle,|0+\rangle\Bigr)\nonumber\\
&&+\Phi^*_1\psi_2\Bigl(|0+\rangle,|0-\rangle\Bigr)+
\Phi_2^*\psi_1\Bigl(|0-\rangle,|0+\rangle\Bigr)+\Phi_2^*\psi_2\Bigl(|0-\rangle,|0-\rangle\Bigr)=\nonumber\\
&=&\Phi_1^*\psi_2+\Phi_2^*\psi_1
\label{grass.tre-ventidue}
\end{eqnarray}
%%%
where in the last step we have used the scalar products (\ref{grass.tre-quindici}). 
According to (\ref{grass.tre-ventidue}), the norm of the state $\psi_1+\alpha\psi_2$ is given by:
%%%
\begin{equation}
\langle \psi|\psi\rangle=\psi_1^*\psi_2+\psi_2^*\psi_1=2 \,\textrm{Re}\;\psi_1^*\psi_2. \label{grass.tre-ventitre}
\end{equation}
%%%
Differently from the SvH norms (\ref{grass.due-venticinque})
the gauge ones (\ref{grass.tre-ventitre}) are not always positive definite. 
For example the zero-forms  (i.e.
states like $\psi=\psi_1$) have zero norm and the same happens for the one-forms $\psi=\alpha\psi_2$.
It is also easy to build negative norm states like for example:
%%%
\begin{equation}
\psi=\psi_1-\psi_1\alpha ,\label{grass.tre-ventiquattro}
\end{equation}
%%%
for which $\|\psi\|^2=-2|\psi_1|^2$. 

Having worked out all the details for the case of one Grassmann
variable, we should now turn to the CPI where, for
one degree of freedom, we have two Grassmann variables ($c^q,c^p$). 
We can proceed along the same lines we followed for the SvH case: we derive all the scalar products from 
one choice of normalization, e.g. $\Bigl(|0-,0-\rangle,|0+,0+\rangle\Bigr)=i$, 
using the hermiticity conditions (\ref{grass.tre-due}) and the anticommutators (\ref{grass.tre-tre-a}) 
among the Grassmann variables. For example: 
%%%
\begin{eqnarray}
&&\Bigl(|0+,0+\rangle,|0-,0-\rangle\Bigr)=\Bigl(-\widehat{\bar{c}}_q\widehat{\bar{c}}_p|0-,0-\rangle,
\widehat{c}^q\widehat{c}^p|0+,0+\rangle
\Bigr)=\nonumber\\
&&=\Bigl(|0-,0-\rangle,-\widehat{\bar{c}}_p\widehat{\bar{c}}_q\widehat{c}^q\widehat{c}^p|0+,0+\rangle\Bigr)=
\Bigl(|0-,0-\rangle,-[\widehat{\bar{c}}_q,\widehat{c}^q][\widehat{\bar{c}}_p,\widehat{c}^p]|0+,0+\rangle\Bigr)=
\nonumber\\
&&=-\Bigl(|0-,0-\rangle,|0+,0+\rangle\Bigr)=-i .
\end{eqnarray}
%%%
Using the same kind of calculation it is possible to prove that the only non-zero 
scalar products are:
%%%
\begin{equation}
\label{grass.tre-ventinove}
	 \left\{
		\begin{array}{l}
		\displaystyle 
		\Bigl(|0-,0-\rangle,|0+,0+\rangle\Bigr)=i
		\smallskip\\
		\Bigl(|0+,0+\rangle,|0-,0-\rangle\Bigr)=-i
		\smallskip\\
		\Bigl(|0-,0+\rangle,|0+,0-\rangle\Bigr)=-i
		\smallskip \\
		\Bigl(|0+,0-\rangle,|0-,0+\rangle\Bigr)=i.
		\end{array}
		\right.
\end{equation}
%%%
Then, as in the SvH case, we can build the following states 
%%%
\begin{equation}
|\alpha_q-,\alpha_p-\rangle\equiv \textrm{exp}[-\alpha_q\widehat{\bar{c}}_q-\alpha_p\widehat{\bar{c}}_p]|0-,0-\rangle \label{grass.tre-trenta}
\end{equation}
%%%
which are eigenstates of $\widehat{c}^q$ and $\widehat{c}^p$ with eigenvalues $\alpha_q$ and $\alpha_p$:
%%%
\begin{equation}
\label{grass.tre-trentuno}
	 \left\{
		\begin{array}{l}
		\displaystyle 
		\widehat{c}^q|\alpha_q-,\alpha_p-\rangle=\alpha_q|\alpha_q-,\alpha_p-\rangle
		\smallskip \\
		\displaystyle
		\widehat{c}^p|\alpha_q-,\alpha_p-\rangle=\alpha_p|\alpha_q-,\alpha_p-\rangle.
		\end{array}
		\right.
\end{equation}
%%%
The proof is the same as in the SvH case because it is based only on the commutation relations which are the same in both
cases. Analogously we can build the states 
%%%
\begin{equation}
|\beta_q+,\beta_p+\rangle=\textrm{exp}[-\beta_q\widehat{c}^q-\beta_p\widehat{c}^p]|0+,0+\rangle \label{grass.tre-trentadue}
\end{equation}
%%%
which are eigenstates of $\widehat{\bar{c}}_q,\widehat{\bar{c}}_p$
%%%
\begin{equation}
\label{grass.tre-trentatre}
	 \left\{
		\begin{array}{l}
		\displaystyle 
		\widehat{\bar{c}}_q|\beta_q+,\beta_p+\rangle=\beta_q|\beta_q+,\beta_p+\rangle
		\smallskip \\
		\displaystyle
		\widehat{\bar{c}}_p|\beta_q+,\beta_p+\rangle=\beta_p|\beta_q+,\beta_p+\rangle
		\end{array}
		\right.
\end{equation}
%%%
and the states 
%%%
\begin{equation}
\label{grass.tre-trentaquattro}
	 \left\{
		\begin{array}{l}
		\displaystyle 
		|\alpha_q-,\beta_p+\rangle=\textrm{exp}[-\alpha_q\widehat{\bar{c}}_q-\beta_p\widehat{c}^p]|0-,0+\rangle
		\smallskip \\
		\displaystyle
		|\beta_q+,\alpha_p-\rangle=\textrm{exp}[-\beta_q\widehat{c}^q-\alpha_p\widehat{\bar{c}}_p]|0+,0-\rangle
		\end{array}
		\right.
\end{equation}
%%%
which are respectively eigenstates of the following operators:
%%%
\begin{equation}
\label{grass.tre-trentasei}
	 \left\{
		\begin{array}{l}
		\displaystyle 
		\widehat{c}^q|\alpha_q-,\beta_p+\rangle=\alpha_q|\alpha_q-,\beta_p+\rangle
		\smallskip \\
		\displaystyle
		\widehat{\bar{c}}_p|\alpha_q-,\beta_p+\rangle=\beta_p|\alpha_q-,\beta_p+\rangle
		\end{array}
		\right.
\end{equation}
%%%
%%%
\begin{equation}
\label{grass.tre-trentasette}
	 \left\{
		\begin{array}{l}
		\displaystyle 
		\widehat{\bar{c}}_q|\beta_q+,\alpha_p-\rangle=\beta_q|\beta_q+,\alpha_p-\rangle
		\smallskip \\
		\displaystyle
		\widehat{c}^p|\beta_q+,\alpha_p-\rangle=\alpha_p|\beta_q+,\alpha_p-\rangle.
		\end{array}
		\right.
\end{equation}
%%%
Next, we can calculate the gauge scalar products among all these states using (\ref{grass.tre-ventinove}) and 
the hermiticity relations (\ref{grass.tre-due}). The results are:
%%%
\begin{eqnarray}
\label{grass.tre-trentotto}
&&\Bigl(|\alpha_q-,\alpha_p-\rangle,|\beta_q+,\beta_p+\rangle\Bigr)=i\cdot \textrm{exp}(\alpha_q^*\beta_q+\alpha_p^*\beta_p)\nonumber\\
&&\Bigl(|\alpha_q-,\alpha_p-\rangle,|\alpha_q^{\prime}-,\beta_p+\rangle\Bigr)=i\cdot \delta(\alpha_q^*-\alpha_q^{\prime})
\textrm{exp}(\alpha_p^*\beta_p)\nonumber\\
&&\Bigl(|\alpha_q-,\alpha_p-\rangle,|\beta_q+,\alpha_p^{\prime}-\rangle\Bigr)=i\cdot \delta(\alpha_p^*-\alpha_p^{\prime})
\textrm{exp}(\alpha_q^*\beta_q)\nonumber\\
&&\Bigl(|\alpha_q-,\alpha_p-\rangle,|\alpha_q^{\prime}-,\alpha_p^{\prime}-\rangle\Bigr)=i\cdot
\delta(\alpha_q^*-\alpha_q^{\prime})\delta
(\alpha_p^*-\alpha_p^{\prime})\\
&&\Bigl(|\beta_q+,\beta_p+\rangle,|\alpha_q-,\beta_p^{\prime}+\rangle\Bigr)=
i\cdot\delta(\beta_p^*-\beta_p^{\prime})\textrm{exp}(\beta_q^*\alpha_q)
\nonumber\\
&&\Bigl(|\beta_q+,\beta_p+\rangle,|\beta_q^{\prime}+,\alpha_p-\rangle\Bigr)=
i\cdot\delta(\beta_q^{\prime}-\beta_q^*)\textrm{exp}(\beta_p^*\alpha_p)
\nonumber\\
&&\Bigl(|\beta_q+,\beta_p+\rangle,
|\beta_q^{\prime}+,\beta_p^{\prime}+\rangle\Bigr)=i\cdot\delta(\beta_q^*-\beta_q^{\prime})\delta(\beta_p^*-\beta_p^{\prime})\nonumber\\
&&\Bigl(|\beta_q+,\beta_p+\rangle,|\alpha_q-,\alpha_p-\rangle\Bigr)=-i\cdot
\textrm{exp}(\beta_q^*\alpha_q+\beta_p^*\alpha_p)\nonumber.
\end{eqnarray}
%%%
Via these scalar products it is easy to prove the following resolutions of the identity:
%%%
\begin{equation}
i\int d\alpha_q d\alpha_p|\alpha_q\pm,\alpha_p\pm\rangle\langle\pm\alpha_p^*,\pm\alpha_q^*|={\bf 1} \label{grass.tre-trentanove}
\end{equation}
%%%
which should be compared with the SvH ones (\ref{grass.due-quarantuno}). 

Using (\ref{grass.tre-trentanove}) we can derive the scalar product between two
generic states as in the SvH case, see 
(\ref{grass.due-quarantotto})-(\ref{grass.svhprod}). Let us first include in
(\ref{grass.tre-trentanove}) also the variables $\varphi$ or $\lambda$ as in
(\ref{grass.due-cinquantasei-a})-(\ref{grass.due-cinquantasei-b}). 
Written in terms of $c$ and $\bar{c}$ these relations are\footnote[3]{We have used the variable
$c$ instead of $\alpha$ in the ket $|\alpha -\rangle$ to indicate that this last one is an eigenstate of $\widehat{c}$
with eigenvalue $c$. In the same way we have replaced $|\alpha+\rangle$ with $|\bar{c} +\rangle$.} 
%%%
\begin{eqnarray}
&&\label{grass.tre-quaranta-a}
i\int dqdpdc^qdc^p|q,p,c^q-,c^p-\rangle\langle-c^{p*},-c^{q*},p,q|={\bf 1} \\
&&\displaystyle \label{grass.tre-quaranta-b} i\int d\lambda_q d\lambda_p d\bar{c}_qd\bar{c}_p|\lambda_q,\lambda_p,\bar{c}_q+,\bar{c}_p+\rangle
\langle+\bar{c}_p^*,+\bar{c}_q^*,\lambda_p,\lambda_q|={\bf 1}. 
\end{eqnarray}
%%%
Inserting the resolution of the identity (\ref{grass.tre-quaranta-a}) into the scalar product
$\langle\psi|\Phi\rangle$ we obtain:
%%%
\begin{eqnarray}
\label{grass.reality-1}
\langle\psi|\Phi\rangle&=&i\int dqdpdc^qdc^p\langle\psi|q,p,c^q-,c^p-\rangle
\langle-c^{p*},-c^{q*},p,q|\Phi\rangle=\nonumber\\
&=&i\int dq dp dc^qdc^p\,\psi_+^*(q,p,c^q,c^p)\Phi_-(q,p,c^q,c^p) 
\end{eqnarray}
%%%
where 
%%%
\begin{equation}
	\left\{
		\begin{array}{l}
		\displaystyle 
		\psi_+(q,p,c^q,c^p)\equiv\langle -c^p,-c^q,p,q|\psi\rangle
		\smallskip \\
		\displaystyle
		\Phi_-(q,p,c^q,c^p)\equiv\langle
		-c^{p*},-c^{q*},p,q|\Phi\rangle\equiv\Phi_0+\Phi_qc^q+\Phi_pc^p+\Phi_2c^qc^p.
		\end{array}
		\right.
\end{equation}
%%%
Note that:
%%%
\begin{eqnarray}
&&\psi_+(q,p,c^q,c^p)=\langle -c^p,-c^q, p,q|\psi\rangle=\nonumber\\
&&=i\int dq^{\prime}dp^{\prime}dc^{q\prime}dc^{p\prime}\langle -c^p,-c^q,p,q|q^{\prime},p^{\prime},c^{q\prime}-,
c^{p\prime}-\rangle\langle -c^{p*\prime},-c^{q*\prime},p^{\prime}, q^{\prime}|\psi\rangle=\nonumber\\
&& =\int dc^{q\prime}dc^{p\prime}
\delta(c^{p\prime}-c^{p*})\delta(c^{q\prime}-c^{q*})
\cdot(\psi_0+\psi_qc^{q\prime}+\psi_pc^{p\prime}+\psi_2c^{q\prime}c^{p\prime})=\nonumber\\
&&=\psi_0+\psi_qc^{q*}+\psi_pc^{p*}-\psi_2c^{p*}c^{q*}.
\end{eqnarray}
%%%
Therefore the complex conjugate of $\psi_+(q,p,c^q,c^p)$ is
%%%
\begin{equation}
\psi^*_+(q,p,c^q,c^p)=\psi_0^*+\psi_q^*c^{q}+\psi_p^*c^{p}-\psi_2^*c^{q}c^{p} \label{grass.reality0}
\end{equation}
%%%
and inserting (\ref{grass.reality0}) into (\ref{grass.reality-1}) we obtain:
%%%
\begin{eqnarray}
\langle \psi|\Phi\rangle&=&i\int d\varphi dc^qdc^p[\psi_0^*+c^q\psi_q^*+c^p\psi_p^*-c^qc^p\psi_2^*]\cdot
[\Phi_0+\Phi_qc^q+\Phi_pc^p+\Phi_2c^qc^p]=\nonumber\\ 
&=&i\int d\varphi\Bigl[\psi_2^*\Phi_0-\Phi_2\psi^*_0+\psi_p^*\Phi_q-\psi_q^*\Phi_p\Bigr] .\label{grass.tre-ventotto-x}
\end{eqnarray}
The presence of the factor ``$i$" in (\ref{grass.tre-ventotto-x}) is crucial in order to have a real norm. In fact:
%%%
\begin{eqnarray}
\langle\psi|\psi\rangle&=&i\int d\varphi\,\Bigl[\psi_2^*\psi_0-\psi_2\psi_0^*+\psi_p^*\psi_q-\psi_q^*\psi_p\Bigr]=\nonumber\\
&=&2\,\textrm{Im}\int d\varphi\Bigl[\psi_2\psi_0^*+\psi_q^*\psi_p\Bigr] .\label{grass.mart}
\end{eqnarray}
%%%
From (\ref{grass.mart}) we see
that this is not a positive definite scalar product. Moreover the zero-forms have zero-norm. This means that the scalar
product of
KvN for the zero-forms is definitely not of the gauge type because the zero-forms in KvN had positive
norm. 

The resolutions of the identity (\ref{grass.tre-quaranta-a})-(\ref{grass.tre-quaranta-b}) are useful also in the derivation of the path
integral. In the SvH case we proved that the 
resolutions of the identity (\ref{grass.due-cinquantasei-a}) and (\ref{grass.due-cinquantasei-b}) 
turned the operatorial formalism into the classical path
integral. Let us now see what happens for the gauge scalar product. 
In this case the transition amplitude $K(f|i)$ between an initial configuration
$(\varphi_i,c^q_i,c^p_i)$ and a final one $(\varphi_f,c^q_f,c^p_f)$ becomes:
%%%
\begin{equation}
\displaystyle K(f|i)=\langle -c_f^{p*},-c_f^{q*}, \varphi_f|e^{-i\widehat{\HT}(t_f-t_i)}
|\varphi_i,c^q_i-,c^p_i-\rangle.
\label{grass.tre-quarantuno}
\end{equation}
%%%
The difference with respect to (\ref{grass.due-cinquantasette}) is that the final bra is 
$\langle-c_f^{p*},-c_f^{q*},\varphi_f|$
and not $\langle +c_f^{p*},+c_f^{q*},\varphi_f|$. The reason is that the dual of the 
ket $|c^q-,c^p-\rangle$ in the SvH case 
is a bra of the form $\langle +c^{p*},+c^{q*}|$ while in the gauge case is a bra of the form $\langle
-c^{p*},-c^{q*}|$.
In fact starting from the relations:
%%%
\begin{equation}
\label{grass.tre-quarantuno-a}
	  \left\{
		\begin{array}{l}
		\displaystyle 
		\widehat{c}^q|\varphi_f,c_f^{q*}-,c_f^{p*}-\rangle=c_f^{q*}|\varphi_f,c^{q*}_f-,c^{p*}_f-\rangle
		\smallskip \\
		\displaystyle
		\widehat{c}^p|\varphi_f,c_f^{q*}-,c_f^{p*}-\rangle=c_f^{p*}|\varphi_f,c_f^{q*}-,c_f^{p*}-\rangle
		\end{array}
		\right.
\end{equation}
%%%
and performing their Hermitian conjugation according to the gauge scalar product,
we get
%%%
\begin{equation}
\label{grass.tre-quarantuno-b}
	 \left\{
		\begin{array}{l}
		\displaystyle 
		\langle-c_f^{p*},-c_f^{q*},\varphi_f|\widehat{c}^q=\langle-c_f^{p*},-c_f^{q*},\varphi_f|c^q_f
		\smallskip \\
		\displaystyle
		\langle-c_f^{p*},-c_f^{q*},\varphi_f|\widehat{c}^p=\langle-c_f^{p*},-c_f^{q*},\varphi_f|c^p_f
		\end{array}
		\right.
\end{equation}
%%%
i.e. the bra $\langle-c_f^{p*},-c_f^{q*},\varphi_f|$ is just an eigenstate of 
$\widehat{c}^q$ and $\widehat{c}^p$ with eigenvalues
$c^q_f$ and $c^p_f$.  

Let us now slice the time interval $(t_f-t_i)$ into $N$ intervals
of length $\epsilon$ like in (\ref{grass.due-cinquantotto}) and insert a relation like 
(\ref{grass.tre-quaranta-b}) on the left of 
each operator
$\displaystyle e^{-i\widehat{\HT}\epsilon}$ and one like (\ref{grass.tre-quaranta-a}) on the right. What we obtain is
%%%
\begin{eqnarray}
\displaystyle
\label{grass.tre-quarantadue}
K(f|i)&\hspace{-0.2cm}=&\hspace{-0.2cm}\langle-c_f^{p*},-c_f^{q*},p_f,q_f|
\biggl\{\prod_{j=1}^{N}i^2\int d\lambda_{q_j} d\lambda_{p_j} 
d\bar{c}_{q_j} d\bar{c}_{p_j}dq_jdp_jdc^{q}_jdc^{p}_j\nonumber\\
&&|\lambda_{q_j},\lambda_{p_j},\bar{c}_{q_j}+,\bar{c}_{p_j}+\rangle\cdot
\langle +\bar{c}_{p_j}^*,+\bar{c}_{q_j}^*,\lambda_{p_j},\lambda_{q_j}|\textrm{exp}[-i\epsilon\widehat{\HT}]
|q_{\scriptscriptstyle j},p_{\scriptscriptstyle j},c^q_{\scriptscriptstyle j}-,
c^p_{\scriptscriptstyle j}-\rangle\nonumber\\
&&\cdot\langle -c_j^{p*},-c_j^{q*},p_{\scriptscriptstyle j},q_{\scriptscriptstyle j}|\biggr\}|q_i,p_i,c^q_i-,
c^p_i-\rangle .
\end{eqnarray}
%%%
Using (\ref{grass.tre-trentotto}) the last term in the expression above gives
%%%
\begin{equation}
\langle
-c_{\scriptscriptstyle 1}^{p*},-c_{\scriptscriptstyle 1}^{q*},
p_{\scriptscriptstyle 1},q_{\scriptscriptstyle 1}|q_i,p_i,c^q_i-,c^p_i-\rangle=
i\delta(q_{\scriptscriptstyle 1}-q_i)\delta(p_{\scriptscriptstyle 1}-p_i)\delta(c^q_{\scriptscriptstyle 1}-
c^q_i)\delta(c^p_{\scriptscriptstyle 1}-c^p_i).
\label{grass.en1}
\end{equation}
%%%
Via the integration in $q_{\scriptscriptstyle 1},p_{\scriptscriptstyle 1},c^q_{\scriptscriptstyle 1}
,c^p_{\scriptscriptstyle 1}$ the Dirac deltas
identify $q_{\scriptscriptstyle 1}=q_i$, $p_{\scriptscriptstyle 1}=p_i$, $c^q_{\scriptscriptstyle 1}=c^q_i$ and
$c^p_{\scriptscriptstyle 1}=c^p_i$. Using again the scalar products (\ref{grass.tre-trentotto}) it is easy to evaluate
the other terms in (\ref{grass.tre-quarantadue}):
%%%
\begin{eqnarray}
\displaystyle 
\label{grass.en2}
&&\langle -c^{p*}_{j},-c^{q*}_{j},p_j,q_j|\lambda_{q_{j-1}},\lambda_{p_{j-1}},\bar{c}_{q_{j-1}}+,
\bar{c}_{p_{j-1}}+\rangle=\nonumber\\
&&=i \,	\textrm{exp}\bigl[ip_j\lambda_{p_{j-1}}+iq_j\lambda_{q_{j-1}}
\bigr] \textrm{exp}\bigl[c^q_j\bar{c}_{q_{j-1}}+c^p_j\bar{c}_{p_{j-1}}\bigr],\bigskip\medskip\\
\displaystyle &&\langle +\bar{c}_{p_j}^{*},+\bar{c}_{q_j}^{*},\lambda_{p_j},\lambda_{q_j}\bigl|
\textrm{exp}[-i\epsilon\widehat{\HT}]
\bigr|q_j,p_j,c_j^q-,c_j^p-\rangle=\nonumber\\
&&=-i \, \textrm{exp}\bigl[-i\epsilon\HT(j)\bigr]	\textrm{exp}[-i\lambda_{q_j}q_j-i\lambda_{p_j}p_j+
\bar{c}_{q_j}c^q_j+\bar{c}_{p_j}c^p_j].
\end{eqnarray}
%%%
The $N$ factors ``$-i$" and the $N$ factors ``$+i$" 
which appear in the expressions above will get compensated by the $2N$ factors ``$+i$"
in the resolutions of the identity (\ref{grass.tre-quaranta-a})
and (\ref{grass.tre-quaranta-b}) we have used.
The $N$ resulting ``$-$" signs can be
absorbed by turning around the integrations in $c^q$ and $c^p$:
%%%
\begin{equation}
-\int dq_jdp_jdc^q_jdc^p_j=\int dq_jdp_jdc^p_jdc^q_j.
\end{equation}
%%%
So, combining all the pieces, the final result is\footnote[4]{The factor ``$i$" 
in front of the RHS of (\ref{grass.tre-quarantadue-b}) is due to the normalization we have chosen:
in fact for $t_f=t_i$ and $f=i$
the $K(f|i)$ in  (\ref{grass.tre-quarantuno})
is equal to ``$i$" instead of ``${\bf 1}$".}:
%%%
\begin{equation}
K(f|i)=i \int {\mathcal D}\mu \;	\textrm{exp}\biggl[i\epsilon\biggl(\sum_{j=1}^N\lambda_j
\frac{\varphi_{j+1}-\varphi_j}{\epsilon}+
i\bar{c}_j\frac{c_{j+1}-c_j}{\epsilon}-\HT(j)\biggr)\biggr] \label{grass.tre-quarantadue-b}
\end{equation}
%%%
where the boundary conditions are:
%%%
\begin{equation}
\varphi_0=\varphi_i,\;\;\;\;\;\;\varphi_{\scriptscriptstyle N+1}=\varphi_f,\;\;\;\;\;\;c_0=c_i,\;\;\;\;\;\;
c_{\scriptscriptstyle{N+1}}=c_f
\end{equation}
%%%
and the measure is
%%%
\begin{equation}
\displaystyle \displaystyle {\mathcal D}\mu= \biggl(\prod_{j=2}^Nd\varphi_jd\lambda_jd\bar{c}_jdc_j\biggr)
d\lambda_1d\bar{c}_1
\end{equation}
%%%
which is identical to the measure of the SvH case in (\ref{grass.due-sessantatre}).  
Both measures can be turned, without any sign change, into the following one:
%%%
\begin{equation}
\displaystyle {\mathcal
D}\mu=d\lambda_{q_1}d\lambda_{p_1}d\bar{c}_{q_1}d\bar{c}_{p_1}
\prod_{j=2}^Ndq_jdp_jd\lambda_{q_j}d\lambda_{p_j}dc_j^qd\bar{c}_{q_j}
dc^p_jd\bar{c}_{p_j}
\end{equation}
%%%
which is the one which originally appeared in the CPI \cite{Goz89}. 
Furthermore the discretized Lagrangian appearing in (\ref{grass.tre-quarantadue-b}) goes into the usual $\LT$ 
of (\ref{ann.suplag}) in the continuum limit.
All this confirms that also via the gauge scalar product we
can reproduce the classical path integral. 

We can summarize this section by saying that 
in the gauge scalar
product $\widehat{\HT}$ is {\it Hermitian}, the CPI can be obtained from the operatorial formalism but the
scalar product is {\it not positive definite}. 

%%%%%%%%%%%%%%%%%%%%%%%%%%%%%%%%%%%%%%%%%%%%%%%%%%%%%%%%%%%%%%%%%%%%%%%%%%%%%%%%%%%%%%%%%%%%%%%%%%%%%%%%%%%%%%%%%%%%%%%%%%%%%%%

\bigskip

\section{The Symplectic Scalar Product}
\noindent
The gauge scalar product which we explored in the previous section is not the only one under which $\widehat{\HT}$
is Hermitian. In this section we will explore another one which has the same feature and whose hermiticity conditions are:
%%%
\begin{equation}
\label{grass.quattro-uno}
	 \left\{
		\begin{array}{l}
		\displaystyle 
		(\widehat{c}^a)^{\dagger}=i\omega^{ab}\widehat{\bar{c}}_b
		\smallskip \\
		\displaystyle
		(\widehat{\bar{c}}_d)^{\dagger}=i\omega_{df}\widehat{c}^f
		\end{array}
		\right.
\end{equation}
%%%
%%%
\begin{equation}
\label{grass.quattro-due}
	 \left\{
		\begin{array}{l}
		\displaystyle 
		\widehat{\varphi}^{a\dagger}=\widehat{\varphi}^a
		\smallskip \\
		\displaystyle
		\widehat{\lambda}_a^{\dagger}=\widehat{\lambda}_a.
		\end{array}
		\right.
\end{equation}
%%%
Because of the presence of the symplectic matrix $\omega^{ab}$ in (\ref{grass.quattro-uno}) 
we will call ``symplectic" the  scalar product which produces the hermiticity conditions above. 
Under these conditions the bosonic part of $\widehat{\HT}$ turns out to be Hermitian as it was in the SvH case
(\ref{grass.due-cinquantuno-b}). If we take $\displaystyle H=\frac{p^2}{2}+V(q)$, 
the fermionic part (\ref{grass.due-cinquantuno-xx}) can be written
as $\widehat{\HT}_{ferm}=i\widehat{\bar{c}}_q\widehat{c}^p-i\widehat{\bar{c}}_pV^{\prime\prime}\widehat{c}^q$ 
where $\displaystyle
V^{\prime\prime}=\frac{\partial^2V}{\partial q^2}$. Applying the hermiticity conditions (\ref{grass.quattro-uno}) we get
%%%
\begin{eqnarray}
\widehat{\HT}^{\dagger}_{ferm}&=&(i\widehat{\bar{c}}_q\widehat{c}^p)^{\dagger}-(i\widehat{\bar{c}}_p
V^{\prime\prime}\widehat{c}^q)^{\dagger}=
-i\widehat{c}^{p\dagger}\widehat{\bar{c}}_q^{\dagger}+i\widehat{c}^{q\dagger}V^{\prime\prime}
\widehat{\bar{c}}_p^{\dagger}=\nonumber\\
&=&-i(-i\widehat{\bar{c}}_q)(-i\widehat{c}^p)+i\cdot i\widehat{\bar{c}}_pV^{\prime\prime}i\widehat{c}^q=
i\widehat{\bar{c}}_q\widehat{c}^p-i\widehat{\bar{c}}_pV^{\prime\prime}\widehat{c}^q=\widehat{\HT}_{ferm},
\label{grass.quattro-tre}
\end{eqnarray}
%%%
i.e. also the fermionic part of $\widehat{\HT}$ is Hermitian.

As usual let us now proceed by constructing a resolution of the identity and to find
out the expression of the symplectic scalar product in terms of the components of the wave functions, as in
(\ref{grass.due-cinquantuno}) for the SvH case and in (\ref{grass.tre-ventotto-x}) for the gauge case. The
vector space is the same as in the SvH and the gauge cases and is spanned by
the four states $\bigl\{|0-,0-\rangle$, $|0-,0+\rangle$, $|0+,0-\rangle$, $|0+,0+\rangle\bigr\}$ defined in the usual way:
%%%
\begin{equation}
\label{grass.quattro-quattro}
	 \left\{
		\begin{array}{l}
		\displaystyle 
		\widehat{c}^q|0-,0-\rangle=\widehat{c}^p|0-,0-\rangle=0
		\smallskip \\
		\displaystyle
		|0+,0-\rangle=\widehat{\bar{c}}_q|0-,0-\rangle
	 \smallskip \\
	 \displaystyle 
	 |0-,0+\rangle=-\widehat{\bar{c}}_p|0-,0-\rangle
	 \smallskip \\
	 |0+,0+\rangle =\widehat{\bar{c}}_q|0-,0+\rangle=\widehat{\bar{c}}_p\widehat{\bar{c}}_q|0-,0-\rangle.
		\end{array}
		\right.
\end{equation}
%%%
As in the previous cases we can choose how to normalize one of 
these states. This time we choose the following normalization:
%%%
\begin{equation}
\Bigl(|0+,0+\rangle,|0+,0+\rangle\Bigr)=1. \label{grass.quattro-quattro-a}
\end{equation}
%%%
From this and (\ref{grass.quattro-quattro}) one can easily obtain that the only non-zero scalar
products are
%%%
\begin{equation}
\label{grass.quattro-cinque}
	 \left\{
		\begin{array}{l}
		\displaystyle
		\Bigl(|0-,0+\rangle,|0+,0-\rangle\Bigr)=i
	 \smallskip \\
	 \displaystyle 
	 \Bigl(|0+,0-\rangle,|0-,0+\rangle\Bigr)=-i
	 \smallskip \\
	 \Bigl(|0-,0-\rangle,|0-,0-\rangle\Bigr)=-1.
		\end{array}
		\right.
\end{equation}
%%%
In (\ref{grass.quattro-cinque}) we notice the presence of negative norm states like $|0-,0-\rangle$.
The states which are needed in the decomposition of the identity are the ones we have already built 
in the SvH and gauge cases:
%%%
\begin{equation}
\label{grass.quattro-sei}
	 \left\{
		\begin{array}{l}
		\displaystyle  
		|\alpha_q-,\alpha_p-\rangle\equiv e^{-\alpha_q\widehat{\bar{c}}_q-\alpha_p\widehat{\bar{c}}_p}|0-,0-\rangle
		\smallskip \\
		\displaystyle
		|\beta_q+,\alpha_p-\rangle\equiv e^{-\beta_q\widehat{c}^q-\alpha_p\widehat{\bar{c}}_p}|0+,0-\rangle
	 \smallskip \\
	 \displaystyle 
	 |\alpha_q-,\beta_p+\rangle\equiv e^{-\alpha_q\widehat{\bar{c}}_q-\beta_p\widehat{c}^p}|0-,0+\rangle
	 \smallskip \\
	 |\beta_q+,\beta_p+\rangle\equiv e^{-\beta_q\widehat{c}^q-\beta_p\widehat{c}^p}|0+,0+\rangle.
		\end{array}
		\right.
\end{equation}
%%%
From (\ref{grass.quattro-quattro-a})-(\ref{grass.quattro-cinque}) and the commutation relations 
(\ref{ijmpa.comm}) we can easily
derive the scalar products among the states (\ref{grass.quattro-sei}). They are
%%%
\begin{eqnarray}
\label{grass.quattro-sette}
&&\Bigl(|\alpha_q-,\alpha_p-\rangle,|\alpha_q^{\prime}-,\alpha_p^{\prime}-\rangle\Bigr)=
-	\textrm{exp}(-i\alpha_q^*\alpha_p^{\prime}+i\alpha_p^*\alpha_q^{\prime})\nonumber\\
&&\Bigl(|\alpha_q-,\beta_p+\rangle,|\beta_q+,\beta_p^{\prime}+\rangle\Bigr)=
i\delta(\beta_p^{\prime}+i\alpha_q^*)	\textrm{exp}(i\beta_q\beta_p^*)\nonumber\\
&&\Bigl(|\alpha_q-,\alpha_p-\rangle,|\beta_q+,\alpha_p^{\prime}-\rangle\Bigr)=
\delta(\beta_q-i\alpha_p^*)	\textrm{exp}(i\alpha_p^{\prime}\alpha_q^*)\nonumber\\
&&\Bigl(|\alpha_q-,\alpha_p-\rangle,|\beta_q+,\beta_p+\rangle\Bigr)=\delta(\beta_q-i\alpha^*_p)
\delta(\beta_p+i\alpha_q^*)\nonumber\\
&&\Bigl(|\alpha_q-,\alpha_p-\rangle,|\alpha_q^{\prime}-,\beta_p+\rangle\Bigr)=
-\delta(\beta_p+i\alpha_q^*)	\textrm{exp}(-i\alpha_q^{\prime}\alpha_p^*)\nonumber\\
&&\Bigl(|\beta_q+,\alpha_p-\rangle,|\beta_q^{\prime}+,\beta_p+\rangle\Bigr)=
-i\delta(\beta_q^{\prime}-i\alpha_p^*)	\textrm{exp}(-i\beta_p\beta_q^*)\\
&&\Bigl(|\beta_q+,\alpha_p-\rangle,|\alpha_q-,\beta_p+\rangle\Bigr)=
-i\,\textrm{exp}(i\beta_q^*\beta_p+i\alpha_p^*\alpha_q)\nonumber\\
&&\Bigl(|\beta_q+,\alpha_p-\rangle,|\beta_q^{\prime}+,\alpha_p^{\prime}-\rangle\Bigr)=
\delta(\alpha_p^*+i\beta_q^{\prime})\delta(\alpha_p^{\prime}-i\beta_q^*)\nonumber\\
&&\Bigl(|\alpha_q-,\beta_p+\rangle,|\alpha_q^{\prime}-,\beta_p^{\prime}+\rangle\Bigr)=
\delta(\alpha_q^{\prime}+i\beta_p^*)\delta(i\beta_p^{\prime}-\alpha_q^*)\nonumber\\
&&\Bigl(|\alpha_q^{\prime}-,\alpha_p^{\prime}-\rangle,|\alpha_q-,\alpha_p-\rangle\Bigr)=
-	\textrm{exp}(-i\alpha_q^{\prime*}\alpha_p+i\alpha_p^{\prime*}\alpha_q)\nonumber\\
&&\Bigl(|\beta_q+,\beta_p+\rangle,|\beta_q^{\prime}+,\beta_p^{\prime}+\rangle\Bigr)=
\textrm{exp}(i\beta_q^*\beta_p^{\prime}-i\beta_p^*\beta_q^{\prime})\nonumber .
\end{eqnarray}
%%%
These relations should be compared with their analogs in the SvH (\ref{grass.due-trentotto-c})
and in the gauge case (\ref{grass.tre-trentotto}). Before proceeding let us remember what we discussed after (\ref{grass.due-quaranta}). There
we proved that in passing from the $|ket\rangle$ to the $\langle bra|$ the order of the entries had to be reversed: if in the $|ket
\rangle$ the first entry was $q$, in the corresponding $\langle bra|$ it had to be $p$. This was the case for  the SvH and gauge scalar
products but it is no longer the case here. In fact let us remember for example the relation 
%%%
\begin{equation}
|0_q-,0_p-\rangle=\widehat{c}^q\widehat{c}^p|0_q+,0_p+\rangle \label{grass.quattro-otto-a}
\end{equation}
%%%
and let us perform the Hermitian conjugation (\ref{grass.quattro-uno}) of the operators appearing in (\ref{grass.quattro-otto-a}); we obtain 
the following relation:
%%%
\begin{equation}
\langle 0_q+,0_p+|\widehat{\bar{c}}_q\widehat{\bar{c}}_p=\langle 0_q-,0_p-| \label{grass.quattro-otto-b}
\end{equation}
%%%
which is perfectly consistent.
This consistency indicates that we can keep the same order in the entries of the
$\langle bra|$ and the $|ket\rangle$. 

The resolutions of the identity that we get for the symplectic scalar product are 
%%%
\begin{eqnarray}
\displaystyle \label{grass.quattro-nove-a} 	
&&\int d\alpha_p d\alpha_q|\alpha_q-,\alpha_p-\rangle\langle i\alpha_p^*+,(-i\alpha_q^*)+|={\bf 1} \\
\displaystyle \label{grass.quattro-nove-b}
&&\int d\alpha_p d\alpha_q|\alpha_q+,\alpha_p+\rangle\langle (-i\alpha_p^*)-,i\alpha_q^*-|={\bf 1}.
\end{eqnarray}
%%%
These resolutions should be compared with those of the SvH 
(\ref{grass.due-quarantuno}) and the gauge case (\ref{grass.tre-trentanove}). The strange form of
the $\langle bra|$ appearing in  (\ref{grass.quattro-nove-a}) is actually necessary in order to make it an eigenstate of
$\widehat{c}^p$ and $\widehat{c}^q$. In fact if we start from
%%%
\begin{equation}
\widehat{\bar{c}}_q|i\alpha_p^*+,(-i\alpha_q^*)+\rangle=i\alpha_p^*|i\alpha_p^*+,(-i\alpha_q^*)+\rangle
\end{equation}
%%%
and perform the Hermitian conjugation, we get:
%%%
\begin{equation}
\langle i\alpha_p^*+,(-i\alpha_q^*)+|\widehat{\bar{c}}_q^{\dagger}=\langle i\alpha_p^*+,(-i\alpha_q^*)+|(i\alpha_p^*)^*.
\end{equation}
%%%
Using the hermiticity conditions (\ref{grass.quattro-uno}), we obtain that 
%%%
\begin{eqnarray}
&&\langle i\alpha_p^*+,(-i\alpha_q^*)+|(-i\widehat{c}^p)=\langle i\alpha_p^*+,(-i\alpha_q^*)+|(-i\alpha_p)\Rightarrow\nonumber\\
&&\Rightarrow \langle i\alpha_p^*+, (-i\alpha_q^*)+|\widehat{c}^p=\langle i\alpha_p^*+,(-i\alpha_q^*)+|\alpha_p.
\label{grass.quattro-nove-c}
\end{eqnarray}
%%%
The same can be shown for $\widehat{c}^q$. Let us start from 
%%%
\begin{equation}
\widehat{\bar{c}}_p|i\alpha_p^*+,(-i\alpha_q^*)+\rangle=-i\alpha_q^*|i\alpha_p^*+,(-i\alpha_q^*)+\rangle
\end{equation}
%%%
and perform its Hermitian conjugation. What we get is:
%%%
\begin{eqnarray}
&&\langle i\alpha_p^*+,(-i\alpha_q^*)+|i\widehat{c}^q=\langle i\alpha_p^*+, (-i\alpha_q^*)+|i\alpha_q \Rightarrow\nonumber\\
&&\Rightarrow \langle i\alpha_p^*+,(-i\alpha_q^*)+|\widehat{c}^q=\langle i\alpha_p^*+,(-i\alpha_q^*)+|\alpha_q
\label{grass.quattro-nove-d}
\end{eqnarray}
%%%
which proves our claim.
Via (\ref{grass.quattro-nove-a}) it is now easy to write the symplectic scalar product in terms of
the components of the states as in (\ref{grass.due-cinquantuno}) for the SvH case and in (\ref{grass.tre-ventotto-x}) for the gauge case. Let us
consider two states
$|\psi\rangle$ and $|\Phi\rangle$ and their scalar product $\langle\Phi|\psi\rangle$. By inserting the resolution of the
identity (\ref{grass.quattro-nove-a}), we obtain:
%%%
\begin{eqnarray}
\langle \Phi|\psi\rangle&=&\int d\alpha_p d\alpha_q\langle\Phi|\alpha_q-,\alpha_p-\rangle
\langle i\alpha_p^*+,(-i\alpha_q^*)+|\psi\rangle=\nonumber\\
&=&\int d\alpha_pd\alpha_q\,\Phi^*_+(\alpha_q,\alpha_p)\psi_-(\alpha_q,\alpha_p) \label{grass.quattro-dieci-a}
\end{eqnarray}
%%%
where:
%%%
\begin{equation}
\psi_-(\alpha_q,\alpha_p)\equiv\langle i\alpha_p^*+,(-i\alpha_q^*)+|\psi\rangle\label{grass.quattro-dieci-b}
\end{equation}
%%%
and 
%%%
\begin{equation}
\Phi_+(\alpha_q,\alpha_p)\equiv\langle\alpha_q-,\alpha_p-|\Phi\rangle .\label{grass.quattro-undici}
\end{equation}
%%%
Since $\alpha_q$ and $\alpha_p$
are Grassmann variables and $\psi_-(\alpha_q,\alpha_p)$ is a function of them 
we can write it as 
%%%
\begin{equation}
\psi_-(\alpha_q,\alpha_p)=\psi_0+\psi_q\alpha_q+\psi_p\alpha_p+\psi_2\alpha_q\alpha_p. \label{grass.quattro-dodici}
\end{equation}
%%%
Let us now find out the expression of $\Phi_+$ in terms of its components:
%%%
\begin{eqnarray}
\label{grass.quattro-tredici}
\Phi_+(\alpha_q,\alpha_p)&\hspace{-0.cm}=&\hspace{-0.2cm}
\langle\alpha_q-,\alpha_p-|\Phi\rangle=\nonumber\\
&\hspace{-0.2cm}=&\hspace{-0.2cm}\int d\alpha_p^{\prime}d\alpha_q^{\prime}\langle\alpha_q-,\alpha_p-|\alpha_q^{\prime}-,
\alpha_p^{\prime}-\rangle\langle i\alpha_p^{\prime *}+,(-i\alpha_q^{\prime *})+|\Phi\rangle=\nonumber\\
&\hspace{-0.2cm}=&\hspace{-0.2cm}\int d\alpha_p^{\prime}d\alpha_q^{\prime}\langle\alpha_q-,\alpha_p-|\alpha_q^{\prime}-,
\alpha_p^{\prime}-\rangle\Phi_-(\alpha_q^{\prime},\alpha_p^{\prime})=\\
&\hspace{-0.2cm}=&\hspace{-0.2cm}-\int
d\alpha_p^{\prime}d\alpha_q^{\prime}\,
\textrm{exp}(-i\alpha_q^*\alpha_p^{\prime}+i\alpha_p^*\alpha_q^{\prime})\cdot
(\Phi_0+\Phi_q\alpha_q^{\prime}+\Phi_p\alpha_p^{\prime}+\Phi_2\alpha_q^{\prime}\alpha_p^{\prime})\nonumber\\
&\hspace{-0.cm}=&\hspace{-0.2cm}
-\Phi_2-i\Phi_p\alpha_p^*-i\Phi_q\alpha_q^*+\Phi_0\alpha_p^*\alpha_q^*\nonumber.
\end{eqnarray}
%%%
Inserting the expressions (\ref{grass.quattro-dodici}) and (\ref{grass.quattro-tredici}) in (\ref{grass.quattro-dieci-a}) we get:
%%%
\begin{eqnarray}
\langle \Phi|\psi\rangle&=&\int d\alpha_p d\alpha_q(-\Phi_2^*+i\Phi_p^*\alpha_p+i\Phi_q^*\alpha_q+\Phi_0^*\alpha_q\alpha_p)
\cdot \\
&&\cdot(\psi_0+\psi_q\alpha_q+\psi_p\alpha_p+\psi_2\alpha_q\alpha_p)=
\Phi_0^*\psi_0+i(\Phi_q^*\psi_p-\Phi_p^*\psi_q)-\Phi_2^*\psi_2. \nonumber
\end{eqnarray}
%%%
If we include also the bosonic variables $\varphi$ we obtain:
%%%
\begin{equation}
\displaystyle \langle \Phi|\psi\rangle
=\int d\varphi\biggl[\Phi_0^*\psi_0+i(\Phi_q^*\psi_p-\Phi_p^*\psi_q)-\Phi_2^*\psi_2\biggr] \label{grass.quattro-quattordici}
\end{equation}
%%%
and the hermiticity conditions (\ref{grass.quattro-uno}) are correctly reproduced since
%%%
\begin{equation}
\label{grass.quattro-quattordici-a}
	 \left\{
		\begin{array}{l}
		\displaystyle  
		\langle\Phi|\widehat{c}^q\psi\rangle=\langle i\widehat{\bar{c}}_p\Phi|\psi\rangle, \quad \;\;\;\;\;\;\;\;\;
	 \langle \Phi|i\widehat{\bar{c}}_p\psi\rangle=\langle \widehat{c}^q\Phi|\psi\rangle 
		\smallskip \\
		\displaystyle
		\langle \Phi|\widehat{c}^p\psi\rangle=\langle -i\widehat{\bar{c}}_q\Phi|\psi\rangle, \;\;\;\;\;\;\;\;\;\;
	 \langle \Phi|-i\widehat{\bar{c}}_q\psi\rangle=\langle \widehat{c}^p\Phi|\psi\rangle.
	 \smallskip
		\end{array}
		\right.
\end{equation}
%%%
Let us also notice that the scalar product (\ref{grass.quattro-quattordici}) reproduces the KvN one for the zero-forms,
%%%
\begin{equation}
\displaystyle \langle \Phi|\psi\rangle=\int d\varphi \;\Phi_0^*\psi_0 .\label{grass.quattro-quattordici-a1}
\end{equation}
%%%
This happened also in the SvH but not in the gauge case.
Unfortunately the symplectic scalar product is not
positive definite. In fact from (\ref{grass.quattro-quattordici}) we notice that
there are states, like the two-forms $\Phi=\Phi_2\alpha_q\alpha_p$, with negative norm: 
%%%
\begin{equation}
\displaystyle \langle \Phi|\Phi\rangle=-\int d\varphi|\Phi_2|^2 \label{grass.quattro-quattordici-b}
\end{equation}
%%%
and states, like the one-forms with real coefficients $\psi=\psi_q\alpha^q+\psi_p\alpha^p$, with zero norm:
%%%
\begin{equation}	
\displaystyle \langle\psi|\psi\rangle=i\int d\varphi (\psi_q\psi_p-\psi_p\psi_q)=0 .\label{grass.quattro-quattordici-c}
\end{equation}
%%%

The symplectic scalar product can be easily generalized to the case of $n$ degrees of freedom. 
First of all let us rewrite (\ref{grass.quattro-quattordici}) as:
%%%
\begin{equation}
\displaystyle
\langle\Phi|\psi\rangle=\int d\varphi\biggl[\Phi_0^*\psi_0
+i\Phi_a^*\omega^{ab}\psi_b+\frac{i^2}{2!}\Phi_{a_1a_2}^*\omega^{a_1b_1}\omega^{a_2b_2}\psi_{b_1b_2}	
\biggr] \label{grass.quattro-sedici}
\end{equation}
%%%
where the notation has the following meaning:
%%%
\begin{equation}
\varphi^a=(q,p),\;\;\;\;\Phi_a=(\Phi_q,\Phi_p),\;\;\;\;\Phi_{qp}=\Phi_2,\;\;\;\;\Phi_{qq}=0,\;\;\;\;\Phi_{pp}=0. 
\end{equation}
%%%
Then it is a long but straightforward calculation to prove
that two generic states of the form:
%%%
\begin{equation}
\displaystyle
\psi=\sum_{m=0}^{2n}\frac{1}{m!}\psi_{a_1\ldots a_m}c^{a_1}\ldots c^{a_m};\;\;\;\;\;
\Phi=\sum_{m=0}^{2n}\frac{1}{m!}\Phi_{b_1\ldots b_m}c^{b_1}\ldots c^{b_m} \label{grass.quattro-diciassette}
\end{equation}
%%%
have the following symplectic scalar product:
%%%
\begin{equation}
\displaystyle
\langle\Phi|\psi\rangle=\sum_{m=0}^{2n}\frac{i^m}{m!}\int d\varphi \;
\Phi^*_{a_1\ldots a_m}\omega^{a_1b_1}\ldots
\omega^{a_mb_m}\psi_{b_1\ldots b_m}. \label{grass.quattro-diciotto}
\end{equation}
%%%

The last issue we want to explore is whether we can reproduce, via the symplectic scalar product and its
decompositions of the identity, the CPI starting from the operatorial formalism,
as in the SvH and gauge cases. 
Let us first include in the resolutions of the identity
(\ref{grass.quattro-nove-a})-(\ref{grass.quattro-nove-b}) the bosonic variables. What we get is: 
%%%
\begin{eqnarray}
&&\int d\varphi\, dc^pdc^q|\varphi,\,c^q-,c^p-\rangle\langle ic^{p*}+,(-ic^{q*})+,\varphi|={\bf 1}
\label{grass.quattro-diciannove-a}\\
&&\int d\lambda\,d\bar{c}_p\,d\bar{c}_q|\lambda,\,\bar{c}_q+,\bar{c}_p+\rangle\langle(-i\bar{c}_p^*)-,
i\bar{c}_q^*-,\lambda|={\bf 1}. \label{grass.quattro-diciannove-b}
\end{eqnarray}
%%%
The kernel $K(f|i)$ in the symplectic case takes the
form:
%%%
\begin{equation}
\displaystyle 
K(f|i)=\langle ic^{p*}_f+,(-ic^{q*}_f)+,\varphi_f\Bigl|	
e^{-i\widehat{\HT}(t_f-t_i)}\Bigr|\varphi_i,\,c^q_i-,c^p_i-\rangle.
\label{grass.quattro-venti}
\end{equation}
%%%
Note that the difference with respect to (\ref{grass.due-cinquantasette}) and (\ref{grass.tre-quarantuno}) 
is in the initial $\langle bra|$.
Since $K(f|i)\equiv K(\varphi_f,c^q_f,c^p_f|\varphi_i,c^q_i,c^p_i)$ is the transition amplitude 
between an initial configuration $(\varphi_i,c^q_i,c^p_i)$ and a final one $(\varphi_f,c^q_f,c^p_f)$, the $\langle bra|$
which appears in (\ref{grass.quattro-venti}), i.e. $\langle ic_f^{p*}+,(-ic_f^{q*})+,\varphi_f|$, 
must be an eigenstate of $\widehat{c}^p$
and $\widehat{c}^q$ with eigenvalues $c^p_f$ and $c^q_f$. This is actually the case as it has been proved in 
(\ref{grass.quattro-nove-c}) and (\ref{grass.quattro-nove-d}). 
The procedure then continues, as in the previous cases, by splitting the time interval $(t_f-t_i)$ in
(\ref{grass.quattro-venti}) into $N$ intervals of length $\epsilon$ 
%%%
\begin{equation}
\displaystyle 
K(f|i)=\langle ic^{p*}_f+,(-ic^{q*}_f)+,
\varphi_f|\underbrace{e^{-i\epsilon\widehat{\HT}}e^{-i\epsilon\widehat{\HT}}\ldots 
e^{-i\epsilon\widehat{\HT}}}_{N\; \textrm{terms}}|\varphi_i,c^q_i-,c^p_i-\rangle.
\end{equation}
%%%
Inserting (\ref{grass.quattro-diciannove-b}) on the left and (\ref{grass.quattro-diciannove-a}) on the right of each term
$e^{-i\epsilon\widehat{\HT}}$, we get the same discretized expression as in
(\ref{grass.due-sessantadue}). We skip the details here because they are very similar to the SvH and
gauge cases and the result is the same. Therefore also in the symplectic case we can reproduce the classical path integral 
(\ref{ann.prob3}).

We can say that up to now in this chapter we have shown that there is more than one extension of the KvN scalar product, 
the SvH and the symplectic one, but both
of them have defects: for the first one $\widehat{\HT}$ is not Hermitian, while for the second one the scalar
product is not positive definite. The gauge scalar product instead is not even an extension of the KvN theory because, 
differently from the KvN case, the zero-forms
states have zero-norms, see (\ref{grass.tre-ventotto-x}). 

%%%%%%%%%%%%%%%%%%%%%%%%%%%%%%%%%%%%%%%%%%%%%%%%%%%%%%%%%%%%%%%%%%%%%%%%%%%%%%%%%%%%%%%%%%%%%%%%%%%%%%%%%%%%%%%%%%%%%%%%

\bigskip

\section{``Physical" Hilbert Space} 
\noindent
In this section we will address the issue of what is the {\it physical} subspace of the full Hilbert space 
underlying the CPI.
With a little abuse of notation, we call {\it physical} subspace the one made of positive norm states
on which $\widehat{\HT}$ is Hermitian. We shall perform this analysis for all the three scalar products
studied in the previous sections starting from the SvH one of
Sec. {\bf 4.1}. 

\subsection{Salomonson-van Holten Case}
In this case all the states in the Hilbert space have positive definite norm but
$\widehat{\HT}$ is not always Hermitian. This is an unacceptable feature because it would lead to the non-conservation of the
norm creating in this way difficulties in assigning the meaning of probability to the norm of a generic state, 
differently from what 
happens in the zero-form case. The linear subspace of the full Hilbert space
on which $\widehat{\HT}$ is Hermitian is defined by the following condition:
%%%
\begin{equation}
\Bigl(\widehat{\HT}-\widehat{\HT}^{\dagger}\Bigr)|\psi\rangle=0. \label{grass.cinque-uno}
\end{equation}
%%%
The next thing we have to guarantee is that the vector subspace defined by (\ref{grass.cinque-uno}) be closed
under time evolution. This means that  a state $|\psi^{\prime}\rangle$, obtained via an infinitesimal time evolution 
from a physical
state $|\psi\rangle$: 
%%%
\begin{equation}
\displaystyle |\psi^{\prime}\rangle=e^{-i\epsilon\widehat{\HT}}|\psi\rangle \label{grass.cinque-tre},
\end{equation}
%%%
must still be physical, i.e.:
%%%
\begin{equation}
(\widehat{\HT}-\widehat{\HT}^{\dagger})
|\psi^{\prime}\rangle=0. \label{grass.cinque-quattro}
\end{equation}
%%%
Inserting (\ref{grass.cinque-tre}) into (\ref{grass.cinque-quattro}) we get
%%%
\begin{eqnarray}
&&(\widehat{\HT}-\widehat{\HT}^{\dagger})
|\psi^{\prime}\rangle=(\widehat{\HT}-\widehat{\HT}^{\dagger})|
\psi\rangle-i\epsilon (\widehat{\HT}-\widehat{\HT}^{\dagger})\widehat{\HT}|\psi\rangle=\nonumber\\ 
&&=-i\epsilon \Bigl[\widehat{\HT},(\widehat{\HT}-\widehat{\HT}^{\dagger})\Bigr]|\psi\rangle=
i\epsilon \Bigl[\widehat{\HT},\widehat{\HT}^{\dagger}\Bigr]|\psi\rangle
\end{eqnarray}
%%%
that implies that for $|\psi^{\prime}\rangle$ to be physical the following condition must also be satisfied 
%%%
\begin{equation}
\Bigl[\widehat{\HT},\widehat{\HT}^{\dagger}\Bigr]|\psi\rangle=0. \label{grass.cinque-cinque}
\end{equation}
%%%
Let us analyse the commutator structure of (\ref{grass.cinque-cinque}). If we write
$\widehat{\HT}=\widehat{\HT}_{bos}+\widehat{\HT}_{ferm}$ 
we get that the commutator contained in the LHS of (\ref{grass.cinque-cinque}) turns into the following expression:
%%%
\begin{equation}
[\widehat{\HT},\widehat{\HT}^{\dagger}]=
[\widehat{\HT}_{ferm},\widehat{\HT}^{\dagger}_{ferm}]+[\widehat{\HT}_{bos},\widehat{\HT}_{ferm}^{\dagger}
]+[\widehat{\HT}_{ferm},\widehat{\HT}_{bos}] .\label{grass.cinque-sette}
\end{equation}
%%%
Let us look at the first term on the RHS of (\ref{grass.cinque-sette}). The general expression of 
$\widehat{\HT}_{ferm}$ was given in (\ref{grass.due-cinquantuno-xx}) and, choosing $H$ to be of the form $\displaystyle
H=\sum_{i=1}^np_i^2/2+V(q_1,\ldots,q_n)$, we get 
%%%
\begin{equation}
\label{grass.cinque-sette-b} \widehat{\HT}_{ferm}=i\widehat{\bar{c}}_{q_i}\widehat{c}^{p_i}-
i\widehat{\bar{c}}_{p_j}\partial_i\partial_jV\widehat{c}^{q_i} .
\end{equation}
%%%
Using the SvH Hermitian conjugation rules (\ref{grass.due-cinquantadue}), we obtain:
%%%
\begin{equation}
\widehat{\HT}^{\dagger}_{ferm}=-i\widehat{\bar{c}}_{p_i}\widehat{c}^{q_i}+i\widehat{\bar{c}}_{q_i}
\partial_i\partial_jV\widehat{c}^{p_j}.
\end{equation}
%%%
So the first term in (\ref{grass.cinque-sette}) turns out to be 
%%%
\begin{equation}
[\widehat{\HT}_{ferm},\widehat{\HT}^{\dagger}_{ferm}]=\widehat{\bar{c}}_{q_i}\widehat{c}^{q_i}-
\widehat{\bar{c}}_{p_i}\widehat{c}^{p_i}+(\partial_i\partial_jV)(\partial_l\partial_mV)
[\widehat{\bar{c}}_{p_j}\widehat{c}^{p_m}\delta^i_l-\widehat{\bar{c}}_{q_l}\widehat{c}^{q_i}\delta_j^m]
\label{grass.cinque-otto}
\end{equation}
%%%
while the second and the third term in (\ref{grass.cinque-sette}) contain third order derivatives in the potential $V$. To find solutions
$|\psi\rangle$ of  (\ref{grass.cinque-cinque}), whose form is independent of the potential, we should impose that 
$|\psi\rangle$ be annihilated separately by the terms in (\ref{grass.cinque-sette}) which contain no derivative in $V,$ next by
those which contain first derivatives of $V$ and so on. By looking at
(\ref{grass.cinque-sette}) and (\ref{grass.cinque-otto}) the term with no derivative of $V$ is
$(\widehat{\bar{c}}_{q_i}\widehat{c}^{q_i}-\widehat{\bar{c}}_{p_i}\widehat{c}^{p_i})$; 
imposing it on $|\psi\rangle$ we get
\begin{equation}
(\widehat{\bar{c}}_{q_i}\widehat{c}^{q_i}-\widehat{\bar{c}}_{p_i}\widehat{c}^{p_i})|\psi\rangle=0
\end{equation}
%%%
which implies 
%%%
\begin{equation}
\displaystyle c^{q_i}\frac{\partial}{\partial c^{q_i}}|\psi\rangle=c^{p_i}\frac{\partial}{\partial c^{p_i}}|\psi\rangle.
\label{grass.cinque-nove}
\end{equation}
If we represent $|\psi\rangle$ as
%%%
\begin{equation}
\displaystyle \psi(\varphi,c)=\sum_{j=0}^{2n}\frac{1}{j!}\psi_{a,b,\ldots,j}(\varphi)c^ac^b\ldots c^j 
\label{grass.cinque-nove-b}
\end{equation}
then (\ref{grass.cinque-nove}) is satisfied by those $\psi(\varphi,c)$ which contain the same number of $c^q$ and 
$c^p$. Clearly these forms are Grassmannian even, which implies immediately that odd forms cannot be physical.   
Before going on to check whether also the terms in (\ref{grass.cinque-otto}) with second derivatives in $V$ 
annihilate these forms, let us remember that we must also satisfy the condition (\ref{grass.cinque-uno}). 
The operator $\widehat{\HT}-\widehat{\HT}^{\dagger}$ with $H$
of the form $\displaystyle H=\sum_{i=1}^n\frac{p_i^2}{2}+V(q_1,\ldots,q_n)$ has the expression 
%%%
\begin{equation}
\widehat{\HT}-\widehat{\HT}^{\dagger}=(i\widehat{\bar{c}}_{q_i}\widehat{c}^{p_i}+i\widehat{\bar{c}}_{p_i}
\widehat{c}^{q_i})-i
(\widehat{\bar{c}}_{p_j}\partial_i\partial_jV\widehat{c}^{q_i}+\widehat{\bar{c}}_{q_i}\partial_i\partial_jV
\widehat{c}^{p_j}) .\label{grass.cinque-dieci}
\end{equation}
%%%
Again, a physical form must be annihilated separately by the terms independent of $V$ and by those depending on it. So,
using (\ref{grass.cinque-dieci}),  (\ref{grass.cinque-uno}) gives the following two conditions:
%%%
\begin{eqnarray}
&&\label{grass.cinque-undici-a} (\widehat{\bar{c}}_{q_i}\widehat{c}^{p_i}+\widehat{\bar{c}}_{p_i}
\widehat{c}^{q_i})|\psi\rangle=0 \\
&&\label{grass.cinque-undici-b} (\widehat{\bar{c}}_{p_j}\partial_i\partial_jV\widehat{c}^{q_i}+\widehat{\bar{c}}_{q_i}
\partial_i\partial_jV\widehat{c}^{p_j})|\psi\rangle=0.
\end{eqnarray}
%%%
Let us now remember  that, because of  (\ref{grass.cinque-nove}), the state $|\psi\rangle$ must contain the same 
number of $c^q$
and $c^p$. Therefore it is easy to realize that (\ref{grass.cinque-undici-a}) implies that the following two relations 
must hold separately:
%%%
\begin{eqnarray}
&&\widehat{\bar{c}}_{q_i}\widehat{c}^{p_i}|\psi\rangle=0 \label{grass.cinque-dodici-a}\\
&&\widehat{\bar{c}}_{p_i}\widehat{c}^{q_i}|\psi\rangle=0. \label{grass.cinque-dodici-b}
\end{eqnarray}
%%%
Analogously (\ref{grass.cinque-undici-b}) implies that each
term in it separately annihilates $|\psi\rangle$:
%%%
\begin{eqnarray}
\displaystyle 
\label{grass.cinque-tredici-a}	
&&\widehat{\bar{c}}_{p_j}(\partial_i\partial_jV)\widehat{c}^{q_i}|\psi\rangle=0
\smallskip \\
\displaystyle
\label{grass.cinque-tredici-b}	
&&\widehat{\bar{c}}_{q_i}(\partial_i\partial_jV)\widehat{c}^{p_j}|\psi\rangle=0.
\end{eqnarray}
%%%
Let us now construct a linear combination of (\ref{grass.cinque-dodici-a}) and (\ref{grass.cinque-tredici-a}) of the 
following form:
%%%
\begin{equation}
(i\widehat{\bar{c}}_{q_i}\widehat{c}^{p_i}-i\widehat{\bar{c}}_{p_j}\partial_i\partial_jV\widehat{c}^{q_i})|\psi\rangle=0.
\end{equation}
%%%
It is easy to realize, looking at (\ref{grass.cinque-sette-b}), that this is equivalent to:
%%%
\begin{equation}
\widehat{\HT}_{ferm}|\psi\rangle=0. \label{grass.cinque-diciotto}
\end{equation}
%%%
Performing instead a linear combination of (\ref{grass.cinque-dodici-b}) and 
(\ref{grass.cinque-tredici-b}) of the form 
%%%
\begin{equation}
(-i\widehat{\bar{c}}_{p_i}\widehat{c}^{q_i}+i\widehat{\bar{c}}_{q_j}\partial_i\partial_jV\widehat{c}^{p_i})|\psi\rangle=0
\end{equation}
%%%
we immediately realize that this is equivalent to:
%%%
\begin{equation}
\widehat{\HT}^{\dagger}_{ferm}|\psi\rangle=0. \label{grass.cinque-diciannove}
\end{equation}
These are the two relations which complete our proof. In fact, using
(\ref{grass.cinque-diciotto})-(\ref{grass.cinque-diciannove}), we have that (\ref{grass.cinque-otto}) becomes:
%%%
\begin{eqnarray}
[\widehat{\HT},\widehat{\HT}^{\dagger}]|\psi\rangle &=&
[\widehat{\HT}_{ferm},\widehat{\HT}^{\dagger}_{ferm}]|\psi\rangle+
[\widehat{\HT}_{bos},\widehat{\HT}_{ferm}^{\dagger}]|\psi\rangle+
[\widehat{\HT}_{ferm},\widehat{\HT}_{bos}]|\psi\rangle=\nonumber\\
&=&-\widehat{\HT}^{\dagger}_{ferm}\widehat{\HT}_{bos}|\psi\rangle+\widehat{\HT}_{ferm}\widehat{\HT}_{bos}|
\psi\rangle=-\widehat{\HT}_{ferm}^{\dagger}|\psi^{\prime}
\rangle+\widehat{\HT}_{ferm}|\psi^{\prime}\rangle=0.\nonumber\\
\end{eqnarray}
%%%
The last step is based on the fact that $|\psi^{\prime}\rangle\equiv\widehat{\HT}_{bos}|\psi\rangle$
is  still a physical state. In fact $\widehat{\HT}_{bos}$ acts only on the bosonic coefficients of the states 
and so it does not modify their Grassmannian structure which
determines whether a state is physical or not. 

Up to now we have proved that a state, to be physical, must be annihilated by the fermionic part of the Hamiltonian 
$\widehat{\HT}$. 
The next step is to find the explicit form of such states. We want to start with an example. Let us take  a 
two-form with $n$ degrees of freedom. In order to satisfy
(\ref{grass.cinque-nove}) the two-form must contain one
variable $c^q$ and one variable $c^p$ and so it must be of the form:
%%%
\begin{equation}
\psi=\psi_{q_ip_k}c^{q_i}c^{p_k} .\label{grass.gio1}
\end{equation}
%%%
If we impose (\ref{grass.cinque-dodici-b}) on the state (\ref{grass.gio1}) we obtain:
%%%
\begin{equation}
\displaystyle c^{q_{\alpha}}\frac{\partial}{\partial c^{p_{\alpha}}}\psi=0 \Longrightarrow \psi_{q_ip_{\alpha}}c^{q_{\alpha}}
c^{q_i}=0 .\label{grass.gio2}
\end{equation}
%%%
For the properties of the Grassmann variables the previous relation is satisfied if we take $\alpha=i$. This
means that we have to take a two-form of the type:
%%%
\begin{equation}
\psi=\psi_{q_1p_1}c^{q_1}c^{p_1}+\psi_{q_2p_2}c^{q_2}c^{p_2}+\ldots+ \psi_{q_np_n}c^{q_n}c^{p_n} \label{grass.cinque-quattordici-a}
\end{equation} 
%%%
i.e. a form in which each $c^{q_i}$ is coupled with the relative $c^{p_i}$. 
Let us indicate, for simplicity, the various components $\psi_{q_jp_j}$ as $\psi_{\scriptscriptstyle (j)}
(\varphi)$. Then (\ref{grass.cinque-quattordici-a}) 
can be written as 
%%%
\begin{equation}
\displaystyle \psi=\sum_j\psi_{\scriptscriptstyle (j)}(\varphi)c^{q_j}c^{p_j} .\label{grass.cinque-quattordici-b}
\end{equation} 
%%%
Inserting (\ref{grass.cinque-quattordici-b}) into  (\ref{grass.cinque-tredici-a})-(\ref{grass.cinque-tredici-b}), 
it is easy to prove that they can be satisfied only if all the
coefficients $\psi_{\scriptscriptstyle (j)}(\varphi)$ in (\ref{grass.cinque-quattordici-b}) are the same
%%%
\begin{equation}
\psi_{\scriptscriptstyle (j)}(\varphi)=K(\varphi) .\label{grass.cinque-quindici-b}
\end{equation}
%%%
So (\ref{grass.cinque-quattordici-a}) turns out to be 
%%%
\begin{equation}	
\psi=K(\varphi)[c^{q_1}c^{p_1}+c^{q_2}c^{p_2}+\ldots+c^{q_n}c^{p_n}] .\label{grass.cinque-sedici}
\end{equation}
%%%
One sees that somehow the dependence on the indices of the coefficients of the
two-forms has disappeared. 
In general the coefficients 
$K(\varphi)$ will be the same for forms of the same rank but they will change with the rank. 
So for example a ``{\it physical}"
inhomogeneous form of rank up to 4 will be:
%%%
\begin{equation}
\displaystyle \psi=\psi_0(\varphi)+K(\varphi)\sum_ic^{q_i}c^{p_i}+S(\varphi)\sum_{i,j}(c^{q_i}c^{p_i})(c^{q_j}c^{p_j})
+\ldots
\label{grass.cinque-diciassette}
\end{equation}
%%%
Anyhow all our construction proves only that states of the form
(\ref{grass.cinque-diciassette}) are physical but not that they are the only ones. We feel anyhow very confident that they are actually
the only ones. From the physical point of view the homogeneous physical forms, like (\ref{grass.cinque-sedici}), 
are ``somehow" isomorphic to the zero-forms. In fact $\widehat{\HT}_{ferm}$ annihilates them, see
(\ref{grass.cinque-diciotto}), and this is the same that happens on the zero-forms. Basically $\widehat{\HT}_{ferm}$
acts on the Grassmann variables in (\ref{grass.cinque-sedici}) annihilating them; so we are left with only
$K(\varphi)$ changing under the time evolution and this $K(\varphi)$ evolves like a zero-form. 
Instead an inhomogeneous state like
(\ref{grass.cinque-diciassette}) is made of a sum of terms, each isomorphic to a zero-form; so we can say that it is 
like a linear superposition of the various zero-forms
$\psi_0(\varphi),K(\varphi),S(\varphi)$.
It is also easy to realize that among these physical states we always have the zero-forms and the $2n$ or volume-forms.
Before concluding we should point out that the physical condition (\ref{grass.cinque-uno}) limits the forms to be of the type
(\ref{grass.cinque-diciassette}) only if we do not put any restriction on the potential $V$. If we put restriction instead, 
for example choosing a harmonic oscillator potential or a separable potential, then the condition (\ref{grass.cinque-uno}) is satisfied by a
wider class of forms than the ones in (\ref{grass.cinque-diciassette}). This concludes the analysis of the SvH case.

\subsection{Symplectic Case}

Let us now turn to the other scalar products and in particular to the symplectic one of Sec. {\bf 4.3}. The
Hamiltonian $\widehat{\HT}$ in this case is Hermitian but not all the states of the Hilbert space have positive
norm. So the ``physical" Hilbert space, which we will indicate with $\mathbf{H}_{phys}$, should be a vector
subspace of the full Hilbert space, made only of positive norm states. 
Anyhow this subspace $\mathbf{H}_{phys}$ cannot be identified with the set $\mathbf{H}^{\scriptscriptstyle (+)}$
of {\it all} the positive norm states. In fact it is easy to realize that 
$\mathbf{H}^{\scriptscriptstyle (+)}$ is not a vector space because the linear combination of two states with positive norm,
$\psi\equiv\alpha\psi_+^{(1)}+\beta\psi_+^{(2)}$
%%%
where $\alpha$ and $\beta$ are complex coefficients, does not necessarily belong to $\mathbf{H}^{\scriptscriptstyle (+)}$. We will provide
an explicit example of this fact in (\ref{grass.counter}). So
$\mathbf{H}_{phys}$ can only be a particular subspace of $\mathbf{H}^{\scriptscriptstyle (+)}$. In order to build it, 
it is better to change the
variables, and pass from the set $(\widehat{q}_i,\widehat{p}_i,\widehat{\lambda}_{q_i},\widehat{\lambda}_{p_i}, 
\widehat{c}^{q_i},\widehat{c}^{p_i},\widehat{\bar{c}}_{q_i},\widehat{\bar{c}}_{p_i})$ 
to the following one:
%%%
\begin{equation}
 \left\{
		\begin{array}{l}
		\displaystyle 
	        \label{grass.cinque-venti-a}	
	        \widehat{z}_i\equiv\frac{1}{\sqrt{2}}(\widehat{q}_i+i\widehat{p}_i), \qquad\qquad\;\;\;\;\;\; 
	        \widehat{\bar{z}}_i\equiv\frac{1}{\sqrt{2}}(\widehat{q}_i-i\widehat{p}_i)
		\smallskip \\
		\displaystyle
		\widehat{l}_i\equiv\frac{1}{\sqrt{2}}(\widehat{\lambda}_{q_i}-i\widehat{\lambda}_{p_i}), 
		\qquad\qquad \;\;\;\;
	        \widehat{\bar{l}}_i\equiv\frac{1}{\sqrt{2}}(\widehat{\lambda}_{q_i}+i\widehat{\lambda}_{p_i}) 	
	        \smallskip \\
	        \displaystyle
		\widehat{\xi}^i\equiv\frac{1}{\sqrt{2}}(\widehat{c}^{q_i}+i\widehat{c}^{p_i}), \qquad\qquad\;\;\;\;
	        \widehat{\bar{\xi}}_i\equiv\frac{1}{\sqrt{2}}(-\widehat{\bar{c}}_{q_i}+i\widehat{\bar{c}}_{p_i})
	        \smallskip \\
	        \displaystyle
		\widehat{\xi}^{i*}\equiv\frac{1}{\sqrt{2}}(\widehat{c}^{q_i}-i\widehat{c}^{p_i}), \qquad\qquad\;\;\;
	        \widehat{\bar{\xi}}^*_i\equiv\frac{1}{\sqrt{2}}(\widehat{\bar{c}}_{q_i}+i\widehat{\bar{c}}_{p_i}).
		\end{array}
		\right.
\end{equation}
%%%
From (\ref{ijmpa.comm}) it is easy to work out the graded commutators among the new variables
(\ref{grass.cinque-venti-a}). In particular we will be
interested in the following ones:
%%%
\begin{equation}
 \left\{
		\begin{array}{l}
		\displaystyle 
		\label{grass.cinque-ventuno}	[\widehat{\xi}^i,\widehat{\bar{\xi}}_j]=-\delta^i_{j}, \qquad\qquad\; 
		[\widehat{\xi}^i,\widehat{\bar{\xi}}_j^*]=0
		\smallskip \\
		\displaystyle
		[\widehat{\xi}^{i*},\widehat{\bar{\xi}}^*_j]=+\delta_{j}^i, \qquad\qquad 
		[\widehat{\xi}^{i*},\widehat{\bar{\xi}}_j]=0.
		\end{array}
	\right.
\end{equation}
%%%
Under the symplectic Hermitian conjugation (\ref{grass.quattro-uno}), we get 
%%%
\begin{equation}
\widehat{\xi}^{i\dagger}=\widehat{\bar{\xi}}_i,
\qquad\qquad\qquad \widehat{\xi}^{i*\dagger}=\widehat{\bar{\xi}}_i^*. \label{grass.cinque-ventidue}
\end{equation}
%%%
Note that this ``hermiticity" properties for the Grassmann variables $(\widehat{\xi},\widehat{\xi}^*), 
(\widehat{\bar{\xi}},\widehat{\bar{\xi}}^*)$ are the same
as the SvH one (\ref{grass.due-cinquantadue}) for the variables $\widehat{c}^a,\widehat{\bar{c}}_a$. 
The crucial difference 
is in the anticommutator 
%%%
\begin{equation}
[\widehat{\xi}^i,\widehat{\bar{\xi}}_j]=-\delta_{j}^i
\end{equation}
%%%
which, for the analog SvH variables, had the opposite sign on the RHS: $[\widehat{c}^{q_i},\widehat{\bar{c}}_{q_j}]=
\delta_{j}^i$. 
We shall show  that it is just this difference in sign which 
gives rise to negative norm states in the symplectic case. 
Let us define in the case $n=1$ the state:
%%%
\begin{equation}
\widehat{\xi}|0-,0-\rangle=\widehat{\xi}^*|0-,0-\rangle=0. \label{grass.cinque-ventiquattro}
\end{equation}
%%%
Applying the other operators on $|0-,0-\rangle$ we obtain easily the other basic states of the theory:
%%%
\begin{equation}
\left\{
		\begin{array}{l}
		\displaystyle 
	\label{grass.cinque-venticinque-a}	|0+,0-\rangle=\widehat{\bar{\xi}}|0-,0-\rangle
		\smallskip \\
		\displaystyle
		|0-,0+\rangle=-\widehat{\bar{\xi}}^*|0-,0-\rangle
	        \smallskip \\
	        \displaystyle 
	    |0+,0+\rangle=\widehat{\bar{\xi}}^*\widehat{\bar{\xi}}|0-,0-\rangle.
		\end{array}
		\right.
\end{equation}
%%%
These states $|0\pm,0\pm\rangle$ will be different from those defined in 
(\ref{grass.quattro-quattro}) via the operators $\widehat{c}$ and $\widehat{\bar{c}}$
because they are eigenstates of different operators. Besides 
the hermiticity conditions, let us choose, as usual, a normalization for one of the states $|0\pm,0\pm\rangle$. 
In particular let us impose:
%%%
\begin{equation}
\Bigl(|0-,0-\rangle,|0-,0-\rangle\Bigr)=-1. \label{grass.cinque-ventisei-a}
\end{equation}
%%%
Via the definitions (\ref{grass.cinque-venticinque-a}) and the anticommutation relations 
(\ref{grass.cinque-ventuno}), we easily obtain
the following normalization conditions for the other states:
%%%
\begin{equation}
\left\{
		\begin{array}{l}
		\displaystyle 
		\label{grass.cinque-ventisei}	\Bigl(|0+,0-\rangle,|0+,0-\rangle\Bigr)=1
		\smallskip \\
		\displaystyle
		\Bigl(|0-,0+\rangle,|0-,0+\rangle\Bigr)=-1
	        \smallskip \\
	        \displaystyle 
	        \Bigl(|0+,0+\rangle,|0+,0+\rangle\Bigr)=1.
		\end{array}
		\right.
\end{equation}
%%%
From the definition (\ref{grass.cinque-venticinque-a}) 
we could represent the states as follows:
%%%
\begin{equation}
    \left\{
		\begin{array}{l}
		\displaystyle 
		\label{grass.cinque-ventisette}	
		\displaystyle
		|0-,0-\rangle=\xi\xi^*,\qquad\qquad |0-,0+\rangle=\xi
	        \smallskip \\
	        \displaystyle 
	        |0+,0-\rangle=\xi^*,\qquad\qquad \;\;|0+,0+\rangle=1.
		\end{array}
		\right.
\end{equation}
%%%
From this representation one sees that $|0+,0+\rangle$ is the basis of the zero-forms. This explains the reason for the
normalization choice $\Bigl(|0-,0-\rangle,|0-,0-\rangle\Bigr)=-1$: in this way the normalization of $|0+,0+\rangle$
turns out to be +1
and, as a consequence, the zero-form-states have positive norm as in the KvN case. According to
(\ref{grass.cinque-ventisette}) the scalar products
(\ref{grass.cinque-ventisei}) can be written as
%%%
\begin{equation}
	 \left\{
		\begin{array}{l}
		\displaystyle 
		\label{grass.cinque-ventisette-b}	
		\displaystyle
		(\xi\xi^*,\xi\xi^*)=-1,\qquad\qquad\; (\xi,\xi)=-1
	        \smallskip \\
	        \displaystyle 
	        (\xi^*,\xi^*)=1,\qquad\qquad\qquad (1,1)=1
		\end{array}
		\right.
\end{equation}
%%%
and so a generic state 
%%%
\begin{equation}
\psi(\xi,\xi^*)=\psi_0+\psi_{\xi}\xi+\psi_{\xi^*}\xi^*+\psi_2\xi\xi^*
\end{equation}
%%%
has the following norm: 
%%%
\begin{equation}
\langle\psi|\psi\rangle=|\psi_0|^2-|\psi_{\xi}|^2+|\psi_{\xi^*}|^2-|\psi_2|^2 .\label{grass.cinque-ventotto}
\end{equation}
%%%
The differences of sign between (\ref{grass.cinque-ventotto}) and (\ref{grass.due-cinquantuno}) 
are due to the minus signs appearing in (\ref{grass.cinque-ventisei-a})-(\ref{grass.cinque-ventisei}) 
instead of the plus signs appearing in (\ref{grass.due-trentasette-a})-(\ref{grass.due-trentasette-c}). 

We will now generalize our treatment to the case of
$n=2$. If we still want that $|0+0+0+0+\rangle$ be the basis of the zero-forms and have positive norm like in the KvN
case, then we should not choose the analog of the normalization (\ref{grass.cinque-ventisei-a}) but rather:
%%%
\begin{equation}
\Bigl(|0-0-0-0-\rangle,|0-0-0-0-\rangle\Bigr)=1 .\label{grass.cinque-ventinove}
\end{equation}
%%%
With this choice it is easy to prove that we get
%%%
\begin{equation}
\Bigl(|0+0+0+0+\rangle,|0+0+0+0+\rangle\Bigr)=1 \label{grass.cinque-trenta}
\end{equation}
%%%
and, as a consequence, the zero-forms have positive norm. In general, for $n$ degrees of freedom, in order to have positive norm
for the zero-forms we should make the following choice for the normalization of the state
$|0-0-\ldots 0-0-\rangle$:
%%%
\begin{equation}
\langle 0-0-\ldots 0-0-|0-0-\ldots 0-0-\rangle=(-1)^n. \label{grass.cinque-trentuno}
\end{equation}
%%%
We have now all the ingredients to start looking for the physical states. From the norms in (\ref{grass.cinque-ventisette-b}) we infer that in
general a homogeneous form $\displaystyle \psi=\frac{1}{l!}\psi_{ij\ldots l}\xi^i\xi^{j*}\ldots \xi^l$ 
has positive norm if the number of
$\xi$ variables is odd. This rule  
holds not only for $n=1$ but also for higher $n$. For example,
the representation of $|0-0-0-\ldots 0-\rangle$ in $n$ dimensions is 
%%%
\begin{equation}
|0-0-0-\ldots 0-\rangle=\xi^1\xi^{1*}\xi^2\xi^{2*}\ldots \xi^n\xi^{n*}
\end{equation}
%%%
and its norm is (\ref{grass.cinque-trentuno}), i.e. $(-1)^n$, which is $+1$ if $n$ (number of $\xi$ contained) is
even and $-1$ if $n$ is odd. 
So we have a criterion to look for  positive norm states: if a generic homogeneous state $\displaystyle 
\frac{1}{l!}\psi_{ab\ldots l} c^ ac^b\ldots
c^l$ is given, we first transform the $c^ a$ variables into $\xi^ i,\xi^{i*}$ variables via
(\ref{grass.cinque-venti-a}):
%%%
\begin{equation}
\displaystyle \frac{1}{l!} \psi_{ab\ldots l}c^ac^b\ldots c^l\;\Longrightarrow\; 
\frac{1}{l!}\widetilde{\psi}_{ij\ldots}\xi^i\xi^{j*}\ldots
\end{equation}
%%%
and then we count the number of $\xi$: if they are even, the state has positive norm; if they are odd, the state
has negative norm. Of
course this is a sufficient and necessary condition for homogeneous states but not for non-homogeneous ones. For
example the state 
%%%
\begin{equation}
\displaystyle \psi=\frac{1}{2!}\psi_{ab}\xi^a\xi^b+\psi_a\xi^a \label{grass.cinque-trentadue}
\end{equation}
%%%
is made of two parts, a two-form $\psi_{ab}\xi^a\xi^b$, and a one-form
$\psi_a\xi^a$. From what we said above the two-form has positive
 norm because  it
contains two $\xi$, while the one-form has a negative one. Still the overall norm 
%%%
\begin{equation}
\displaystyle \langle\psi|\psi\rangle=\sum_{a<b}\psi_{ab}\psi^*_{ab}-\sum_a\psi_a\psi_a^* \label{grass.counter}
\end{equation}
%%%
could well be positive. Indeed this is the statement that the subspace $\mathbf{H}^{\scriptscriptstyle (+)}$ of 
$\mathbf{H}$ 
is not a vector space. In fact  in the example (\ref{grass.cinque-trentadue}) we have summed a vector of 
$\mathbf{H}^{\scriptscriptstyle (+)}$
with one of $\mathbf{H}^{\scriptscriptstyle (-)}$ and ended up in a vector of
$\mathbf{H}^{\scriptscriptstyle (+)}$. Anyhow it is possible to find a subspace of
$\mathbf{H}^{\scriptscriptstyle (+)}$ which is a vector space. 

Let us stick to the homogeneous positive forms and let us check what happens under time evolution. First we  rewrite
the Hamiltonian $\widehat{\HT}$ in terms of the new variables
(\ref{grass.cinque-venti-a}) as:
%%%
\begin{equation}
\widehat{\HT}=i\partial_aH\widehat{l}_a-i\bar{\partial}_aH\widehat{\bar{l}}_a+
(\widehat{\xi}^k\widehat{\bar{\xi}}_a+\widehat{\xi}^{a*}\widehat{\bar{\xi}}_k^*)
\partial_k\bar{\partial}_aH
+\widehat{\xi}^{a*}\widehat{\bar{\xi}}_k\bar{\partial}_a\bar{\partial}_kH+
\widehat{\xi}^a\widehat{\bar{\xi}}_k^*\partial_a\partial_kH 
\label{grass.cinque-trentatre}
\end{equation}
%%%
where $\displaystyle \bar{\partial}_i=\frac{\partial}{\partial \bar{z}_i}$ and $\displaystyle
\partial_i=\frac{\partial}{\partial z_i}$. If we now take a generic homogeneous state of positive norm, i.e., with an even
number of $\xi$:
%%%
\begin{equation}
\displaystyle \psi=\frac{1}{q!}\psi_{ij\ldots q}\xi^i\xi^{j*}\ldots \xi^q, \label{grass.cinque-trentaquattro}
\end{equation}
%%%
in general the time evolution will turn it into a positive norm state because $\widehat{\HT}$ is Hermitian and the
evolution is unitary. Nevertheless the final state will be the sum of two terms, the first with an even number of 
$\xi$ and the second with an odd
number. In fact the last two terms in
(\ref{grass.cinque-trentatre}) change the number of $\xi$ and $\xi^*$ factors in the state
(\ref{grass.cinque-trentaquattro}). For example the first of these two terms removes a $\xi$ from
(\ref{grass.cinque-trentaquattro}) and injects a $\xi^*$ into it. So the resulting state is an {\it inhomogeneous} 
form in $\xi$.
This is not surprising because, even if $\widehat{\HT}$ conserves the form number in $c^a$, 
it does not conserve the form number in $\xi^i$
and $\xi^{i*}$ separately. If we restrict our space of homogeneous positive norm states to those
which are annihilated by the last two terms of (\ref{grass.cinque-trentatre}), then the time evolution will occur only via the first
four terms of (\ref{grass.cinque-trentatre}) which will not modify the number of $\xi$ and $\xi^*$
contained in the state. It is easy to check that the states of the form:
%%%
\begin{equation}
\displaystyle \psi_{phys}\equiv\psi_0(\varphi)+B(\varphi)\sum_{i,j}\xi^i\xi^{i*}\xi^j\xi^{j*}
+C(\varphi)\sum_{i,j,k,l}\xi^i\xi^{i*}\xi^j\xi^{j*}\xi^k\xi^{k*}\xi^l\xi^{l*}+\ldots \label{grass.cinque-trentacinque}
\end{equation}
%%%
are annihilated by the last two terms of $\HT$. The features of these states 
are: 
\begin{itemize}
\item[{\bf 1)}] each homogeneous form contained in them is made of products 
of an even number of $\xi^i$ and $\xi^{i*}$;
\item[{\bf 2)}] all indices are summed over;
\item[{\bf 3)}] in the homogeneous forms each term has the same coefficient:
in our example $B(\varphi)$ is the coefficient of the 4-form, $C(\varphi)$ is the coefficient 
of the 8-form.
\end{itemize}
The states (\ref{grass.cinque-trentacinque}) have positive norm because they are the sum of 
orthogonal positive norm states. 
Moreover the time evolution turns them in states with
the same features because the last two terms in $\widehat{\HT}$, which could break the pairs
$\xi^i\xi^{i*}$, give zero on states of the form (\ref{grass.cinque-trentacinque}). 
So this family of states is closed under time
evolution. Last but not least, differently than generic positive norm states, those of the form
(\ref{grass.cinque-trentacinque}) make a vector space: the sum of two forms with arbitrary coefficients is 
still a form which has the
properties {\bf 1), 2), 3)} which define this family. 
So these states have all the features to be physical states: they have positive norm, they are closed under 
time evolution and
they make a vector space. Furthermore not only the last two terms of $\widehat{\HT}$,
but also the previous two containing second derivatives of $H$, annihilate the states (\ref{grass.cinque-trentacinque}):
%%%
\begin{equation}
\Bigl[(\widehat{\xi}^k\widehat{\bar{\xi}}_a+\widehat{\xi}^{a*}\widehat{\bar{\xi}}_k^*)
\partial_k\bar{\partial}_aH\Bigr]\psi_{phys}=0.
\label{grass.cinque-trentasei}
\end{equation}
%%%
The four terms containing  second derivatives of $H$ are what we called $\widehat{\HT}_{ferm}$ 
in the first part of this section. So (\ref{grass.cinque-trentasei}) implies that
%%%
\begin{equation}
\widehat{\HT}_{ferm}\psi_{phys}=0. \label{grass.cinque-trentasette}
\end{equation}
%%%
This feature is preserved under time evolution because $\Bigl[\widehat{\HT},
\widehat{\HT}_{ferm}\Bigr]\psi_{phys}=0$. Note that  (\ref{grass.cinque-trentasette}) is the same equation we  
obtained in the SvH case (\ref{grass.cinque-diciotto}). Therefore also for the states 
(\ref{grass.cinque-trentacinque}) there is no evolution
of the Grassmann variables. Only $\widehat{\HT}_{bos}$ evolves the states acting on the
coefficients $\psi_0(\varphi),B(\varphi),C(\varphi)$ just like the Liouvillian on the zero-forms. 
In this sense also the symplectic physical states,
like the SvH ones, are ``isomorphic" to a set of zero-forms. Nevertheless the SvH physical states are many more than
the symplectic physical ones. 
In fact if we take for example a 4-form in (\ref{grass.cinque-trentacinque}), 
by turning the $\xi^i,\xi^{i*}$ into $c^q,c^p$
variables via (\ref{grass.cinque-venti-a}), we get :
%%%
\begin{eqnarray}
\displaystyle 
A(z,\bar{z})\xi^i\xi^{i*}\xi^j\xi^{j*}&\hspace{-0.2cm}=&\hspace{-0.2cm}
\frac{\widetilde{A}(\varphi)}{4}\Bigl[(c^{q_i}+ic^{p_i})(c^{q_i}-ic^{p_i})
(c^{q_j}+ic^{p_j})(c^{q_j}-ic^{p_j})\Bigr]=\nonumber\\
&\hspace{-0.2cm}=&\hspace{-0.2cm}
\frac{\widetilde{A}(\varphi)}{4}\Bigl[(-ic^{q_i}c^{p_i}+ic^{p_i}c^{q_i})(-ic^{q_j}c^{p_j}+ic^{p_j}c^{q_j})\Bigr]
=\nonumber\\
&\hspace{-0.2cm}=&\hspace{-0.2cm}
\frac{\widetilde{A}(\varphi)}{4}2ic^{p_i}c^{q_i}c^{p_j}c^{q_j}2i=
-\widetilde{A}(\varphi)c^{p_i}c^{q_i}c^{p_j}c^{q_j}
\label{grass.cinque-trentotto}
\end{eqnarray}
%%%
and this is a physical 4-form also in the SvH case. But if we take a 6-form an analogous calculation gives:
%%%
\begin{equation}
\displaystyle A(z,\bar{z})\:\xi^i\xi^{i*}\xi^j\xi^{j*}\xi^k\xi^{k*}=
-i\widetilde{A}(\varphi)\:c^{p_i}c^{q_i}c^{p_j}c^{q_j}c^{p_k}c^{q_k}.
\end{equation}
%%%
A 6-form like this is physical in the  SvH case  because it is annihilated by $\widehat{\HT}_{ferm}$ and so
$\widehat{\HT}$ is Hermitian on it. Nevertheless in the symplectic case 
it cannot be a physical form since it has negative norm because it contains an odd number of $\xi$. 
So the class of physical states is wider in the SvH case than in the symplectic one. 

\subsection{Connection between the Symplectic and the Gauge Case}

What about the gauge scalar product we analysed in Sec. {\bf 4.2}? Actually we are less interested in it because the
zero-forms have zero norm violating in this way the main feature of the KvN scalar product which we wanted to
maintain. Nevertheless in order to find the physical Hilbert space also in this
case the way to proceed is the following. Let us define the new Grassmann variables:
%%%
\begin{equation}
 \left\{
		\begin{array}{l}
		\displaystyle 
		\label{grass.cinque-trentanove}	
		\displaystyle
		\widehat{\psi}^a\equiv\frac{\widehat{c}^a+i\omega^{ab}\widehat{\bar{c}}_b}{\sqrt{2}}
	        \smallskip \\
	        \displaystyle 
	        \widehat{\bar{\psi}}_a\equiv\frac{\widehat{\bar{c}}_a+i\omega_{ab}\widehat{c}^b}{\sqrt{2}}.
	        \end{array}
		\right.
\end{equation}
%%%
Using the symplectic hermiticity conditions (\ref{grass.quattro-uno}) for the variables
($\widehat{c}^a,\widehat{\bar{c}}_a$), we get that
%%%
%%%
\begin{equation}
\widehat{\psi}^{a\dagger}=\widehat{\psi}^a, \qquad\qquad \widehat{\bar{\psi}}_a^{\dagger}=\widehat{\bar{\psi}}_a
\label{grass.cinque-trentanove-a}
\end{equation}
%%%
which means that $\widehat{\psi}^a$ and $\widehat{\bar{\psi}}_a$ are Hermitian like the Grassmann variables in the gauge
scalar product (\ref{grass.tre-due}). This is an interesting connection between the symplectic 
and the gauge scalar product which can be used to find the physical subspace.
In fact it is easy to prove that the anticommutation relations among ($\widehat{\psi}^a,\widehat{\bar{\psi}}_a$) 
are the same as the ones among the variables $\widehat{c}^a$:
%%%
\begin{equation}
[\widehat{\psi}^a,\widehat{\bar{\psi}}_b]=\delta_b^a,\;\;\;\; [\widehat{\psi}^a,\widehat{\psi}^b]=0,\;\;\;\; 
[\widehat{\bar{\psi}}_a,\widehat{\bar{\psi}}_b]=0.
\end{equation}
%%%
Furthermore the inverse transformations of (\ref{grass.cinque-trentanove}) are:
%%%
\begin{equation}
	 \left\{
		\begin{array}{l}
		\displaystyle 
	\label{grass.cinque-quaranta}	
		\displaystyle
		\widehat{c}^a=\frac{\widehat{\psi}^a-i\omega^{ab}\widehat{\bar{\psi}}_b}{\sqrt{2}}
	        \smallskip \\
	        \displaystyle 
	        \widehat{\bar{c}}_a=\frac{\widehat{\bar{\psi}}_a-i\omega_{ab}\widehat{\psi}^b}{\sqrt{2}}.
	        \end{array}
		\right.
\end{equation}
%%%
Having proved all this we can introduce, as in (\ref{grass.cinque-venti-a}), the Grassmann variables: 
%%%
\begin{eqnarray}
&&\displaystyle \widehat{\xi}^i=\frac{1}{\sqrt{2}}(\widehat{c}^{q_i}+i\widehat{c}^{p_i})=
\frac{1}{2}(\widehat{\psi}^{q_i}+i\widehat{\psi}^{p_i}-i
\widehat{\bar{\psi}}_{p_i}-\widehat{\bar{\psi}}_{q_i})\nonumber\\
&&\displaystyle
\widehat{\xi}^{i*}=\frac{1}{\sqrt{2}}(\widehat{c}^{q_i}-i\widehat{c}^{p_i})=
\frac{1}{2}(\widehat{\psi}^{q_i}-i\widehat{\psi}^{p_i}-
i\widehat{\bar{\psi}}_{p_i}+\widehat{\bar{\psi}}_{q_i})\nonumber\\
&&\displaystyle \widehat{\bar{\xi}}_i=\frac{1}{\sqrt{2}}(-\widehat{\bar{c}}_{q_i}+i\widehat{\bar{c}}_{p_i})=
\frac{1}{2}(-\widehat{\bar{\psi}}_{q_i}-i\widehat{\psi}^{p_i}+i\widehat{\bar{\psi}}_{p_i}+\widehat{\psi}^{q_i})
\nonumber\\
&&\displaystyle \widehat{\bar{\xi}}_i^*=\frac{1}{\sqrt{2}}(\widehat{\bar{c}}_{q_i}+i\widehat{\bar{c}}_{p_i})=
\frac{1}{2}(\widehat{\bar{\psi}}_{q_i}+i\widehat{\psi}^{p_i}+i\widehat{\bar{\psi}}_{p_i}+\widehat{\psi}^{q_i}).
\end{eqnarray}
%%%
It is easy to realize that, if $\widehat{\psi}$ and $\widehat{\bar{\psi}}$ satisfy the algebra and 
the anticommutation relations
of the gauge scalar product, then the set of operators $\widehat{\xi}, \widehat{\xi}^*, \widehat{\bar{\xi}},
\widehat{\bar{\xi}}^*$ satisfy exactly (\ref{grass.cinque-ventuno})-(\ref{grass.cinque-ventidue}). 
Therefore, even starting from the
gauge scalar product, we can repeat the same kind of considerations made in the symplectic case in order to find
out which is the subset of physical states. 

%%%%%%%%%%%%%%%%%%%%%%%%%%%%%%%%%%%%%%%%%%%%%%%%%%%%%%%%%%%%%%%%%%%%%%%%%%%%%%%%%%%%%%%%%%%%%%%%%%%%%%%%%%%%%%%

\bigskip

\section{Generalized Scalar Products}
\noindent
From the previous sections we can conclude that all the three scalar products we have analysed have 
either $\widehat{\HT}$ non-Hermitian or the scalar product non-positive definite. 
We shall show in this section that this is not a feature of the particular scalar products that we have 
introduced: even the most
general one cannot have both $\widehat{\HT}$ Hermitian and no negative norm states.

Let us limit our analysis to the case of $n=1$ and $\displaystyle H=\frac{p^2}{2}+V(q)$. What we want to do now
is to find out the most general hermiticity conditions for ($\widehat{c}^a,\widehat{\bar{c}}_a$) under which 
$\widehat{\HT}$ is Hermitian\footnote[5]{For $\widehat{\varphi}$ and $\widehat{\lambda}$ we will stick to the standard 
hermiticity conditions (\ref{grass.due-uno}).}. After this
we will analyse whether any of the associated scalar product is positive definite. 

The bosonic part of $\widehat{\HT}$ is always Hermitian and therefore we should only care about the fermionic part
which is
%%%
\begin{equation}
\widehat{\HT}_{ferm}=i\widehat{\bar{c}}_q\widehat{c}^p-i\widehat{\bar{c}}_pV^{\prime\prime}(q)\widehat{c}^q. 
\label{grass.sei-uno}
\end{equation}
%%%
For this to be Hermitian the two pieces on the RHS of (\ref{grass.sei-uno}) must be separately Hermitian since the
second one, differently from the first, contains the potential $V$. So we must have:
%%%
\begin{eqnarray}
&& \label{grass.sei-due} (i\widehat{\bar{c}}_q\widehat{c}^p)^{\dagger}=i\widehat{\bar{c}}_q\widehat{c}^p \\
&& \label{grass.sei-tre} (-i\widehat{\bar{c}}_p\widehat{c}^q)^{\dagger}=-i\widehat{\bar{c}}_p\widehat{c}^q.
\end{eqnarray}
%%%
Let us now see which are the most general hermiticity conditions on $\widehat{\bar{c}}_q$ and $\widehat{c}^p$ which 
satisfy (\ref{grass.sei-due}). Imposing a general condition of the form:
%%%
\begin{equation}
	 \left\{
		\begin{array}{l}
		\displaystyle 
	        \label{grass.sei-quattro}	
		\displaystyle
		\widehat{c}^{p\dagger}=\alpha\widehat{c}^p+\beta\widehat{\bar{c}}_q
	        \smallskip \\
	        \displaystyle 
	        \widehat{\bar{c}}_q^{\dagger}=\gamma\widehat{c}^p+\delta\widehat{\bar{c}}_q
	        \end{array}
		\right.
\end{equation}
%%%
and inserting it in (\ref{grass.sei-due}) we get
%%%
\begin{equation}
\alpha\delta-\beta\gamma=1. \label{grass.sei-cinque}
\end{equation}
%%%
Besides (\ref{grass.sei-due}) $\widehat{c}^p$ and $\widehat{\bar{c}}_q$ have to satisfy also the relations 
$\bigl((\widehat{c}^p)^{\dagger}\bigr)^{\dagger}=\widehat{c}^p,
\; \bigl((\widehat{\bar{c}}_q)^{\dagger}\bigr)^{\dagger}=\widehat{\bar{c}}_q$. Inserting (\ref{grass.sei-quattro}) 
into them
we get the following further conditions on the coefficients $\alpha,\beta,\gamma,\delta$:
%%%
\begin{eqnarray}
\label{grass.sei-sei-a}	
		\displaystyle
		&&\alpha^*\alpha+\beta^*\gamma=1
	        \smallskip \\
	        \label{grass.sei-sei-b}
	        \displaystyle 
	        &&\alpha^*\beta+\beta^*\delta=0
			\bigskip \\
        \label{grass.sei-sette-a}	
		\displaystyle
		&&\alpha\gamma^*+\delta^*\gamma=0
	        \smallskip\\
	        \label{grass.sei-sette-b}
	        \displaystyle 
	        &&\gamma^*\beta+\delta^*\delta=1.
\end{eqnarray}
%%%
From (\ref{grass.sei-sei-a}) and (\ref{grass.sei-sette-b}) we get 
%%%
\begin{equation}
	|\alpha|^2-|\delta|^2=-2i \,\textrm{Im}(\beta^*\gamma) \,\Longrightarrow\, \left\{
		\begin{array}{l}
	        \label{grass.sei-nove-a}	
			\displaystyle
			|\alpha|=|\delta|
	        \smallskip \\
	        \displaystyle 
	        \textrm{Im}(\beta^*\gamma)=0.
	        \end{array}
		\right.
\end{equation}
%%%
So $\beta^*\gamma$ is a real number and from (\ref{grass.sei-sei-a}) we get that 
$\beta^*\gamma \leq 1$.
We will have to analyse three different cases:
%%%
\begin{equation}
	 \left\{
		\begin{array}{l}
	        \label{grass.sei-undici}	
			\displaystyle
			\beta^*\gamma=1
	        \smallskip \\
	        \displaystyle 
	        \beta^*\gamma=0
	        \smallskip \\
	        \beta^*\gamma=z, 
	        \end{array}
		\right.
\end{equation}
%%%
where $z$ is a real number different from 0 and 1. It is possible to show that the hermiticity conditions
(\ref{grass.sei-quattro}) in the three cases (\ref{grass.sei-undici}) become respectively:
%%%
\begin{eqnarray}
\beta^*\gamma=1 & \Rightarrow & \left\{
		\begin{array}{l}
	        \label{grass.sei-dodici-1}	
		\displaystyle
		\widehat{c}^{p\dagger}=ib\widehat{\bar{c}}_q
	        \smallskip \\
	        \displaystyle 
	        \widehat{\bar{c}}_q^{\dagger}=\frac{i}{b}\widehat{c}^p
	        \end{array}
		\right. \\
		\medskip
	\beta^*\gamma=0 & \Rightarrow & \left\{
		\begin{array}{l}
	        \label{grass.sei-dodici-2}	
		\displaystyle
		\widehat{c}^{p\dagger}=e^{i\theta_{\alpha}}\widehat{c}^p
	        \smallskip \\
	        \displaystyle 
	        \widehat{\bar{c}}_q^{\dagger}=i\gamma_{\scriptscriptstyle 			
			I}\widehat{c}^p+e^{-i\theta_{\alpha}}\widehat{\bar{c}}_q
	        \end{array}
		\right. \\
		\medskip
	\beta^*\gamma=z & \Rightarrow & \left\{
		\begin{array}{l}
	        \label{grass.sei-dodici-3}	
		\displaystyle
		\widehat{c}^{p\dagger}=e^{i\theta_{\alpha}}\widehat{c}^p+ib\widehat{\bar{c}}_q
	        \smallskip \\
	        \displaystyle 
	        \widehat{\bar{c}}_q^{\dagger}=e^{-i\theta_{\alpha}}\widehat{\bar{c}}_q
	        \end{array}
		\right.
\end{eqnarray}
%%%
where $b$, $\gamma_{\scriptscriptstyle I}$ are the imaginary part of $\beta$, $\gamma$ and the variable 
$\theta_\alpha$ is the phase of $\alpha$: 
$\alpha=e^{i\theta_{\alpha}}$. Note that, if $b=-1$, (\ref{grass.sei-dodici-1}) gives part of the hermiticity 
conditions of
the symplectic scalar product (\ref{grass.quattro-uno}). If instead $\theta_{\alpha}=0$ and 
$\gamma_{\scriptscriptstyle I}=0$, (\ref{grass.sei-dodici-2})
gives part of the gauge scalar product hermiticity conditions (\ref{grass.tre-due}). Analogously if $\theta_{\alpha}=0$ 
and
$b=0$, (\ref{grass.sei-dodici-3}) also gives part of the gauge scalar product hermiticity conditions
(\ref{grass.tre-due}). So, via our procedure, we got some generalizations of either the symplectic or the gauge scalar
product. We did not get generalizations of the SvH one because in that case $\widehat{\HT}$ can be 
non-Hermitian while here we are searching for all the scalar products under which $\widehat{\HT}$ is
Hermitian. Let us remember that up to now we have only satisfied (\ref{grass.sei-due}). In order to satisfy also the 
hermiticity conditions
(\ref{grass.sei-tre}) we can note that
they can be obtained from (\ref{grass.sei-due}), replacing $p$ with $q$ and vice versa. So the associated hermiticity
conditions can be derived from (\ref{grass.sei-dodici-1}), (\ref{grass.sei-dodici-2}),
(\ref{grass.sei-dodici-3}), replacing $p$ with $q$ and vice versa: 
%%%
\begin{eqnarray}
	 &&\left\{
		\begin{array}{l}
	        \label{grass.sei-tredici-1}	
		\displaystyle
		\widehat{c}^{q\dagger}=ia\widehat{\bar{c}}_p
	        \smallskip \\
	        \displaystyle 
	        \widehat{\bar{c}}_p^{\dagger}=\frac{i}{a}\widehat{c}^q
	        \end{array}
		\right. \\
		\medskip
	&&\left\{
		\begin{array}{l}
	        \label{grass.sei-tredici-2}	
		\displaystyle
		\widehat{c}^{q\dagger}=e^{i\theta_{\beta}}\widehat{c}^q
	        \smallskip \\
	        \displaystyle 
	        \widehat{\bar{c}}_p^{\dagger}=i\gamma_{\scriptscriptstyle I}^{\prime}\widehat{c}^q+
	        e^{-i\theta_{\beta}}\widehat{\bar{c}}_p
	        \end{array}
		\right. \\
		\medskip
	&&\left\{
		\begin{array}{l}
	        \label{grass.sei-tredici-3}	
		\displaystyle
		\widehat{c}^{q\dagger}=e^{i\theta_{\beta}}\widehat{c}^q+ia\widehat{\bar{c}}_p
	        \smallskip \\
	        \displaystyle 
	        \widehat{\bar{c}}_p^{\dagger}=e^{-i\theta_{\beta}}\widehat{\bar{c}}_p.
	        \end{array}
		\right.
\end{eqnarray}
%%%
In the formulae above the variables $a,\theta_{\beta},\gamma_{\scriptscriptstyle I}^{\prime}$ are real parameters 
that can vary like $b,\theta_{\alpha},\gamma_{\scriptscriptstyle I}$ 
in (\ref{grass.sei-dodici-1}), (\ref{grass.sei-dodici-2}), (\ref{grass.sei-dodici-3}). Now, having three conditions
(\ref{grass.sei-dodici-1}), (\ref{grass.sei-dodici-2}), (\ref{grass.sei-dodici-3}) which satisfy (\ref{grass.sei-due}), 
and three
(\ref{grass.sei-tredici-1}), (\ref{grass.sei-tredici-2}), (\ref{grass.sei-tredici-3}) which satisfy (\ref{grass.sei-tre}), 
we have nine combinations which satisfy
both (\ref{grass.sei-due}) and (\ref{grass.sei-tre}). 
The next step in our procedure is to see whether some of the scalar products associated to these nine hermiticity 
conditions are positive definite. In order to perform this analysis in a
neater form we shall introduce in each scalar product a metric $g^{ij}$. If we write the states as:
%%%
\begin{equation}
	 \left\{
		\begin{array}{l}
	        \label{grass.sei-quattordici}	
		\displaystyle
		\psi=\psi_0+\psi_1c^q+\psi_2c^p+\psi_3c^qc^p
	        \smallskip \\
	        \displaystyle 
	        \Phi=\Phi_0+\Phi_1c^q+\Phi_2c^p+\Phi_3c^qc^p
	        \end{array}
		\right.
\end{equation}
%%%
then we can define the matrix $g^{ij}$ by the following equation:
%%%
\begin{equation}
\langle \Phi|\psi\rangle\equiv\int d\varphi\,\Phi_i^*g^{ij}\psi_j  \label{grass.sei-quindici}
\end{equation}
%%%
where $i,j$ can be $(0,1,2,3)$. It is easy to get convinced that all the three
scalar products, SvH (\ref{grass.due-cinquantuno}), gauge (\ref{grass.tre-ventotto-x}) 
and symplectic one (\ref{grass.quattro-quattordici}), can be
written in the form (\ref{grass.sei-quindici}) with different choices of the metric. 

Let us now see which metric we obtain out of the hermiticity conditions 
(\ref{grass.sei-dodici-1})-(\ref{grass.sei-tredici-1}) which are:
%%%
\begin{equation}
	 \left\{
		\begin{array}{l}
	        \label{grass.sei-sedici}	
		\displaystyle
		\widehat{c}^{p\dagger}=ib\widehat{\bar{c}}_q
	        \smallskip \\
	        \displaystyle 
	        \widehat{\bar{c}}_q^{\dagger}=\frac{i}{b}\widehat{c}^p
	        \smallskip \\
	        \widehat{c}^{q\dagger}=ia\widehat{\bar{c}}_p
	        \smallskip \\
	        \displaystyle \widehat{\bar{c}}_p^{\dagger}=\frac{i}{a}\widehat{c}^q.
	        \end{array}
		\right.
\end{equation}
%%%
The first two equations of (\ref{grass.sei-sedici}) can be written as  
\begin{equation}
	 \left\{
		\begin{array}{l}
	        \label{grass.sei-diciassette}	
		\displaystyle
		\langle \widehat{c}^p\Phi|\psi\rangle=\langle \Phi|ib\frac{\partial}{\partial c^q}\psi\rangle
	        \smallskip \\
	        \displaystyle 
	        \langle ib\frac{\partial}{\partial c^q}\Phi|\psi\rangle=\langle \Phi|\widehat{c}^p\psi\rangle .
	        \end{array}
		\right.
\end{equation}
%%%
Similarly the last two equations of (\ref{grass.sei-sedici}) are equivalent to 
\begin{equation}
	 \left\{
		\begin{array}{l}
	        \label{grass.sei-diciotto}	
		\displaystyle
		\langle \widehat{c}^q\Phi|\psi\rangle=\langle\Phi|ia\frac{\partial}{\partial c^p}\psi\rangle
	        \smallskip \\
	        \displaystyle 
	        \langle ia\frac{\partial}{\partial c^p}\Phi|\psi\rangle=\langle\Phi|\widehat{c}^q\psi\rangle.
	        \end{array}
		\right.
\end{equation}
%%%
Inserting (\ref{grass.sei-quindici}) into (\ref{grass.sei-diciassette}) and (\ref{grass.sei-diciotto})
we get that the real parameters $a$ and $b$ in (\ref{grass.sei-sedici}) must be one the opposite of the other: $a=-b$.
With the choice $g^{00}=1$ the whole metric turns out to be
%%%
\begin{equation}
g^{ij}=\begin{pmatrix}1 & 0 & 0 & 0\\ 0 & 0 & -ib & 0\\ 0 & ib & 0 & 0\\ 0 & 0 & 0 & -b^2\end{pmatrix}.
\label{grass.sei-diciannove}
\end{equation}
%%%
Let us first notice that with the
choice $b=-1$ the metric (\ref{grass.sei-diciannove}) reproduces, via (\ref{grass.sei-quindici}), 
the usual symplectic scalar product
(\ref{grass.quattro-quattordici}) which is not positive definite.
In general, to check whether (\ref{grass.sei-diciannove}) gives positive definite
scalar products, we should calculate the eigenvalues of (\ref{grass.sei-diciannove}) and see 
if they are all positive. These eigenvalues are:
%%%
\begin{equation}
	 \left\{
		\begin{array}{l}
	        \label{grass.sei-venti}	
		\displaystyle
		\lambda_1=1
	        \smallskip \\
	        \displaystyle 
	        \lambda_2=+b
	        \smallskip \\
	        \displaystyle
	        \lambda_3=-b 
	        \smallskip \\
	        \displaystyle 
	        \lambda_4=-b^2.
	        \end{array}
		\right.
\end{equation}
%%%
So we see that there are always two negative eigenvalues. This ultimately confirms that the scalar product
associated to (\ref{grass.sei-diciannove}) is not positive definite. Let us now turn 
to another of the nine possible hermiticity
conditions, in particular the one obtained combining (\ref{grass.sei-dodici-1}) with (\ref{grass.sei-tredici-2}):
%%%
\begin{equation}
	 \left\{
		\begin{array}{l}
	        \label{grass.sei-ventuno}	
		\displaystyle
		\widehat{c}^{p\dagger}=ib\widehat{\bar{c}}_q
	        \smallskip \\
	        \displaystyle 
	        \widehat{\bar{c}}_q^{\dagger}=\frac{i}{b}\widehat{c}^p
	        \smallskip \\
	        \displaystyle
	        \widehat{c}^{q\dagger}=e^{i\theta_{\beta}}\widehat{c}^q
	        \smallskip \\
	        \displaystyle 
	        \widehat{\bar{c}}_p^{\dagger}=i\gamma_{\scriptscriptstyle I}^{\prime} 
	        \widehat{c}^q+e^{-i\theta_{\beta}}\widehat{\bar{c}}_p.
	        \end{array}
		\right.
\end{equation}
%%%
It is easy to realize that this choice of hermiticity conditions is not consistent. 
In fact from the standard anticommutation
relation $[\widehat{c}^q,\widehat{\bar{c}}_q]=1$ we get this other one 
%%%
\begin{equation}
[\widehat{c}^{q\dagger},\widehat{\bar{c}}_q^{\dagger}]=1 \label{grass.sei-ventidue}
\end{equation}
%%%
and replacing in it the expression obtained from (\ref{grass.sei-ventuno}) we get: 
%%%
\begin{equation}
\displaystyle [\widehat{c}^{q\dagger},\widehat{\bar{c}}_q^{\dagger}]=
[e^{i\theta_{\beta}}\widehat{c}^q,(i/b)\widehat{c}^p]=0
\end{equation}
%%%
which contradicts (\ref{grass.sei-ventidue}). 
We have a similar problem with the hermiticity conditions obtained combining (\ref{grass.sei-dodici-1}) 
and (\ref{grass.sei-tredici-3}):
%%%
\begin{equation}
	 \left\{
		\begin{array}{l}
	        \label{grass.sei-ventitre}	
		\displaystyle
		\widehat{c}^{p\dagger}=ib\widehat{\bar{c}}_q
	        \smallskip \\
	        \displaystyle 
	        \widehat{\bar{c}}_q^{\dagger}=\frac{i}{b}\widehat{c}^p
	        \smallskip \\
	        \displaystyle
	        \widehat{c}^{q\dagger}=e^{i\theta_{\beta}}\widehat{c}^q+ia\widehat{\bar{c}}_p
	        \smallskip \\
	        \displaystyle 
	        \widehat{\bar{c}}_p^{\dagger}=e^{-i\theta_{\beta}}\widehat{\bar{c}}_p.
	        \end{array}
		\right.
\end{equation}
%%%
In fact from $[\widehat{c}^p,\widehat{\bar{c}}_p]$=1 we get:
%%%
\begin{equation}
[\widehat{c}^{p\dagger},\widehat{\bar{c}}_p^{\dagger}]=1 \label{grass.sei-ventiquattro}
\end{equation}
%%%
but inserting (\ref{grass.sei-ventitre}) in (\ref{grass.sei-ventiquattro}) we obtain 
%%%
\begin{equation}
[\widehat{c}^{p\dagger},\widehat{\bar{c}}_p^{\dagger}]=[ib\widehat{\bar{c}}_q, e^{-i\theta_{\beta}}\widehat{\bar{c}}_p]=0
\end{equation}
%%%
which contradicts (\ref{grass.sei-ventiquattro}).

Another hermiticity condition which is inconsistent is the one obtained 
combining  (\ref{grass.sei-dodici-2}) with (\ref{grass.sei-tredici-1}). In that case in fact, instead of 
(\ref{grass.sei-ventiquattro}), we would get
%%%
\begin{equation}
\displaystyle [\widehat{c}^{p\dagger},\widehat{\bar{c}}_p^{\dagger}]=[e^{i\theta_{\alpha}}\widehat{c}^p,
(i/a)\widehat{c}^q]=0.
\end{equation}
%%%
Analogously the hermiticity conditions obtained combining (\ref{grass.sei-dodici-3}) 
and (\ref{grass.sei-tredici-1}) are inconsistent.
In fact in this case we get
%%%
\begin{equation}
[\widehat{c}^{q\dagger},\widehat{\bar{c}}_q^{\dagger}]=[ia\widehat{\bar{c}}_p,e^{-i\theta_{\alpha}}\widehat{\bar{c}}_q]=0
\end{equation}
%%%
which contradicts (\ref{grass.sei-ventidue}). Let us now analyse the case 
(\ref{grass.sei-dodici-2})-(\ref{grass.sei-tredici-2}) which gives
the following hermiticity conditions:
%%%
\begin{equation}
	 \left\{
		\begin{array}{l}
	        \label{grass.sei-ventiquattro-b}	
			\displaystyle
			\widehat{c}^{p\dagger}=e^{i\theta_{\alpha}}\widehat{c}^p
	        \smallskip \\
	        \displaystyle 
	        \widehat{\bar{c}}_q^{\dagger}=i\gamma_{\scriptscriptstyle I}
	        \widehat{c}^p+e^{-i\theta_{\alpha}}\widehat{\bar{c}}_q
	        \smallskip \\
	        \displaystyle
	        \widehat{c}^{q\dagger}=e^{i\theta_{\beta}}\widehat{c}^q
	        \smallskip \\
	        \displaystyle 
	        \widehat{\bar{c}}_p^{\dagger}=i\gamma_{\scriptscriptstyle I}^{\prime} 				 
			\widehat{c}^q+e^{-i\theta_{\beta}}\widehat{\bar{c}}_p.
	        \end{array}
		\right.
\end{equation}
%%%
The parameters $\theta_{\alpha},\theta_{\beta},\gamma_{\scriptscriptstyle I},
\gamma_{\scriptscriptstyle I}^{\prime}$ have to satisfy the following conditions: 
$\theta_{\alpha}=\theta_{\beta},\,\gamma_{\scriptscriptstyle I}=-\gamma_{\scriptscriptstyle I}^{\prime}$
which lead to the following metric $g^{ij}$: 
%%%
\begin{equation}
g^{ij}=\begin{pmatrix} ig^{03}e^{i\theta_{\alpha}}\gamma_{\scriptscriptstyle I} & 0 & 0 & g^{03}\\
0 & 0 & g^{03}e^{i\theta_{\alpha}} & 0\\ 
0 & -g^{03}e^{i\theta_{\alpha}} & 0 & 0\\
-g^{03}e^{2i\theta_{\alpha}} & 0 & 0 & 0\end{pmatrix}. \label{grass.sei-ventisei}
\end{equation}
%%%
The eigenvalue equation associated to this metric is
%%%
\begin{equation}
\bigl(\lambda^2+(g^{03})^2e^{2i\theta_{\alpha}}\bigr)
\bigl(\lambda^2-ig^{03}e^{i\theta_{\alpha}}\gamma_{\scriptscriptstyle I}\lambda+(g^{03})^2
e^{2i\theta_{\alpha}}\bigr)=0.
\end{equation}
%%%
Two of its eigenvalues are
%%%
\begin{equation}
\lambda=\pm ig^{03}e^{i\theta_{\alpha}} \label{grass.sei-ventisette}
\end{equation}
%%%
and one sees that, for example, the choice $\theta_{\alpha}=0$ and $g^{03}=i$ leads to a 
positive and a negative eigenvalue. In general it is impossible to obtain from 
(\ref{grass.sei-ventisette}) two real eigenvalues. This proves that even the scalar 
product which is associated to the hermiticity
conditions (\ref{grass.sei-ventiquattro-b}) is not positive definite. This case is a 
generalization of the gauge scalar product 
(\ref{grass.tre-ventotto-x}); in fact it reduces to it with the choice 
$\gamma_{\scriptscriptstyle I}=0$, $g^{03}=-i$, $\theta_{\alpha}=0$. 
Let us now proceed to other cases like for example the one whose hermiticity conditions are the combination 
of (\ref{grass.sei-dodici-2}) and (\ref{grass.sei-tredici-3}). From these relations we get 
%%%
\begin{equation}
	 \left\{
		\begin{array}{l}
	        \label{grass.sei-ventotto}	
		\displaystyle
		\widehat{c}^{p\dagger}=e^{i\theta_{\alpha}}\widehat{c}^p
	        \smallskip \\
	        \displaystyle 
	        \widehat{c}^{q\dagger}=e^{i\theta_{\beta}}\widehat{c}^q+ia\widehat{\bar{c}}_p
	        \smallskip \\
	        \displaystyle 
	        \widehat{\bar{c}}_q^{\dagger}=i\gamma_{\scriptscriptstyle I}
	        \widehat{c}^p+e^{-i\theta_{\alpha}}\widehat{\bar{c}}_q
	        \smallskip \\
	        \displaystyle 
	        \widehat{\bar{c}}_p^{\dagger}=e^{-i\theta_{\beta}}\widehat{\bar{c}}_p.
	        \end{array}
		\right.
\end{equation}
%%%
Using these expressions and imposing the following relations:
 %%%
\begin{equation}
	 \left\{
		\begin{array}{l}
	        \label{grass.sei-ventinove}	
		\displaystyle
		[\widehat{c}^q,\widehat{c}^p]^{\dagger}=0
	        \smallskip \\
	        \displaystyle 
	        [\widehat{\bar{c}}_q,\widehat{\bar{c}}_p]^{\dagger}=0,
	        \smallskip \\
	        \end{array}
		\right.
\end{equation}
%%%
we get that in (\ref{grass.sei-ventotto}) 
we must choose $a=\gamma_{\scriptscriptstyle I}=0$. Therefore
the hermiticity conditions are the same as those which led to the metric 
(\ref{grass.sei-ventisei}) with $\gamma_{\scriptscriptstyle I}=0$. Therefore also the scalar product
(\ref{grass.sei-dodici-2})-(\ref{grass.sei-tredici-3}) is not positive definite. 
A similar analysis can be carried out for 
the hermiticity conditions given by the combination of (\ref{grass.sei-dodici-3}) 
and (\ref{grass.sei-tredici-2}).
Also in this case we get that, in order to satisfy (\ref{grass.sei-ventinove}), 
we must put
$b=\gamma_{\scriptscriptstyle I}^{\prime}=0$ in (\ref{grass.sei-dodici-3}) and 
(\ref{grass.sei-tredici-2}). 
These relations lead again to the metric (\ref{grass.sei-ventisei}) with 
$\gamma_{\scriptscriptstyle I}=0$ 
which is not positive definite. The last of the nine hermiticity conditions that we have to examine 
is the combination of (\ref{grass.sei-dodici-3}) and (\ref{grass.sei-tredici-3}).
This leads to the metric
%%%
\begin{equation}
g^{ij}=\begin{pmatrix}0 & 0 & 0 & g^{03}\\
0 & 0 & g^{03}e^{i\theta_{\alpha}} & 0\\ 
0 & -g^{03}e^{i\theta_{\alpha}} & 0 & 0\\
-g^{03}e^{2i\theta_{\alpha}} & 0 & 0 & -ig^{03}e^{i\theta_{\alpha}}b\end{pmatrix} \label{grass.sei-trenta}
\end{equation}
%%%
whose eigenvalues are given by the solutions
of the equation
%%%
\begin{equation}
\bigl(\lambda^2+(g^{03})^2e^{i\theta_{\alpha}}\bigr)\bigl(\lambda^2+
ig^{03}e^{i\theta_{\alpha}}b\lambda+(g^{03})^2e^{2i\theta_{\alpha}}\bigr)=0.
\end{equation}
%%%
Two solutions are given by
%%%
\begin{equation}
\lambda=\pm ig^{03}e^{i\theta_{\alpha}} \label{grass.sei-trentuno}
\end{equation}
%%%
and again it is impossible to obtain two real and positive eigenvalues from them.
Since the metric (\ref{grass.sei-trenta}) have negative eigenvalues, the associated scalar product is
not positive definite. It should be noticed that also (\ref{grass.sei-trenta}) is a generalization of the gauge scalar product
which is obtained with the following choice of parameters: $b=0$, $\theta_{\alpha}=0$, $g^{03}=-i$.

To summarize what we have done in this section, we can say that the whole set of consistent scalar products under which 
$\widehat{\HT}$ is Hermitian is given by the three metrics (\ref{grass.sei-ventisei}),
(\ref{grass.sei-trenta}) and (\ref{grass.sei-diciannove}). 
The first two are generalizations of the gauge scalar product while the last one is a generalization of the
symplectic case. None of these three generalizations leads to a positive definite scalar product. 
So we have proved that if
$\widehat{\HT}$ {\it is Hermitian for every choice of the Hamiltonian} $H$ 
then {\it the associated scalar product is not positive definite}. 
As a consequence, based on standard rules of logic, we have that if
{\it the associated scalar product is positive definite} then
$\widehat{\HT}$ {\it can turn out to be non-Hermitian}. 
This second case is exemplified in the SvH scalar product. 

Of course the theorem above holds if we work in the full
Hilbert space. It is easy to see that, even with the generalized scalar products 
(\ref{grass.sei-diciannove}), (\ref{grass.sei-ventisei}) and (\ref{grass.sei-trenta}),
the subspace of positive norm states, closed under time evolution and with the feature of being a
vector space, is isomorphic to the space of zero-forms. We will skip the proof because it is very similar to that
presented in Sec. {\bf 4.4}.

%%%%%%%%%%%%%%%%%%%%%%%%%%%%%%%%%%%%%%%%%%%%%%%%%%%%%%%%%%%%%%%%%%%%%%%%%%%%%%%%%%%%%%%%%%%%%%%%%%%%%%%%%%%%%%%%%%%%%%%%%%%%%%

\bigskip

\section{Grassmann Variables and Jacobi Fields}
\noindent
Despite the detailed mathematical analysis contained in this chapter, the reader may still be puzzled by the results we have
gotten. In fact it is difficult to accept that in CM we cannot have at the same time a positive definite
scalar product and a Hermitian Hamiltonian. In this section we would like to give some tentative physical explanations of this result.

Let us for example analyse chaotic systems, i.e. systems which have trajectories flying away
exponentially as time passes by. The variables which describe this behaviour better are the so-called Jacobi fields which
are defined as
%%%
\begin{equation}
\delta\varphi^a(t,\varphi_0)=
\varphi_2^a(t,\varphi_0+\delta\varphi_0)-\varphi_1^a(t,\varphi_0) \label{grass.sette-uno}
\end{equation}
%%%
where $\varphi_1^a(t)$ and $\varphi_2^a(t)$ are the two trajectories which start at time $t=0$ very close to each other
respectively in $\varphi_0$ and $\varphi_0+\delta\varphi_0$.
The evolution of these Jacobi fields $\delta\varphi^a$ is the same as that of the Grassmann
variables $c^a$:
%%%
\begin{equation}
\displaystyle \biggl[\delta_b^a\partial_t-\omega^{al}\frac{\partial^2H}{\partial\varphi^l\partial\varphi^b}\biggr]
\delta\varphi^b=0 .\label{grass.sette-due}
\end{equation}
%%%
The Euclidean square distance between the two trajectories in phase space is given by
%%%
\begin{equation}
D(\varphi_0,t)\equiv\|\delta\varphi^a\|^2 \label{grass.sette-tre}
\end{equation}
%%%
and it is a function of $t$ and $\varphi_0$. In more precise mathematical terms the inequality 
which defines a system as chaotic is the following one:
%%%
\begin{equation}
\displaystyle \lim_{t\to\infty}\frac{1}{t}\;\ln\int d\varphi_0D(\varphi_0,t)>0. \label{grass.sette-quattro}
\end{equation}
%%%
One immediately infers from (\ref{grass.sette-quattro}) that the phase space of chaotic systems has regions of non-zero
measure such that the trajectories which originate from there fly away exponentially as time passes by.

Now we want to show that, in those regions, also the components $\psi_a$ 
of any one-form $\psi=\psi_ac^a$ behave as the Jacobi fields. 
Let us first turn to the ``momentum" representation for the Grassmann variables replacing $c$ with $\bar{c}$. 
In the $\bar{c}$ representation it is natural to use, instead of the notation 
(\ref{grass.due-cinquanta-a}):
%%%
\begin{equation}
\langle +c^{p*},+c^{q*},\varphi|\psi\rangle=\psi_0+\psi_qc^q+\psi_pc^p+\psi_2c^qc^p \label{grass.i-44-1}
\end{equation}
%%%
the following one:
%%%
\begin{equation}
\langle -\bar{c}_p^*,-\bar{c}_q^*,\varphi|\psi\rangle=\psi^0+\psi^q\bar{c}_q+\psi^p\bar{c}_p+\psi^2\bar{c}_q\bar{c}_p.
\label{grass.i-44-2}
\end{equation}
%%%
In this basis a completeness relation, analog to the first one of (\ref{grass.due-quarantuno}), is then
%%%
\begin{equation}
\int d\bar{c}_qd\bar{c}_p|\bar{c}_q+,\bar{c}_p+\rangle\langle-\bar{c}_p^*,-\bar{c}_q^*|={\bf 1}
\end{equation}
%%%
and inserting it into the LHS of (\ref{grass.i-44-1}) we get the relation between (\ref{grass.i-44-1}) and (\ref{grass.i-44-2}), i.e.:
%%%
\begin{eqnarray}
\langle +c^{p*},+c^{q*}|\psi\rangle&\hspace{-0.2cm}=&\hspace{-0.2cm}
\int d\bar{c}_qd\bar{c}_p\langle +c^{p*},+c^{q*}|\bar{c}_q+,\bar{c}_p+\rangle
\cdot \langle -\bar{c}_p^*, -\bar{c}_q^*|\psi\rangle=\nonumber\\
&\hspace{-0.2cm}=&\hspace{-0.2cm}\int d\bar{c}_qd\bar{c}_p (1+c^q\bar{c}_q+c^p\bar{c}_p-c^qc^p\bar{c}_q\bar{c}_p)\cdot
(\psi^0+\psi^q\bar{c}_q+\psi^p\bar{c}_p+\psi^2\bar{c}_q\bar{c}_p)=\nonumber\\
&\hspace{-0.2cm}=&\hspace{-0.2cm}-\psi^2-\psi^pc^q+\psi^qc^p+\psi^0c^qc^p.
\end{eqnarray}
%%%
Comparing this with the RHS of (\ref{grass.i-44-1}) we get that:
%%%
\begin{equation}
	 \left\{
		\begin{array}{l}
		\displaystyle
		\label{grass.i-45-1}
		\psi_0=-\psi^2
	        \smallskip \\
	        \displaystyle 
	        \psi_q=-\psi^p
	        \smallskip \\
	        \displaystyle 
	        \psi_p=\psi^q
	        \smallskip \\
	        \displaystyle 
	        \psi_2=\psi^0.
	        \end{array}
		\right.
\end{equation}
%%%
The reason why we have introduced the $\bar{c}$ representation is that, as we will show below, 
the components $\psi^q$ and $\psi^p$ transform, under time evolution, as the Jacobi fields 
$\delta q,\delta p$ of (\ref{grass.sette-uno}).
The evolution of the wave functions (\ref{grass.i-44-2}) is given by the Hamiltonian $\widehat{\HT}$ expressed
in the ``momentum" representation, i.e.:
%%%
\begin{equation}
\displaystyle \widehat{\HT}_{ferm}=i\bar{c}_aM^a_d\frac{\partial}{\partial\bar{c}_d}. \label{grass.i-46-2}
\end{equation}
%%%
where $M^a_d=\omega^{ab}\partial_b\partial_dH$.
%%%
Restricting ourselves to the part of the wave function (\ref{grass.i-44-2}) which is 
linear in the $\bar{c}$ variables, we have that the infinitesimal evolution gives:
%%%
\begin{equation}
\displaystyle \psi^a(\epsilon)\bar{c}_a(\epsilon)=e^{-i\widehat{\HT}\epsilon}\,\biggl(\psi^a(0)\bar{c}_a(0)
\biggr).
\label{grass.i-46-3}
\end{equation}
%%%
Inserting (\ref{grass.i-46-2}) into (\ref{grass.i-46-3}) we get easily:
%%%
\begin{equation}
\psi^a(\epsilon) =\psi^a(0)+\epsilon M^a_d\psi^d(0). \label{grass.i-47-1}
\end{equation}
%%%
If we now solve, for an infinitesimal time, the equation of motion (\ref{grass.sette-due}) 
for the Jacobi fields we get:
%%%
\begin{equation}
\delta\varphi^a(\epsilon)=\delta\varphi^a(0)+\epsilon M^a_d\delta\varphi^d(0). \label{grass.i-47-2}
\end{equation}
%%%
This proves that the variables $\psi^a$ and the Jacobi fields $\delta\varphi^a$ evolve in the same way and if the
latter diverge
exponentially with $t$, the same happens to $\psi^a$. This implies that the behaviour with $t$ of the distance
$D(\varphi_0,t)$ (\ref{grass.sette-tre}) and of $\displaystyle \sum_a |\psi_a(\varphi_0,t)|^2$
are the same. Furthermore if
(\ref{grass.sette-quattro}) holds, also the following inequality holds:
%%%
\begin{equation}
\displaystyle \lim_{t\to\infty}\frac{1}{t}\textrm{ln}\int d\varphi_0\sum_a|\psi^a(\varphi_0,t)|^2>0.
\label{grass.i-47-3}
\end{equation}
%%%
Then via (\ref{grass.i-45-1}) we can replace (\ref{grass.i-47-3}) with 
%%%
\begin{equation}
\lim_{t\to\infty}\frac{1}{t}\textrm{ln}\int d\varphi_0\sum_a|\psi_a(\varphi_0,t)|^2>0
\end{equation}
%%%
and the argument of the logarithm is exactly the SvH norm of the one-form $\psi_a(\varphi_0,t)c^a$. 
So in the case of chaotic systems the norm of the states $\psi=\psi_ac^a$ in the SvH scalar product
%%%
\begin{equation}
\displaystyle \int d\varphi_0\sum_a|\psi_a(\varphi_0,t)|^2 \label{grass.sette-cinque}
\end{equation}
%%%
diverges exponentially, just like (\ref{grass.sette-quattro}).
Now if we take the sum of a zero-form $\psi_0$ and a one-form $\psi_ac^a$
%%%
\begin{equation}
\widetilde{\psi}\equiv\psi_0+\psi_ac^a.
\end{equation}
%%%
the SvH norm of $\widetilde{\psi}$ becomes: 
%%%
\begin{equation}
\displaystyle \|\widetilde{\psi}\|^2=\int d\varphi_0|\psi_0(\varphi_0,t)|^2+\int d\varphi_0\sum_a|\psi_a
(\varphi_0,t)|^2 .\label{grass.sette-sei}
\end{equation}
%%%
Let us suppose that in the SvH scalar product $\widehat{\HT}$ were Hermitian. Then the norm of
$\widetilde{\psi}$ would be conserved under the time evolution. Anyhow we know that the second piece in
(\ref{grass.sette-sei}), i.e. $\displaystyle \biggl(\int d\varphi_0\sum_a |\psi_a|^2\biggr)$, 
increases for chaotic systems and this implies that, for
$\|\widetilde{\psi}\|^2$ to be conserved, the first term in (\ref{grass.sette-sei}), i.e. $\displaystyle 
\int d\varphi_0|\psi_0|^2$, cannot be conserved. The non-conservation of this piece
implies a violation of the Liouville theorem but this is an absurd. Therefore $\widehat{\HT}$ must be non-Hermitian in the
SvH scalar product.
To put things in simpler terms: if we have a chaotic system we must be able to
produce, from the operator of evolution $\displaystyle e^{-i\widehat{\HT} t}$, an exponential diverging behaviour like:
%%%
\begin{equation}
\displaystyle e^{-i\widehat{\HT} t}\;\;\longrightarrow K e^{+lt} \label{grass.sette-sette}
\end{equation}
%%%
with $l$ a real number.
This happens only if $\widehat{\HT}$ is not Hermitian. In fact only in this case $\widehat{\HT}$ can have 
complex eigenvalues producing something like (\ref{grass.sette-sette}). 
This same kind of behaviour can be produced also in the gauge and symplectic case
where $\widehat{\HT}$ is Hermitian but the scalar product is not positive definite. In this case
even Hermitian operators can have complex eigenvalues and the proof goes as follows. 
Let us start from the hermiticity of $\widehat{\HT}$, i.e.:
%%%
\begin{equation}
\langle \psi| \widehat{\HT}\psi\rangle=\langle \widehat{\HT} \psi|\psi\rangle. \label{grass.sette-otto}
\end{equation}
%%%
If $|\psi\rangle$ is an eigenstate of $\widehat{\HT}$ with eigenvalue $\lambda$ then
(\ref{grass.sette-otto}) can be written as
%%%
\begin{equation}
\lambda\langle\psi|\psi\rangle=\lambda^*\langle\psi|\psi\rangle. \label{grass.sette-nove}
\end{equation}
%%%
From this relation we cannot deduce that $\lambda=\lambda^*$ because, in a non-positive definite scalar product, the state
$|\psi\rangle$ could be of zero norm and satisfy (\ref{grass.sette-nove}) whatever is the value, real or complex,
of $\lambda$. 

This analysis explains the reason why for a {\it generic potential}, as we have assumed 
throughout this paper, we
have either a positive definite Hilbert space or $\widehat{\HT}$ Hermitian but never both of them. 
This last possibility can
happen only for {\it specific potentials}. For example in the SvH case the $\widehat{\HT}$ associated
to a Hamiltonian of the form
%%%
\begin{equation}
\displaystyle H=\frac{1}{2}p^2+\frac{1}{2}q^2 
\end{equation}
%%%
is Hermitian, see (\ref{grass.due-cinquantatre})-(\ref{grass.due-cinquantaquattro}). 
As this is a harmonic
oscillator one may be tempted to generalize the result and jump to the conclusion that, 
for integrable systems, differently than for chaotic ones, 
we could have both $\widehat{\HT}$ Hermitian and the scalar product positive definite. It is
actually not so: even for harmonic oscillators only some of them have the features above.
In fact let us insert the mass and the frequency in $H$:
%%%
\begin{equation}
\displaystyle H=\frac{1}{2m}p^2+\frac{1}{2}m\omega^2q^2 .\label{grass.i-49-2}
\end{equation}
%%%
and check if the associated $\widehat{\HT}$ is Hermitian under the SvH hermiticity conditions
%%%
\begin{equation}
\label{grass.i-49-3}
	 \left\{
		\begin{array}{l}
		\displaystyle
		(\widehat{c}^q)^{\dagger}=\widehat{\bar{c}}_q
	        \smallskip \\
	        \displaystyle 
	        (\widehat{c}^p)^{\dagger}=\widehat{\bar{c}}_p.
	        \end{array}
		\right.
\end{equation}
%%%
The only part of $\widehat{\HT}$ which can encounter problems is $\widehat{\HT}_{ferm}$ which, with the
Hamiltonian (\ref{grass.i-49-2}), turns out to be
%%%
\begin{equation}
\displaystyle \widehat{\HT}_{ferm}=i\frac{\widehat{\bar{c}}_q\widehat{c}^p}{m}-i\widehat{\bar{c}}_p
\widehat{c}^qm\omega^2. \label{grass.i-49-4}
\end{equation}
%%%
Using (\ref{grass.i-49-3}) its Hermitian conjugate is
%%%
\begin{equation}
\displaystyle \widehat{\HT}_{ferm}^{\dagger}=i\widehat{\bar{c}}_q\widehat{c}^pm\omega^2-i\frac{\widehat{\bar{c}}_q
\widehat{c}^p}{m}. \label{grass.i-49-5}
\end{equation}
%%%
Comparing (\ref{grass.i-49-4}) and (\ref{grass.i-49-5}) one sees that $\widehat{\HT}_{ferm}$ is Hermitian only if 
%%%
\begin{equation}
\displaystyle \frac{1}{\omega^2}=m^2. \label{grass.i-50-1}
\end{equation}
So, given the mass of the system, only the harmonic oscillators whose frequency is given by (\ref{grass.i-50-1}) are
Hermitian. 
%%%
\setcounter{figure}{0}
\begin{figure}
\centering
\includegraphics{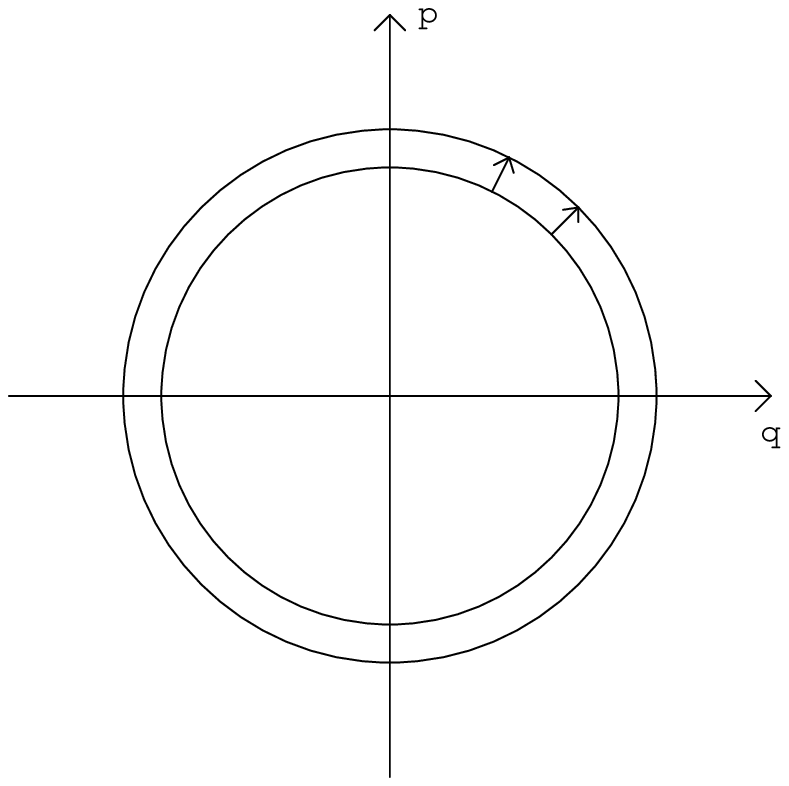}
\caption{{\rm{Phase space trajectories for a harmonic oscillator with}} 
$\displaystyle
\frac{1}{\omega^2}=m^2$.} 
\label{grass.circus}
\end{figure}
\begin{figure}
\centering
\includegraphics{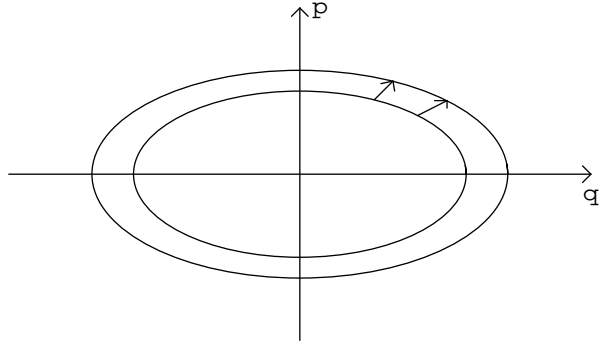}
\caption{{\rm{Phase space trajectories for a harmonic oscillator with}} 
$\displaystyle
\frac{1}{\omega^2}\neq m^2$.}
\label{grass.ellipse}
\end{figure}
%%%
It is easy to realize that all the oscillators satisfying (\ref{grass.i-50-1}) have trajectories in
phase space which are circles (see Fig. {\bf \ref{grass.circus}}). Note that in this case, since 
$\widehat{\HT}$ is Hermitian, the norm of the Jacobi fields, i.e. 
the arrows in Fig. {\bf \ref{grass.circus}}, does not change with time.
If instead $\displaystyle \frac{1}{\omega^2}\neq m^2$ then
the trajectories are ellipses, see Fig. {\bf \ref{grass.ellipse}}, and the norm of the Jacobi fields is not preserved
during the time evolution just as a consequence of the non-hermiticity of $\widehat{\HT}$. 

The reader may object
that the relation (\ref{grass.i-50-1}) can be disrupted by just changing the system of units which we
use to measure $\omega$ and $m$. Anyhow we should note that, if we change the units, we have to change also the hermiticity 
conditions (\ref{grass.i-49-3}) for
dimensional reasons and, under the new hermiticity conditions, $\widehat{\HT}$ is again Hermitian. 
In fact, looking at the Lagrangian (\ref{ann.suplag}), we notice that the kinetic term of the action is $\displaystyle \int
dt\,\bar{c}_q\dot{c}^q$ and so the dimensions of $c^q$ are the inverse of $\bar{c}_q$:
$[c^q]=[\bar{c}_q]^{-1}$.
The same happens for $c^p$: $[c^p]=[\bar{c}_p]^{-1}$. 
From the interaction term $\displaystyle \int dt\,\bar{c}_p\partial_q\partial_qHc^q$ of the action associated to the 
$\LT$
of (\ref{ann.suplag}) we get that
%%%
\begin{equation}
\displaystyle [c^p]=\frac{M}{T}[c^q]. \label{grass.i-52-3}
\end{equation}
So the hermiticity conditions (\ref{grass.i-49-3}) should be written as
%%%
\begin{equation}
(\widehat{c}^q)^{\dagger}=\widehat{\bar{c}}_q\cdot {\bf 1}_q,\;\;\;\;\;\;\; 
(\widehat{c}^p)^{\dagger}=\widehat{\bar{c}}_p\cdot {\bf 1}_p
\end{equation}
%%%
where ${\bf 1}_q$ and ${\bf 1}_p$ are dimensionful quantities. 
Of course we could choose $\widehat{c}^q$ to be dimensionless but if $\widehat{c}^q$ is dimensionless 
then $\widehat{c}^p$ must have
dimensions because of (\ref{grass.i-52-3}). Consequently ${\bf 1}_p$ would have the dimension of $[c^p]^2$ and so
it would change with the system of units.
Let us for example choose the S.I. system of units and the hermiticity conditions 
\begin{equation}
	 \left\{
		\begin{array}{l}
		\displaystyle
		\label{grass.i-53-1}
		(\widehat{c}^q)^{\dagger}=\widehat{\bar{c}}_q
	        \smallskip \\
	        \displaystyle 
	        (\widehat{c}^p)^{\dagger}=\widehat{\bar{c}}_p.
	        \end{array}
		\right.
\end{equation}
%%%
The harmonic oscillators which are Hermitian are those satisfying:
%%%
\begin{equation}
	 \left\{
		\begin{array}{l}
		\displaystyle
		\label{grass.i-53-2}
		m=\alpha\cdot \textrm{Kg}
	        \smallskip \\
	        \displaystyle 
	        \omega=\frac{1}{\alpha} \textrm{sec}^{-1}.
	        \end{array}
		\right.
\end{equation}
%%%
If we now pass to the CGS system the relation (\ref{grass.i-53-2}) becomes 
%%%
\begin{equation}
	 \left\{
		\begin{array}{l}
		\displaystyle
		\label{grass.i-53-3}
		m=\alpha\cdot10^3 \textrm{g}
	        \smallskip \\
	        \displaystyle 
	        \omega=\frac{1}{\alpha} \textrm{sec}^{-1}
	        \end{array}
		\right.
\end{equation}
%%%
and $m\omega=10^3\neq 1$. In any case also the hermiticity conditions (\ref{grass.i-53-1}), in the new units, have to be
changed:
%%%
\begin{equation}
	 \left\{
		\begin{array}{l}
		\displaystyle
		\label{grass.i-53-4}
		(\widehat{c}^q)^{\dagger}=\widehat{\bar{c}}_q
	        \smallskip \\
	        \displaystyle 
	        (\widehat{c}^p)^{\dagger}=10^6\,\widehat{\bar{c}}_p
	        \end{array}
		\right.
\end{equation}
%%%
and with these new hermiticity conditions $\widehat{\HT}$ is Hermitian even under the condition 
(\ref{grass.i-53-3}). 

This concludes this chapter. In the next one we will analyse the same issues 
by using an entirely bosonic $\widehat{\HT}$ first introduced in \cite{Regini}. 
We shall also analyse what happens when we change the representation for the Grassmann variables 
\cite{metaplectic}.

%% file: chapter5.tex
\def \HT{{\mathcal H}}
\def \LT{{\mathcal L}}
\def \ET{{\widetilde{\mathcal E}}}
\def \HCT{\hat{\mathcal H}}
\def \s{\scriptscriptstyle}

\def \I{1{\hspace*{-.32cm} |}}

\pagestyle{fancy}
\chapter*{\begin{center}
5. Hilbert Space Structure with Forms: II
\end{center}}
\addcontentsline{toc}{chapter}{\numberline{5}Hilbert Space Structure with
Forms: II}
\setcounter{chapter}{5}
\setcounter{section}{0}
\markboth{{\it{5. Hilbert Space Structure with Forms: II}}}{}

\begin{quote}
{\it{
What is proved by impossibility proofs is lack of imagination.}}\medskip\\
-{\bf John Bell}, 1982.
\end{quote}

\bigskip

\noindent As we have seen in the previous chapter the introduction 
of the Grassmann variables creates some problems at the level 
of the CPI, since it becomes impossible to endow the associated Hilbert 
space with a positive definite scalar product and, at the same time, 
to have a unitary evolution. To overcome these difficulties 
we want to analyse in this chapter two other functional formulations of CM. 
In the first one the Jacobi fields are represented by {\it bosonic}
variables and belong to the vector (or its dual) representation of the symplectic 
group. In the second formulation the Jacobi fields are given as {\it condensates}
of Grassmann variables belonging to the spinor representation of the metaplectic 
group. For both formulations we shall show that, differently from what happens
in the case presented in the previous chapter, it is possible to endow the associated Hilbert space
with a positive definite scalar product and to describe the dynamics via a Hermitian Hamiltonian.
Unfortunately both these new formulations present some difficulties at the geometrical 
level which make them less pleasant and easy to use than the previous one.
The content of this chapter with further calculational details can be found in \cite{7P}.

\bigskip

\section{Bosonic Functional Approach (BFA)}

As we have already seen the starting point of 
the CPI formulation of CM \cite{Goz89} is given by the transition probability (\ref{CPI3}):
%%%
\begin{equation}
P(\varphi^a,t|\varphi^a_i,t_i)=\int {\mathcal D}\varphi \,
\widetilde{\delta}\Bigl[\varphi^a-\phi^a_{cl}(t;\varphi_i)\Bigr]
\label{bos.2-1}
\end{equation}
%%%
where the variables $\varphi^a\equiv(q^i,p^i)$ 
are the phase space coordinates of the symplectic manifold ${\cal M}$ and 
$\phi^a_{cl}(t;\varphi_i)$ are the solutions of the Hamiltonian
equations of motion
%%%
\begin{equation}
\displaystyle \dot{\varphi}^a=\omega^{ab}\frac{\partial H}{\partial \varphi^b}
\end{equation}
%%%
with initial conditions $\varphi_i$. Eq. (\ref{bos.2-1}) can be rewritten as:
%%%
\begin{equation}
\displaystyle P(\varphi^a,t|\varphi^a_i,t_i)=\int {\cal
D}\varphi\,\widetilde{\delta}\biggl[\dot{\varphi}^a-\omega^{ab}\frac{\partial
H}{\partial
\varphi^b}\biggr]\textrm{det}\biggl[\delta_l^a\partial_t-\omega^{ab}\frac{\partial^2H}{\partial
\varphi^b\partial\varphi^l}\biggr] \label{bos.2-2}
\end{equation}
%%%
where the determinant appearing in (\ref{bos.2-2}) is the functional determinant 
needed
to pass from the zeroes of the function 
$\displaystyle F(\varphi,\dot{\varphi})\equiv \dot{\varphi}^a-\omega^{ab}\frac{\partial H}
{\partial \varphi^b}$ to the function itself.
This functional determinant is positive definite \cite{schulman} and this {\it crucial}
property is based on the fact that between two phase space points there is at most only
one trajectory. This property does not hold between two points of the configuration space
where the associated functional determinant in general is not positive definite.

As we have seen in Sec {\bf 1.1}, in the CPI formulation \cite{Goz89} of CM the next step was to 
``exponentiate" the determinant in (\ref{bos.2-2}) via Grassmann variables like 
in the Faddeev-Popov (FP) method of gauge theories. In Ref. \cite{Regini} a
different strategy was adopted. In particular the determinant in (\ref{bos.2-2})
was replaced with the inverse determinant:
%%%
\begin{equation}
\displaystyle \textrm{det}\biggl[\delta_l^a\partial_t-\omega^{ab}\frac{\partial^2H}{\partial
\varphi^b\partial\varphi^l}\biggr]=\biggl\{\textrm{det}\biggl[\delta_l^a\partial_t+\omega^{ab}
\frac{\partial^2H}{\partial\varphi^b\partial\varphi^l}\biggr]\biggr\}^{-1}.
\label{bos.2-3}
\end{equation}
%%%
The next steps in \cite{Regini}
were to insert (\ref{bos.2-3}) in 
(\ref{bos.2-2}) and to ``exponentiate" the inverse matrix via bosonic variables using 
the well-known formula
%%%
\begin{equation}
\int dx^idy_j \, \textrm{exp}\,(ix^iA^j_iy_j) \propto \Bigl\{\textrm{det}[A_i^j]\Bigr\}^{-1}. \label{bos.2-4}
\end{equation}
%%%
This formula of Gaussian integration applies only to matrices with positive determinant but this is just
our case as we explained above. Note that this is no longer the case for the FP 
determinant which, as signalled by the Gribov problem, is not positive definite. This is the
reason why the FP determinant could not be exponentiated via bosonic variables. Various attempts
exist in the literature to write fermionic determinants via bosonic variables \cite{slavnov} but they
are all different from the one we have presented here.

Let us now insert (\ref{bos.2-3}) and (\ref{bos.2-4}) into (\ref{bos.2-2}). 
The result is the following path integral\footnote[1]{The acronym BFA stands for Bosonic Functional Approach.}:
%%%
\begin{eqnarray}
\displaystyle P_{\scriptscriptstyle BFA}&=&\int {\mathcal D} \varphi\;\widetilde{\delta}\biggl[\dot{\varphi}^a-
\omega^{ab}
\frac{\partial H}{\partial \varphi^b}\biggr]\biggl\{\textrm{det}\biggl[\delta^a_l\partial_t
+\omega^{ab}\frac{\partial^2H}{\partial\varphi^b\partial\varphi^l}\biggr]\biggr\}^{-1} \nonumber\\
&=&\int {\mathcal D}\varphi{\mathcal D}\lambda{\mathcal D}\pi{\mathcal D}\xi\,\textrm{exp}\,
\biggl(i\int dt \,{\cal L}_{\scriptscriptstyle BFA}\biggr)
\label{bos.2-5-a}
\end{eqnarray}
%%%
where 
%%%
\begin{equation}
\label{bos.2-5-b}
{\cal L}_{\scriptscriptstyle BFA}=\lambda_a\biggl[\dot{\varphi}^a-
\omega^{ab}\frac{\partial H}{\partial \varphi^b}\biggr]+\pi^l
\biggl[\delta_l^a\partial_t+\omega^{ab}\frac{\partial^2H}{\partial \varphi^b\partial\varphi^l}\biggr]\xi_a.
\end{equation}
%%%
The variables $\lambda_a$ are the same as in the CPI \cite{Goz89} formulation of CM
and they are needed to perform the Fourier transform of the Dirac delta 
$\displaystyle \widetilde{\delta}\biggl(\dot{\varphi}^a-
\omega^{ab}\frac{\partial H}{\partial\varphi^b}\biggr)$. The variables $\lambda_a$ are {\it bosonic}
like the variables $\pi^l,\xi_a$ which were introduced to exponentiate the matrix 
$\displaystyle \biggl[\delta_l^a\partial_t+\omega^{ab}\frac{\partial^2H}{\partial\varphi^b\partial\varphi^l}
\biggr]$ and to produce its inverse determinant. The variables $\pi^l$ and $\xi_a$ 
are the analog of the 
$x^i$ and $y_j$ of (\ref{bos.2-4}). Let us remember that in the original CPI formulation of CM 
\cite{Goz89} the Lagrangian obtained was
%%%
\begin{equation}
{\cal L}_{\scriptscriptstyle CPI}=\lambda_a\biggl[\dot{\varphi}^a-\omega^{ab} \frac{\partial H}{\partial\varphi^b}
\biggr]+i\bar{c}_a\biggl[\delta_l^a\partial_t
-\omega^{ab}\frac{\partial^2H}{\partial\varphi^b\partial\varphi^l}\biggr]c^l \label{bos.2-6}
\end{equation}
%%%
which can be compared with the ${\cal L}_{\scriptscriptstyle BFA}$ of (\ref{bos.2-5-b}) if, in this last one, we interchange
$\pi^l,\xi_a$:
%%%
\begin{equation}
{\cal L}_{\scriptscriptstyle BFA}=\lambda_a\biggl[\dot{\varphi}^a-\omega^{ab}\frac{\partial H}{\partial \varphi^b}\biggr]-
\xi_a\biggl[\delta_l^a\partial_t-
\omega^{ab}\frac{\partial^2H}{\partial\varphi^b\partial\varphi^l}\biggr]\pi^l+(\textrm{s.t.}).
\label{bos.2-7}
\end{equation}
%%%
From (\ref{bos.2-7}) we see that, modulo the surface terms (s.t.), we get ${\cal L}_{\scriptscriptstyle BFA}$
from ${\cal L}_{\scriptscriptstyle CPI}$ by replacing 
the Grassmann variables $i\bar{c}_a$ and $c^l$ with
the bosonic ones $-\xi_a$ and $\pi^l$.

\bigskip

\section{Operatorial Formalism of the BFA}

We should now build the operatorial formalism associated to the BFA. 
The commutators among the basic variables ($\widehat{\varphi}^a,\widehat{\lambda}_a,\widehat{\pi}^a,\widehat{\xi}_a$) 
can be straightforwardly
derived from the path integral (\ref{bos.2-5-a}) by inspecting the kinetic term of (\ref{bos.2-5-b}). They turn
out to be:
%%%
\begin{eqnarray}
&&[\widehat{\varphi}^a,\widehat{\lambda}_b]=i\delta_b^a \nonumber\\
&&[\widehat{\xi}_a,\widehat{\pi}^b]=i\delta_a^b. \label{bos.3-1}
\end{eqnarray}
%%%
Next we choose the ``Schr\"odinger" 
representation in which we realize $\widehat{\varphi}^a$ and $\widehat{\pi}^a$ as multiplicative operators while 
$\widehat{\lambda}_a$ and $\widehat{\xi}_a$ as derivative ones: 
%%%
\begin{equation}
\left\{
	\begin{array}{l}
	\displaystyle \label{bos.3-2}
	\widehat{\lambda}_a=-i\frac{\partial}{\partial\varphi^a} \smallskip \\
	\displaystyle \widehat{\xi}_a= i\frac{\partial}{\partial\pi^a}.
	\end{array}
	\right.
\end{equation}
%%%
So in this representation the associated Hilbert space is made of the functions $\psi(\varphi^a,\pi^a)$ of the
$4n$ coordinates $(\varphi^a,\pi^a)$. A very natural, and 
{\it positive definite}, scalar product that can be introduced in this space is
%%%
\begin{equation}
\displaystyle \langle \psi|\psi^{\prime}\rangle\equiv \int d\varphi d\pi\,\psi^*(\varphi^a,\pi^a)
\psi^{\prime}(\varphi^a,\pi^a). \label{bos.3-3}
\end{equation}
%%%
It is extremely easy to check that the $8n$ operators $\widehat{\varphi}^a,\widehat{\lambda}_a,
\widehat{\pi}^a,\widehat{\xi}_a$ are all Hermitian under the scalar product (\ref{bos.3-3}):
%%%
\begin{equation}
\left\{
	\begin{array}{l}
	\displaystyle \label{bos.3-4}
	\widehat{\varphi}^{a\dagger}=\widehat{\varphi}^a \smallskip \\
	\widehat{\lambda}_a^{\dagger}=\widehat{\lambda}_a \smallskip \\
	\widehat{\xi}_a^{\,\dagger}=\widehat{\xi}_a \smallskip \\
	\widehat{\pi}^{a\dagger}=\widehat{\pi}^a.
	\end{array}
	\right.
\end{equation}
%%%
Let us now derive from the Lagrangian (\ref{bos.2-5-b}) the associated Hamiltonian:
%%%
\begin{equation}
{\cal H}_{\scriptscriptstyle BFA}=\lambda_a\omega^{ab}\partial_bH-\pi^l\omega^{ab}
\partial_b\partial_lH\xi_a. \label{bos.3-5-a}
\end{equation}
%%%
Turning the variables ($\varphi^a,\lambda_a,\pi^a,\xi_a$) into operators, the Hamiltonian itself becomes:
%%%
\begin{equation}
\widehat{\cal H}_{\scriptscriptstyle BFA}=\widehat{\lambda}_a\omega^{ab}\partial_bH-\widehat{\pi}^l\omega^{ab}
\partial_b\partial_lH\widehat{\xi}_a.
\label{bos.3-5}
\end{equation}
%%%
It is straightforward to check that this Hamiltonian is Hermitian under the hermiticity conditions (\ref{bos.3-4})
%%%
\begin{eqnarray}
\widehat{\cal H}_{\scriptscriptstyle BFA}^{\dagger}&\hspace{-0.2cm}=&\hspace{-0.2cm}
(\widehat{\lambda}_a\omega^{ab}\partial_bH-\widehat{\pi}^l\omega^{ab}\partial_b\partial_lH
\widehat{\xi}_a)^{\dagger}=
(\partial_bH)\omega^{ab}\widehat{\lambda}_a-\widehat{\xi}_a\omega^{ab}\partial_b\partial_lH\widehat{\pi}^l=\nonumber\\
&\hspace{-0.2cm}=&\hspace{-0.2cm}\widehat{\lambda}_a\omega^{ab}\partial_bH+i\omega^{ab}\partial_a\partial_bH
-\widehat{\pi}^l\omega^{ab}\partial_b\partial_lH\widehat{\xi}_a-i\delta_a^l
\omega^{ab}\partial_b\partial_lH=\widehat{\cal H}_{\scriptscriptstyle BFA}. \label{bos.3-6}
\end{eqnarray}
%%%
In the last two steps
we have used the commutation relations $[\widehat{\lambda}_a,\partial_bH]=-i\partial_a\partial_bH$, 
$[\widehat{\xi}_a,\widehat{\pi}^l]=i\delta_a^l$ and the fact that $\omega^{ab}\partial_b\partial_aH$ is zero because of
the antisymmetry of $\omega^{ab}$. So we can say that, differently than in the CPI 
case analysed in the previous chapter, in the
BFA case we can have both a positive definite Hilbert space and a Hermitian Hamiltonian. 

The reader may remember that in the previous chapter we gave some {\it physical} reasons of why we could not have 
both a positive 
definite Hilbert space and a Hermitian Hamiltonian in the CPI case: basically 
in a chaotic system the Jacobi fields $c^a$ grow exponentially and, as a consequence, some 
of the one-forms
%%%
\begin{equation}
\psi(\varphi,c)=\psi_a(\varphi)c^a \label{bos.3-7}
\end{equation}
%%%
have a norm which also grows exponentially with time. Therefore such norms are not
conserved and this happens only if the evolution is not unitary 
or equivalently if the Hamiltonian is not Hermitian.
This kind of reasoning cannot be applied in the BFA case. In fact here the role of the Jacobi fields is taken by the 
variables $\pi^a$ whose equations of motion can be derived from the Lagrangian ${\cal L}_{\scriptscriptstyle BFA}$ 
of (\ref{bos.2-5-b}):
%%%
\begin{equation}
\dot{\pi}^a=\omega^{ad}\partial_d\partial_bH\pi^b. \label{bos.3-8-b}
\end{equation}
%%%
So the analog of the wave function (\ref{bos.3-7}) is:
%%%
\begin{equation}
\psi(\varphi,\pi)=\psi_a(\varphi)\pi^a. \label{bos.3-8}
\end{equation}
%%%
Unfortunately this wave function is not normalizable according to the scalar product (\ref{bos.3-3}). 
So an exponential 
increase in $\pi^a$ would not lead to the conclusion that the evolution is not unitary because 
the state (\ref{bos.3-8}) 
itself is not part of the Hilbert space already at $t=0$ since it is not normalizable.
If the reader is not immediately  convinced by our arguments, he should remember that the line of reasoning we 
followed 
in the CPI case to motivate our physical understanding was crucially based on the use 
of wave functions linear in the Jacobi fields. These functions no longer belong to our Hilbert space in the BFA approach. 

\bigskip

\section{Geometrical Analysis and Symmetries in the BFA}

In Ref. \cite{Regini} a first geometrical analysis of the bosonic formalism was proposed: 
basically the variables $\pi^a,\xi_a$ were identified with the {\it components} of vectors and forms whose 
{\it basis} were respectively $\bar{c}_a$ and $c^a$:
%%%
\begin{equation}
\left\{
	\begin{array}{l}
	\displaystyle \label{bos.4-1}
	V=\pi^a\bar{c}_a \smallskip\\
	F=\xi_ac^a.
	\end{array}
	\right.
\end{equation}
%%%
In fact it is easy to prove that, under the infinitesimal diffeomorphism generated by 
$\widehat{{\cal H}}_{\scriptscriptstyle BFA}$ over the original phase space:
%%%
\begin{equation}
\varphi^{\prime a}=\varphi^a+\epsilon\omega^{ab}\partial_bH,
\end{equation}
%%%
the variables $\pi^a$ and $\xi_a$ transform in the following way:
%%%
\begin{eqnarray}
\displaystyle &&\pi^{\prime a}=\pi^a+\epsilon\omega^{ac}\partial_c\partial_bH\pi^b=
\frac{\partial \varphi^{\prime a}}{\partial\varphi^b}\pi^b\nonumber\\
&&\xi_a^{\prime}=\xi_a-\epsilon\omega^{bc}\partial_c\partial_aH\xi_b=
\frac{\partial\varphi^b}{\partial\varphi^{\prime a}}\xi_b,
\end{eqnarray}
%%%
i.e. just as components of vectors and forms.
With this identification 
the Hamiltonian $\widehat{\cal H}_{\scriptscriptstyle BFA}$ of (\ref{bos.3-5}) 
cannot be given the meaning of a Lie derivative.
In fact we know that the Lie derivative ${\cal L}_{(dH)^{\sharp}}$ \cite{Marsd} 
changes the components of a vector 
as follows:
%%%
\begin{equation}
\delta \pi^l=(\partial_a\pi^l)\omega^{ab}\partial_bH-(\partial_a\omega^{lb}\partial_bH)\pi^a \label{bos.4-2}
\end{equation}
%%%
while the $\widehat{\cal H}_{\scriptscriptstyle BFA}$ 
of (\ref{bos.3-5}) induces on $\widehat{\pi}^l$, via the commutators (\ref{bos.3-1}), the following transformation
%%%
\begin{equation}
\delta \widehat{\pi}^l=\Bigl[\widehat{\pi}^l,i\widehat{\cal H}_{\scriptscriptstyle BFA}\Bigr]=
(-\partial_a\omega^{lb}\partial_bH)\widehat{\pi}^a \label{bos.4-3}
\end{equation}
%%%
which is clearly different from (\ref{bos.4-2}).
So if we insist in the analysis presented in \cite{Regini}, we should {\it first} 
abandon the interpretation of $\widehat{\cal H}_{\scriptscriptstyle BFA}$ 
as the Lie derivative along the Hamiltonian flow.
{\it Second}, if we interpret $\widehat{\pi}^a$ and $\widehat{\xi}_a$ 
as components, we should make them dependent on $\widehat{\varphi}$ 
and consequently we should give a connection to glue the fibers of $T^*(T^*{\cal M})$ of which 
$\widehat{\pi}^a$ and $\widehat{\xi}_a$ are coordinates \cite{Regini}. This connection does not appear in a 
natural way in our formalism.
So, in order to bypass these {\it two} problems, we will interpret $\widehat{\pi}^a$ and $\widehat{\xi}_a$ as 
{\it basis} respectively of forms and vector fields. 
One-forms and vector fields are then given by 
%%%
\begin{equation}
\left\{
	\begin{array}{l}
	\displaystyle \label{bos.4-4}
	F=F_a(\widehat{\varphi})\widehat{\pi}^a \smallskip\\
	V=V^a(\widehat{\varphi})\widehat{\xi}_a.
	\end{array}
	\right.
\end{equation}
%%%
As a consequence, it is easy to check that the $\widehat{\cal H}_{\scriptscriptstyle BFA}$ 
of (\ref{bos.3-5}) can be interpreted, up to a constant factor, as 
the Lie derivative. 
In fact the commutator between $i\widehat{\cal H}_{\scriptscriptstyle BFA}$ and 
the vector $V$ of (\ref{bos.4-4}) gives:
%%%
\begin{equation}
\Bigl[i\widehat{\cal H}_{\scriptscriptstyle BFA}, V^e(\widehat{\varphi})\widehat{\xi}_e\Bigr]=
\Bigl[(\partial_aV^e)\omega^{ab}
\partial_bH-(\partial_a\omega^{eb}
\partial_bH)V^a\Bigr]\widehat{\xi}_e \label{bos.4-5}
\end{equation}
%%%
and this is exactly how vector components $V^e$ change \cite{Marsd} under the Lie 
derivative of the Hamiltonian flow:
%%%
\begin{equation}
\delta V^e(\varphi)\equiv V^{\prime e}(\varphi)-V^e(\varphi)=
(\partial_aV^e)\omega^{ab}\partial_bH-(\partial_a\omega^{eb}\partial_bH)V^a. \label{bos.4-6}
\end{equation}
%%%
Analogously, on the one-forms $F=F_e(\widehat{\varphi})\widehat{\pi}^e$ of (\ref{bos.4-4}), $\widehat{\cal
H}_{\scriptscriptstyle BFA}$ acts as follows:
%%%
\begin{equation}
[i\widehat{\cal H}_{\scriptscriptstyle BFA}, F_e(\widehat{\varphi})\widehat{\pi}^e]=
\Bigl[(\partial_aF_e)\omega^{ab}\partial_bH+(\partial_e\omega^{ab}\partial_bH)
F_a\Bigr]\pi^e \label{bos.4-7}
\end{equation}
%%%
and this is exactly how one-forms components $F_e$ transform \cite{Marsd} under the Lie derivative:
%%%
\begin{equation}
\delta F_e(\varphi)=F_e^{\prime}(\varphi)-F_e(\varphi)=(\partial_aF_e)\omega^{ab}\partial_bH+
(\partial_e\omega^{ab}\partial_bH)F_a. \label{bos.4-8}
\end{equation}
%%%
To give to $\widehat{\cal H}_{\scriptscriptstyle BFA}$ the meaning of
a Lie derivative, another check we should do is the following. The
commutator of two Lie derivatives has the property \cite{Marsd}:
%%%
\begin{equation}
\displaystyle \Bigl[{\cal L}_{(dH_{\scriptscriptstyle 1})^{\sharp}},{\cal L}_{(dH_{\scriptscriptstyle 2})^{\sharp}}\Bigr]=
{\cal L}_{[(dH_{\scriptscriptstyle 1})^{\sharp},(dH_{\scriptscriptstyle 2})^{\sharp}]_{Lb}}
\label{bos.4-9}
\end{equation}
%%%
where $H_{\scriptscriptstyle 1}$ and $H_{\scriptscriptstyle 2}$ are the Hamiltonians entering the 
Lie derivative and $[(dH_{\scriptscriptstyle 1})^{\sharp},(dH_{\scriptscriptstyle 2})^{\sharp}]_{Lb}$ 
are the {\it Lie brackets} (Lb) between the associated Hamiltonian vector fields. According
to our conventions the Lie brackets can be related to the Poisson brackets between $H_{\scriptscriptstyle 1}$ and 
$H_{\scriptscriptstyle 2}$ in the following way \cite{Marsd}:
%%%
\begin{eqnarray}
\displaystyle \Bigl[(dH_{\scriptscriptstyle 1})^{\sharp},(dH_{\scriptscriptstyle 2})^{\sharp}\Bigr]_{Lb}&=&
\Bigl[\omega^{bc}\partial_cH_{\scriptscriptstyle 1}\partial_b\omega^{ad}\partial_dH_{\scriptscriptstyle 2}
-\omega^{bc}\partial_cH_{\scriptscriptstyle 2}(\partial_b\omega^{ad}\partial_dH_{\scriptscriptstyle 1})\Bigr]\xi_a=
\nonumber\\
&=&-\Bigl[\omega^{ad}\partial_d(\partial_bH_{\scriptscriptstyle 1}
\omega^{bc}\partial_cH_{\scriptscriptstyle 2})\Bigr]\xi_a=
-\Bigl(d\{H_{\scriptscriptstyle 1},H_{\scriptscriptstyle 2}\}\Bigr)^{\sharp} \label{bos.correspondence}.
\end{eqnarray}
%%%
Therefore (\ref{bos.4-9}) can be rewritten as:
%%%
\begin{equation}
\displaystyle \Bigl[{\cal L}_{(dH_1)^{\sharp}},{\cal L}_{(dH_2)^{\sharp}}\Bigr]=
{\cal L}_{-(d\{H_{\scriptscriptstyle 1},H_{\scriptscriptstyle 2}\})^{\sharp}}. \label{bos.4-9-bis}
\end{equation}
%%%
If we want to identify each Lie derivative ${\cal L}_{(dH)^{\sharp}}$ with an operator 
$i\widehat{\cal H}_{\scriptscriptstyle H}$,
the relation (\ref{bos.4-9-bis}) has to be satisfied also between the correspondent $\widehat{\cal H}$:
%%%
\begin{equation}
\Bigl[i\widehat{\cal H}_{H_1},i\widehat{\cal H}_{H_2}\Bigr]=-i\widehat{\cal H}_{\{H_1,H_2\}} \label{bos.4-10}
\end{equation}
%%%
where we have put on the $\widehat{\cal H}$ 
of (\ref{bos.3-5}) the label $H_{\scriptscriptstyle 1}$, $H_{\scriptscriptstyle 2}$ or 
$\{H_{\scriptscriptstyle 1},H_{\scriptscriptstyle 2}\}$ to indicate
the function entering each $\widehat{\cal H}_{\scriptscriptstyle BFA}$. It is easy to prove that 
(\ref{bos.4-10}) is satisfied using the commutators (\ref{bos.3-1}). This confirms that it is consistent 
to assign to $\widehat{\cal H}_{\scriptscriptstyle BFA}$ the meaning of a Lie derivative.

As we have seen in Chapter {\bf 3}, in the CPI there are various 
conserved universal charges \cite{Goz89}
which somehow signal the redundancy of the $8n$ variables $(\varphi^a,\lambda_a,c^a,\bar{c}_a)$ used in 
describing CM. Also in the bosonic case we have many extra variables $(\lambda_a,\pi^a,\xi_a)$ 
besides the $2n$ phase space ones $\varphi^a$ and so we expect to find various symmetries like in the CPI. 

The way we start our search for the symmetries in the bosonic case is rather naive but it is one of the few 
we could think of. Basically, as the variables $\widehat{\pi}^a,\widehat{\xi}_a$ 
take the place - in the bosonic case - of the Grassmann 
variables $\widehat{c}^a,\widehat{\bar{c}}_a$, we simply rewrite the conserved charges of the CPI replacing in them
$\widehat{c}^a$ and $\widehat{\bar{c}}_a$ with $\widehat{\pi}^a$ and $\widehat{\xi}_a$. 
If in the CPI the conserved charges were 
\cite{Goz89}\cite{Deotto}
%%%
\begin{equation}
\left\{
	\begin{array}{l}
	\displaystyle \label{bos.4-11}
	\widehat{Q}_f=i\widehat{c}^a\widehat{\bar{c}}_a \;\;\;\;\;\;\; \smallskip\\
	\widehat{N}=\widehat{c}^a\partial_aH \smallskip\\
	\widehat{\overline{N}}=\widehat{\bar{c}}_a\omega^{ab}\partial_bH
	\end{array}
	\right.
\end{equation}
%%%
and
%%%
\begin{equation}
\left\{
	\begin{array}{l}
	\displaystyle \label{bos.4-12}
	\widehat{Q}=i\widehat{c}^a\widehat{\lambda}_a \;\;\;\;\;\;\;\;\;\;\;\;{\textrm{BRS charge}} \smallskip\\
	\widehat{\overline{Q}}=i\widehat{\bar{c}}_a\omega^{ab}\widehat{\lambda}_b \;\;\;\;\;\;\;{\textrm{anti-BRS charge}} \smallskip\\
	\widehat{Q}_{\scriptscriptstyle H}=\widehat{Q}-\widehat{N} \;\;\;\;\;\;\;{\textrm{susy charge}} \smallskip\\
	\widehat{\overline{Q}}_{\scriptscriptstyle H}=\widehat{\overline{Q}}+
	\widehat{\overline{N}} \;\;\;\;\;\;\;{\textrm{susy charge}},
	\end{array}
	\right.
\end{equation}
%%%
then in the bosonic case we get the following two set of charges:
%%%
\begin{equation}
\left\{
	\begin{array}{l}
	\displaystyle \label{bos.4-13}
	\widehat{Q}_f^{\scriptscriptstyle (B)}=i\widehat{\pi}^a\widehat{\xi}_a \smallskip\\
	\widehat{N}^{\scriptscriptstyle (B)}=\widehat{\pi}^a\partial_aH \smallskip\\
	\widehat{\overline{N}}^{\scriptscriptstyle (B)}=\widehat{\xi}_a\omega^{ab}\partial_bH
	\end{array}
	\right.
\end{equation}
%%%
and 
%%%
\begin{equation}
\left\{
	\begin{array}{l}
	\displaystyle \label{bos.4-14}
	\widehat{Q}^{\scriptscriptstyle (B)}=i\widehat{\pi}^a\widehat{\lambda}_a \smallskip\\
	\widehat{\overline{Q}}^{\scriptscriptstyle (B)}=i\widehat{\xi}_a\omega^{ab}\widehat{\lambda}_b \smallskip\\
	\widehat{Q}_{\scriptscriptstyle H}^{\scriptscriptstyle (B)}=
	\widehat{Q}^{\scriptscriptstyle (B)}-\widehat{N}^{\scriptscriptstyle (B)} \smallskip\\
	\widehat{\overline{Q}}_{\scriptscriptstyle H}^{\scriptscriptstyle (B)}=
	\widehat{\overline{Q}}^{\scriptscriptstyle (B)}+\widehat{\overline{N}}^{\scriptscriptstyle (B)}
	\end{array}
	\right.
\end{equation}
%%%
where the superscript $(B)$ on the charge indicates that it refers to the BFA case. 
We note that, by replacing 
$\widehat{c}^a$ with $\widehat{\pi}^a$ and $\widehat{\bar{c}}_a$ with $\widehat{\xi}_a$, 
we have not really done the replacement which would bring the 
$\widehat{\HT}$ of the CPI into the $\widehat{\cal H}$ of the BFA (\ref{bos.3-5})
but the difference is only in some multiplicative 
factors $(\pm i)$ which cannot modify the conservation of the charges. 
The careful reader may also notice that in the CPI there were other two conserved charges \cite{Goz89}:
%%%
\begin{equation}
\left\{
	\begin{array}{l}
	\displaystyle 
	\widehat{K}=\frac{1}{2}\widehat{c}^a\omega_{ab}\widehat{c}^b \smallskip\\
	\displaystyle \widehat{\overline{K}}=\frac{1}{2}\widehat{\bar{c}}_a\omega^{ab}\widehat{\bar{c}}_b.
	\end{array}
	\right.
\end{equation}
%%%
We did not list them because, via our substitutions, the corresponding charges in the BFA case
would be zero because of the bosonic character of ($\widehat{\pi}$, $\widehat{\xi}$) 
and the antisymmetry of $\omega_{ab}$:
%%%
\begin{equation}
\left\{
	\begin{array}{l}
	\displaystyle 
	\widehat{K}^{\scriptscriptstyle (B)}=\frac{1}{2}\widehat{\pi}^a\omega_{ab}\widehat{\pi}^b=0 \smallskip\\
	\displaystyle \widehat{\overline{K}}^{\scriptscriptstyle (B)}=\frac{1}{2}\widehat{\xi}_a\omega^{ab}
	\widehat{\xi}_b=0.
	\end{array}
	\right.
\end{equation}
%%%
Now, turning back to the set of charges in (\ref{bos.4-13}), it is easy to check that they are all conserved, i.e.:
%%%
\begin{equation}
[\widehat{Q}_f^{\scriptscriptstyle (B)},\widehat{\cal H}_{\scriptscriptstyle BFA}]=
[\widehat{N}^{\scriptscriptstyle (B)},\widehat{\cal H}_{\scriptscriptstyle BFA}]=
[\widehat{\overline{N}}^{\scriptscriptstyle (B)},\widehat{\cal H}_{\scriptscriptstyle BFA}]=0. \label{bos.4-15}
\end{equation}
%%%
On the other hand
the charges present in (\ref{bos.4-14}) are apparently not conserved. In fact the 
bosonic analog of the BRS charge has the following commutator with 
$\widehat{\cal H}_{\scriptscriptstyle BFA}$: 
%%%
\begin{equation}
[\widehat{Q}^{\scriptscriptstyle (B)},\widehat{\cal H}_{\scriptscriptstyle BFA}]=
-\widehat{\pi}^l\omega^{ab}\partial_b\partial_l\partial_cH\widehat{\pi}^c\widehat{\xi}_a \label{bos.4-16}
\end{equation}
%%%
while for the anti-BRS charge we get:
%%%
\begin{equation}
[\widehat{\overline{Q}}^{\scriptscriptstyle (B)},\widehat{\cal H}_{\scriptscriptstyle BFA}]=
-\widehat{\xi}_a\omega^{ab}\widehat{\xi}_s\omega^{st}(\partial_b\partial_t\partial_lH)\widehat{\pi}^l.
\label{bos.4-17}
\end{equation}
%%%
Analogously the bosonic 
charges $\widehat{Q}_{\scriptscriptstyle H}^{\scriptscriptstyle(B)},
\widehat{\overline{Q}}_{\scriptscriptstyle H}^{\scriptscriptstyle(B)}$ of (\ref{bos.4-14}) cannot be conserved because 
they are a linear combination of $\widehat{Q}^{\scriptscriptstyle (B)}$ and 
$\widehat{\overline{Q}}^{\scriptscriptstyle(B)}$,
which are not conserved, with $\widehat{N}^{\scriptscriptstyle (B)}$ and $\widehat{\overline{N}}^{\scriptscriptstyle (B)}$
which are conserved. 

Let us now look at the RHS of (\ref{bos.4-16}) and (\ref{bos.4-17}) which indicate by 
``{\it how much}" the conservation law is violated. 
It is easy to notice that these RHS do not contain $\widehat{\lambda}_a$ and so they commute
with the original phase space operators $\widehat{\varphi}^a$.
As a consequence the infinitesimal transformations generated by $\widehat{Q}^{\scriptscriptstyle (B)}$
and by the Hamiltonian $\widehat{\cal H}_{\scriptscriptstyle BFA}$ {\it commute} when they are applied on 
$\widehat{\varphi}$. 
In fact the infinitesimal BRS transformations generated by $\widehat{Q}^{\scriptscriptstyle (B)}$ 
on a field $\widehat{A}$ are given
by the commutator of $\widehat{Q}^{\scriptscriptstyle (B)}$ with the field: $\delta_{Q^{\scriptscriptstyle (B)}}A=
[\epsilon \widehat{Q}^{\scriptscriptstyle (B)}, \widehat{A}]$ where $\epsilon$ is an infinitesimal parameter. 
The same happens for the transformations generated by the Hamiltonian:
$\delta_{\cal H}\widehat{A}=[\bar{\epsilon}\,\widehat{\cal H}_{\scriptscriptstyle BFA},\widehat{A}]$. 
Let us take for $\widehat{A}$ the original phase space variables
$\widehat{\varphi}^a$. If we perform first an infinitesimal time evolution and then a BRS transformation we obtain
%%%
\begin{equation}
\delta_{Q^{\scriptscriptstyle (B)}}\delta_{\cal H}\widehat{\varphi}^a=
\epsilon\bar{\epsilon}\Bigl[\widehat{Q}^{\scriptscriptstyle (B)}, 
[\widehat{\cal H}_{\scriptscriptstyle BFA},\widehat{\varphi}^a]\Bigr]
\end{equation} 
%%%
while, if we perform the transformations in the inverse order, we get:
%%%
\begin{equation}
\delta_{\cal H}\delta_{Q^{\scriptscriptstyle (B)}}\widehat{\varphi}^a=
\epsilon\bar{\epsilon}\Bigl[\widehat{\cal H}_{\scriptscriptstyle BFA}, [\widehat{Q}^{\scriptscriptstyle (B)},
\widehat{\varphi}^a]\Bigr].
\end{equation} 
%%%
Now we can use the Jacobi identities to obtain
%%%
\begin{eqnarray}
\delta_{Q^{\scriptscriptstyle (B)}}\delta_{\cal H}\widehat{\varphi}^a-
\delta_{\cal H}\delta_{Q^{\scriptscriptstyle (B)}}
\widehat{\varphi}^a&=&\epsilon\bar{\epsilon}\biggl(\Bigl[\widehat{Q}^{\scriptscriptstyle (B)},
[\widehat{\cal H}_{\scriptscriptstyle BFA},
\widehat{\varphi}^a]\Bigr]-\Bigl[\widehat{\cal H}_{\scriptscriptstyle BFA},[\widehat{Q}^{\scriptscriptstyle (B)},
\widehat{\varphi}^a]\Bigr]\biggr)=\nonumber\\
&=&-\epsilon\bar{\epsilon}\Bigl[\widehat{\varphi}^a,[\widehat{Q}^{\scriptscriptstyle (B)},
\widehat{\cal H}_{\scriptscriptstyle BFA}]\Bigr]=0 
\label{bos.zip}
\end{eqnarray}
%%%
where in the last step we have used the fact that the RHS of (\ref{bos.4-16}) commutes with $\widehat{\varphi}^a$. 
``Somehow" we can say that the transformations generated by 
$\widehat{Q}^{\scriptscriptstyle (B)}$ and $\widehat{\cal H}_{\scriptscriptstyle BFA}$ 
commute on the original phase space. Of course
the same happens for the anti-BRS charge $\widehat{\overline{Q}}^{\scriptscriptstyle (B)}$ and for the 
supersymmetry charges $\widehat{Q}_{\scriptscriptstyle H}^{\scriptscriptstyle (B)},
\widehat{\overline{Q}}_{\scriptscriptstyle H}^{\scriptscriptstyle (B)}$. 

Now we want to provide a geometrical
interpretation of this fact at least for the BRS-charge. Let us perform 
an infinitesimal BRS transformation on $\widehat{\varphi}^a$:
%%%
\begin{equation}
\delta_{Q^{\scriptscriptstyle (B)}}\widehat{\varphi}^a=[\epsilon \widehat{Q}^{\scriptscriptstyle (B)},
\widehat{\varphi}^a]=
[\epsilon i\widehat{\pi}^b\widehat{\lambda}_b,\widehat{\varphi}^a]=\epsilon\widehat{\pi}^a
\label{bos.4-20}
\end{equation}
%%%
where $\epsilon$ is an infinitesimal commuting parameter. The new phase space point $\varphi^{\prime a}$ reached after 
this transformation is:
%%%
\begin{equation}
\varphi^{\prime a}=\varphi^a+\epsilon\pi^a. \label{bos.4-21}
\end{equation}
%%%
Remember now that $\pi^a$ is a Jacobi field, i.e. it satisfies the same equation of motion as the first variations
(\ref{bos.3-8-b}). So if $\varphi^a$ is a point on a trajectory then 
$\varphi^{\prime a}$ is a point on a nearby trajectory
as indicated in the following picture:

\begin{center}
\begin{picture}(240,120)
\pscurve[]{c-c}(-0.9,2.4)(0.9,2.9)(2.9,3.2)(4.9,3.4)(6.9,3.5)
\pscurve[]{c-c}(-0.9,0.4)(0.9,0.7)(2.9,0.9)(4.9,1)(6.9,1.1)
\psline[]{->}(0.9,0.7)(0.9,1.8)
\psline[](0.9,1.8)(0.9,2.9)

\rput(0.9,3.4){$\varphi^{\prime a}=\varphi^a+\epsilon\pi^a$}
\rput(0.9,0.4){$\varphi^a$}
\end{picture}
\end{center}

\noindent From (\ref{bos.zip}) 
we expect that we could move from the point $\varphi^a$ along its trajectory via the Hamiltonian 
$\widehat{\cal H}_{\scriptscriptstyle BFA}$ for an interval of time $\Delta t$, reach a point $\varphi_{\scriptscriptstyle (1)}^a$ and 
from there jump, via a
BRS transformation to a point $\varphi^{\prime a}_{\scriptscriptstyle(1)}$ on the nearby trajectory. 
Moving then back on this 
second trajectory for an interval of time $\Delta t$ we should reach the point $\varphi^{\prime a}$ that we originally
reached via a simple BRS transformation from $\varphi^a$. All this is illustrated in the picture below: 

\begin{center}
\begin{picture}(240,120)
\pscurve[]{c-c}(-0.9,2.4)(0.9,2.9)(2.9,3.2)(4.9,3.4)(6.9,3.5)
\pscurve[]{c-c}(-0.9,0.4)(0.9,0.7)(2.9,0.9)(4.9,1)(6.9,1.1)
\psline[](0.9,0.7)(0.9,1.8)
\psline[](0.9,1.8)(0.9,2.9)
\psline[](0.9,1.8)(0.8,1.7)
\psline[](0.9,1.8)(1.0,1.7)
\psline[](4.9,1)(4.9,2.2)
\psline[](4.9,2.2)(4.9,3.4)
\psline[](2.9,3.2)(3.0,3.1)
\psline[](2.9,3.2)(3.0,3.3)
\psline[](4.9,2.2)(4.8,2.1)
\psline[](4.9,2.2)(5.0,2.1)
\psline[](2.9,0.9)(2.8,1.0)
\psline[](2.9,0.9)(2.8,0.8)
\rput(1.4,1.8){\small{BRS}}
\rput(2.9,2.9){\small{$\widehat{\cal H}\Delta t$}}
\rput(2.9,0.5){\small{$\widehat{\cal H}\Delta t$}}

\rput(0.9,3.4){$\varphi^{\prime a}=\varphi^a+\epsilon\pi^a$}
\rput(0.9,0.4){$\varphi^a$}
\rput(4.9,3.8){$\varphi_{\scriptscriptstyle (1)}^{\prime a}=\varphi_{\scriptscriptstyle (1)}^a+\epsilon\pi^a$}
\rput(4.9,0.7){$\varphi_{\scriptscriptstyle (1)}^{a}$}
\rput(5.4,2.2){\small{BRS}}
\end{picture}
\end{center}

\noindent This diagram illustrates what Eq. (\ref{bos.zip}) tells us:
{\it in the} $\varphi$-{\it space} the BRS charge $\widehat{Q}^{\scriptscriptstyle (B)}$ 
and the Hamiltonian $\widehat{\cal H}_{\scriptscriptstyle BFA}$ commute.

Let us now turn to the bosonic analogs of the susy charges 
$\widehat{Q}_{\scriptscriptstyle H}^{\scriptscriptstyle (B)},
\widehat{\overline{Q}}_{\scriptscriptstyle H}^{\scriptscriptstyle (B)}$. As they are linear combinations
of $\widehat{Q}^{\scriptscriptstyle (B)},\widehat{\overline {Q}}^{\scriptscriptstyle (B)},
\widehat{N}^{\scriptscriptstyle (B)},
\widehat{\overline{N}}^{\scriptscriptstyle (B)}$ and these last two 
charges commute 
with $\widehat{\cal H}_{\scriptscriptstyle BFA}$, from (\ref{bos.4-16}) and (\ref{bos.4-17}) we get 
%%%
\begin{eqnarray}
&&[\widehat{Q}_{\scriptscriptstyle H}^{\scriptscriptstyle (B)},\widehat{\cal H}_{\scriptscriptstyle BFA}]=
-\widehat{\pi}^l\omega^{ab}\partial_b\partial_l\partial_cH\widehat{\pi}^c\widehat{\xi}_a\nonumber\\
&&[\widehat{\overline{Q}}_{\scriptscriptstyle H}^{\scriptscriptstyle (B)},\widehat{\cal H}_{\scriptscriptstyle BFA}]=
-\widehat{\xi}_a\omega^{ab}\widehat{\xi}_s\omega^{st}(\partial_b\partial_t\partial_lH)\widehat{\pi}^l.
\label{bos.4-22}
\end{eqnarray}
%%%
So also the transformations generated by $\widehat{Q}_{\scriptscriptstyle H}^{\scriptscriptstyle (B)}$,
$\widehat{\overline{Q}}_{\scriptscriptstyle H}^{\scriptscriptstyle (B)}$
and those generated by $\widehat{\cal H}_{\scriptscriptstyle BFA}$ commute on the phase space variables 
$\widehat{\varphi}^a$.
It would be interesting to check whether they behave as true supersymmetry
charges\footnote[2]{Since
$\widehat{Q}_{\scriptscriptstyle H}^{\scriptscriptstyle (B)}$ and 
$\widehat{\overline{Q}}_{\scriptscriptstyle H}^{\scriptscriptstyle (B)}$ 
are bosonic charges in all the following formulae there will be commutators instead of the usual 
anticommutators of the supersymmetry algebra.}, i.e. whether:
%%%
\begin{equation}
[\widehat{Q}_{\scriptscriptstyle H}^{\scriptscriptstyle (B)},
\widehat{\overline{Q}}^{\scriptscriptstyle (B)}_{\scriptscriptstyle H}]
=2\widehat{\cal H}_{\scriptscriptstyle BFA}.
\end{equation}
%%%
It is actually easy to work out the following commutator between 
$\widehat{Q}_{\scriptscriptstyle H}^{\scriptscriptstyle (B)}$ and
$\widehat{\overline{Q}}_{\scriptscriptstyle H}^{\scriptscriptstyle (B)}$: 
%%%
\begin{equation}
[\widehat{Q}_{\scriptscriptstyle H}^{\scriptscriptstyle (B)},
\widehat{\overline{Q}}_{\scriptscriptstyle H}^{\scriptscriptstyle (B)}]=
2\widehat{\cal H}_{\scriptscriptstyle BFA}+4\widehat{\pi}^a\omega^{de}\partial_e\partial_aH\widehat{\xi}_d. 
\label{bos.4-23}
\end{equation}
%%%
We see that we can get the standard supersymmetry algebra if the last term on the RHS of (\ref{bos.4-23}) were zero. 
Again this term does not contain $\widehat{\lambda}_a$ and so on the $\widehat{\varphi}^a$ variables we have that
%%%
\begin{equation}
\delta_{Q_{\scriptscriptstyle H}^{\scriptscriptstyle (B)}}
\delta_{\overline{Q}_{\scriptscriptstyle H}^{\scriptscriptstyle (B)}}
\widehat{\varphi}^a-\delta_{\overline{Q}_{\scriptscriptstyle H}^{\scriptscriptstyle (B)}}
\delta_{Q_{\scriptscriptstyle H}^{\scriptscriptstyle (B)}}\widehat{\varphi}^a=
2\delta_{\widehat{\cal H}}\widehat{\varphi}^a
\end{equation}
%%%
i.e. the supersymmetry algebra holds. 

Usually supersymmetry is described as the ``{\it square root}" of the time translations. Let us find out whether 
there is anything
like that in our bosonic case. Instead of the two charges $\widehat{Q}_{\scriptscriptstyle H}^{\scriptscriptstyle (B)}$ 
and $\widehat{\overline{Q}}_{\scriptscriptstyle H}^{\scriptscriptstyle (B)}$, let us build the following
ones:
%%%
\begin{equation}
\left\{
	\begin{array}{l}
	\displaystyle 
	\widehat{Q}_{1}^{\scriptscriptstyle (B)}=\widehat{Q}^{\scriptscriptstyle (B)}-
	\widehat{\overline{N}}^{\scriptscriptstyle (B)} \smallskip\\
	\widehat{Q}_{2}^{\scriptscriptstyle (B)}=\widehat{\overline{Q}}^{\scriptscriptstyle (B)}
	+\widehat{N}^{\scriptscriptstyle (B)}.
	\label{bos.4-24}
	\end{array}
	\right.
\end{equation}
%%%
The transformations under $\widehat{Q}_{1}^{\scriptscriptstyle (B)}$ can be easily worked out. The result is:
%%%
\begin{equation}
\left\{
	\begin{array}{l}
	\displaystyle 
	\delta_{Q_1^{\scriptscriptstyle (B)}}\widehat{\varphi}^a=\epsilon\widehat{\pi}^a \smallskip\\
	\delta_{Q_1^{\scriptscriptstyle (B)}}\widehat{\xi}_a=\epsilon\widehat{\lambda}_a \smallskip\\
	\delta_{Q_1^{\scriptscriptstyle (B)}}\widehat{\pi}^a=-i\epsilon\omega^{ae}\partial_eH \smallskip\\
	\delta_{Q_1^{\scriptscriptstyle (B)}}\widehat{\lambda}_a=
	-i\epsilon \widehat{\xi}_b\omega^{be}(\partial_e\partial_aH)
	\label{bos.4-25}
	\end{array}
	\right.
\end{equation}
%%%
where $\epsilon$ is an infinitesimal commuting parameter. Let us check whether, by performing these transformations twice, 
we get a time translation. Using (\ref{bos.4-25}) and restricting ourselves 
on the original phase space $\widehat{\varphi}^a$, we get
%%%
\begin{equation}
\delta^2_{Q_1^{\scriptscriptstyle (B)}}\widehat{\varphi}^a=\delta_{Q_1^{\scriptscriptstyle (B)}}(\epsilon
\widehat{\pi}^a)=
-i\epsilon^2\omega^{ae}\partial_eH=-i\epsilon^2\dot{\widehat{
\varphi}}^a \label{bos.4-26}
\end{equation}
%%%
where in the last step above we have used the equations of motion. The result seems to confirm that 
$\widehat{Q}_1^{\scriptscriptstyle (B)}$ is the ``{\it
square root}" of the time translations. In fact (\ref{bos.4-26}) is an infinitesimal time translation if we equate
$\epsilon^2=\Delta t$. So we could say that, in order
to do an infinitesimal time translation, we could perform two $\widehat{Q}_1^{\scriptscriptstyle (B)}$-transformations 
in a row each with 
``infinitesimal" parameter $\epsilon$ related to the ``square root" of $\Delta t$. 
We think that it is curious that, at least on some hypersurfaces of our $8n$-dimensional space we could,
without introducing Grassmann variables and
via purely bosonic charges, get a square root of a time translation. 

Let us now go back to geometry and to the bosonic BRS charge $\widehat{Q}^{\scriptscriptstyle (B)}$. 
As we have seen in Chapter {\bf 3} in the Grassmann case 
the BRS charge can be identified with the exterior derivative and one of its properties 
is that it commutes with the Lie derivative. This is no longer the case for our
$\widehat{Q}^{\scriptscriptstyle (B)}$ as it is proved in (\ref{bos.4-16}). 
Even if $\widehat{Q}^{\scriptscriptstyle (B)}$ and  $\widehat{\cal H}_{\scriptscriptstyle BFA}$
commute in the $\varphi$-space, it is not enough. In fact the exterior
derivative must commute with the Lie derivative in the full space of forms which is somehow an extension
of the ordinary phase space. Actually it is the space of higher forms which has to be properly defined in 
the BFA case and this is what we will do in the next section. 

\bigskip

\section{Higher Forms in the BFA}

In (\ref{bos.4-4}) we have seen that it is possible 
to build one-forms using the operators $\widehat{\pi}^a$ instead of the Grassmann variables $c^a$. 
The problem arises when we want to build higher forms:
in fact it is difficult to define a {\it wedge} product $\wedge$ so that, for example, the basis
$d\varphi^a\wedge d\varphi^b$ for two-forms is antisymmetric in the interchange $a\leftrightarrow b$. This 
operation was naturally
incorporated in the Grassmann formalism \cite{Goz89}: in fact, by representing the forms $d\varphi^a$ 
with the anticommuting variables
$c^a$, the antisymmetry of $d\varphi^a\wedge d\varphi^b$ was
automatically reproduced by the antisymmetry 
of the product $c^ac^b$:
%%%
\begin{equation}
	\begin{array}{c}
	\displaystyle 
	d\varphi^a\wedge d\varphi^b=-d\varphi^b\wedge d\varphi^a \smallskip\\
	\Updownarrow \smallskip\\
	c^ac^b=-c^bc^a.
	\label{bos.5-1}
	\end{array}
\end{equation}
%%%
In the bosonic case we do not have Grassmann variables and the forms $\widehat{\pi}^a$ commute among themselves. 
Therefore we cannot represent a two-form $d\varphi^a\wedge d\varphi^b$ as $\widehat{\pi}^a\widehat{\pi}^b$, 
because in this way we would loose its anticommuting nature.

The way out seems to be the standard procedure used in the literature on differential geometry \cite{Eguchi}, i.e.
to introduce a tensor product among the cotangent spaces whose basis are $d\varphi^a$ and define the wedge product 
$\wedge$ as
%%%
\begin{equation}
\displaystyle 
d\varphi^a\wedge d\varphi^b\equiv \frac{1}{2}(d\varphi^a\otimes d\varphi^b-d\varphi^b\otimes d\varphi^a).
\end{equation}
%%%
In our case the role of $d\varphi^a$ is played by the operators $\widehat{\pi}^a$ and, since we want to build tensor 
products among them, we have to enlarge our Hilbert space. Originally it was made of the functions 
$\psi(\varphi,\pi)$ belonging to the tensor product of the two Hilbert spaces spanned by $|\varphi^a\rangle$
and $|\pi^a\rangle$:
%%%
\begin{equation}
{\mathbf H}\equiv {\mathbf H}_{\varphi}\otimes {\mathbf H}_{\pi}. \label{bos.5-2}
\end{equation}
%%%
From now on we will use instead the following new Hilbert space: 
%%%
\begin{equation}
{\mathbf H}_{2n}\equiv {\mathbf H}_{\varphi}\otimes {\mathbf H}_{\pi_{(1)}}
\otimes {\mathbf H}_{\pi_{(2)}}\ldots\otimes {\mathbf H}_{\pi_{(2n)}}. \label{bos.5-3}
\end{equation}
%%%
where we have made the tensor products of $2n$ copies of the space ${\mathbf H}_{\pi}$ and labeled them 
with ${\mathbf H}_{\pi_{(1)}}$, $\ldots {\mathbf H}_{\pi_{(2n)}}$.
If we limit ourselves to the case $n=1$ the space (\ref{bos.5-3}) reduces to: 
%%%
\begin{equation}
{\mathbf H}_2\equiv {\mathbf H}_{\varphi}\otimes {\mathbf H}_{\pi_{(1)}}
\otimes {\mathbf H}_{\pi_{(2)}} \label{bos.5-5}
\end{equation}
%%%
and, for example, a two-form can be represented as 
%%%
\begin{equation}
\displaystyle \widehat{F}
=F_{ab}(\widehat{\varphi})\otimes \frac{1}{2}\Bigl[\widehat{\pi}_{\scriptscriptstyle (1)}^a\otimes
\widehat{\pi}^b_{\scriptscriptstyle (2)}-\widehat{\pi}_{\scriptscriptstyle (1)}^b\otimes
\widehat{\pi}^a_{\scriptscriptstyle (2)}\Bigr]. \label{bos.5-4}
\end{equation}
%%%
The operator (\ref{bos.3-5}) was good to represent the Lie derivative but only in the space (\ref{bos.5-2}).
To represent the Lie derivative on the new space (\ref{bos.5-5}) we should generalize 
$\widehat{\cal H}_{\scriptscriptstyle BFA}$ to the following operator:
%%%
\begin{equation}
\widehat{\cal H}\equiv \widehat{\lambda}_a\omega^{ab}\partial_bH(\widehat{\varphi})\otimes 
{\bf 1}_{\scriptscriptstyle (1)}\otimes {\bf 1}_{\scriptscriptstyle (2)}
-\omega^{be}\partial_e\partial_a H(\widehat{\varphi})\otimes \Bigl(\widehat{\pi}^a_{\scriptscriptstyle (1)}
\widehat{\xi}_b^{\scriptscriptstyle (1)}\otimes {\bf 1}_{\scriptscriptstyle (2)}
+{\bf 1}_{\scriptscriptstyle (1)}\otimes 
\widehat{\pi}^a_{\scriptscriptstyle (2)}\widehat{\xi}_b^{\scriptscriptstyle (2)}\Bigr). \label{bos.5-6}
\end{equation}
%%%
Using the following commutators:
%%%
\begin{eqnarray}
&&[\widehat{\pi}_{\scriptscriptstyle (i)}^a,\widehat{\pi}_{\scriptscriptstyle (j)}^b]=0\nonumber\\
&&[\widehat{\xi}^{\scriptscriptstyle (i)}_a,\widehat{\xi}^{\scriptscriptstyle (j)}_b]=0\nonumber\\
&&[\widehat{\xi}_a^{\scriptscriptstyle (i)},\widehat{\pi}_{\scriptscriptstyle (j)}^b]=
i\delta_a^b\delta^{\scriptscriptstyle (i)}_{\scriptscriptstyle (j)} \label{bos.5-7}\\
&&[\widehat{\varphi}^a,\widehat{\pi}_{\scriptscriptstyle (i)}^b]=0 \nonumber\\
&&[\widehat{\varphi}^a,\widehat{\xi}_b^{\scriptscriptstyle (i)}]=0 \nonumber
\end{eqnarray}
%%%
it is easy to see that the action of the $\widehat{\cal H}$ of (\ref{bos.5-6}) 
on the two-form $\widehat{F}$ of (\ref{bos.5-4})
is:
%%%
\begin{equation}
\displaystyle [i\widehat{\cal H},\widehat{F}]=
\omega^{ab}\Bigl[\partial_bH\partial_aF_{de}+\partial_b\partial_dHF_{ae}
+\partial_b\partial_eHF_{da}\Bigr]\otimes \frac{1}{2}(\widehat{\pi}^d_{\scriptscriptstyle (1)}\otimes
\widehat{\pi}^e_{\scriptscriptstyle (2)}-
\widehat{\pi}^e_{\scriptscriptstyle (1)}
\otimes\widehat{\pi}^d_{\scriptscriptstyle (2)}) \label{bos.5-8}
\end{equation}
%%%
and this is exactly the manner how the two-forms transform under the Lie derivative \cite{Marsd}. 
In the case $n=1$ we have only zero-, one- and two-forms; if the two-forms are represented
by (\ref{bos.5-4}), the zero-forms $\widehat{G}$
and the one-forms $\widehat{C}$ are respectively given by:
%%%
\begin{equation}
\widehat{G}
=G(\widehat{\varphi})\otimes \Bigl[{\bf 1}_{\scriptscriptstyle (1)}\otimes {\bf 1}_{\scriptscriptstyle (2)}\Bigr],
\end{equation}
%%%
\begin{equation}
\widehat{C}
=C_d(\widehat{\varphi})\otimes \Bigl[\widehat{\pi}_{\scriptscriptstyle (1)}^d\otimes {\bf 1}_{\scriptscriptstyle (2)}+
{\bf 1}_{\scriptscriptstyle (1)}\otimes \widehat{\pi}^d_{\scriptscriptstyle (2)}\Bigr]. \label{bos.oneforms}
\end{equation}
%%%
The commutator of $i\widehat{\cal H}$ with $\widehat{C}$
gives the correct action of the Lie derivative on one-forms:
%%%
\begin{equation}
[i\widehat{\cal H},\widehat{C}]=
(\partial_aC_d\omega^{ab}\partial_bH+\omega^{ae}\partial_e\partial_dHC_a)
\otimes\Bigl[\widehat{\pi}^d_{\scriptscriptstyle (1)}\otimes{\bf 1}_{\scriptscriptstyle (2)}
+{\bf 1}_{\scriptscriptstyle (1)}
\otimes\widehat{\pi}^d_{\scriptscriptstyle (2)}\Bigr]. \label{bos.comm1}
\end{equation}
%%%
So we can conclude that, in the case $n=1$, the operator (\ref{bos.5-6}) represents a good extension of the Lie derivative on 
the entire space of differential forms.

It is easy to generalize this operator
to the Lie derivative which acts in a space with an arbitrary number $n$ of degrees of freedom. 
It is the following one:
%%%
\begin{equation}
\widehat{\cal H}\equiv\lambda_a\omega^{ab}\partial_bH\otimes{\bf 1}^{\otimes 2n}-\omega^{be}\partial_e\partial_aH\otimes
{\bf S}\Bigl[\widehat{\pi}^a\widehat{\xi}_b\otimes{\bf 1}^{\otimes (2n-1)}\Bigr] \label{bos.5-9}
\end{equation}
%%%
where by ${\bf 1}^{\otimes 2n}$ we indicate the tensor product of $2n$ identity operators, and with ${\bf S}$ the 
symmetrizer of the operators contained in the square brackets. So for example for $n=2$ we have
%%%
\begin{eqnarray}
{\bf S}\Bigl[\widehat{\pi}^a\widehat{\xi}_b\otimes{\bf 1}^{\otimes 3}\Bigr]&=&
\widehat{\pi}_{\scriptscriptstyle (1)}^a\widehat{\xi}_b^{\scriptscriptstyle (1)}
\otimes {\bf 1}_{\scriptscriptstyle (2)}\otimes {\bf 1}_{\scriptscriptstyle (3)}
\otimes {\bf 1}_{\scriptscriptstyle (4)}+{\bf 1}_{\scriptscriptstyle (1)} \otimes 
\widehat{\pi}^a_{\scriptscriptstyle (2)}\widehat{\xi}_b^{\scriptscriptstyle (2)}
\otimes {\bf 1}_{\scriptscriptstyle (3)}\otimes {\bf 1}_{\scriptscriptstyle (4)}+\nonumber\\
&&+{\bf 1}_{\scriptscriptstyle (1)}\otimes{\bf 1}_{\scriptscriptstyle (2)}
\otimes\widehat{\pi}^a_{\scriptscriptstyle (3)}\widehat{\xi}_b^{\scriptscriptstyle (3)}
\otimes{\bf 1}_{\scriptscriptstyle (4)}+{\bf 1}_{\scriptscriptstyle (1)}\otimes {\bf 1}_{\scriptscriptstyle (2)}\otimes
{\bf 1}_{\scriptscriptstyle (3)}\otimes 
\widehat{\pi}^a_{\scriptscriptstyle (4)}\widehat{\xi}_b^{\scriptscriptstyle (4)}.\nonumber \label{bos.5-10}
\end{eqnarray}
%%%
Let us remember that the indices $(1),(2),\ldots (2n)$ always indicate on which Hilbert space
${\mathbf H}_{\pi_{(i)}}$ in (\ref{bos.5-3}) the operators $\widehat{\pi}_{\scriptscriptstyle (i)}$,
$\widehat{\xi}^{\scriptscriptstyle (i)}$ and ${\bf 1}_{\scriptscriptstyle (i)}$ act.
In the same way it is possible to generalize the concept of differential form. An $m$-form in a $2n$-dimensional space
is given by:
%%%
\begin{equation}
\displaystyle \widehat{P}\equiv{\bf S}\biggl\{\frac{1}{m!}P_{a_1\ldots a_m}(\widehat{\varphi})
\otimes
{\bf A}\Bigl\{\widehat{\pi}_{\scriptscriptstyle (1)}^{a_1}\otimes\widehat{\pi}_{\scriptscriptstyle (2)}^{a_2}\otimes\ldots
\widehat{\pi}_{\scriptscriptstyle (m)}^{a_m}\Bigr\}\otimes{\bf 1}^{\otimes (2n-m)}\biggr\}, \label{bos.genform}
\end{equation}
%%%
where ${\bf A}$ indicates the antisymmetrizer of the basis $\widehat{\pi}_{\scriptscriptstyle (i)}^{a_i}$
of the $m$ cotangent spaces needed to build an $m$-form. The position of this $m$ operators $\widehat{\pi}$
inside the string of the $2n$ Hilbert spaces is completely arbitrary. Therefore if we do not want to choose
a particular position we can symmetrize the $2n-m$ identity operators with the $m$ operators $\widehat{\pi}$
by means of the symmetrizer ${\bf S}$ as we did in (\ref{bos.oneforms}) for the one-forms.
Then the commutator
between (\ref{bos.5-9}) and (\ref{bos.genform}) 
reproduces the correct action of the Lie derivative on an arbitrary differential form 
$P$:
%%%
\begin{equation}
[i\widehat{\cal H},\widehat{P}]={\cal L}_{(dH)^{\sharp}}
P. \label{bos.genlie}
\end{equation} 
%%%

Before concluding this section let us notice that, in the BFA approach, the higher forms are not
represented by {\it wave functions} of the theory like $\psi(\varphi,c)$ in the CPI case, 
but by {\it operators} like in
(\ref{bos.genform}). In fact the wave functions of the BFA approach do not have the structure of the Grassmannian
ones:
%%%
\begin{equation}
\psi(\varphi,c)=\psi_0(\varphi)+\psi_a(\varphi)c^a+\psi_{ab}(\varphi)
c^ac^b+\ldots +\psi_{abc\ldots l}c^ac^bc^c\ldots c^l.
\end{equation}
%%%
In the CPI case
it was this structure which allowed us  to identify $\psi_0(\varphi)$ with zero-forms, $\psi_a(\varphi)c^a$ with one-forms, 
$\psi_{ab}c^ac^b$ with 
two-forms etc. In the bosonic case instead the wave functions $\psi(\varphi,\pi)$ are generic functions of $\pi$ and this 
forbids their identification with forms. Moreover, as we said previously, a one-form would be represented by 
$\psi(\varphi,\pi)=\psi_a(\varphi)\pi^a$ which would be not an acceptable wave function because it is not normalizable. 
Of course this does not mean that in the formalism given by (\ref{bos.5-9}) 
we cannot introduce wave functions. It simply means that such wave functions cannot  
have the meaning of higher forms, differently than operators like (\ref{bos.genform}). 

The wave functions associated to the multi-form formalism of the Hamiltonian (\ref{bos.5-9}) basically make up
the Hilbert space (\ref{bos.5-3}) and so they are given by $\psi(\varphi, \pi_{\scriptscriptstyle (1)},
\ldots,\pi_{\scriptscriptstyle (2n)})$. It is possible
to introduce also in this space the following positive definite scalar product:
%%%
\begin{equation}
\langle \psi_1|\psi_2\rangle\equiv \int d\varphi^a\prod_{i=1}^{2n}d\pi_{\scriptscriptstyle
(i)}^a\;\psi_1^*\psi_2
\label{bos.5-11}
\end{equation}
%%%
and it is easy to prove that with this product 
the Hamiltonian (\ref{bos.5-9}) is Hermitian. 
The reader may remember that our original $\widehat{\cal H}_{\scriptscriptstyle BFA}$ (\ref{bos.3-5}) 
was derived from the path integral formalism
(\ref{bos.2-5-a}). Also the multi-form Hamiltonian (\ref{bos.5-9}) can be derived from the following path integral:
%%%
\begin{equation}
\displaystyle 
Z=\int {\mathcal D}\varphi{\mathcal D}\lambda\prod_{i=1}^{2n}{\mathcal D}\pi_{\scriptscriptstyle (i)}{\mathcal D}
\xi^{\scriptscriptstyle (i)} \,\textrm{exp}\Bigl[i\int dt \,{\cal L}_{\scriptscriptstyle{MF}}\Bigr]
\label{bos.5-12-a}
\end{equation}
%%%
where the multiform (MF) Lagrangian ${\cal L}_{\scriptscriptstyle{MF}}$ is given by:
%%%
\begin{equation}
\displaystyle \label{yesyes}
{\cal L}_{\scriptscriptstyle{MF}}=
\lambda_a\dot{\varphi}^a+\sum_{i=1}^{2n}\pi^a_{\scriptscriptstyle (i)}\dot{\xi}_a^{\scriptscriptstyle (i)}
-{\cal H}_{\scriptscriptstyle{MF}} 
\end{equation}
%%%
with
%%%
\begin{equation}
{\cal H}_{\scriptscriptstyle{MF}}=
\lambda_a\omega^{ab}\partial_bH-\sum_{i=1}^{2n}\pi^a_{\scriptscriptstyle (i)}
\omega^{be}\partial_e\partial_aH\xi_b^{\scriptscriptstyle (i)}.
\label{bos.5-12}
\end{equation}
%%%
At this level the proof of the hermiticity of ${\cal H}_{\scriptscriptstyle MF}$ under the scalar product 
(\ref{bos.5-11}) is identical to the proof given for $\widehat{\cal H}_{\scriptscriptstyle BFA}$
in (\ref{bos.3-6}). 
Basically in the Hamiltonian (\ref{bos.5-12}) we have a set of extra variables $(\pi_{\scriptscriptstyle (i)},
\xi^{\scriptscriptstyle (i)})$
for each extra Hilbert space ${\mathbf{H}}_{\pi_{(i)}}$ appearing in (\ref{bos.5-3}).
It is actually then easier to work with the Hamiltonian ${\cal H}_{\scriptscriptstyle{MF}}$ 
than with the one in (\ref{bos.5-9}). We can in fact turn 
the $\pi_{\scriptscriptstyle (i)}^a, \xi_a^{\scriptscriptstyle (i)}$ into operators by just looking at the 
kinetic term of (\ref{yesyes}) and deriving from it the commutators 
that we introduced by hand in (\ref{bos.5-7}) plus the usual one 
$[\widehat{\varphi}^a,\widehat{\lambda}_b]=i\delta_b^a$. 
This confirms that the path integral (\ref{bos.5-12-a})
is basically the one behind the operatorial
formalism (\ref{bos.5-9}). Unfortunately this path integral does not have a ``natural" interpretation differently than  
the one in (\ref{bos.2-5-a}), in the sense that the latter is naturally related to (\ref{bos.2-2}) 
and (\ref{bos.2-1}) which
are just Dirac deltas on the classical paths. These Dirac deltas are natural objects in a functional approach 
to CM because they just give weight one to classical paths and weight zero to non-classical ones. Nothing like that
can be done for the path integral (\ref{bos.5-12-a}) which can be turned into a Dirac delta of the equations of motion
like in (\ref{bos.2-2}) but it gets multiplied not by one determinant but by $2n$ of them. This structure does not allow
us to pass to the Dirac deltas of the classical trajectories appearing in (\ref{bos.2-1}). So somehow the path integral 
(\ref{bos.5-12-a}) does not have a simple intuitive understanding. This is the price we pay: we have a formalism
with a positive definite scalar product and a Hermitian Hamiltonian but a physical understanding is lacking. 
If on the contrary we keep the intuitive single particle path integral associated to the Hilbert space (\ref{bos.5-2})
then the tensor product structure $\otimes$, needed to build higher forms like in (\ref{bos.genform}), 
has to be given from outside 
and is not provided directly by the path integral.
On the contrary in the original CPI \cite{Goz89} the whole formalism, even for higher forms, has 
a nice and intuitive 
understanding and construction (because it can be reduced to a Dirac delta on classical paths), and no extra structure
has to be brought in from outside, but 
we have to give up one of the two conditions: either the positive definiteness of the scalar
product or the hermiticity of the Hamiltonian. 

\bigskip

\section{Metaplectic Representation}

So far we have widely used the concept of Lie derivative \cite{Marsd}. In particular we have seen how it acts on 
vector fields (\ref{bos.4-6}), on forms (\ref{bos.4-8}) or on general
tensors in the case of {\it symplectic} manifolds. The notion of Lie
derivative can be extended to {\it general} manifolds ${\cal M}_n$ with $\textrm{Diff}({\cal M}_n)$
as group of diffeomorphisms and with ${\cal G}$ as {\it structure group} of the associated
(co-)tangent bundle \cite{Eguchi}.
Under the action of an element of $\textrm{Diff}({\cal
M}_n)$, which generates an infinitesimal displacement $\delta\varphi^a=-h^a(\varphi)$,
an arbitrary tensor field ${\cal X}$ on ${\cal M}_n$ is transformed as
follows:
%%%
\begin{equation}
{\cal X}^{\prime}(\varphi)-{\cal X}(\varphi)={\cal L}_h{\cal X}(\varphi) \label{bos.6-1}
\end{equation}
%%%
where ${\cal L}_h$ is the Lie derivative associated to the vector field $h=h^a\partial_a$. 
The general abstract expression of ${\cal L}_h$ is \cite{dewitt}:
%%%
\begin{equation}
{\cal L}_h=h^a\partial_a-\partial_bh^aG^b_{\; a} \label{bos.6-2}
\end{equation}
%%%
where $G^b_{\;a}$ are the generators of the structure group ${\cal G}$ in the representation to which ${\cal X}$ belongs.
We will indicate with $a,b$ the group indices and with $\alpha,\beta$ the representation ones. So the matrix
representation of $(G^a_{\;b})$ will be $(G^a_{\;b})^{\alpha}_{\;\beta}$ where $\alpha$ are also the indices of ${\cal X}$, if we
organize it as a vector. For a generic manifold we have that 
${\cal G}=\textrm{GL}(n,R)$, for a Riemann manifold ${\cal G}=O(n)$,
and for a symplectic manifold ${\cal G}=\textrm{Sp}(2N)$. It is easy to 
check that, if ${\cal X}$ is a vector or a form and 
${\cal M}_{2N}$ a symplectic manifold, (\ref{bos.6-1}) reduces to the usual transformations (\ref{bos.4-6}) and
(\ref{bos.4-8})
\cite{metaplectic}. As we have already said the coefficients $G^b_{\; a}$ are the generators of the structure group ${\cal G}$ in the
representation to which ${\cal X}$ belong. Now ${\cal G}$ could have also spinor representations like in the case of
${\cal G}=O(n)$. This means that we can introduce the concept of Lie derivative also for spinors 
but not along all the vector fields $h^a$. We have to restrict ourselves to the Lie derivatives along Killing vector
fields in the Riemann case \cite{dewitt} and along Hamiltonian vector fields $h^a=\omega^{ab}\partial_bH$
in the symplectic one.
We will not give the reasons why we have to restrict ourselves to these
particular vector fields for spinors ${\cal X}$ but we refer the reader to the literature \cite{dewitt}. 
Basically it is only
for those fields that the usual commutator structure of the Lie derivatives (\ref{bos.4-9}) is preserved even for spinors.

Before proceeding let us rewrite (\ref{bos.6-2}) in a slightly modified form, by introducing the following objects:
%%%
\begin{eqnarray}
&& K_{ab}(\varphi)\equiv\partial_a\partial_bH(\varphi)\nonumber\\
&& \Sigma^{ab}\equiv i(\omega^{ca}G^b_{\;c}+\omega^{cb}G^a_{\;c}).
\end{eqnarray}
%%%
Both are symmetric in $a,b$ and $K_{ab}\Sigma^{ab}=2i\partial_bh^cG^b_{\; c}$;
so (\ref{bos.6-2}) can be rewritten as
%%%
\begin{equation}
{\cal L}_h=h^a\partial_a+\frac{i}{2}K_{ab}\Sigma^{ab}. \label{bos.6-4}
\end{equation}
This is the classical Lie derivative for fields whose components transform, under infinitesimal $\textrm{Sp}(2N)$
transformations of the tangent space, via the operator
%%%
\begin{equation}
\displaystyle S=1-\frac{i}{2}K_{ab}\Sigma^{ab}. \label{bos.6-5}
\end{equation}
%%%
We want now to apply this formalism to spinors, i.e. we want to use for $\Sigma^{ab}$ in (\ref{bos.6-5}) the spinorial
representation of $\textrm{Sp}(2N)$. To do this we have to pass to the universal covering group of $\textrm{Sp}(2N)$
which is the metaplectic group $\textrm{Mp}(2N)$ \cite{littlejohn}\cite{konstant}\cite{dewitt2}. In analogy to the
spinorial representation of the Lorentz group, we first have to build the counterpart of
the representation of the Clifford algebra
%%%
\begin{equation}
\gamma^{\mu}\gamma^{\nu}+\gamma^{\nu}\gamma^{\mu}=2g^{\mu\nu} \label{bos.6-6}
\end{equation}
%%%
which, in the case of the metaplectic group \cite{dewitt2}, is 
%%%
\begin{equation}
\gamma^a\gamma^b-\gamma^b\gamma^a=2i\omega^{ab}. \label{bos.6-7}
\end{equation}
%%%
This algebra, because of the crucial difference in sign on the LHS of (\ref{bos.6-7}) with respect to (\ref{bos.6-6}), does not
admit finite dimensional unitary representations. The reason is the same as the one for which we
cannot find any finite dimensional representation for the operators $\widehat{q},\widehat{p}$ in QM: 
they obey the algebra
$\widehat{q}\widehat{p}-\widehat{p}\widehat{q}=i\hbar$ and, if they were represented by finite dimensional matrices, 
by taking the trace on both sides, we would get
a contradictory result. The only representations are infinite dimensional. We will indicate with ${\cal V}$
this infinite dimensional
Hilbert space and with ``$x$" the indices of its vectors. The generators of the
metaplectic group are operators in this Hilbert space and the matrices $\gamma^a$ in 
(\ref{bos.6-7}) can be represented by the infinite matrices 
$(\gamma^a)^x_{\;y}$.  As $\textrm{Mp}(2N)$ is the covering group of
$\textrm{Sp}(2N)$ there are two elements $M(S)$ and $-M(S)$ of $\textrm{Mp}(2N)$ associated to each $S\in
\textrm{Sp}(2N)$. Correspondingly the multiplication rules will be 
%%%
\begin{equation}
M(S_1)M(S_2)=\pm M(S_1S_2).
\end{equation}
%%%
Following the procedure used for spinors in the case of the Lorentz group, we now try to find an operator $M(S)$ on
${\cal V}$ such that 
%%%
\begin{equation}
M(S)^{-1}\gamma^aM(S)=S^a_{\;b}\gamma^b. \label{bos.6-8}
\end{equation}
%%%
We choose $S^a_{\;b}$ infinitesimally close to the identity and parameterize it like in (\ref{bos.6-5}). For $M(S)$ we 
make the ansatz
%%%
\begin{equation}
\displaystyle M(S)=1-\frac{i}{2}K_{ab}\Sigma^{ab}_{\textrm{meta}} \label{bos.6-9}
\end{equation}
%%%
where $\Sigma^{ab}_{\textrm{meta}}$ are the operators $\Sigma^{ab}$ in the metaplectic representation. 
If we insert (\ref{bos.6-9}) into (\ref{bos.6-8}) we get:
%%%
\begin{equation}
\displaystyle \Sigma_{\textrm{meta}}^{ab}=\frac{1}{4}(\gamma^a\gamma^b+\gamma^b\gamma^a). \label{bos.6-9-a}
\end{equation}
%%%
So a representation for the matrix $\gamma^a$ gives rise to an associated representation for the generators
$\Sigma^{ab}_{\textrm{meta}}$. We will consider only representations in which $\gamma^a$ is Hermitian with respect 
to the inner product in ${\cal V}$. As a consequence also $\Sigma_{\textrm{meta}}^{ab}$ is Hermitian and 
$M(S)$ turns out to be unitary:
%%%
\begin{eqnarray}
&&\displaystyle (\gamma^a)^{\dagger}=\gamma^a\nonumber\\
&&\displaystyle (\Sigma^{ab}_{\textrm{meta}})^{\dagger}=\Sigma^{ab}_{\textrm{meta}} \label{bos.em}\\
&&\displaystyle M(S)^{\dagger}=M(S)^{-1}.\nonumber
\end{eqnarray}
%%%
Explicit representations of the symplectic operators 
$\gamma^a$ have been worked out and can be found in the literature \cite{metaplectic}.
We will briefly review one of them here.
The Clifford algebra (\ref{bos.6-7}) is isomorphic to the standard Heisenberg algebra made of $N$ positions $\widehat{x}^k$ and $N$
momenta operators $\widehat{p}^j$:
%%%
\begin{eqnarray}
&&[\widehat{x}^k,\widehat{p}^j]=i\hbar\delta^{kj} \label{bos.6-10}\nonumber\\
&&[\widehat{x}^k,\widehat{x}^j]=0\qquad\qquad\quad k,j=1,\ldots, N. \\
&&[\widehat{p}^k,\widehat{p}^j]=0\nonumber
\end{eqnarray}
%%%
In fact, combining $\widehat{x}^k$ and $\widehat{p}^k$ into a single variable
%%%
\begin{equation}
\widehat{\phi}^a=(\widehat{p}^k,\widehat{x}^k) \qquad\qquad a=1,\ldots, 2N,
\end{equation}
%%%
we get that the algebra (\ref{bos.6-10}) can be written as:
%%%
\begin{equation}
\widehat{\phi}^a\widehat{\phi}^b-\widehat{\phi}^b\widehat{\phi}^a=i\hbar\omega^{ab} \label{bos.comm}
\end{equation}
%%%
which is isomorphic to the metaplectic analog of the Clifford algebra (\ref{bos.6-7}); so we have the following 
representation for $\gamma^a$:
%%%
\begin{equation}
\displaystyle \gamma^a=\biggl(\frac{2}{\hbar}\biggr)^{\frac{1}{2}}\widehat{\phi}^a.
\end{equation}
%%%
In the ``Schr\"odinger" representation in which the operators $\widehat{x}^k$ are diagonal, we have:
%%%
\begin{eqnarray}
&& (\gamma^k)^x_{\;y}=\biggl(\frac{2}{\hbar}\biggr)^{\frac{1}{2}}\langle x|\widehat{x}^k|y\rangle=
\biggl(\frac{2}{\hbar}\biggr)^{\frac{1}{2}}x^k\delta^N(x-y)\nonumber\\
&& (\gamma^{N+k})^x_{\;y}=\biggl(\frac{2}{\hbar}\biggr)^{\frac{1}{2}}\langle x|\widehat{p}^k|y\rangle=
-i(2\hbar)^{\frac{1}{2}}\partial_k\delta^N(x-y)
\end{eqnarray}
%%%
where $x,y$ are the Hilbert space indices. 
With the representation above and using 
(\ref{bos.6-9-a}), we get the following expression for the $K_{ab}\Sigma^{ab}_{\textrm{meta}}$ entering the 
matrix $M(S)$ of
(\ref{bos.6-9}):
%%%
\begin{eqnarray}
\displaystyle
\biggl(\frac{1}{2}K_{ab}\Sigma^{ab}_{\textrm{meta}}\biggr)^x_{\;y}&\hspace{-0.2cm}=&\hspace{-0.2cm}
\biggl[-\frac{1}{2}K_{kj}\partial^k\partial^j
-\frac{1}{2}iK_{N+k,j}(x^k\partial^j+\partial^jx^k)+\nonumber\\
&\hspace{-0.2cm}&\hspace{-0.2cm}+\frac{1}{2}K_{N+k,N+j}x^kx^j\biggr]\delta^N(x-y) \label{bos.6-11-b}
\end{eqnarray}
%%%
where we have put $\hbar=1$.

The geometrical picture we have so far is the following: the base space is the phase space ${\cal M}_{2N}$
and on its fibers the structure group $\textrm{Sp}(2N)$ acts no longer in the {\it vector}
representation like in Ref. \cite{Goz89} or in Secs. {\bf 5.1}-{\bf 5.4}, 
but in the {\it spinor} representation. In this way we get a bundle similar to the ``spin-bundle" \cite{Eguchi} but 
whose fibers are
the Hilbert spaces ${\cal V}$. So we end up in a Hilbert bundle which we call ${\cal V}_{\varphi}$ to indicate that there
is a fiber ${\cal V}$ at each point ${\cal \varphi}$ of the phase space ${\cal M}_{2N}$. In each
fiber a state $|\psi\rangle$ can be represented in its basis $\langle x|$ as:
%%%
\begin{equation}
\psi^x\equiv\langle x|\psi\rangle \label{bos.fiber}
\end{equation}
%%%
while its dual state $\langle\psi|\in{\cal V}^*$ is given by:
%%%
\begin{equation}
\langle\psi|x\rangle=(\psi^x)^*.
\end{equation}
%%%
The dual pairing is then the usual inner product
%%%
\begin{equation}
\langle {\cal X}|\psi\rangle\equiv\int d^Nx\;({\cal X}^x)^*\psi^x
\end{equation}
%%%
on $L^2(R^N,d^Nx)$. To avoid possible mistakes we want to underline 
that the bra $\langle x|$ entering (\ref{bos.fiber}) or the operators $\widehat{\phi}$ entering (\ref{bos.comm})
have nothing to do with the variables ${\cal \varphi}$ parameterizing the phase space ${\cal M}_{2N}$.

Going back to the Hilbert bundle ${\cal V}_{\cal \varphi}$ a section is locally given by a function $\psi$
%%%
\begin{eqnarray}
\psi:{\cal M}_{2N} &\rightarrow & {\cal V}\nonumber\\
{\cal \varphi} &\rightarrow  & |\psi;{\cal \varphi}\rangle\in{\cal V}_{\cal \varphi}. \label{bos.6-12}
\end{eqnarray}
%%%
Here the notation $|\psi;{\cal \varphi}\rangle$ indicates that the vector $\psi$
lives in the local Hilbert space (fiber) ${\cal V}$
associated to the point ${\cal \varphi}$ of the base manifold. At the level of matrix elements the function (\ref{bos.6-12}) is
defined by its components:
%%%
\begin{equation}
\psi^x({\cal \varphi})=\langle x|\psi;{\cal \varphi}\rangle.
\end{equation}
%%%
By replacing ${\cal V}$ with its Hilbert dual ${\cal V}^*$ we can construct:
%%%
\begin{equation}
{\cal X}_x({\cal \varphi})=\langle {\cal X;\varphi}|x\rangle, \qquad \langle {\cal X;\varphi}|\in {\cal V}^*_{\cal
\varphi}.
\end{equation}
%%%
In our formalism it is natural to consider also ``multispinor" fields
%%%
\begin{equation}
{\cal \varphi}\rightarrow {\cal X}^{x_1\ldots x_q}_{y_1\ldots y_p}({\cal \varphi}) \label{bos.6-13-a}
\end{equation}
%%%
which assume values in the tensor product:
%%%
\begin{equation}
\underbrace{{\cal V}_{\cal \varphi}^*\otimes {\cal V}_{\cal \varphi}^*\otimes\ldots\otimes {\cal V}_{\cal \varphi}^*}_
{p \;\textrm{factors}}
\otimes\underbrace{{\cal V}_{\cal \varphi}\otimes{\cal V}_{\cal \varphi}\otimes\ldots\otimes{\cal V}_{\cal \varphi}}_{q\; 
\textrm{factors}}. \label{bos.6-13-b}
\end{equation}
%%%
The symplectic spinors and multispinors have been first studied in detail in Ref. \cite{konstant}. Restricting
ourselves to a spinor, its evolution equation under the Hamiltonian vector field
$h^a=\omega^{ab}\partial_bH$ is:
%%%
\begin{eqnarray}
\displaystyle \partial_t{\cal X}_x({\cal \varphi},t)&=&-{\cal L}_h{\cal X}_x({\cal \varphi},t)=\nonumber\\
&=&-\int dy \biggl[\delta(x-y)h^a\partial_a+\frac{i}{2}K_{ab}({\cal \varphi})(\Sigma_{\textrm{meta}}^{ab})^y_{\;x}\biggr]
{\cal X}_y({\cal
\varphi}, t) \label{bos.6-13}
\end{eqnarray}
%%%
where $K_{ab}(\Sigma^{ab}_{\textrm{meta}})^y_{\;x}$ is given by (\ref{bos.6-11-b}). 
In general if we indicate the representation indices with Greek letters $(\alpha,\beta)$ and the
group (or manifold) indices with Latin ones $(a,b)$ Eq. (\ref{bos.6-13}) can be replaced by 
%%%
\begin{eqnarray}
\displaystyle \partial_t{\cal X}_{\alpha}({\cal \varphi}, t)&=&-{\cal L}_h{\cal X}_{\alpha}({\cal
\varphi},t)=\nonumber\\
&=& -\biggl[\delta_{\alpha}^{\beta}h^a\partial_a+\frac{i}{2}K_{ab}({\cal \varphi})(\Sigma^{ab})_{\;\alpha}^{\beta}\biggr]
{\cal X}_{\beta}({\cal \varphi},t). \label{bos.6-14}
\end{eqnarray}
%%%
Note also that we have not put the label ``meta" on the $\Sigma^{ab}$
matrix just because (\ref{bos.6-14}) is the equation of evolution of the spinor $\chi$ in any representation. 

\bigskip

\section{Metaplectic Hamiltonian and Scalar Product}

Up to now we have used the abstract formalism of differential geometry 
that one can find in the literature \cite{Marsd}\cite{Eguchi}\cite{konstant}\cite{dewitt2}, 
but we would like to put it in the same kind of language of the CPI
\cite{Goz89}. The procedure is straightforward \cite{metaplectic}.
Let us extend ${\cal M}_{2N}$ to a new space ${\cal M}_{\textrm{ext}}$ labeled by the coordinates $({\cal
\varphi}^a,\lambda_a,\eta^{\alpha},\bar{\eta}_{\alpha})$ where $\lambda_a$ are the same kind of variables we used 
in the first part of this chapter while $\eta^{\alpha},\bar{\eta}_{\alpha}$ are Grassmann variables
and they are as many as the indices $\alpha$. Note that in the CPI of 
Ref. \cite{Goz89}, since the vector (or form) representation has the same dimension as the manifold
${\cal M}_{2N}$, the Grassmann variables $c^a,\bar{c}_a$ 
were as many as the variables ${\cal \varphi}^a$ (or $\lambda_a$). Here instead the number 
of indices $\alpha$ is equal to 
the dimension $M$ of the representation that we are using. 
So the dimension of ${\cal M}_{\textrm{ext}}$ is not $8N$, as in the CPI case, but $4N+2M$ where
$M$ is the dimension of the representation. 

Next let us endow ${\cal M}_{\textrm{ext}}$ with the following extended Poisson structure (epb):
%%%
\begin{eqnarray}
&&\{{\cal\varphi}^a,\lambda_b\}_{epb}=\delta_b^a,\qquad\quad\; \{{\cal \varphi}^a,{\cal
\varphi}^b\}_{epb}=\{\lambda_a,\lambda_b\}_{epb}=0\nonumber\\
&&\{\bar{\eta}_{\beta},\eta^{\alpha}\}_{epb}=-i\delta^{\alpha}_{\beta},\qquad
\{\eta^{\alpha},\eta^{\beta}\}_{epb}=\{\bar{\eta}_{\alpha},\bar{\eta}_{\beta}\}_{epb}=0 \label{bos.7-1}
\end{eqnarray}
%%%
and with the following Hamiltonian
%%%
\begin{equation}
\displaystyle {\cal H}_{\scriptscriptstyle MFA}=h^a({\cal \varphi})
\lambda_a+\frac{1}{2}\bar{\eta}_{\alpha}
K_{ab}({\cal \varphi})(\Sigma^{ab}_{\textrm{meta}})^{\alpha}_{\;\beta}\eta^{\beta} \label{bos.7-2}
\end{equation}
%%%
where the acronym MFA means ``Metaplectic Functional 
Approach". We have used it 
to identify also this Hamiltonian because it is the one appearing in the functional approach which we will present 
later on.
As last ingredient let us build the following hat ``$\wedge$" map between the multispinor fields
of the abstract formalism (\ref{bos.6-13-a}) and the variables belonging to ${\cal M}_{\textrm{ext}}$:
%%%
\begin{equation}
\displaystyle {\cal X}({\cal \varphi})\;\hat{\longrightarrow}\;
\widehat{\cal X}\equiv \frac{1}{p!q!}{\cal X}^{\beta_1\ldots\beta_q}_{\alpha_1\ldots\alpha_p}({\cal \varphi})
\bar{\eta}_{\beta_1}\ldots\bar{\eta}_{\beta_q}\eta^{\alpha_1}\ldots\eta^{\alpha_p}. \label{bos.7-3}
\end{equation}
%%%
It is then a very long but straightforward calculation to show that the action of the Lie derivative on
${\cal X}$ can be realized via the extended Poisson brackets and the Hamiltonian
${\cal H}_{\scriptscriptstyle MFA}$ as:
%%%
\begin{equation}
({\cal L}_h{\cal X})\;\hat{\longrightarrow}\;-\{{\cal H}_{\scriptscriptstyle MFA},
\widehat{\cal X}\}_{epb}.
\label{bos.7-4}
\end{equation}
%%%
It is easy to show that the standard equation of motion (\ref{bos.6-14}) for the spinor field
$\partial_t{\cal X}_{\alpha}=-{\cal L}_h{\cal X}_{\alpha}$ can be written in terms of the Hamiltonian
${\cal H}_{\scriptscriptstyle MFA}$ as:
%%%
\begin{equation}
\partial_t\widehat{\cal X}=\{{\cal H}_{\scriptscriptstyle MFA},\widehat{\cal X}\}_{epb}
\end{equation}
%%%
where $\widehat{\cal X}$ is given by (\ref{bos.7-3}).
Via ${\cal H}_{\scriptscriptstyle MFA}$ and the extended Poisson brackets it is easy to obtain
the evolution of all the variables $({\cal \varphi}^a,\lambda_a,\eta^{\alpha},
\bar{\eta}_{\alpha})$ of the extended manifold ${\cal M}_{\textrm{ext}}$. The 
equations of motion for $\varphi^a$ are the standard ones of CM:
%%%
\begin{equation}
\dot{\cal \varphi}^a=h^a({\cal \varphi}(t)) \label{bos.7-6}
\end{equation}
%%%
while the equations for the Grassmann variables are: 
\begin{equation}
\left\{
	\begin{array}{l}
	\displaystyle 
	\dot{\eta}^{\alpha}=-\frac{i}{2}K_{ab}(\Sigma^{ab}_{\textrm{meta}})^{\alpha}_{\;\beta}\eta^{\beta} \smallskip\\
	\displaystyle \dot{\bar{\eta}}_{\alpha}=\frac{i}{2}K_{ab}\bar{\eta}_{\beta}
	(\Sigma^{ab}_{\textrm{meta}})^{\beta}_{\;\alpha}
	\label{bos.7-7}.
	\end{array}
	\right.
\end{equation}
%%%
Let us notice that the last two equations are quite different from the one of the Jacobi field $\delta{\cal \varphi}^a$
%%%
\begin{equation}
\displaystyle \dot{(\delta{\cal \varphi}^a)}=\partial_lh^a({\cal \varphi})(\delta{\cal \varphi}^l).
\label{bos.7-8}
\end{equation}
%%%
So we cannot identify $\eta^{\alpha}$ with the Jacobi fields of CM. Instead it is easy to
show that they are a sort of ``square root" of the Jacobi fields \cite{metaplectic} in the sense that the composite objects
${\mathcal P}^a(t)$ defined as 
%%%
\begin{equation}
{\mathcal P}^a(t)\equiv\bar{\eta}_{\alpha}(\gamma^a)^{\alpha}_{\;\beta}\bar{\eta}^{\beta} \label{bos.7-9}
\end{equation}
%%%
have the same equations of motion as the Jacobi fields.

It is possible to give also a path integral version of the formalism presented in this section, 
as explained in detail in Ref. \cite{metaplectic}. The associated generating functional is
%%%
\begin{equation}
\displaystyle Z_{\scriptscriptstyle MFA}=\int {\mathcal D}{\cal \varphi}{\mathcal D}\lambda{\mathcal D}
\eta{\mathcal D}\bar{\eta}
\;\textrm{exp}\,i\int dt\bigl[\lambda_a\dot{\varphi}^a+i\bar{\eta}_{\alpha}\dot{\eta}^{\alpha}-
{\cal H}_{\scriptscriptstyle MFA}\bigr].
\label{bos.7-10}
\end{equation}
%%%
As for the  CPI case \cite{Goz89} it is easy to derive the ``operatorial" version of this MFA path
integral. From the kinetic term in (\ref{bos.7-10}) one gets the following graded commutators:
%%%
\begin{equation}
\left\{
	\begin{array}{l}
	\displaystyle 
	[\widehat{\cal\varphi}^a,\widehat{\lambda}_b]=i\delta_b^a \smallskip\\
	\displaystyle [\widehat{\bar{\eta}}_{\beta},\widehat{\eta}^{\alpha}]=\delta^{\alpha}_{\beta}.
	\label{bos.7-11}
	\end{array}
	\right.
\end{equation}
%%%
All the commutators not indicated in (\ref{bos.7-11})
are zero. 
In a ``Schr\"odinger-type" representation  $\widehat{\cal \varphi}^a$ and $\widehat{\eta}^{\alpha}$ are 
{\it multiplicative} operators while the associated momenta $\widehat{\lambda}_a$, 
$\widehat{\bar{\eta}}_{\alpha}$ have to be 
realized as {\it derivative} operators in order to satisfy the algebra (\ref{bos.7-11}):
%%%
\begin{eqnarray}
\displaystyle &&\widehat{\lambda}_a=-i\frac{\partial}{\partial{{\cal \varphi}^a}}\nonumber\\
&&\widehat{\bar{\eta}}_{\alpha}=\frac{\partial}{\partial\eta^{\alpha}}. \label{bos.7-12}
\end{eqnarray}
%%%
The associated representation space is
given by the set of functions: 
%%%
\begin{equation}
\displaystyle {\cal X}({\cal \varphi},\eta)\equiv \sum_p\frac{1}{p!}{\cal X}_{\alpha_1\alpha_2\ldots\alpha_p}
({\cal \varphi}) \eta^{\alpha_1}\eta^{\alpha_2}\ldots \eta^{\alpha_p} \label{bos.7-13}
\end{equation}
%%%
while the metaplectic Hamiltonian (\ref{bos.7-2}) is turned into the operator
%%%
\begin{equation}
\displaystyle \widehat{\cal H}_{\scriptscriptstyle MFA}={\cal H}_{\scriptscriptstyle MFA}
\biggl(\widehat{\cal \varphi},\widehat{\lambda}=-i\frac{\partial}{\partial{\cal \varphi}},
\widehat{\eta},\widehat{\bar{\eta}}=\frac{\partial}{\partial\eta}\biggr). \label{bos.7-14}
\end{equation}
%%%

The next step is to endow the space of functions (\ref{bos.7-13}) with a scalar product and check if $\widehat{\cal
H}_{\scriptscriptstyle MFA}$ is Hermitian under it. 
We will choose the analog of the SvH scalar product by imposing the following hermiticity
conditions:
%%%
\begin{equation}
\left\{
	\begin{array}{l} 
	\widehat{\eta}^{\alpha\dagger}=\widehat{\bar{\eta}}_{\alpha} \smallskip \\
	\widehat{\bar{\eta}}_{\alpha}^{\dagger}=\widehat{\eta}^{\alpha} \smallskip\\
	\widehat{\varphi}^{a\dagger}=\widehat{\varphi}^a \smallskip\\
	\widehat{\lambda}_a^{\dagger}=\widehat{\lambda}_a.
	\label{bos.7-15}
	\end{array}
	\right.
\end{equation}
%%%
Along the same lines developed in the previous chapter, it is easy to show that the scalar product induced 
by (\ref{bos.7-15}) is given by:
%%%
\begin{equation}
\displaystyle \langle \tau|{\cal X}\rangle=\sum_pK(p)\tau^{*\alpha_1\ldots \alpha_p}({\cal \varphi})
{\cal X}_{\alpha_1\ldots\alpha_p}
({\cal \varphi}). \label{bos.7-16}
\end{equation}
%%%
where $K(p)$ is a positive combinatorial factor.
One immediately notices that this is a {\it positive definite} scalar product. 
Let us now check whether the Hamiltonian $\widehat{\cal H}_{\scriptscriptstyle MFA}$ is Hermitian under this
scalar product. 
As usual the bosonic part of 
(\ref{bos.7-2}), which is the same as
in the CPI case, is Hermitian. So we have to check out only the fermionic part which is
%%%
\begin{equation}
\displaystyle \widehat{\cal
H}_{\scriptscriptstyle MFA}^{\textrm{ferm}}=\frac{1}{2}\partial_a\partial_bH\widehat{\bar{\eta}}_x
(\Sigma^{ab}_{\textrm{meta}})^x_{\; y}\widehat{\eta}^y.
\label{bos.7-17}
\end{equation}
%%%
We have indicated the indices with $x,y$ because in the metaplectic case, 
as explained above, they are a continuous set labeling the infinite states of the Hilbert space ${\cal V}$.
Second let us remember that $\Sigma^{ab}_{\textrm{meta}}$
have to be chosen Hermitian in the metaplectic representation:
%%%
\begin{equation}
(\Sigma^{ab}_{\textrm{meta}})^{\dagger}=\Sigma^{ab}_{\textrm{meta}}. \label{bos.7-18}
\end{equation}
%%%
This hermiticity of course refers to the indices $(x,y)$ and not to $(a,b)$. 
So if we indicate the elements $(\Sigma^{ab}_{\textrm{meta}})^x_{\;y}$ as 
$\langle x|\Sigma^{ab}_{\textrm{meta}}|y\rangle$, then (\ref{bos.7-18}) implies that:
%%%
\begin{equation}
\langle x|\Sigma^{ab}_{\textrm{meta}}|y\rangle^*=\langle y|\Sigma^{ab\;\dagger}_{\textrm{meta}}|x\rangle
=\langle y|\Sigma_{\textrm{meta}}^{ab}|x\rangle
\end{equation}
%%%
which in normal matrix language means
%%%
\begin{equation}
(\Sigma^{ab}_{\textrm{meta}})^{x*}_{\;y}=(\Sigma^{ab}_{\textrm{meta}})^y_{\;x}. \label{bos.7-19}
\end{equation}
%%%
Now it is easy to prove
the hermiticity of $\widehat{\cal H}_{\scriptscriptstyle MFA}^{\textrm{ferm}}$. In fact:
%%%
\begin{eqnarray}
&&(\widehat{\cal H}^{\textrm{ferm}}_{\scriptscriptstyle MFA})^{\dagger}=\biggl(\frac{1}{2}
(\partial_a\partial_bH)\widehat{\bar{\eta}}_x(\Sigma^{ab}_{\textrm{meta}})^x_{\;y}\widehat{\eta}^y\biggr)^{\dagger}=
\frac{1}{2}(\partial_a\partial_bH)\widehat{\eta}^{y\dagger}
(\Sigma^{ab}_{\textrm{meta}})^{x*}_{\;y}\widehat{\bar{\eta}}_x^{\dagger}=
\nonumber\\
&&=\frac{1}{2}(\partial_a\partial_bH)\widehat{\bar{\eta}}_y(\Sigma^{ab}_{\textrm{meta}})^y_{\;x}\widehat{\eta}^x=
\frac{1}{2}(\partial_a\partial_bH)\widehat{\bar{\eta}}_x(\Sigma^{ab}_{\textrm{meta}})^x_{\;y}\widehat{\eta}^y=
\widehat{\cal
H}_{\scriptscriptstyle MFA}^{\textrm{ferm}}.
\end{eqnarray}
%%%
This proves that the full $\widehat{\cal H}_{\scriptscriptstyle MFA}$ 
is Hermitian under the
SvH scalar product. This does not happen for the $\widehat{\HT}$ 
of the CPI \cite{Goz89} and it is easy to understand why. In fact the usual $\widehat{\cal
H}_{\scriptscriptstyle CPI}$ can be given the following form:
%%%
\begin{equation}
\widehat{\HT}_{\scriptscriptstyle CPI}=
h^a\widehat{\lambda}_a+\frac{1}{2}\widehat{\bar{c}}_eK_{ab}({\cal \varphi})(\Sigma^{ab}_{\textrm{vec}})^e_{\;f}
\widehat{c}^f
\end{equation}
%%%
where $(\Sigma^{ab}_{\textrm{vec}})^e_f$ is the $\Sigma$ associated to the transformations of vectors 
under $\textrm{Sp}(2N)$:
%%%
\begin{equation}
(\Sigma^{ab}_{\textrm{vec}})^e_{\; f}=-i(\delta^a_f\omega^{be}+\delta^b_f\omega^{ae}).
\end{equation}
%%%
Then it is easy to check that this $\Sigma$ does not satisfy the analog of (\ref{bos.7-19}), i.e.:
%%%
\begin{equation}
(\Sigma^{ab}_{\textrm{vec}})^{e*}_{\;f}\neq (\Sigma^{ab}_{\textrm{vec}})^f_{\; e}.
\end{equation}
%%%
This explains why $\widehat{\HT}_{\scriptscriptstyle CPI}$ is not Hermitian
with the SvH scalar product. 

Summarizing the work contained in the last two chapters we can conclude that the KvN program  
of introducing a positive
definite Hilbert space in CM {\it fails}
as soon as we try to include the higher forms $\psi({\cal \varphi},c)$ of the CPI, besides the zero-forms
$\psi_0({\cal \varphi})$. It fails in the sense that we cannot
have both a positive definite scalar product and a Hermitian Hamiltonian. 
To disentangle from the problem above we prefer to abandon the Hilbert space picture of the KvN program in the 
CPI case and to interpret the functions $\psi({\cal \varphi},c)$ of the CPI {\it not}
as ``classical wave functions" but only as {\it probability
densities} $\rho({\cal \varphi},c)$, i.e. elements of $L^1$. Doing so there is no need to introduce a scalar product and a Hilbert
space structure. 
Otherwise, if we do not want to abandon the
``classical wave function" picture of KvN, we can use the wave function picture that the
metaplectic case of Secs. {\bf 5.5}-{\bf 5.6} provides. In fact the metaplectic functions and the associated Hilbert spaces
are not affected by problems like the lack of a positive definite scalar product or the lack of hermiticity of 
the Hamiltonian. 
Moreover also the MFA provides
a description of the standard classical Hamiltonian dynamics. In fact from its path integral (\ref{bos.7-10}) we have that the
usual phase space points ${\cal \varphi}^a$ evolve as in CM:
%%%
\begin{equation}
\displaystyle \dot{\cal \varphi}^a=\omega^{ab}\frac{\partial H}{\partial{\cal \varphi}^b}
\end{equation}
%%%
and the equations of the Jacobi fields $\delta{\cal \varphi}^a$, which are crucial to study some
features  of the dynamics such as ergodicity, chaos etc. \cite{Goz89}\cite{Liapunov}\cite{Deotto}, can be reconstructed from the 
dynamics of $\eta$, $\bar{\eta}$ via (\ref{bos.7-9}). So somehow the metaplectic formulation is a different manner
to describe the standard classical Hamiltonian dynamics for a point particle. The next
question is how to describe, via the metaplectic formulation, the standard classical {\it statistical} mechanics.  
In the CPI this is described 
by the classical probability densities $\rho_0(\cal\varphi)$ which evolve via the Liouville operator, or 
(if we include also the Jacobi fields like in Ref. \cite{Goz89}) by the Grassmann-valued densities
%%%
\begin{equation}
\rho({\cal \varphi},c)=\rho_0({\cal \varphi})+\rho_a({\cal \varphi})c^a+\rho_{ab}({\cal \varphi})c^ac^b+\ldots \label{bos.8-1}
\end{equation}
%%%
whose evolution is generated by the standard Lie derivative $\widehat{\cal H}_{\scriptscriptstyle CPI}$ 
\cite{Goz89}.
In the metaplectic representation we have instead functions like (\ref{bos.7-13})
%%%
\begin{equation}
\displaystyle \psi({\cal \varphi},\eta)=\sum_p\frac{1}{p!}\psi_{\alpha_1\ldots\alpha_p}({\cal \varphi})\eta^{\alpha_1}
\eta^{\alpha_2}\ldots\eta^{\alpha_p} \label{bos.8-2}
\end{equation}
%%%
and it is from these wave functions that we should reconstruct the probability densities (\ref{bos.8-1}). 
We feel that somehow the 
metaplectic wave functions (\ref{bos.8-2}) should be related to something like the ``square roots" of the probability densities
(\ref{bos.8-1}). We get that feeling by noticing that in the metaplectic representation 
we have already done something similar in the sense that
$\eta,\bar{\eta}$ are like the ``square roots" of the Jacobi fields $\delta{\cal \varphi}^a$ as shown in (\ref{bos.7-9}).
What we need is a scalar product such that the following relation
holds\footnote[3]{In (\ref{bos.8-3})
we have put the subindex ${\cal \varphi}$  on 
$|\psi\rangle_{\cal \varphi}$ to indicate that they belong to the Hilbert space
${\cal V}$ jetting out of the point ${\cal \varphi}$.}:
%%%
\begin{equation}
_{\cal \varphi}\langle \psi|\eta\rangle\langle\eta|\psi\rangle_{\cal \varphi}=
\rho({\cal \varphi},\bar{\eta}\gamma\eta)=\rho({\cal \varphi},c).
\label{bos.8-3}
\end{equation}
%%%
In other words we would like that the $\eta$, $\bar{\eta}$ on 
the LHS of (\ref{bos.8-3}) get combined by this scalar product 
into the combination $\bar{\eta}\gamma^a\eta$ which are basically the $\delta{\cal \varphi}^a$ (or $c^a$) Jacobi fields,
see (\ref{bos.7-9}). We want that they combine in this way 
because the probability densities (\ref{bos.8-1}) contain the variables $c$
and not $\eta$, or $\bar{\eta}$. 

The scalar product (\ref{bos.8-3}) is not the SvH one that we explored in (\ref{bos.7-16}). 
In fact the SvH scalar product does not pull in 
the $\gamma$ matrices which instead are necessary in (\ref{bos.8-3}) to get the combination $\bar{\eta}\gamma\eta$
inside the $\rho$. 
So far we have not succeeded in building the ``scalar" product (\ref{bos.8-3}), in checking if it is positive definite and if
$\widehat{\cal H}_{\scriptscriptstyle MFA}$ is Hermitian under it. 
However, in order to get some practice in the interplay between the
metaplectic wave functions (\ref{bos.8-2}) and the CPI probability densities (\ref{bos.8-1}), we have asked ourselves how,
from the components of the $\psi$ appearing in (\ref{bos.8-2}), we can build objects which at least have the same
indices and transformation properties as the components of the $\rho$ appearing in (\ref{bos.8-1}). One 
solution we found is the following one
%%%
\begin{equation}
\rho_{\underbrace{\scriptstyle ab\ldots d}_{p}}({\cal \varphi})=\textrm{Tr}\Bigl[|\psi^{\scriptscriptstyle (p)}
\rangle\langle\psi^{\scriptscriptstyle (p)}|\gamma_{[a}\otimes\gamma_b
\otimes\ldots\gamma_{d]}\Bigr] \label{bos.8-4}
\end{equation}
%%%
where with $|\psi^{\scriptscriptstyle (p)}\rangle$ we indicate the components of the states (\ref{bos.7-13}) 
with $p$ indices and with $\otimes$
we indicate the tensor products among the Hilbert spaces like in (\ref{bos.6-13-b}). 

The {\it first} thing we can derive from (\ref{bos.8-4}) is
that if we transform $|\psi^{\scriptscriptstyle (p)}\rangle$ 
according to the metaplectic transformations then the resulting $\rho_{ab\ldots d}$
turns out to transform according to the symplectic ones. 

{\it Second}, the metaplectic
$|\psi^{\scriptscriptstyle (p)}\rangle$ has the following 
components $\psi_{\alpha_1\ldots\alpha_p}$ whose number of indices can
run from zero to $\infty$, while $\rho$ can have at most $N$ indices. This means that we have much more information
stored in the $|\psi^{\scriptscriptstyle (p)}\rangle$ that what is needed to build the $\rho$. What does this imply? 

{\it Third},
while (\ref{bos.8-4}) produces a $\rho$ out of a $\psi$, it is not clear whether the inverse procedure 
is true and unique, i.e. whether,
given a $\rho$ with all its components, it is possible to find a $|\psi\rangle$ such that (\ref{bos.8-4}) or (\ref{bos.8-3})
is satisfied. These are the topics we are currently working on. Our reason to insist with the metaplectic formulation 
is because, as proved in this chapter, this is the only formulation of CM
where a true Hilbert space can be introduced and therefore it is the formulation which can most probably 
be directly compared
with QM. Actually the metaplectic representation made its appearance before in issues related to quantization. 
In particular
it appeared in Geometric Quantization \cite{woodhouse} and in a beautiful approach to quantization developed in 
\cite{reutermeta}. In this last paper quantization was achieved via three postulates and the first was just the necessity 
to go to the metaplectic representation for the Lie derivative. At that time it was not clear why one had to do that 
step first. Now instead it is clear: going to the metaplectic representation is the only way to introduce  
a true Hilbert space in CM. 

The reader may object to the statement we made above that the MFA is the only way to introduce
a true Hilbert space in CM. Actually there is also the BFA case 
presented in Secs. {\bf 5.1}-{\bf 5.4} in which, after all, the Hilbert space has a
positive definite scalar product and a Hermitian Hamiltonian. 
What we do not like in the BFA approach is that higher forms and tensors have to be built by hand 
introducing from outside 
the operation $\otimes$ of tensor product. What we mean is that in the CPI 
case of Ref. \cite{Goz89} the higher
tensors and forms were generated automatically as functions on the extended phase space which is the only ingredient
entering the associated path integral. In the bosonic or BFA 
case instead we have to introduce an extra structure which is the cross product $\otimes$ or use a strange path
integral like (\ref{bos.5-12-a}). Anyhow we will leave to the reader and his taste which of the two programs to pursue in case he is interested 
in carrying further this research.

%% file: conclusions.tex
\def \HT{{\mathcal H}}
\def \LT{{\mathcal L}}
\def \ET{{\widetilde{\mathcal E}}}
\def \HCT{\widehat{\mathcal H}}
\def \s{\scriptscriptstyle}

\pagestyle{fancy}
\chapter*{\begin{center}
6. Conclusions and Outlook
\end{center}}
\addcontentsline{toc}{chapter}{\numberline{6}Conclusions and Outlook}
\setcounter{chapter}{6}
\setcounter{section}{1}
\markboth{{\it{6. Conclusions and Outlook}}}{\it{6. Conclusions and Outlook}}

\begin{quote}
{\it{
The mathematician plays a game in which he himself invents the rules while the physicist 
plays a game in which the rules are provided by Nature, but as time goes on it becomes 
increasingly evident that the rules which the mathematician finds interesting 
are the same as those which Nature has chosen.}}\medskip\\
-{\bf P.A.M. Dirac}, 1939.
\end{quote}

\section*{} \setcounter{equation}{0}
\noindent In this thesis we have studied some aspects of the original KvN
formulation of CM which had never been studied before in the literature, such
as the role played by the phases of the elements of KvN
Hilbert space and the rules needed to couple a point particle with an external gauge
field. Next, following Ref. \cite{Goz89}, we have extended the KvN formalism 
to the space of differential forms where we have analyzed first 
some purely {\it geometrical topics} and second some {\it Hilbert space features}.
The {\it geometrical topics} were centered on the Cartan calculus of differential 
symplectic geometry which, as shown in Ref. \cite{Goz89}, could be reproduced 
in a compact form via the Grassmann variables and other structures present in the 
CPI. In this thesis we showed that the CPI could also provide a unified view 
of those various generalizations of the Lie brackets known as the Schouten-Nijenhuis, 
the Fr\"olicher-Nijenhuis and the Nijenhuis-Richardson brackets. Furthermore 
we showed that all the Cartan calculus could be reproduced not only via the Grassmann
variables of the CPI but also via some suitable tensor products of Pauli matrices.
Regarding the {\it Hilbert space features} mentioned above, in this thesis we have 
shown that, once forms are included in the formalism, the associated Hilbert space 
cannot have at the same time a positive definite scalar product and a Hermitian 
Hamiltonian. Two solutions to this problem have been studied in details 
in Chapter {\bf 5}. 

Having now completed this thesis, what to do next? 
The main problem we are interested in is the
quantization of the theory. We know that usually the quantization of a system is
performed via the Dirac's correspondence rules, which consist in replacing the classical
Poisson brackets $\{\,\cdot\,,\,\cdot\,\}_{pb}$ by commutators according to the following
relation:
%%%
\begin{equation}
\displaystyle \{\,\cdot\,,\,\cdot\,\}_{pb}\;\longrightarrow\;
\frac{[\,\cdot\,,\,\cdot\,]_{\scriptscriptstyle QM}}{i\hbar}.
\end{equation}
%%%
Now, if we formulate CM \`a la KvN, the Poisson brackets are already replaced by the KvN commutators.
So the quantization of the system can be performed: 1) either finding suitable rules to go from 
the KvN commutators to the quantum ones; 2) or finding a way to reproduce the algebra of quantum
observables using KvN commutators and operators.
Up to now we have found a compact way to realize the second road towards a quantization
based on the following map:
%%%
\begin{equation}
\displaystyle {\mathcal Q}:\;f(\widehat{q},\widehat{p})\;\longrightarrow\;
f\Bigl(\widehat{q}-\frac{\hbar}{2}\widehat{\lambda}_p,\widehat{p}+\frac{\hbar}{2}\widehat{\lambda}_q\Bigr)
\label{mapq}
\end{equation}
%%%
where the domain of ${\mathcal Q}$ is made up of all the standard classical observables
$f(\widehat{q},\widehat{p})$ and its image is made up of suitable
Hermitian operators living in the KvN Hilbert space and depending also on $\widehat{\lambda}$.
In particular it is possible to prove that, using KvN commutators, all the functions
of the form $\displaystyle f\Bigl(\widehat{q}-\frac{\hbar}{2}\widehat{\lambda}_p,\widehat{p}+\frac{\hbar}
{2}\widehat{\lambda}_q\Bigr)$ appearing on the RHS of (\ref{mapq}) 
satisfy exactly the algebra of the 
observables of QM. So maybe this is a rule to quantize the KvN classical mechanics
and it might be useful also in clarifying the physical role of the KvN operators which depends on
$\widehat{\lambda}$.

An even more interesting road towards a quantization of the theory starts from 
the CPI and its basic ideas have been explained in \cite{Abrikosov}. Having both classical 
and quantum mechanics formulated via path integrals, the quantization of the system is equivalent
to finding some rules to go from the CPI to the quantum path integral. A very simple 
and elegant trick to connect at least the weights of the path integrals is based on the
concept of superfield introduced in (\ref{ann.super1}). Basically if we take the
Lagrangian $L$ which appears in the weight of the quantum path integral, replace in it
the fields with the superfields and integrate over $\theta,\bar{\theta}$, what we obtain
is just the Lagrangian ${\mathcal L}$ of (\ref{ann.suplag}), which appears in the weight of the CPI,
modulus some surface terms:
%%%
\begin{equation}
\displaystyle
i\int d\theta d\bar{\theta} L(\Phi)=\LT-\frac{d}{dt}(\lambda_pp+i\bar{c}_pc^p).
\label{trentuno}
\end{equation}
%%%%%%
Inserting (\ref{trentuno}) into the kernel of evolution (\ref{zoe}) for the functions
in the $(\varphi,c)$ representation we obtain:
%%%%%%
\begin{eqnarray}
\displaystyle
\label{trentadue}
&&K(\varphi_f,c_f,t_f|\varphi_i,c_i,t_i)=
\langle q_f,p_f,c^q_f,c^p_f,t_f|q_i,p_i,c^q_i,c^p_i,t_i\rangle=\int{\mathcal D}\mu \;\textrm{exp}\; i\int dt\LT
=\nonumber\\
&&=\int {\cal D}\mu \;\textrm{exp}\biggl[i\int i dt d\theta d\bar{\theta}L(\Phi)+
i\int dt\frac{d}{dt}(\lambda_pp+i\bar{c}_pc^p)\biggr]=\\
&&=\int {\cal D}\mu \;\textrm{exp}\biggl[i\int i dt d\theta d\bar{\theta}L(\Phi)+
i\lambda_{p,f}p_f-i\lambda_{p,i}p_i-\bar{c}_{p,f}c^p_f+\bar{c}_{p,i}c^p_i\biggr] \nonumber
\end{eqnarray}
%%%%%%
where we have indicated the functional measure of integration with:
%%%%%%
\begin{equation}
{\cal D}\mu={\cal D}^{\prime\prime}\varphi^a{\cal D}\lambda_a{\cal D}^
{\prime\prime}c^a{\cal D}\bar{c}_a.
\label{trentatre}
\end{equation}
%%%%%%
It is easy to prove that the surface terms appearing in (\ref{trentadue}) can be re-absorbed by
going from the $(\varphi,c)$ representation to the mixed one $(q,\lambda_p,c^q,\bar{c}_p)$.
In fact the kernel of propagation for the functions in this mixed representation is related to the
previous one by the following relation:
%%%%%%
\begin{eqnarray}
\displaystyle
&&\langle q_f,\lambda_{p,f},c^q_f,\bar{c}_{p,f},t_f|q_i,\lambda_{p,i},
c^q_i,\bar{c}_{p,i},t_i\rangle=\int \frac{dp_f}{\sqrt{2\pi}}\frac{dp_i}{\sqrt{2\pi}}
dc^p_fdc^p_i\;\textrm{exp}(-i\lambda_{p,f}p_f+\bar{c}_{p,f}c^p_f)\cdot \nonumber\\
&&\qquad\qquad\qquad\cdot \langle q_f,p_f,c^q_f,c^p_f|q_i,p_i,c^q_i,c^p_i\rangle\; \textrm{exp}(i\lambda_{p,i}p_i
-\bar{c}_{p,i}c^p_i).
\label{trentaquattro}
\end{eqnarray}
%%%%%%
The exponentials of the partial Fourier transforms appearing in (\ref{trentaquattro}) are just equal 
and opposite to
the ones obtained from the surface terms in (\ref{trentadue}). Therefore the kernel of propagation
in the mixed representation can be written in the following very compact form\footnote[1]{The ket
$|q_i,\lambda_{p,i},c^q_i,\bar{c}_{p,i}\rangle$ is an eigenstate for the operators
$\widehat{q}$, $\widehat{\lambda}_p$, $\widehat{c}^q$, $\widehat{\bar{c}}_p$ and, consequently, is an eigenstate
also for the superfield $\widehat{\Phi}^q$. In this way we tentatively identify $|\Phi^q_i\rangle$ with 
$|q_i,\lambda_{p,i},c^q_i,\bar{c}_{p,i}\rangle$. A problem to be solved is the following: 
is this correspondence one-to-one?
All the eigenstates of $\widehat{q}$, $\widehat{\lambda}_p$, $\widehat{c}^q$ and $\widehat{\bar{c}}_p$ are 
eigenstates for the superfield operator $\widehat{\Phi}^q$ but does it make sense to consider, if they exist, 
also eigenstates of $\widehat{\Phi}^q$ which are not of the form $|q_i,\lambda_{p,i},c^q_i,\bar{c}_{p,i}\rangle$?}:
%%%%%%
\begin{equation}
\langle\Phi^q_f,t_f|\Phi^q_i,t_i\rangle=\int {\cal D}\Phi \; \textrm{exp}\,\biggl[i\int i dt d\theta d\bar{\theta}
L(\Phi)\biggr] \label{trentacinque}
\end{equation}
%%%%%%
where the functional integration over the superfield is defined as:
%%%%%%
\begin{equation}
{\cal D}\Phi\equiv {\cal D}^{\prime\prime}q{\cal D}p{\cal D}^
{\prime\prime}\lambda_p{\cal D}\lambda_q{\cal D}^{\prime\prime}c^q{\cal D}c^p{\cal D}^
{\prime\prime}\bar{c}_p{\cal D}\bar{c}_q.
\label{trentasei}
\end{equation}
%%%%%%
The quantization procedure is then equivalent to sending the variables $\theta,\bar{\theta}\to 0$
according to the following scheme:
%%%%%%%%%
\begin{eqnarray}
\label{centoottantanove}
\langle\Phi^{q}_{f},t_f|\Phi^q_i,t_i\rangle=\int{\cal D}\Phi~
&\hspace{-0.2cm}\textrm{exp}&\hspace{-0.2cm}{i\int idtd\theta d{\bar\theta}L[\Phi]}\nonumber\\
% &\hspace{-0.2cm}\Downarrow
\mathrm{quantization:}
&\hspace{-0.2cm}\left \Downarrow\rule[3pt]{0pt}{10pt} \right.
&~\theta,{\bar\theta}\rightarrow 0\\
\langle q_{f},t_f|q_{i},t_i\rangle=\int{\cal D}\varphi
&\hspace{-0.2cm}\textrm{exp}&\hspace{-0.2cm}{{\frac{i}{\hbar}}\int dtL[\varphi]} \nonumber
\end{eqnarray}
%%%%%%
where the limit at the state level has to be intended as:
%%%%%%
\begin{equation}
\lim_{\theta,\bar{\theta}\to 0}|\Phi^q\rangle=|\lim_{\theta,\bar{\theta}\to 0}\Phi^q\rangle=|q\rangle.
\end{equation}
%%%%%%
This road to quantization is interesting because, differently than the usual procedure, it is performed
at the level of space and time, freezing to zero the variables $\theta,\bar{\theta}$ which are the 
Grassmannian partners of time in the superspace. 
What remains to be done in these projects is to study in more details the physics and the geometry 
underlying these procedures of quantization. In any case the work contained in this
thesis may be helpful: in fact we think that 
a better understanding of the physical features of the Hilbert space of CM
is crucial in order to make a comparison with quantum theories and to discover the real meaning
of the associated quantization rules.

As we have already said in the first chapter another problem that we would like to analyze in the future 
is whether the large number of Hermitian operators present in the KvN Hilbert space may be useful
in order to describe those intermediate regimes which are at the interface between quantum and 
classical mechanics and which are becoming more and more important in modern experiments.
In fact, even if we find out the physical meaning and the correct rules to quantize CM \`a la KvN, 
the problem of understanding which are the realms of applicability
of quantum and classical mechanics would remain. What happens in the intermediate region, at the border
between the two theories? Maybe the answer lies just within the KvN Hilbert space which
seventy years after its birth has not revealed yet all its secrets and mysteries.

\newpage

\markboth{{\it{Acknowledgments}}}{}

\section*{Acknowledgments}

First of all I would like to express my gratitude to my supervisor, Ennio Gozzi, for his constant
support and encouragement during all the stages of this thesis. I cannot thank him enough for all the
helpful and enlightening discussions which gave rise to the greatest part of this work.

I would like to thank also Martin Reuter for his many inspiring ideas and suggestions
and for his hospitality in Mainz
in August 2001, M.V. Ioffe for some important technical suggestions, 
G. Marmo, F. Benatti and G. Pastore for having triggered with their questions some of the 
projects developed in this thesis and last, but not least, G.C. Ghirardi, E. Deotto and L. Marinatto
for many valuable discussions. 

This work has been supported by grants from INFN, MURST, University of Trieste and Regione Friuli-Venezia
Giulia.